%% file: Mopra_paper_v4.tex
\documentclass[useAMS,usenatbib]{mn2e}
\usepackage{graphicx}
\usepackage{subfigure}
\usepackage{amsmath}
\usepackage{verbatim}
\usepackage{upgreek}
\usepackage{subfigure}

\usepackage{multirow}
\usepackage{nicefrac}
\usepackage{color}
\title[A molecular line survey of a sample of AGB stars and planetary nebulae]{A molecular line survey of a sample of AGB stars and planetary nebulae}
\author[C. L. Smith, A. A. Zijlstra, G. A. Fuller]{C. L. Smith $^{1}$\thanks{E-mail: chrsmith@yorku.ca \newline Present address for C. L. Smith: \emph{York University, Centre for Research in the Earth and Space Sciences (CRESS), Canada}.}, A. A. Zijlstra $^{1}$, G. A. Fuller $^{1}$ \\
$^{1}$ Jodrell Bank Centre for Astrophysics, University of Manchester, Manchester, M13 9PL, UK}

\begin{document}

\date{Accepted for publication in MNRAS 19th Aug 2015}
\pagerange{\pageref{firstpage}--\pageref{lastpage}} \pubyear{2015?}

\maketitle

\label{firstpage}
\begin{abstract}
A millimeter molecular line survey of three carbon-rich AGB stars and two oxygen-rich planetary nebulae has been carried out over the frequency range 80.5-115.5 GHz. Sixty eight different transitions were detected in the data from 27 different molecular species.  The hyperfine structure of C$_2$H and C$^{13}$CH has been fitted to constrain the optical depth of their transitions. All other transitions have been constrained on the basis of their line profile shapes. Rotation temperatures and column densities have been calculated for all possible species, with adaptations to the methods applied in order to account for the hyperfine structure of various transitions. From the column densities, carbon, silicon and sulphur isotopic ratios have been determined. The results corroborate IRAS 15194-5115 as a J-type star, whilst excluding IRAS 15082-4808 and IRAS 07454-7112 as such.
\end{abstract}

\begin{keywords}
stars: AGB and post-AGB, planetary nebulae: general, stars: abundances, stars: carbon, line: profiles, ISM: abundances
\end{keywords}

\section{Introduction}

Asymptotic giant branch (AGB) stars and planetary nebulae (PNe) are rich sites of dust and molecule formation. As a result, their spectra are full of a variety of spectral lines. A vast quantity of molecular species present in these objects emit in the millimeter region of the spectrum. From these spectral lines, a host of information can be gleaned, including the abundances and temperatures of species present as well as their extent in the envelopes of the host stars. Thus millimeter molecular line surveys are an ideal method of identifying the chemical composition and thermodynamic properties of the regions in which these species exist.

J-type stars are a sub-set of carbon-rich AGB stars with an unusually low $^{12}$C/$^{13}$C ratio ($<10$). It is thought that these could be the origin of presolar A+B grains: silicon carbide (SiC) grains with pre-Solar System origins, found in meteorites which do not have a confirmed origin (see \citealp{Zinner2003} for further details about the SiC presolar grains). By identifying and subsequently studying J-type stars in detail, we can ascertain the isotopic characteristics of these stars and compare them to the measured isotopic abundances found in these mysterious A+B grains.

Isotopic abundance calculations are most easily carried out using optically thin rather than optically thick transitions to measure the relative abundances of two isotopologues. Moderately optically thick lines may be used but would require an optical depth correction or the use of a radiative transfer code. However, often the method used to measure, for example, the optical depth of CO transitions requires a knowledge of the carbon isotopic ratio. Transitions with hyperfine structure alleviate this requirement as the relative intensities of the different components vary with increasing optical depth of the overall transition. By fitting these hyperfine components, the optical depth may be constrained and thus accurate isotopic ratios determined.

This paper details a millimeter molecular line survey carried out over a sample of three carbon-rich AGB stars and two oxygen-rich PNe. The aim of this survey is to use high velocity-resolution spectra to identify molecules present in the targets, resolve hyperfine structure where possible and ultimately determine reliable column densities and isotopic ratios for all sources observed. This line survey is intended to compliment those previously carried out (e.g. \citealp{Woods2003, Edwards2013, Ramstedt2014}) by fully resolving the lines and subsequently analysing both the line profiles and hyperfine structures of the transitions identified.

\section{Observations}

\subsection{Targets}

The sample comprises of five  galactic sources: three carbon stars and two oxygen-rich planetary nebulae. Further details about the sources can be found in Table \ref{tab_source}.

\begin{table}
\renewcommand{\arraystretch}{1.2}
\centering
\setlength{\tabcolsep}{4pt}
\caption{\label{tab_source}Source and coordinate list}
\begin{tabular}{l l c c c}
\hline
IRAS	& Name &	RA & Dec	&	Type		\\
\hline 
15194-5115	& II Lup	&	15 23 05.1 & -51 25 58.7	&	C-AGB      \\
15082-4808	& RAFGL 4211	&	15 11 41.9 & -48 20 01.3	&	C-AGB 	\\
07454-7112	& AI Vol	&	07 45 02.8 & -71 19 43.2 	&	C-AGB	\\
14192-4355	& IC 4406	&	14 22 26.3 & -44 09 04.3 	&	O-PN	\\
18021-1950	& NGC 6537	&	18 05 13.1 & -19 50 34.9 	&	O-PN	\\
\hline
\end{tabular}
\end{table}

\subsubsection{IRAS 15194-5115}

IRAS 15194-5115, also known as II Lup, is the brightest carbon star in the southern hemisphere at 12 $\upmu$m and the third brightest in both hemispheres at 12 $\upmu$m, with only IRC+10216 and CIT 6 being brighter \citep{Nyman1993}.  It has a pulsation period of 575 days \citep{Feast2003}. Distance estimates in literature studies are between 600-1200 pc \citep{Nyman1995, Woods2003}. The mass-loss rate for this star has been found from non-local thermodynamic equilibrium radiative transfer models to be $1\times10^{-5}$ M$_{\sun}$ yr$^{-1}$ at a distance of 600 pc \citep{Ryde1999, Woods2003}.  

\subsubsection{IRAS 15082-4808}

IRAS 15082-4808, also known as RAFGL 4211, is also a mass losing, carbon-rich AGB star. Its mass loss rate has been derived from models to be approximately $1\times10^{-5}$ M$_{\sun}$ yr$^{-1}$ at 850 pc  \citep{Groenewegen2002}. The distance to this star according to the literature ranges between 640 pc and 1500 pc with the general consensus between 640 and 850pc \citep{Woods2003, Nyman1995}.

\subsubsection{IRAS 07454-7112}

IRAS 07454-7112 is a mass losing, carbon-rich Mira variable star \citep{Menzies2006}. As a bright infrared star, this source has been part of several line surveys in the past as well as being studied in-depth by \citet{Nyman1995} and \citet{Woods2003}. The distance of this source varies in the literature from 710 pc to 850 pc \citep{Woods2003, Menzies2006}, and its mass loss rate has been modelled as $\sim10^{-5}$ M$_{\sun}$ yr$^{-1}$ \citep{Nyman1995} at a distance of 1800 pc .

\subsubsection{IC 4406}

IC 4406 is an oxygen-rich, axially symmetric  planetary nebula with a molecular envelope. It has a collimated outflow parallel to the major axis of the nebula, and because of this outflow, there is a strong dependence of the shape of the line profile with changing position across the nebula, as shown by \citet{Sahai1991}. The central star temperature has been found to be $8\times10^4$ K from three dimensional photoionisation modelling \citep{Gruenwald1997}. This nebula has been shown through both observation and modelling to have a cylindrical structure and extends over a region approximately $100\arcsec$ by $30\arcsec$ \citep{Faes2011}.

\subsubsection{NGC 6537}

NGC 6537, also known as the Red Spider Nebula, is a type I (rich in helium and nitrogen), bi-polar planetary nebula with outflows extending out to an angular size of $100\arcsec$ in the direction NE-SW. The central star temperature is $2\times10^5$ K. A small, opaque, dusty torus exists at 2-4$\arcsec$ from the centre with an asymmetric cavity, found by \citet{Matsuura2005} from $HST$ H$\alpha$ and H$\beta$ maps. \citet{Sabin2007} show, through maps taken with the Submillimetre Common-User Bolometer Array (SCUBA), that the central region of the nebula is approximately $20\arcsec$ in diameter and appears to have a toroidal structure.

\subsection{Telescope}

The data were obtained using the 22 m Mopra telescope, situated approximately 450 km from Sydney, Australia. The 3 mm Monolithic Microwave Integrated Circuit (MMIC) receiver in June 2010 was used in conjunction with the Mopra Spectrometer (MOPS, a digital filterbank spectrometer), in broadband mode for all observations. The MMIC receiver has a frequency range of 77-116 GHz and works in dual polarization single-sideband mode\footnote{Technical summary of the Mopra radio telescope: narrabri.atnf.csiro.au/mopra/mopragu.pdf}. 

The use of broadband mode with MOPS gives an observable frequency range of 8 GHz, split into 4 overlapping bands, each with 2048 channels over a 2.2 GHz frequency range. The band edges suffer from noise, but the overlap given in this set up compensates well for this so little data is lost. Three frequency set-ups were used to observe each source: 84.5--92.5 GHz, 90.5--98.5 GHz and 107.5--115.5 GHz, with the region between 98.5 GHz and 107.5 GHz unobserved.

The observations for this project (ID: M530) were taken over ten days: 18/06/2010 to 27/06/2010. The rms noise on a typical spectrum was 0.01 K and the velocity resolution was 0.91 km/s. The efficiency of Mopra at this frequency is $\sim$0.5, the sensitivity is 22 Jy/K and the beam size at 90 GHz is 38\arcsec. 

\subsection{Data reduction and line identification}

The observations were made in position-switching mode. The on-off quotient\footnote{The quotient is defined as: $T_a^*=\frac{ON-OFF}{OFF}T_{\rm{sys}}$.} was created from these pairs and their corresponding T$_{\rm{sys}}$ (system temperature). The resulting spectrum is the antenna temperature corrected for atmospheric and systematic effects. The main beam temperature is obtained by dividing the corrected antenna temperature by the efficiency of the telescope which varies with frequency. The telescope efficiency is documented as 0.49 at 78 GHz and 0.42 at 116 GHz\footnote{Technical summary of the Mopra radio telescope: narrabri.atnf.csiro.au/mopra/mopragu.pdf}. Therefore, for this data, the efficiency was taken to be 0.49 between 84.5 GHz and 98.5 GHz and 0.42 between 107.5 GHz and 115.5 GHz.

The data were reduced using the ATNF Spectral Analysis Package (ASAP). The on/off observations were matched in time and the resultant spectrum formed. The two polarisations of each set of observations were averaged, weighted by the system temperature. Multiple observations of each source in each frequency set-up were taken and thus needed to be coadded. However, some observations were taken during poor weather and/or suffered from severe baseline rippling. To remove the poor data prior to coadding, the spectrum of each individual observation had a polynomial baseline temporarily fitted and subtracted. Those observations that were badly affected by the weather were flagged and removed from all further analysis. The remaining spectra (taken before the polynomial baseline subtraction) were then averaged in time, weighted by the system temperature. A polynomial baseline was fitted to the averaged spectrum and subtracted to remove any remaining baseline rippling. The full spectra from all sources can be found in Appendix A of \citet{Smith2014d}\footnote{https://www.escholar.manchester.ac.uk/item/?pid=uk-ac-man-scw:240595}.

The lines were identified using the NIST databases \citep{NIST, Splatalogue}. In cases where it was unclear as to which species was the source of a particular line, all observable transitions of each of the candidate species were examined for further detections. In most cases, this resulted in a clear identification of the line. 

\section{Line profiles}\label{lineprofiles}

\begin{table}
\renewcommand{\arraystretch}{1.2}
\setlength{\tabcolsep}{4pt}
\centering
\caption[Details of all detected transitions.]{Details of all detected transitions. All molecular transitions relate to the ground vibrational state unless otherwise indicated. The quantum numbers of SiC$_2$ and c-C$_3$H$_2$ are given in the assymmetric top standard format. \smallskip}\label{alltrans}
\begin{tabular}{l l l}
\hline
\multirow{2}{*}{Molecule} & Freq & \multirow{2}{*}{Transition} \\
	  & (GHz) & \\
\hline
$^{30}$SiO & 84.74617 & J=2-1	\\
HC$_5$N   & 85.20160 & J=32-31 \\
C$^{13}$CH$^{b}$  &  85.22933  &  N=1-0, J=\nicefrac{3}{2}\,-\,\nicefrac{1}{2}, F$_1$=2-1, F=\nicefrac{5}{2}\,-\,\nicefrac{3}{2}    \\  
C$^{13}$CH$^{b}$  &  85.23276  &  N=1-0, J=\nicefrac{3}{2}\,-\,\nicefrac{1}{2}, F$_1$=2-1, F=\nicefrac{3}{2}\,-\,\nicefrac{1}{2}    \\ 
C$^{13}$CH$^{b}$  &  85.24771  &  N=1-0, J=\nicefrac{3}{2}\,-\,\nicefrac{1}{2}, F$_1$=1-0, F=\nicefrac{1}{2}\,-\,\nicefrac{1}{2}    \\ 
C$^{13}$CH$^{b}$  &  85.25695  &  N=1-0, J=\nicefrac{3}{2}\,-\,\nicefrac{1}{2}, F$_1$=1-0, F=\nicefrac{3}{2}\,-\,\nicefrac{1}{2}    \\ 
C$^{13}$CH$^{b}$  &  85.30397  &  N=1-0, J=\nicefrac{1}{2}\,-\,\nicefrac{1}{2}, F$_1$=1-1, F=\nicefrac{1}{2}\,-\,\nicefrac{3}{2}    \\
C$^{13}$CH$^{b}$  &  85.30769  &  N=1-0, J=\nicefrac{1}{2}\,-\,\nicefrac{1}{2}, F$_1$=1-1, F=\nicefrac{3}{2}\,-\,\nicefrac{3}{2}    \\ 
C$^{13}$CH$^{b}$  &  85.31438  &  N=1-0, J=\nicefrac{1}{2}\,-\,\nicefrac{1}{2}, F$_1$=0-1, F=\nicefrac{1}{2}\,-\,\nicefrac{1}{2}    \\
c-C$_3$H$_2$   &  85.33891  &  $2_{1,2}$\,-\,$1_{0,1}$   \\ 
C$_4$H   &  85.63400  &  N=9-8, J=\nicefrac{19}{2}\,-\,\nicefrac{17}{2}   \\
C$_4$H   &  85.67257  &  N=9-8, J=\nicefrac{17}{2}\,-\,\nicefrac{15}{2}   \\
$^{29}$SiO &	85.75920 & J=2-1 \\
H$^{13}$CN  &  86.34018  &  J=1-0  \\
SiO  &  86.84700  &  J=2-1  \\
HN$^{13}$C  &  87.09086  &  J=1-0 \\
C$_2$H  &  87.28416  &  N=1-0, J=\nicefrac{3}{2}\,-\,\nicefrac{1}{2}, F=1-1  \\
C$_2$H$^{b}$  &  87.31693  &  N=1-0, J=\nicefrac{3}{2}\,-\,\nicefrac{1}{2}, F=2-1 \\
C$_2$H$^{b}$  &  87.32862  &  N=1-0, J=\nicefrac{3}{2}\,-\,\nicefrac{1}{2}, F=1-0  \\
C$_2$H$^{b}$  &  87.40200  &  N=1-0, J=\nicefrac{3}{2}\,-\,\nicefrac{1}{2}, F=1-1  \\
C$_2$H$^{b}$  &  87.40717  &  N=1-0, J=\nicefrac{3}{2}\,-\,\nicefrac{1}{2}, F=0-1  \\
C$_2$H  &  87.44651  &  N=1-0, J=\nicefrac{3}{2}\,-\,\nicefrac{1}{2}, F=1-0  \\
H$^{13}$CCCN   &  88.16683  &  J=10-9  \\
HCN  &  88.63185  &  J=1-0 \\
CCCN  &  89.04559  &  N=9-8, J=\nicefrac{19}{2}\,-\,\nicefrac{17}{2}  \\
CCCN  &  89.06436  &  N=9-8, J=\nicefrac{17}{2}\,-\,\nicefrac{15}{2}  \\
HCN  &  89.08791  &  J=1-0 F=2-1 v=(0,2,0) \\
HCO$^{+}$ & 89.18853   & 	J=1-0 \\
HC$^{13}$CCN$^{b}$   &  90.59306  &  J=10-9  \\
HCC$^{13}$CN$^{b}$   &  90.60179  &  J=10-9  \\
HNC  &  90.66356  &  J=1-0  \\
SiS  &  90.77156  &  J=5-4  \\
HCCCN   &  90.97899  &  J=10-9  \\
$^{13}$CS  &  92.49427  &  J=2-1  \\
SiC$_2$  &  93.06364  &  $4_{0,4}$\,-\,$3_{0,3}$   \\
SiC$_2$  &  94.24539  &  $4_{2,3}$\,-\,$3_{2,2}$   \\
C$_4$H   &  95.15032  &  N=10-9, J=\nicefrac{21}{2}\,-\,\nicefrac{19}{2}  \\
C$_4$H   &  95.18894  &  N=10-9, J=\nicefrac{19}{2}\,-\,\nicefrac{17}{2}  \\
SiC$_2$  &  95.57938  &  $4_{2,2}$\,-\,$3_{2,1}$   \\
C$^{34}$S  &  96.41295  &  J=2-1  \\
H$^{13}$CCCN & 96.98300 & J=11-10 \\
CS  &  97.98095  &  J=2-1  \\
\multicolumn{3}{c}{--------------------------- Band Gap --------------------------- } \\
unidentified   &  107.973   &   \dots \\
$^{13}$CN & 108.78020 &  N=1-0, J=\nicefrac{3}{2}\,-\,\nicefrac{1}{2}, F$_1$=2-1, F=3-2 \\
\hline  
\end{tabular}

$^b$: blended transitions
\end{table}
\begin{table}
\renewcommand{\arraystretch}{1.2}
\setlength{\tabcolsep}{4pt}
\contcaption{}
\centering
\begin{tabular}{l l l}
\hline
\multirow{2}{*}{Molecule} & Freq & \multirow{2}{*}{Transition} \\
	  & (GHz) & \\
\hline
$^{13}$CN & 108.78237 &  N=1-0, J=\nicefrac{3}{2}\,-\,\nicefrac{1}{2}, F$_1$=2-1, F=2-1 \\
$^{13}$CN & 108.78698 &  N=1-0, J=\nicefrac{3}{2}\,-\,\nicefrac{1}{2}, F$_1$=2-1, F=1-0 \\
$^{13}$CN & 108.78020 &  N=1-0, J=\nicefrac{3}{2}\,-\,\nicefrac{1}{2}, F$_1$=2-1, F=3-2 \\
$^{13}$CN & 108.78237 &  N=1-0, J=\nicefrac{3}{2}\,-\,\nicefrac{1}{2}, F$_1$=2-1, F=2-1 \\
$^{13}$CN & 108.78698 &  N=1-0, J=\nicefrac{3}{2}\,-\,\nicefrac{1}{2}, F$_1$=2-1, F=1-0 \\
CCCN  &  108.83427  &  N=11-10, J=\nicefrac{23}{2}\,-\,\nicefrac{21}{2}  \\
CCCN  &  108.85302  &  N=11-10, J=\nicefrac{21}{2}\,-\,\nicefrac{19}{2}  \\
SiS  &  108.92430  &  J=6-5  \\
HCCCN   &  109.17364  &  J=12-11  \\
C$^{18}$O & 109.78217 & J=1-0 \\
$^{13}$CO  &  110.20135  &  J=1-0  \\
CN  &  113.14419  & N=1-0, J=\nicefrac{1}{2}\,-\,\nicefrac{1}{2}, F=\nicefrac{1}{2}\,-\,\nicefrac{3}{2} \\
CN  &  113.17050  & N=1-0, J=\nicefrac{1}{2}\,-\,\nicefrac{1}{2}, F=\nicefrac{3}{2}\,-\,\nicefrac{1}{2} \\ 
CN  &  113.19128  & N=1-0, J=\nicefrac{1}{2}\,-\,\nicefrac{1}{2}, F=\nicefrac{3}{2}\,-\,\nicefrac{3}{2} \\
CN$^{b}$  &  113.48812  &  N=1-0, J=\nicefrac{3}{2}\,-\,\nicefrac{1}{2}, F=\nicefrac{3}{2}\,-\,\nicefrac{1}{2}  \\
CN$^{b}$  &  113.49097  &  N=1-0, J=\nicefrac{3}{2}\,-\,\nicefrac{1}{2}, F=\nicefrac{5}{2}\,-\,\nicefrac{3}{2}  \\  
CN$^{b}$  &  113.49964  &  N=1-0, J=\nicefrac{3}{2}\,-\,\nicefrac{1}{2}, F=\nicefrac{1}{2}-\nicefrac{1}{2}  \\
CN$^{b}$  &  113.50891  &  N=1-0, J=\nicefrac{3}{2}\,-\,\nicefrac{1}{2}, F=\nicefrac{3}{2}\,-\,\nicefrac{3}{2}  \\
CN$^{b}$  &  113.52043  &  N=1-0, J=\nicefrac{3}{2}\,-\,\nicefrac{1}{2}, F=\nicefrac{1}{2}\,-\,\nicefrac{3}{2}  \\
C$_4$H   &  114.18251  &  N=12-11, J=\nicefrac{25}{2}\,-\,\nicefrac{23}{2}   \\
C$_4$H   &  114.22104  &  N=12-11, J=\nicefrac{23}{2}\,-\,\nicefrac{21}{2}   \\
CO  &  115.27120  &  J=1-0  \\
SiC$_2$	& 115.38239 & $5_{0,5}$\,-\,$4_{0,4}$	\\
\hline
\end{tabular}

$^b$: blended transitions
\end{table}

The spectral lines observed from circumstellar envelopes (CSEs) have distinctive profiles resulting from a combination of the optical depth of the emitting region and the size of the telescope beam compared to the size of the emitting region. From these shapes we can ascertain the extent of the emitting region relative to the beam, the motion of the emitting molecules and the optical depth of the emitting species.

\citet{Morris1975}, using the \citet{Sobolev1960} approximation, derived line profile shapes for circumstellar envelopes under the following assumptions:

\begin{enumerate}
\item{the CSE is spherically symmetric and isothermal;}

\item{mass loss rate is constant with time;}

\item{the expansion velocity is constant with radius;}

\item{the local line width due to thermal motions and turbulence at any point in the envelope is much less than that created from Doppler broadening due to the expansion of the envelope;}

\item{the central star is negligible in size in comparison to the size of the envelope.}
\end{enumerate}

The assumption of constant expansion velocity with radius does not hold in the innermost regions of the CSE where the majority of the wind acceleration occurs. However, this region is extremely small when compared with the overall size of the envelope. Those high energy lines formed only in the innermost region of the envelope will thus not follow the line shapes detailed in this section. Assumptions (i) and (ii) have been examined in the literature and clumpy or anisotropic structures in CSEs have been reported. This should be kept in mind when comparing these theoretical curves to observations. Assumptions (iv) and (v) are commonly found to be true in observations. These assumptions do not hold for the two planetary nebula sources which are not spherically symmetric and have large velocity gradients. 

For optically thin emission in an unresolved source, the profile was shown by \citet{Morris1975} to be flat-topped, whereas optically thick, unresolved emission has a parabolic profile. For resolved emission, these profiles become flattened parabola and horned profiles for optically thick and thin emission respectively.

\section{Results}

Listed in Table \ref{alltrans} are details of all the transitions detected in the sample and a figure for every transition in each source can be found in the Appendix. A total of 64 transitions were detected across all sources, including some blended hyperfine transitions. 

One emission line detected in IRAS 15194-5115 could not be identified. Its observed frequency is approximately 107.973 GHz, assuming no significant velocity offsets with respect to the systemic velocity of the envelope. A number of different candidate species were identified as potentially causing this transition. The Si$^{13}$CC line at 107.971 GHz has been detected in IRC+10216 and could be the origin of this line, however it would have been expected that the lines at 97.295 GHz and 112.593 GHz which have the same and greater intensities respectively in IRC+10216 \citep{Cernicharo1991}. These two lines are not detected in our data, which leaves the identification uncertain.  It is also possible that this transition could be an artefact of baseline ripple removal, although its strength and the stable baseline regions surrounding it make this option less likely. The line profile is unclear and thus constraints about the optical depth of this transition cannot be deduced.

The peak intensities, integrated intensities and full line width were measured for every transition in all five sources. For unblended transitions, the method was as follows. The spectra were smoothed, generally by a factor of two although a factor of four was used for some of the weakest lines,  then a straight-line baseline was fitted to the data on either side of the line, usually using the regions offset from the line by $\pm$50-100 km/s, to correct for any systematic offset of the spectrum from zero. The limits of the line were identified manually, the peak intensity of the line was measured, and the integrated intensity was measured between  the line of the spectrum and background. The line width was measured as the full width at the base of the profile. These parameters are noted for all transitions in all sources in Table \ref{AGBstar_results} (AGB star sample) and Table \ref{PN_results} (PN sample). Absolute flux calibration uncertainties have been estimated at 10\%. The line widths are accurate to 2.5 km/s. The uncertainties listed in the tables include both calibration and measurement uncertainties.

The blended transitions fell into two categories: separable blended transitions and unseparable blended transitions. The unseparable transitions were those that were blended to such an extent that the edges and thus the extent of the overlap of the individual lines were indeterminable. The integrated intensities, peak intensities and line widths for these blends were taken for the overall blended line and are indicated as such in Tables \ref{AGBstar_results} and \ref{PN_results}. Examples include the CN transitions at $\sim113.49$ GHz and the HC$^{13}$CCN and HCC$^{13}$CN transitions at $\sim90.6$ GHz,  shown in Fig. A1 to A3.

The separable blended transitions are those that are on the limit of being blended, as for C$_2$H, or have clearly defined components, as in C$^{13}$CH. These transitions have been fitted to measure the integrated intensity emanating from each component and these integrated intensities are listed individually in Tables \ref{AGBstar_results} and \ref{PN_results}.

\subsection{Line fitting}

In order to fit the line profiles of the spectra to disentangle the hyperfine structure of transitions and to ascertain the optical depth, an optimization routine is required. The routine chosen for this task is the Genetic Algorithm (GA) {\sc{Pikaia}}. GAs are a class of optimization routine based upon the principle of ``survival of the fittest" employed in nature to produce species that adapt to their environment. For the initial population, the parameters are chosen using a random number generator over the entire X-dimensional parameter space. In subsequent generations, the parameters are constructed from the parameters of the parent solutions with a possibility of parameter mutation. The likelihood of a given individual solution being used as a parent for further solutions is proportional to how well it fits the data. The reproduction plan chosen for this work was full generational replacement with elitism included. Further details on this routine may be found in \citet{Charbonneau1995}.

\subsubsection{C$_2$H}\label{c2h_fitting}

C$_2$H has six detected hyperfine components in IRAS 15194-5115 and four in IRAS 15082-4808. The profiles of C$_2$H have an unusual appearance due to the lines being partially blended. The individual hyperfine components have square profiles, with the brightest in IRAS 15194-5115 showing a slight horned profile, indicating that C$_2$H is just resolved in the beam of the telescope and is optically thin. The two-level step profiles are caused by two hyperfine transitions on the verge of blending and  the three-step profiles are caused by two square profile transitions with a significant overlap, resulting in excess emission in the central region of the profile.

The assumption about optical depth from the line profile shape can be confirmed using the method of \citet{Fuller1993}, but in contrast to this method's usage in that paper, the line profile is assumed to be a top-hat function rather than a Gaussian distribution.

\begin{table*}
\renewcommand{\arraystretch}{1.2}
\centering
\caption[Detected transitions in the AGB star sample]{Detected transitions in the AGB star sample. Frequencies are given in GHz, the line peak, $ T_{peak} $, is given in K, the integrated intensity, $ \int T dv$, is in K km/s and the full width of the line at the base of the line, $\Delta v$, is in km/s. All temperatures are given in corrected antenna temperatures. The efficiency of Mopra is 0.49 before the band gap and 0.42 after.  Absolute flux calibration uncertainties have been estimated at 10\%. The line widths are accurate to 2.5 km/s. The uncertainties listed in the table in parentheses after the value indicate the uncertainty on the final digit and include both calibration and measurement uncertainties.\smallskip}\label{AGBstar_results}
\begin{tabular}{l l c c l c c l l l l }
\hline
 \multirow{2}{*}{Molecule} & Frequency &\multicolumn{3}{c}{IRAS\,15194-5115} &\multicolumn{3}{c}{IRAS\,15082-4808} &\multicolumn{3}{c}{IRAS\,07454-7112} \\	
	  				& (GHz)  		& $ T_{peak}$ & $ \int T dv$ & $\Delta v$ & $ T_{peak} $ & $ \int T dv$ & $\Delta v$  & $ T_{peak} $ & $ \int T dv$ & $\Delta v$ \\
\hline
$^{30}$SiO 			&	84.74617	&	$0.014(6)$	&	0.3(1)	&	39.1	&	\dots	&	\dots	&	\dots	&	\dots	&	\dots	&	\dots	\\
HC$_5$N   			&	85.2016		&	$0.018(7)$	&	0.5(2)	&	48.4	&	\dots	&	\dots	&	\dots	&	\dots	&	\dots	&	\dots	\\
C$^{13}$CH$^{b}$  	&	85.22933	&	$0.011(6)$	&	0.5(3)	&	41.5	&	\dots	&	\dots	&	\dots	&	\dots	&	\dots	&	\dots	\\
C$^{13}$CH$^{b}$  	&	85.23276	&	$0.007(6)$	&	0.3(2)	&	41.5	&	\dots	&	\dots	&	\dots	&	\dots	&	\dots	&	\dots	\\
C$^{13}$CH$^{b}$  	&	85.24771	&	$0.003(5)$	&	0.1(2)	&	41.5	&	\dots	&	\dots	&	\dots	&	\dots	&	\dots	&	\dots	\\
C$^{13}$CH$^{b}$  	&	85.25695	&	$0.007(5)$	&	0.3(2)	&	41.5	&	\dots	&	\dots	&	\dots	&	\dots	&	\dots	&	\dots	\\
C$^{13}$CH$^{b}$  	&	85.30397	&	$0.003(5)$	&	0.1(2)	&	41.5	&	\dots	&	\dots	&	\dots	&	\dots	&	\dots	&	\dots	\\
C$^{13}$CH$^{b}$  	&	85.30769	&	$0.006(5)$	&	0.2(2)	&	41.5	&	\dots	&	\dots	&	\dots	&	\dots	&	\dots	&	\dots	\\
C$^{13}$CH$^{b}$  	&	85.31438	&	$0.003(5)$	&	0.1(2)	&	41.5	&	\dots	&	\dots	&	\dots	&	\dots	&	\dots	&	\dots	\\
c-C$_3$H$_2$   		&	85.33891	&	$0.042(9)$	&	1.0(2)	&	43.6	&	\dots	&	\dots	&	\dots	&	\dots	&	\dots	&	\dots	\\
C$_4$H   			&	85.634		&	$0.043(9)$	&	1.4(3)	&	43.4	&	\dots	&	\dots	&	\dots	&	\dots	&	\dots	&	\dots	\\
C$_4$H   			&	85.67257	&	$0.045(10)$	&	1.5(3)	&	41.5	&	\dots	&	\dots	&	\dots	&	\dots	&	\dots	&	\dots	\\
$^{29}$SiO 			&	85.75920 	&	$0.018(7)$	& 	0.5(2)	&	40.5	&	\dots	&	\dots	&	\dots	&	\dots	&	\dots	&	\dots	\\
H$^{13}$CN  		&	86.34018	&	$0.52(6)$	&	17(2)	&	55.2	&	0.07(1)	&	1.8(3)	&	41.2	&	0.062(9)&	1.4(2)	&	34.6	\\
SiO  				&	86.847		&	$0.17(2)$	&	5.3(7)	&	46.5	&	0.08(1)	&	1.8(3)	&	37.3	&	0.039(7)&	0.7(1)	&	27.9	\\
HN$^{13}$C  		&	87.09086	&	$0.024(7)$	&	0.6(2)	&	48.3	&	\dots	&	\dots	&	\dots	&	\dots	&	\dots	&	\dots	\\
C$_2$H  			&	87.28416	&	$0.013(6)$	&	0.5(2)	&	40.1	&	\dots	&	\dots	&	\dots	&	\dots	&	\dots	&	\dots	\\
C$_2$H$^{b}$  		&	87.31693	&	$0.13(2)$	&	5.1(7)	&	40.1	&	0.04(1)	&	1.5(4)	&	36.9	&	\dots	&	\dots	&	\dots	\\
C$_2$H$^{b}$  		&	87.32862	&	$0.06(1)$	&	2.4(4)	&	40.1	&	0.021(8)&	0.8(3)	&	36.9	&	\dots	&	\dots	&	\dots	\\
C$_2$H$^{b}$  		&	87.402		&	$0.08(1)$	&	3.0(5)	&	40.1	&	0.021(8)&	0.8(3)	&	36.9	&	\dots	&	\dots	&	\dots	\\
C$_2$H$^{b}$  		&	87.40717	&	$0.028(8)$	&	1.1(3)	&	40.1	&	0.009(7)&	0.3(2)	&	36.9	&	\dots	&	\dots	&	\dots	\\
C$_2$H  			&	87.44651	&	$0.014(6)$	&	0.5(2)	&	40.1	&	\dots	&	\dots	&	\dots	&	\dots	&	\dots	&	\dots	\\
H$^{13}$CCCN   		&	88.16683	&	$0.028(8)$	&	0.9(2)	&	44.0	&	\dots	&	\dots	&	\dots	&	\dots	&	\dots	&	\dots	\\
HCN  				&	88.63185	&	$0.51(6)$	&	15(2)	&	56.5	&	0.44(5)	&	10(1)	&	43.8	&	0.14(2)	&	2.8(3)	&	35.6	\\
CCCN  				&	89.04559	&	$0.022(7)$	&	0.7(3)	&	44.0	&	0.024(8)&	0.4(1)	&	37.8	&	\dots	&	\dots	&	\dots	\\
CCCN  				&	89.06436	&	$0.023(7)$	&	0.7(3)	&	42.6	&	0.024(8)&	0.6(2)	&	38.7	&	\dots	&	\dots	&	\dots	\\
HCN  				&	89.08791	&	\dots		&	\dots	&	\dots	&	0.67(7)	&	1.1(1)	&	6.3		&	\dots	&	\dots	&	\dots	\\
HC$^{13}$CCN$^{b}$ 	&	\multirow{2}{*}{\hspace{-0.22cm}$\left.\begin{array}{lr}90.59306\\90.60179\end{array}\right\}$}	&	\multirow{2}{*}{$0.043(9)$}	&	\multirow{2}{*}{1.8(4)}	&	\multirow{2}{*}{74.0}	&	\multirow{2}{*}{\dots}	&	\multirow{2}{*}{\dots}	&	\multirow{2}{*}{\dots}	&	\multirow{2}{*}{\dots}	&	\multirow{2}{*}{\dots}	&	\multirow{2}{*}{\dots}	\\
HCC$^{13}$CN$^{b}$ 	& \\ 
HNC  				&	90.66356	&	$0.10(1)$	&	3.2(5)	&	44.6	&	0.05(1)	&	1.4(3)	&	40.1	&	0.010(4)&	0.3(1)	&	69.5	\\
SiS  				&	90.77156	&	$0.06(1)$	&	2.1(4)	&	44.5	&	0.05(1)	&	1.6(4)	&	45.4	&	0.023(5)&	0.5(1)	&	36.5	\\
HCCCN   			&	90.97899	&	$0.10(1)$	&	3.2(5)	&	44.4	&	0.10(1)	&	2.7(4)	&	39.1	&	0.049(8)&	0.9(1)	&	28.4	\\
$^{13}$CS  			&	92.49427	&	$0.05(1)$	&	1.8(3)	&	44.6	&	0.014(7)&	0.16(8)	&	34.1	&	\dots	&	\dots	&	\dots	\\
SiC$_2$  			&	93.06364	&	$0.07(1)$	&	1.9(3)	&	42.5	&	0.045(7)&	0.9(2)	&	38.2	&	0.024(5)&	0.41(9)	&	31.3	\\
SiC$_2$  			&	94.24539	&	$0.040(9)$	&	1.1(2)	&	42.0	&	0.026(9)&	0.6(2)	&	44.1	&	0.014(4)&	0.20(6)	&	30.0	\\
C$_4$H   			&	95.15032	&	$0.045(9)$	&	1.5(3)	&	45.0	&	0.025(9)&	0.5(2)	&	41.6	&	\dots	&	\dots	&	\dots	\\
C$_4$H   			&	95.18894	&	$0.047(9)$	&	1.6(3)	&	43.3	&	0.026(9)&	0.5(2)	&	41.6	&	\dots	&	\dots	&	\dots	\\
SiC$_2$  			&	95.57938	&	$0.034(8)$	&	1.0(2)	&	43.1	&	0.025(9)&	0.6(2)	&	38.0	&	0.012(4)&	0.18(6)	&	26.2	\\
C$^{34}$S  			&	96.41295	&	$0.025(8)$	&	0.7(2)	&	43.6	&	0.012(7)&	0.3(2)	&	36.9	&	\dots	&	\dots	&	\dots	\\
H$^{13}$CCCN 		&	96.983		&	$0.017(7)$	&	0.4(1)	&	43.3	&	\dots	&	\dots	&	\dots	&	\dots	&	\dots	&	\dots	\\
CS  				&	97.98095	&	$0.31(4)$	&	10(1)	&	42.1	&	0.21(2)	&	6.4(8)	&	41.2	&	0.10(1) &	2.0(3)	&	28.9	\\
\multicolumn{11}{c}{--------------------------- Band Gap --------------------------- } \\
unidentified   		&	107.973		&	$0.023(7)$	&	0.3(1)	&	38.2	&	\dots	&	\dots	&	\dots	&	\dots	&	\dots	&	\dots	\\
$^{13}$CN$^{b}$ 	&	\multirow{3}{*}{\hspace{-0.22cm}$\left.\begin{array}{lr}108.78020\\108.78237\\108.78698 \end{array} \right\} $}	&	 \multirow{3}{*}{$0.027(8)$} 	&	\multirow{3}{*}{0.6(2)}	&	\multirow{3}{*}{58.7}	&	\multirow{3}{*}{\dots}	&	\multirow{3}{*}{\dots}	&	\multirow{3}{*}{\dots}	&	\multirow{3}{*}{\dots}	&	\multirow{3}{*}{\dots}	&	\multirow{3}{*}{\dots}	\\
$^{13}$CN$^{b}$ 	\\ 
$^{13}$CN$^{b}$ 	\\ 
CCCN  				&	108.83427	&	$0.018(7)$	&	0.5(2)	&	48.3	&	0.032(9)&	0.5(2)	&	37.9	&	\dots	&	\dots	&	\dots	\\
CCCN  				&	108.85302	&	$0.026(8)$	&	0.8(2)	&	47.5	&	0.027(9)&	0.7(2)	&	36.4	&	\dots	&	\dots	&	\dots	\\
\hline  
\end{tabular}

$^b$: blended transitions
\end{table*}

\begin{table*}
\renewcommand{\arraystretch}{1.2}
\centering
\contcaption{\smallskip}
\begin{tabular}{l l c c l c c l c c l }
\hline
 \multirow{2}{*}{Molecule} & Frequency &\multicolumn{3}{c}{IRAS\,15194-5115} & \multicolumn{3}{c}{IRAS\,15082-4808} & \multicolumn{3}{c}{IRAS\,07454-7112} \\	
	  & (GHz)  & $ T_{peak} $ & $ \int T dv$ & $\Delta v$ & $ T_{peak} $ & $ \int T dv$ & $\Delta v$  & $ T_{peak} $ & $ \int T dv$ & $\Delta v$ \\
\hline
SiS  		&	108.9243	&	0.045(9)&	1.3(3)	&	44.5	&	0.05(1)	&	1.2(2)	&	35.6	&	0.031(6)&	0.5(1)	&	29.7	\\
HCCCN   	&	109.17364	&	0.06(1)	&	1.9(4)	&	44.4	&	0.09(1)	&	2.5(4)	&	40.7	&	0.057(9)&	1.1(2)	&	31.8	\\
$^{13}$CO  	&	110.20135	&	0.27(3)	&	8.5(10)	&	46.2	&	0.06(1)	&	1.4(3)	&	40.3	&	0.08(1)	&	1.6(2)	&	29.3	\\
CN  		&	113.14419	&	0.012(6)&	0.8(4)	& 	31.4	&	0.06(1)	&	1.1(2)	&	38.5	&	0.02(5)	&	0.28(7)	&	32.9	\\
CN  		&	113.1705	&	0.05(1)	&	1.1(2)	&	43.6	&	0.06(1)	&	0.9(2)	&	39.3	&	0.044(7)&	0.6(1)	&	28.6	\\
CN  		&	113.19128	&	0.06(1)	&	0.7(1)	&	30.0	&	0.07(1)	&	0.9(2)	&	32.9	&	0.048(8)&	0.6(1)	&	30.7	\\
CN$^{b}$  	&	\multirow{5}{*}{\hspace{-0.22cm}$\left.\begin{array}{lr}113.48812\\113.49097\\113.49964\\113.50891\\113.52043\end{array}\right\}$}	&	\multirow{5}{*}{0.11(2)}	&	\multirow{5}{*}{4.5(7)}	&	\multirow{5}{*}{112.5}	&	\multirow{5}{*}{0.17(2)}	&	\multirow{5}{*}{8(1)}	&	\multirow{5}{*}{127.4}	&	\multirow{5}{*}{0.11(1)}	&	\multirow{5}{*}{3.3(4)}	&	\multirow{5}{*}{77.6}	\\
CN$^{b}$  	\\
CN$^{b}$  	\\
CN$^{b}$  	\\
CN$^{b}$  	\\
C$_4$H   	&	114.18251	&	0.06(1)	&	1.5(3)	&	39.6	&	0.026(8)&	0.6(2)	&	43.2	&	\dots	&	\dots	&	\dots	\\
C$_4$H   	&	114.22104	&	0.07(1)	&	1.8(3)	&	42.4	&	0.030(9)&	0.7(2)	&	43.2	&	\dots	&	\dots	&	\dots	\\
CO  		&	115.2712	&	0.49(5)	&	19(2)	&	51.2	&	0.60(7)	&	19(2)	&	41.4	&	0.52(5)	&	11(1)	&	28.7	\\
SiC$_2$		&	115.38239	&	0.044(9)&	1.0(2)	&	49.0	&	0.031(9)&	0.6(2)	&	37.8	&	0.027(6)&	0.32(7)	&	25.9	\\
\hline  
\end{tabular}

$^b$: blended transitions

\end{table*}

\begin{table*}
\renewcommand{\arraystretch}{1.2}
\centering
\caption[Detected transitions in the PN sample]{Detected transitions in the PNe sample. Frequencies are given in GHz, the line peak, $ T_{peak} $, is given in K, the integrated intensity, $ \int T dv$, is in K km/s and the full width of the line at the base of the line, $\Delta v$, is in km/s.  Absolute flux calibration uncertainties have been estimated at 10\%. The line widths are accurate to 2.5 km/s. The uncertainties listed in the table in parentheses after the value indicate the uncertainty on the final digit and include both calibration and measurement uncertainties. For the CO isotopologues in NGC 6357, the integrated intensity is that of the major peak only due to severe absorption being present in all these profiles. The linewidth of the CO isotopologues is across the whole profile, including the absorption region and secondary peak. \smallskip}\label{PN_results}
\begin{tabular}{l l c c l c c l}
\hline
 \multirow{2}{*}{Molecule} & Frequency &\multicolumn{3}{c}{IC 4406} & \multicolumn{3}{c}{NGC 6357} \\
	  & (GHz)  & $ T_{peak} $ & $ \int T dv$ & $\Delta v$ & $ T_{peak} $ & $ \int T dv$ & $\Delta v$  \\
\hline
H$^{13}$CN  	&	86.34018	&	0.025(9)&	0.4(1)	&	52.4	&	0.021(8)&	0.3(1)	&	24.3	\\
HCN  			&	88.63185	&	0.04(1)	&	0.9(2)	&	57.4	&	0.08(1)	&	0.9(2)	&	20.1	\\
HCO$^+$ 		&	89.18853	&	0.03(1)	&	0.8(2)	&	55.3	&	0.04(1)	&	0.5(1)	&	18.1	\\
HNC  			&	90.66356	&	0.03(1)	&	0.4(1)	&	46.3	&	0.06(1)	&	0.43(9)	&	14.3	\\
$^{13}$CS  		&	92.49427	&	\dots	&	\dots	&	\dots	&	0.020(8)&	0.18(7)	&	17.5	\\
C$^{34}$S  		&	96.41295	&	\dots	&	\dots	&	\dots	&	0.027(8)&	0.25(8)	&	17.6	\\
CS				&	97.98095	&	\dots	&	\dots	&	\dots	&	0.05(1)	&	0.5(1)	&	18.1	\\
\multicolumn{8}{c}{--------------------------- Band Gap --------------------------- } \\
C$^{18}$O 		&	109.78217	&	\dots	&	\dots	&	\dots	&	0.07(1)	&	0.35(7)	&	49.3	\\
$^{13}$CO  		&	110.20135	&	0.03(1)	&	0.4(1)	&	53.5	&	0.40(5)	&	2.3(3)	&	45.5	\\
CN$^{b}$  		&	\multirow{2}{*}{\hspace{-0.22cm}$\left.\begin{array}{lr}113.14419\\113.1705\end{array}\right\}$}	&	\multirow{2}{*}{0.05(1)}	&	\multirow{2}{*}{0.32(7)}	&	\multirow{2}{*}{24.3}	&	\multirow{2}{*}{\dots}	&	\multirow{2}{*}{\dots}	&	\multirow{2}{*}{\dots}	\\
CN$^{b}$ 	\\
CN$^{b}$  		&	\multirow{4}{*}{\hspace{-0.22cm}$\left.\begin{array}{lr}113.48812\\113.49097\\113.49964\\113.50891\end{array}\right\}$}	&	\multirow{4}{*}{0.07(1)}	&	\multirow{4}{*}{2.1(4)}	&	\multirow{4}{*}{101.8}	&	\multirow{4}{*}{\dots}	&	\multirow{4}{*}{\dots}	&	\multirow{4}{*}{\dots}	\\
CN$^{b}$  	\\
CN$^{b}$  	\\
CN$^{b}$  	\\
CO  			&	115.2712	&	0.16(2)	&	3.2(5)	&	51.9	&	0.50(6)	&	1.3(2)	&	52.6	\\
\hline  
\end{tabular}

$^b$: blended transitions
\end{table*}

The optical depth of the transition, $\tau$, is now found using:

\begin{equation}\label{CD1_a}
\tau(v)=\tau \sum a_i \phi_i(v),
\end{equation}

\noindent where $a_i$ is the relative intensity of hyperfine component $i$ and $\phi_i$ is the line profile of hyperfine component $i$, given by:

\begin{equation*}
\phi_i=
	\begin{cases}
	1 & \text{if } (v_0+v_i-\Delta v)<v<(v_0+v_i+\Delta v) \\
	0 & \text{otherwise},
	\end{cases}
\end{equation*} 

\noindent where $v_i$ is the offset velocity of component $i$ with respect to the reference velocity, $v_0$, and $\Delta v$ is the line half-width. The line may then be fitted using:

\begin{equation}\label{CD2}
T(v)=A\left(1-\rm{e}^{-\tau(v)}\right)+C,
\end{equation}

\noindent where $A$ is related to the peak intensity of the line and $C$ is a constant to fit for any residual baseline offset from zero. It should, however, be noted that in the optically thin limit, this equation reduces to \mbox{$T(v)=A\tau(v)+C$}, thus the amplitude and optical depth cannot be separately calculated in this limit. 

\begin{figure}
\centering
\includegraphics[trim= 2cm 13cm 2cm 3cm,clip=true, width=0.48\textwidth]{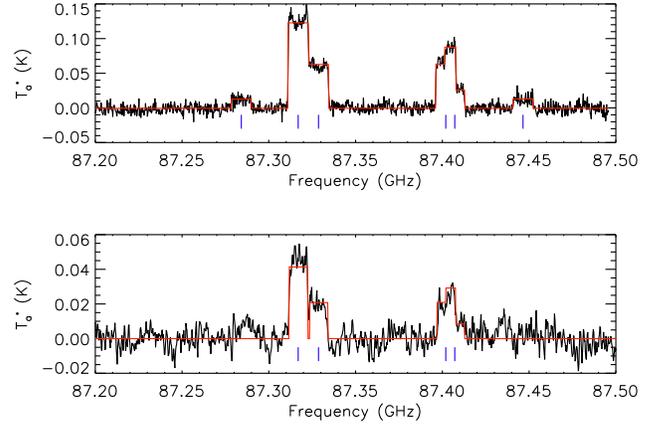}
\caption[C$_2$H fitted for hyperfine structure.]{C$_2$H fitted for hyperfine structure. The upper spectrum is that of IRAS 15194-5115 and the lower is that of IRAS 15082-4808. The red line indicates the best fit to the data and the blue vertical lines indicate the central velocities of each of the fitted components.}\label{c2h_fitted}
\end{figure}

The hyperfine structure was fitted to Equ. \ref{CD1_a} and \ref{CD2} using {\sc{Pikaia}}. The results of the fitting are shown in Fig. \ref{c2h_fitted}. In both cases, the fitting suggests that the transition is optically thin. The resulting integrated intensities of each of the hyperfine components are listed in Table \ref{AGBstar_results}.

\subsubsection{C$^{13}$CH}

\begin{figure}
\includegraphics[trim= 1cm 18.5cm 1cm 3cm,clip=true, width=0.48\textwidth]{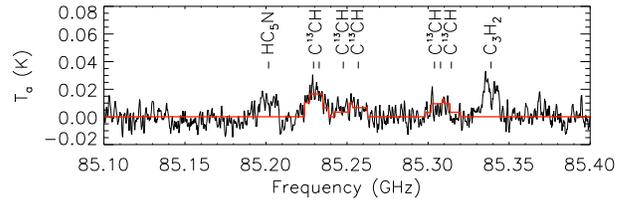}
\caption[C$^{13}$CH fitted for hyperfine structure.]{C$^{13}$CH fitted for hyperfine structure in IRAS 15194-5115.}\label{c13ch_fitted}
\end{figure}

C$^{13}$CH has seven hyperfine components which, in the spectrum of IRAS 15194-5115, are blended into three line-regions. The same approach as outlined above was applied to the C$^{13}$CH spectrum and again, the result was found to be optically thin. The results of the fitting are shown in Fig. \ref{c13ch_fitted}. It should be noted, however, that these lines have a low signal-to-noise ratio so the fits are less certain than those of C$_2$H. The resulting integrated intensities of each of the hyperfine components are listed in Table \ref{AGBstar_results}.

\subsection{Detections in individual sources}

\subsubsection{IRAS 15194-5115}

\begin{figure}
\centering
\subfigure{\includegraphics[trim= 3cm 13cm 2cm 2cm, clip=true, width=0.4\textwidth]{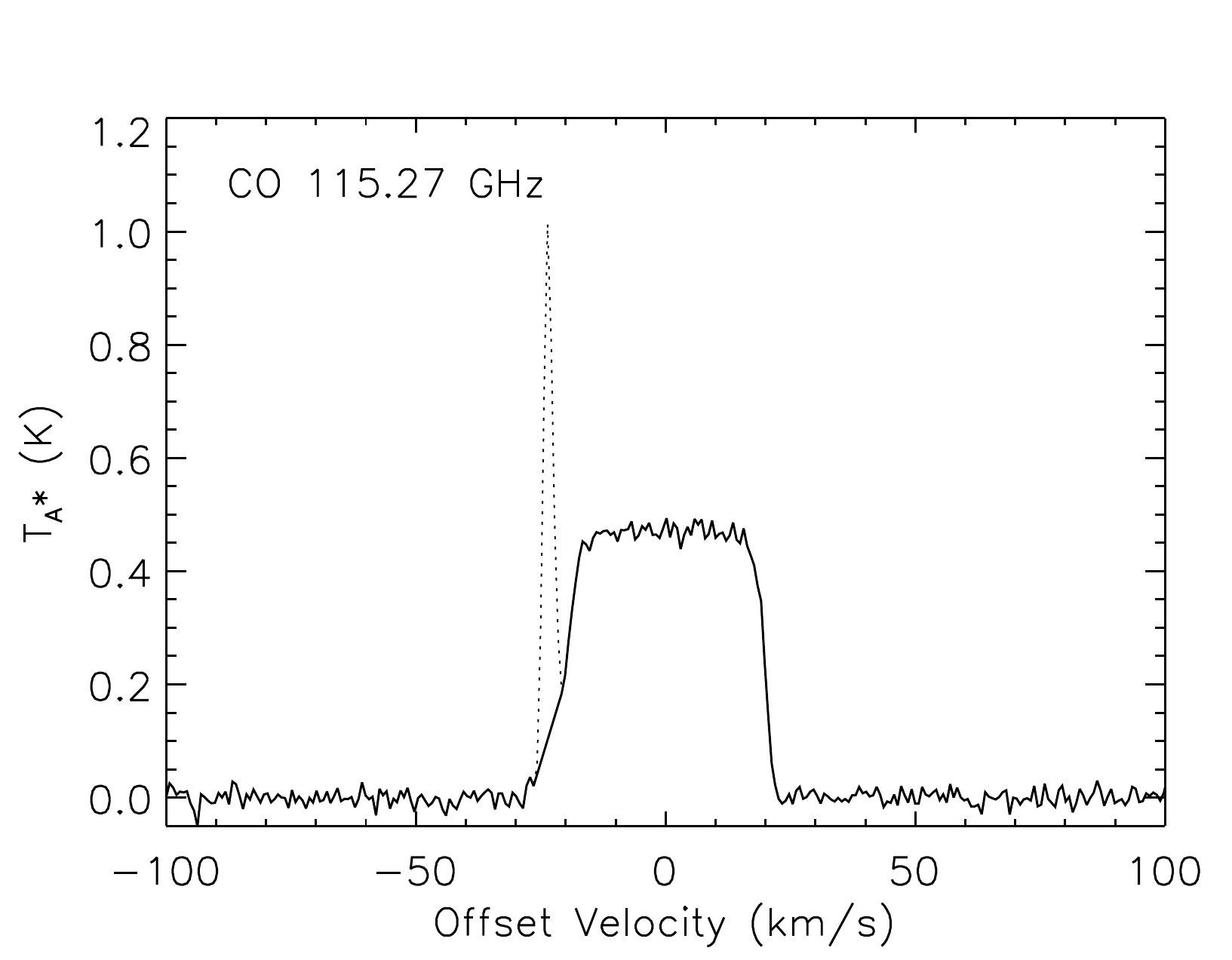}}
\subfigure{\includegraphics[trim= 3cm 13cm 2cm 2cm, clip=true, width=0.4\textwidth]{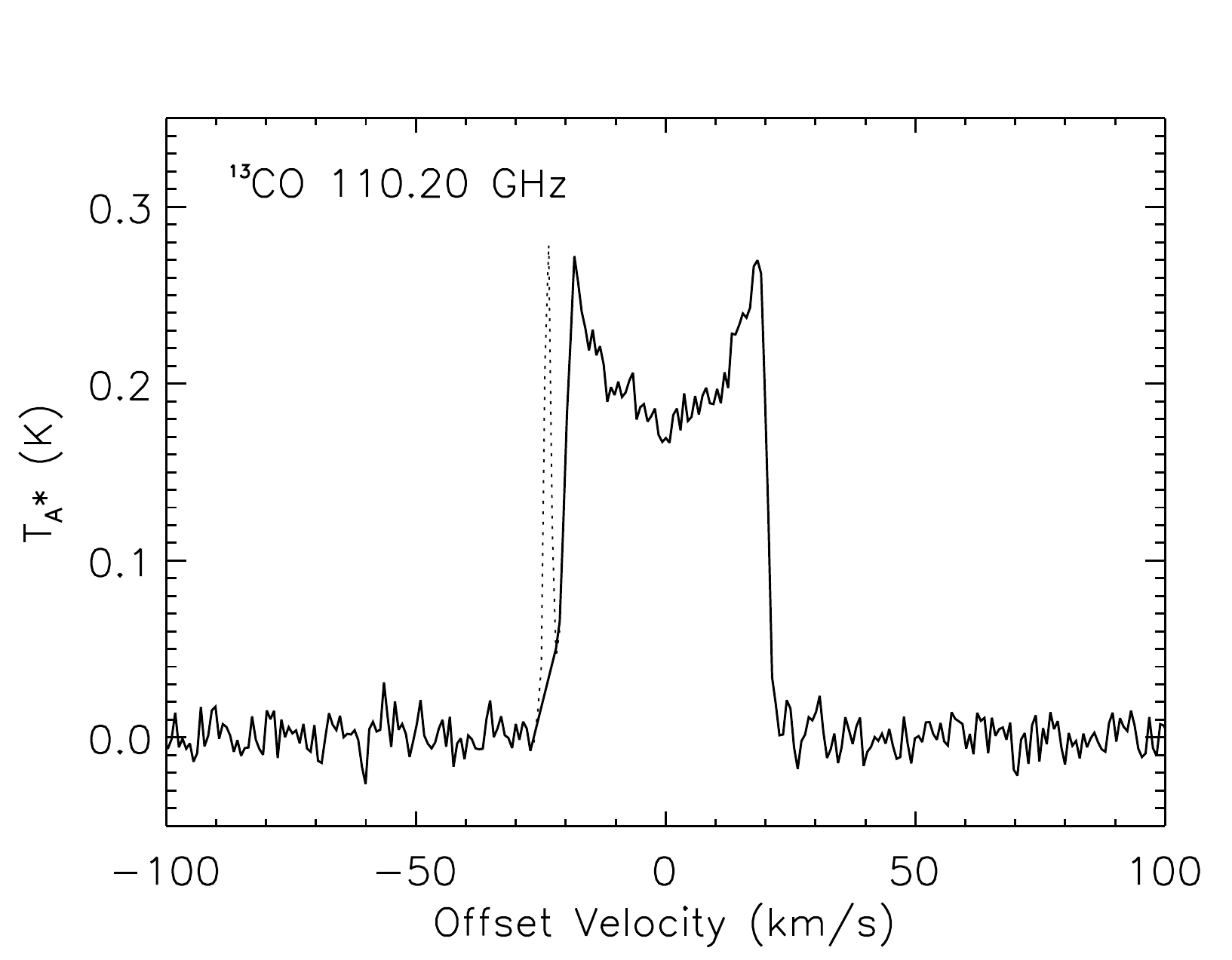}}
\caption{$^{12}$CO (upper) and $^{13}$CO (lower) emission lines in IRAS 15194-5115. Dotted lines indicate the interstellar component. Ordinate axis intensities are in units of corrected antenna temperature, abscissa values are in units of km/s, corrected for LSR velocity of the source, taken as -15.2 km/s.}\label{COfig}
\end{figure}

IRAS 15194-5115 had the largest number of detected transitions of all sources in the sample. CO and $^{13}$CO both show a sharp peak on the blueshifted edge of the line profile, shown in Fig. \ref{COfig}. This is caused by contamination from an interstellar cloud that lies in the same line of sight as IRAS 15194-5115, further details of which can be found in \citet{Nyman1987} and \citet{Nyman1993}. When measuring the peak and integrated intensities of these lines, the contaminant emission was removed and the line interpolated between the remaining points.

Transitions originating from silicon based molecules (e.g. SiO and SiC$_2$) have been detected, including the silicon isotopologues of SiO, allowing silicon isotopic ratios to be calculated.

Multiple $^{13}$C-containing species were detected in this source including, but not limited to, the $^{13}$C isotopologues of HNC, CCH and HCCCN. Emission lines from molecules with a minor isotope of Si or S were also detected.

As explained in Sect. \ref{lineprofiles}, the extent of the emission and the optical depth of the transitions may be constrained by examining line profiles. CO, for example, has a flat topped parabolic profile with the $^{13}$CO counterpart displaying a horned profile, implying optically thick and thin resolved emission respectively. 

HCN displays a Gaussian line shape, implying it does not originate from a region of constant velocity expansion. This line, as shown in \citet{Smith2014}, can be well modelled by a blend of the three hyperfine components, each having a Gaussian velocity distribution.

The hyperfine structure lines in C$_4$H are blended, with their relative intensities impossible to determine from the observed blended transition. Therefore the relative intensities were calculated according to Equ. \ref{wblend}. Only the major hyperfine components, which make up $\sim99\%$ of the  transition intensity have been considered: for example, the N= 9-8, J=$\nicefrac{19}{2}\,-\,\nicefrac{17}{2}$ fine line is considered to be split amongst the F=9-8 and F=10-9 components only.  The blended emission lines are square in profile, implying the assumption that the transitions are optically thin is valid. However, this also implies that the C$_4$H emission region is unresolved, although it is unlikely to be highly unresolved as numerous other species display resolved line profile shapes.

Two transitions of HC$_3$N were detected and these are highly blended hyperfine transitions. Einstein coefficients, upper level energies and statistical weights are available for the overall J=10-9 and J=12-11 transitions, thus alleviating the need to estimate the relative intensities of the unresolved hyperfine lines. In this source, the J=12-11 emission line has a square profile, suggesting optically thin, unresolved emission. The J=10-9 profile is less clearly defined, with straight edges but a more curved central region, suggesting this line could be bordering on optically thick and unresolved.

\subsubsection{IRAS 15082-4808}

\begin{figure}
\subfigure{\includegraphics[trim= 3cm 13cm 2cm 2cm, clip=true, width=0.4\textwidth]{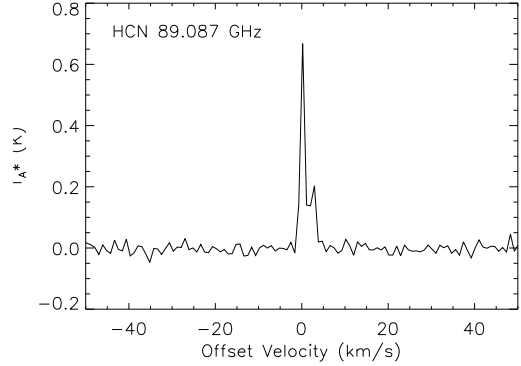}}
\caption{HCN maser line in IRAS 15082-4808. Ordinate axis intensities are in units of corrected antenna temperature, abscissa values are in units of km/s, corrected for LSR velocity of the source, taken as -3 km/s from \citet{Groenewegen2002}.}\label{maserline}
\end{figure}

IRAS 15082-4808 had the second highest number of transitions of the sample. The most unusual detection made is that of the vibrational (v$_2$=2) HCN maser emission at 89.087 GHz. Only ten of these masers, including this detection, have been found to date \citep{Smith2014, Lucas1986, Guilloteau1987, Lucas1988, Lucas1989}. An in-depth study of this maser can be found in \citet{Smith2014} and the maser line is shown in Fig. \ref{maserline}.

The CN line at 113.499 GHz shows interesting structure. The blue shifted part of the spectrum shows a significantly shallower drop than that on the red. This is due to a blend of hyperfine lines causing the appearance of a shallower drop. Similarly the HC$_3$N 109.173 GHz line has this feature, but on the red shifted region of the spectrum and is also likely to be caused by a hyperfine blend. The hyperfine structure lines in C$_4$H are blended and treated in the same fashion as in IRAS 15194-5115. The C$_4$H lines have a low signal-to-noise ratio which makes it difficult to classify the line profiles. They appear to have noisy square profiles, but this is not certain.

\begin{figure}
\subfigure{\includegraphics[trim= 3cm 13cm 2cm 2cm, clip=true, width=0.4\textwidth]{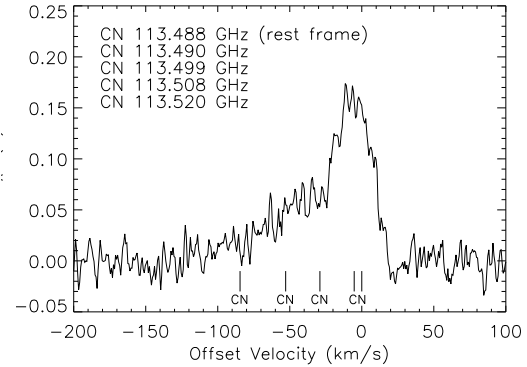}}
\caption{CN emission lines in IRAS 15082-4808. Ordinate axis intensities are in units of corrected antenna temperature, abscissa values are in units of km/s, corrected for LSR velocity of the source, taken as -3 km/s from \citet{Groenewegen2002}.}
\end{figure}

The line profile of these HC$_3$N transitions in IRAS 15082-4808 are parabolic, implying that these lines are optically thick and unresolved. 

Isotopologues of a number of species were detected in this source, including the $^{12}$C and $^{13}$C isotopologues of CS, CO and HCN and the $^{32}$S and $^{34}$S isotopologues of CS. 

\subsubsection{IRAS 07454-7112}

IRAS 07454-7112 has the lowest number of detected transitions in the AGB star portion of the sample. The lines have significantly lower intensities than in the previous two AGB stars. Many of the detected transitions have parabolic shapes and the remainder have too low a signal-to-noise ratio to reliably ascertain the shape of the profile. Despite having similar T$_{\rm{peak}}$ for CO, HCN and CN lines to the aforementioned stars, C$_3$N and the polyyne radicals C$_2$H and C$_4$H  were not detected down to an RMS noise level of 0.01 K. However, three silicon bearing molecules were detected, including three transitions of SiC$_2$.  Similarly to IRAS 15082-4808, two transitions of HC$_3$N were detected as blended transitions, each with parabolic profiles implying optically thin emission..

\subsubsection{IC 4406}

IC 4406 has the lowest number of detections of all of the sample, with transitions from only seven species detected. This is to be expected as planetary nebulae are less molecule-rich than AGB stars due to the high levels of ionising radiation.

CO has a double-peaked appearance in this target; the blue-shifted peak is more intense than the red-shifted peak. This nebula is not spherically symmetric, thus it is difficult to infer information about optical depth and whether or not the nebula is resolved from the line profiles. The variation in line profiles of CO with position on the nebula were mapped by \citet{Sahai1991} and this shows the positional dependence on the profile shape, caused by different gas motions. As the double-peaked appearance and relative intensities of the peaks are strongly dependent on observation position on the nebula, it appears that the double-peaked appearance is likely due to morphology. The $^{13}$CO line shows only the blue-shifted component, explained by its significantly less intense lines and the exact position of the telescope on the source. The integrated intensity of CO in Table \ref{PN_results} is the sum of the integrated intensities of the two components.

The remaining transitions also show the double-peaked appearance, although in the isotopologues of HCN and in HCO$^+$, the red-shifted component is more intense than the blue shifted peak.

\subsubsection{NGC 6537}

NGC 6537 has transitions detected from 10 species, although all three CO isotopologues ($^{12}$CO, $^{13}$CO and C$^{18}$O) show probable interstellar contamination - these line regions can be seen in Fig. A5.

Two isotopologues of HCN were seen (HCN and H$^{13}$CN) along with HNC. Three isotopologues of CS were detected: CS, $^{13}$CS and C$^{34}$S. HCO$^+$ was also detected.

\section{Analysis}

The analysis carried out on the data consists of: the measurement of the rotational temperatures of different species using the population diagram method, the calculation of source-averaged column densities of each species and isotopic ratios from a number of sets of isotopologues. For the following analysis, the antenna temperature has been converted to main beam temperature, assuming an efficiency of 0.49 before the band-gap and 0.42 afterwards. All the constants used in the analysis originate from the JPL and CDMS catalogues \citep{Pickett1998,Muller2005}. Unless otherwise stated, lines are assumed to be optically thin.

\subsection{Excitation temperatures}\label{RD}

For those species where multiple transitions were detected (excluding hyperfine transitions), it was possible to use the population diagram method, as presented in \citet{Goldsmith1999}, to measure the rotational temperatures of different species. It has been assumed that the excitation temperature can be taken as the rotational temperature in the following analysis. The details of the method used are outlined below, including adaptations to account for hyperfine structure.

\subsubsection{Standard population diagram method}

The population of the upper level of a transition, $N_u$, can be expressed as:

\begin{equation}\label{RD3}
N_u=\frac{N}{Q}g_u\exp\left(-\frac{E_u}{{\rm{k_B}}T_{\rm{ex}}}\right),
\end{equation}

\noindent where $N$ is the column density of the molecule, $E_u$ is the energy of the upper level, $g_u$ is the statistical weight of the upper level and $T_{{\rm{ex}}}$ is the excitation temperature. The partition function, $Q$ is given by:

\begin{equation}
Q=\sum N_i=\sum g_i \exp \left(-\frac{E_{ui}}{{\rm{k_B}}T_{{\rm{ex}}}}\right),
\end{equation}

\noindent where the sum is over all energy levels, $i$.

Thus, rearranging and taking logs of Equ. \ref{RD3} leads to:

\begin{equation}\label{RD4}
\ln\left(\frac{N_u}{g_u}\right)=-\frac{1}{T_{{\rm{ex}}}}\left(\frac{E_u}{\rm{k_B}}\right)+\left( \ln N - \ln Q\right),
\end{equation}

\noindent which, when plotting $\ln\left(\frac{N_u}{g_u}\right)$ against $\left(\frac{E_u}{\rm{k_B}}\right)$, gives a straight line with gradient equal to $-\frac{1}{T_{{\rm{ex}}}}$ and y intercept equal to $\left( \ln N - \ln Q\right)$.  The population of the upper level, assuming the source fills the beam and the transition is optically thin, is given by:

\begin{equation}\label{RD2}
N_u=\frac{8\pi {\rm{k}_B} \nu^2 W}{{\rm{hc^3}} A_{ul}},
\end{equation}

\noindent where $W$ is the integrated intensity of the line and $A_{ul}$ is the Einstein A coefficient of transition $u$-$l$.

For some species with hyperfine structure, the input parameters (e.g. $A_{ul}$) for the population diagram method are available for the overall N or J transition (e.g. HCCCN). In these cases, Equ. \ref{RD2} may be directly applied, taking $W$ as the total integrated intensity of all hyperfine transitions.

In many cases, the input parameters are only available for the individual hyperfine components. In this case, the method outlined above must be adapted and these adaptations are detailed in the following subsections.

\subsubsection{Hyperfine structure: separable}

If all hyperfine components of the overall N or J transition have been detected and they are either separated or separable (as is the case for C$_2$H in IRAS 15194-5115, see Sect. \ref{c2h_fitting}) so the integrated intensities of all individual hyperfine components are known, the population of the upper level can be taken as the sum over all components:

\begin{equation}
N_u=\frac{8\pi {\rm{k}_{\rm{B}}}}{\rm{hc^3}}\displaystyle\sum_{i}\frac{\nu_i^2 W_i}{A_{i}}\label{HF_all},
\end{equation}

\noindent where the summation is over all hyperfine transitions, $i$.

If all components are not detected, but those that are detected are separated or separable, the method of calculating the population of the upper level must be adapted to compensate for those components not seen. Thus the upper level population becomes:

\begin{equation}\label{HF_some}
N_u=\frac{8\pi {\rm{k}_{\rm{B}}}}{\rm{hc^3}}\frac{1}{\sum a_{i,{\rm{obs}}}}\displaystyle\sum_{i,{\rm{obs}}}\frac{\nu_{i,{\rm{obs}}}^2 W_{i,{\rm{obs}}}}{A_{i,{\rm{obs}}}},
\end{equation}

\noindent with the summations running over all observed hyperfine transitions ($i,{\rm{obs}}$) and $a_{i,{\rm{obs}}}$ are the relative intensities of the observed hyperfine transitions. For this work, the relative intensities of the lines have been taken as the relative intrinsic line strengths, taken from the JPL or CDMS databases and normalised \citep{Pickett1998,Muller2005}. These parameters can also be found in Appendix B of \citet{Smith2014d}\footnote{https://www.escholar.manchester.ac.uk/item/?pid=uk-ac-man-scw:240595}.

It is important to note that there is a distinct difference between the upper level, $u$, which is the unsplit upper energy level, and that of the individual hyperfine transitions, $i$. In cases where $E_u$, the energy of the unsplit upper level, is not known, the mean of the energies of the upper levels of the hyperfine transitions, $\frac{\sum E_{i,{\rm{upper}}}}{\sum i}$, has been taken in its place.

\subsubsection{Hyperfine structure: unseparable}

For those transitions that are blended to the extent that they cannot be disentangled, yet the required input parameters for the population diagram method are only available for the individual components, a further adaptation is required. It has been assumed that the integrated intensities of the blended transitions are divided amongst their components according to the intrinsic line strength:

\begin{equation}\label{wblend}
W_i=W_{\rm{blend}} \frac{a_i}{\displaystyle\sum_{j=0}^{n}a_j},
\end{equation}

\noindent where $W_i$ is the integrated intensity of hyperfine component $i$, $W_{\rm{{blend}}}$ is the integrated intensity of the blend of $n$ hyperfine transitions including hyperfine transition $i$ and $a_i$ is the relative intrinsic line strength of hyperfine component $i$. $\displaystyle\sum_{j=0}^{n}a_j$ is the sum of the intrinsic line strengths of all hyperfine components that make up the blend. The integrated intensities of the individual transitions may now be used in either Equ. \ref{HF_all} or \ref{HF_some} for those where all hyperfine components or some hyperfine components are detected, respectively.

\subsubsection{Line profiles and excitation temperature results}

\begin{figure}
\centering
\subfigure{\includegraphics[trim= 0cm 0cm 0cm 13cm, clip=true, width=0.4\textwidth]{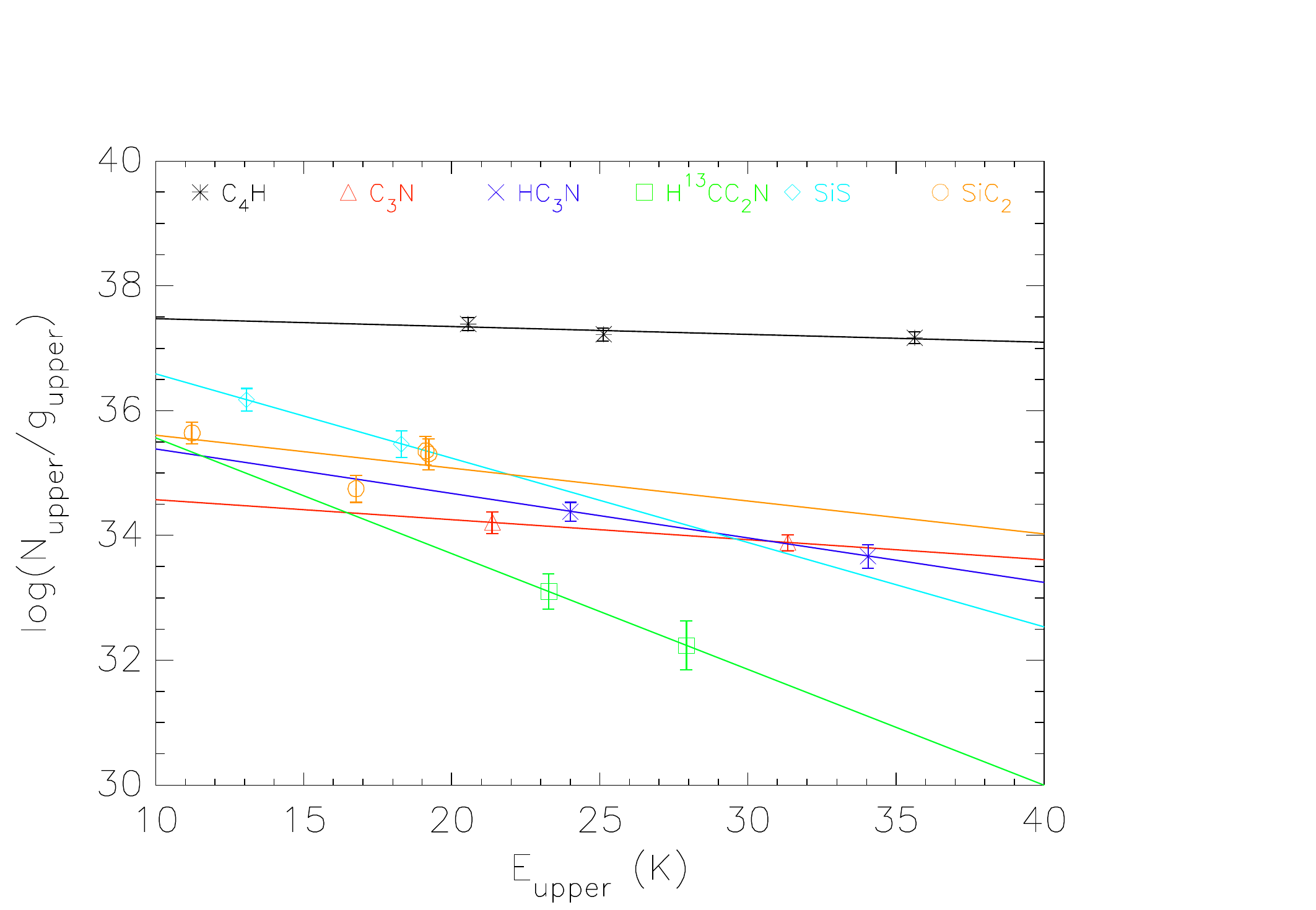}}
\subfigure{\includegraphics[trim= 0cm 0cm 0cm 13cm, clip=true, width=0.4\textwidth]{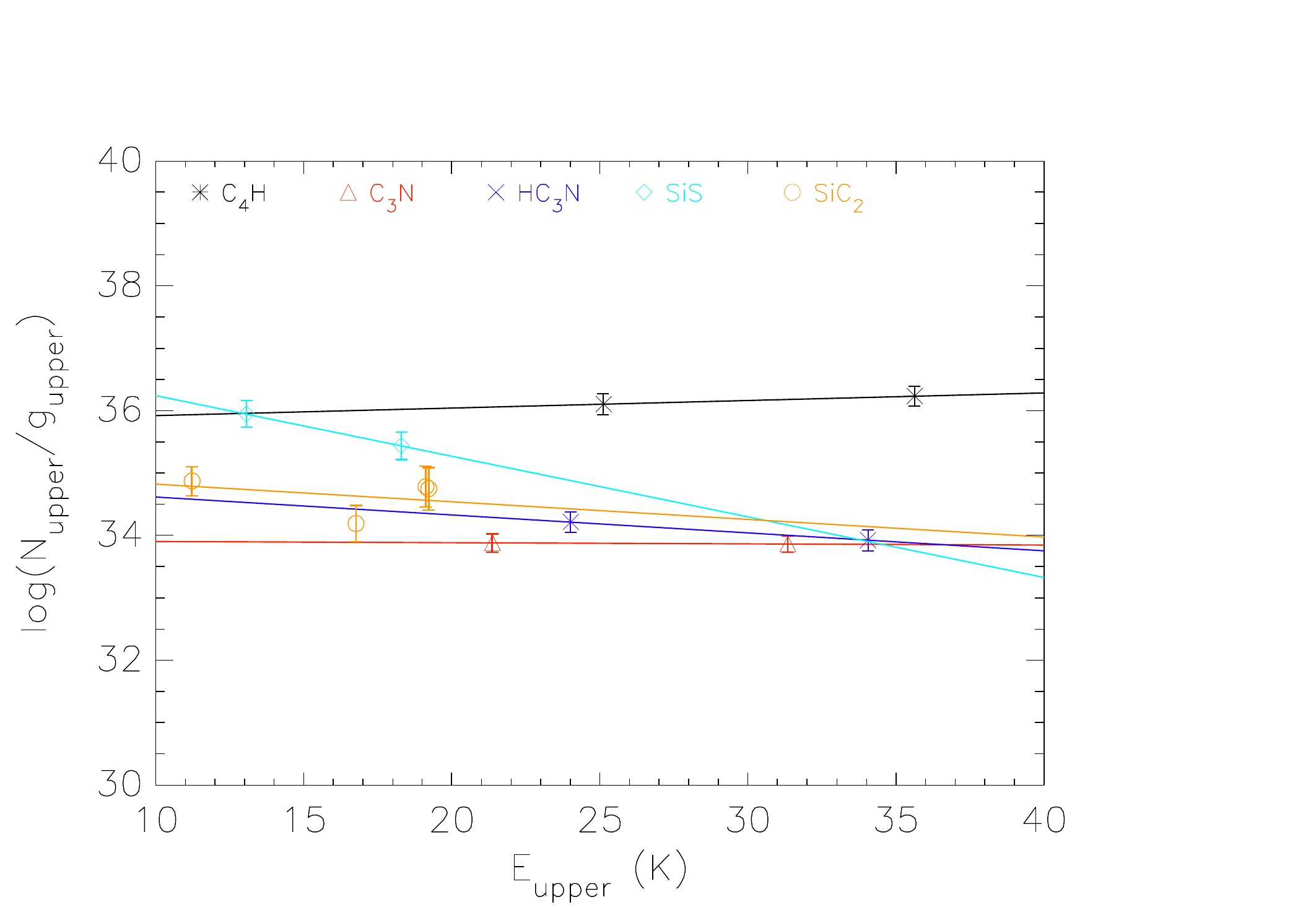}}
\subfigure{\includegraphics[trim= 0cm 0cm 0cm 13cm, clip=true, width=0.4\textwidth]{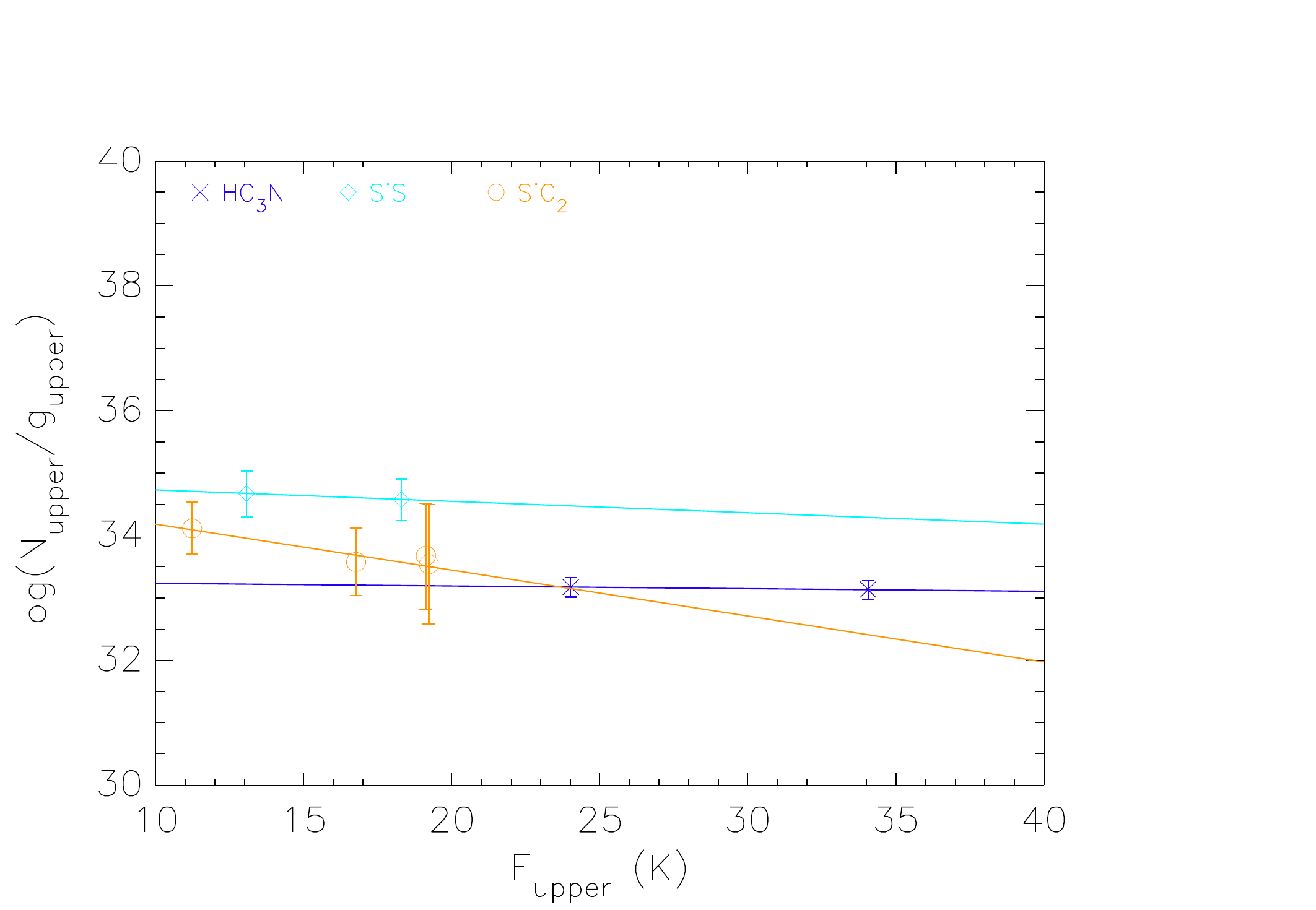}}
\caption{Population diagrams for IRAS $15194-5115$ (upper plot), IRAS $15082-4808$ (middle plot) and IRAS $07454-7112$ (lower plot). All molecules are shown and identidified by their symbols and colours. The black dashed line indicates the fit to all data.}\label{popdias}
\end{figure}

The calculated excitation temperatures from the population diagram method are given in Table \ref{Tex_tab}. Population diagrams are shown for IRAS 15194-5115, IRAS 15082-4808 and IRAS 07454-7112 in Fig. \ref{popdias}.

 Additionally, as only two transitions of this molecule are observed, the reliability of this method is also significantly decreased. The calculated excitation temperatures are given in Table \ref{Tex_tab}. 

\begin{table}
\renewcommand{\arraystretch}{1.2}
\centering
\caption[Excitation temperatures (K)]{Calculated excitation temperatures for a variety of species (K).}\label{Tex_tab}
\medskip
\begin{tabular}{c c c c}
\hline
\multirow{2}{*}{Species}&	IRAS 			&	IRAS 		&	IRAS  \\
				& 15194-5115				&	15082-4808	&	07474-7112 \\
\hline
C$_4$H			&	$79\pm{56}$				&	$\phantom{0}-82\pm{147}\phantom{0.5}$	&	\dots			\\
HC$_3$N			&	$14\pm{5\phantom{0}}$	&	$35\pm{28}$								&	$\phantom{0}240\pm{1260}$ 	\\
H$^{13}$CCCN 	& 	$5\pm{3}$				&	\dots									&	\dots			\\
SiS				&	$7\pm{3}$				&	$10\pm{6\phantom{0}}$					&	$\phantom{0}55\pm{288}$	\\
SiC$_2$			&	$19\pm{11}$				&	$35\pm{51}$								&	$14\pm{16}$		\\
C$_3$N			&	$31\pm{21}$				&	$\phantom{0}\phantom{0}543\pm{5760}$	&	\dots			\\	
\hline
\end{tabular}
\end{table}

Two transitions of SiS (J=5-4 and J=6-5) were detected in all three of the AGB stars in the sample. These lines have no fine or hyperfine splitting associated with them. The two lines have a relatively square profile shape in IRAS 15194-5115, although they show a degree of asymmetry, implying that these transitions are optically thin. In IRAS 15082-4808, the J=6-5 line has a square profile and the J=5-4 transition has a less clear line profile, with resemblance to both square and parabolic, implying these are likely to be optically thin. Both lines of SiS have parabolic profiles in IRAS 07454-7112, implying that both transitions are optically thick.

Up to four SiC$_2$ transition lines were detected in the spectra of each of the AGB stars in the sample. The lines in IRAS 15194-5115 and IRAS 15082-4808 appear to all be square, with only the $\sim95$ GHz transition in IRAS 15082-4808 potentially being parabolic. The lines in the spectra of IRAS 07454-7112 have too low a signal-to-noise ratio to be clearly classified.

C$_3$N was detected in two objects in the sample (IRAS 15194-5115 and IRAS 15082-4808) with two transitions. In both cases, the observed line profiles are square. As with C$_4$H, the lines are blends of unseparable hyperfine components with major components and essentially negligible minor components. Thus the same method for integrated intensity determination has been applied to C$_3$N as was applied to C$_4$H.

\subsection{Column densities}

Two different methods of calculating the column densities of molecules have been applied. These both use the same assumptions: the source fills the beam, the emission is optically thin and LTE applies.

\subsubsection{Column densities: population diagram method}

Column densities can be calculated from the y-intercept, $C$, of the population diagram  (Equ. \ref{RD4}):

\begin{equation}
C=\left( \ln N - \ln Q\right),
\end{equation}

\noindent where $N$ is the source-averaged column density and $Q$ is the partition function of the molecule. The partition functions have been taken from the Jet Propulsion Laboratory (JPL) database or the Cologne Database for Molecular Spectroscopy (CDMS) via Splatalogue \citep{Remijan2007} and are calculated for a variety of excitation temperatures. The partition function evaluated at the closest temperature to that found from the population diagram analysis was adopted for each molecule, barring C$_4$H in IRAS 15082-4808 for which a temperature of 37.5 K was assumed as the measured excitation temperature was unphysical. The values were not interpolated as they are generally within the measured uncertainties of the excitation temperatures. 

The calculated column densities for all molecules for which population diagrams could be constructed are shown in Table \ref{col_density}.

\subsubsection{Column densities: single-line method}\label{CD-slm}

Population diagrams, as described above, could not be created for all molecules, thus the column densities were calculated for the remaining molecules by combining Equ. \ref{RD3} and \ref{RD2}:

\begin{equation}
N=\frac{8\pi{\rm{k}_{\rm{B}}} \nu^2 W}{{\rm{hc^3 A}}_{ul}}\frac{Q}{g_u}\exp{\left(\frac{E_u}{{\rm{k}_B} T_{{\rm{ex}}}}\right)},\label{CD1}
\end{equation}

\noindent where $N$ is the column density of the species, $\nu$ is the frequency of transition, $W$ is the integrated intensity of the emission line, $A_{\rm{ul}}$ is the Einstein coefficient of the transition from the upper level, u, to the lower level, l. Blended lines and  hyperfine transitions were treated in the same fashion as Sect. \ref{RD}, similar to that of \citet{Greaves1992}. If the parameters are available for the overall transition, Equ. \ref{CD1} becomes simply:

\begin{equation}
N=\frac{8\pi{\rm{k}_{\rm{B}}} \nu^2}{{\rm{hc^3 A}}_{ul}}\frac{Q}{g_u}\exp{\left(\frac{E_u}{{\rm{k}_{\rm{B}}} T_{{\rm{ex}}}}\right)}\displaystyle \sum_i W_i.
\end{equation}

\noindent If parameters are only available for the individual hyperfine transitions, Equ. \ref{CD1} becomes:

\begin{equation}
N= \frac{8\pi {\rm{k}_{\rm{B}}}}{\rm{hc^3}}\frac{Q}{g_u}\exp{\left(\frac{E_u}{{\rm{k}_{\rm{B}}} T_{{\rm{ex}}}}\right)}\displaystyle\sum_{i}\frac{\nu_i^2 W_i}{A_{i}},
\end{equation}

\noindent if all hyperfine components are observed, or if only some hyperfine components are observed:

\begin{equation}
N=\frac{8\pi {\rm{k}_B}}{\rm{hc^3}}\frac{Q}{g_u}\exp{\left(\frac{E_u}{{\rm{k}_{\rm{B}}} T_{{\rm{ex}}}}\right)\frac{1}{\sum a_{i,{\rm{obs}}}}}\displaystyle\sum_{i,{\rm{obs}}}\frac{\nu_{i,{\rm{obs}}}^2 W_{i,{\rm{obs}}}}{A_{i,{\rm{obs}}}}.
\end{equation}

\noindent The summations run over either all hyperfine transitions, $i$, or all {\emph{observed}} hyperfine transitions ($i,{\rm{obs}}$). 

The excitation temperature was assumed as: 18.75 K for IRAS 15194-5115 and 37.5 K for all other sources. These were selected as closest to the mean of the calculated excitation temperatures for each of the AGB stars (excluding outliers) and were simply assumed for the PN as no excitation temperatures could be calculated. 

The aforementioned excitation temperatures are given to a higher degree of precision than is realistic for our results from the population diagram analysis. The temperatures have been specified at that level of precision solely due to the fact that the partition functions, available from CDMS, have been calculated at these very specific temperatures. For the effect of the excitation temperature uncertainties on the isotopic ratios and column density calculations, see Sect. \ref{uncertainties}.

The results of these calculations are shown in Table \ref{col_density}. 

\begin{table*}
\renewcommand{\arraystretch}{1.2}
\centering
\caption[Column densities of all detected species.]{Column densities of all detected species in all sources. Quoted values are in cm$^{-2}$. Uncertainties quoted here are the formal errors propagated from calibration and measurement uncertainties. A discussion on the effects of excitation temperature and other assumptions can be found in the text.  Those values calculated from the rotation diagram method are shown in bold, the remaining values used the single-line method. The lower limits in this table originate from transitions which are likely optically thick..}\label{col_density} 
\bigskip
\begin{tabular}{l c c c c c}
\hline
Species	& IRAS 15194-5115 	& IRAS 15082-4808	& IRAS 07454-7112 & IC 4406 & NGC 6357 \\ 
\hline
CO   			&   $>(5.1\pm{0.5})\times10^{16}\phantom{01}$&  $>(8.6\pm{0.9})\times10^{16}\phantom{01}$&   $>(4.8\pm{0.5})\times10^{16}\phantom{01}$   &   $(1.5\pm{0.3})\times10^{16}$   &   \dots   \\
$^{13}$CO   	&   $(5.0\pm{0.6})\times10^{16}$   			&   $(1.2\pm{0.3})\times10^{16}$ 			&   $(1.4\pm{0.2})\times10^{16}$   				&   $(4.4\pm{1.0})\times10^{15}$   &   \dots   \\
HCN   			&   $>(2.2\pm{0.3})\times10^{14}\phantom{01}$& $>(2.7\pm{0.3})\times10^{14}\phantom{01}$&   $>(7.1\pm{0.8})\times10^{13}\phantom{01}$   &   $(2.3\pm{0.5})\times10^{13}$   &  $(2.3\pm{0.5})\times10^{13}$   \\
H$^{13}$CN  	&   $>(2.7\pm{0.3})\times10^{14}\phantom{01}$& $>(5.0\pm{0.8})\times10^{13}\phantom{01}$&   $(3.7\pm{0.5})\times10^{13}$    			&   $(1.1\pm{0.3})\times10^{13}$   &   $(9\pm{3})\times10^{12}$   \\
HNC   			&   $(1.5\pm{0.2})\times10^{13}$   			&   $(1.1\pm{0.2})\times10^{13}$   			&   $(2.6\pm{0.8})\times10^{12}$   				&   $(3.5\pm{0.8})\times10^{12}$   &    $(3.4\pm{0.7})\times10^{12}$   \\
HN$^{13}$C  	&   $(4\pm{1})\times10^{12}$     			&   \dots   								&   \dots   &   \dots   &   \dots    \\
C$_2$H   		&   $(4.1\pm{0.4})\times10^{15}$   			&   $(1.5\pm{0.3})\times10^{15}$    		&   \dots   &   \dots   &   \dots  \\
C$^{13}$CH		&  	$(1.1\pm{0.4})\times10^{15}$			&   \dots 									&   \dots 	&   \dots 	& 	\dots  \\
C$_4$H   		&   ${\mathbf{(2.8\pm{0.7})\times10^{15}}}$   		&   ${\mathbf{(2\pm{2})\times10^{14}}}$	    &   \dots   &   \dots 	&   \dots \\
C$_3$N   		&   ${\mathbf{(1.4\pm{0.8})\times10^{14}}}$   			&   ${\mathbf{(8\pm{4})\times10^{14}}}$ 				&   \dots   &   \dots   &   \dots   \\
HC$_3$N   		&   ${\mathbf{>(2\pm{1})\times10^{13}\phantom{01}}}$	&   ${\mathbf{>(3\pm{2})\times10^{13}\phantom{01}}}$	&   ${\mathbf{>(3\pm{2})\times10^{13}\phantom{01}}}$      &   \dots   &   \dots  \\
H$^{13}$CCCN 	&	${\mathbf{\phantom{0}(8\pm{20})\times10^{13}}}$	&   \dots 									&   \dots   &   \dots   &   \dots   \\
HC$^{13}$CCN 	&	$(1.3\pm{0.3})\times10^{13}$			&   \dots 									&   \dots   &   \dots   &   \dots   \\
HCC$^{13}$CN 	&	$(1.3\pm{0.3})\times10^{13}$			&   \dots 									&   \dots   &   \dots   &   \dots   \\
HC$_5$N			&	$(4\pm{1})\times10^{13}$  				&   \dots 									&   \dots 	& 	\dots 	&   \dots \\
HCO$^{+}$		&   \dots 									&   \dots 									&	\dots 					& $(4\pm{1})\times10^{12}$ & $(2.3\pm{0.5})\times10^{12}$\\
CN   			&   $(1.2\pm{0.1})\times10^{15}$   			&   $(2.9\pm{0.2})\times10^{15}$  			&   $(1.4\pm{0.1})\times10^{15}$   &   $(6.3\pm{0.6})\times10^{14}$   &   \dots   \\
$^{13}$CN    	&   $(4.3\pm{0.9})\times10^{14}$  			&   \dots  									&   \dots  	& \dots 	&   \dots \\
CS   			&   $>(1.1\pm{0.1})\times10^{14}\phantom{01}$&$>(1.1\pm{0.1})\times10^{14}\phantom{01}$ &   $>(3.6\pm{0.5}\times10^{13}\phantom{01}$    &   \dots    &     $(9\pm{2})\times10^{12}$    \\
$^{13}$CS   	&   $(4.2\pm{0.7})\times10^{13}$    		&   $(6\pm{3})\times10^{12}$   				&   \dots	&   \dots	&     $<(7\pm{3})\times10^{12}\phantom{01}$   \\
C$^{34}$S   	&   $(8\pm{2})\times10^{12}$   				&  	$(5\pm{3})\times10^{12}$    			&   \dots   &   \dots  	&      $(5\pm{1})\times10^{12}$   \\
C$_3$H$_2$   	&   $(2.3\pm{0.5})\times10^{13}$ 			&   \dots      								&   \dots   &   \dots   &   \dots  \\
SiO   			&   $>(2.7\pm{0.4})\times10^{13}\phantom{01}$&$>(1.6\pm{0.3})\times10^{13}\phantom{01}$ &   $>(6.2\pm{0.9})\times10^{12}\phantom{01}$      &   \dots &   \dots    \\
$^{29}$SiO   	&   $(2.4\pm{0.9})\times10^{12}$			&   \dots   								&   \dots	&   \dots   &   \dots     \\
$^{30}$SiO   	&   $(1.6\pm{0.7})\times10^{12}$			&   \dots      								&   \dots  	&   \dots       &   \dots \\
SiS   			&   ${\mathbf{(7\pm{5})\times10^{13}}}$  				&   ${\mathbf{(3\pm{3})\times10^{13}}}$				&   ${\mathbf{(1\pm{2})\times10^{13}}}$     &   \dots   &   \dots     \\
SiC$_2$   		&   ${\mathbf{(4\pm{2})\times10^{13}}}$      			&   ${\mathbf{(4\pm{3})\times10^{13}}}$   				&   ${\mathbf{(4\pm{5})\times10^{12}}}$     &   \dots &   \dots   \\
\hline 
\end{tabular}
\end{table*}

\subsection{Isotopic ratios}

Isotopic ratios are most directly measured using observations of optically thin lines. The optical depth can be somewhat constrained by examining the line profiles and making the assumptions described in Sect. \ref{lineprofiles}. 

The isotopic ratios have been calculated by comparing the calculated source-averaged column densities. The results are shown in Table \ref{isotopic_ratios}.

\begin{table*}
\renewcommand{\arraystretch}{1.2}
\centering
\caption[Measured isotopic ratios.]{Isotopic ratios derived from calculated column densities. Uncertainties quoted here are the formal errors propagated from calibration and measurement uncertainties. A discussion on the effects of excitation temperature and other assumptions can be found in the text. \medskip}\label{isotopic_ratios}
\begin{tabular}{l l c c c c c}
\hline
Ratio 				& Species 					& IRAS 15194-5115 				& IRAS 15082-4808 				& IRAS 07454-7112 	& IC 4406 	& NGC 6537 \\
\hline
$^{12}$C/$^{13}$C 	& CO, $^{13}$CO 			& 	$>1.1\pm{0.2}\phantom{01}$	&	$>7\pm{2}\phantom{01}$		&	$>3.4\pm{0.6}$	& 	$3.4\pm{1.0}$	& \dots	\\
					& HCN, H$^{13}$CN			&	$>0.8\pm{0.1}\phantom{01}$	&	$>5\pm{1}\phantom{01}$		&	$>1.9\pm{0.3}$	&	$2.1\pm{0.7}$	& $2.6\pm{1.0}$ 	\\
					& HNC, HN$^{13}$C			&	$3.8\pm{1.1}$				& 	\dots						&	\dots			&	\dots			& \dots \\
					& CS, $^{13}$CS				&	$>2.6\pm{0.5}\phantom{01}$	& 	$18\pm{9}\phantom{0}$		&	\dots			&	\dots			& $>1.3\pm{0.6}\phantom{01}$\\
					& C$_2$H, C$^{13}$CH		& 	$3.7\pm{1.4}$				& 	\dots						&	\dots			&	\dots			& \dots \\
					& HC$_3$N, H${13}$CCCN 		&	$>0.3\pm{0.6}\phantom{01}$	&	\dots						&	\dots			&	\dots			& \dots \\
					& HC$_3$N, HC${13}$CCN 		&	$>1.5\pm{0.8}\phantom{01}$	&	\dots						&	\dots			&	\dots			& \dots \\
					& HC$_3$N, HCC${13}$CN 		&	$>1.5\pm{0.8}\phantom{01}$	&	\dots						&	\dots			&	\dots			& \dots \\
					& CN, $^{13}$CN				&	$>2.8\pm{0.6}\phantom{01}$	&	\dots						&	\dots			&	\dots			& \dots \\
$^{32}$S/$^{34}$S 	& C$^{34}$S, $^{13}$CS		&	$\,21\pm{6}^a$				&	$\,\,\,22\pm{17}^b$			&	\dots			&	\dots			& $\,\,\,4\pm{2}^c$\\
					& CS, C$^{34}$S				&	$>13.8\pm{3.7}\phantom{0}$	&	$>22\pm{13}$					& 	\dots			&	\dots			& $1.8\pm{0.5}$ 	\\
$^{28}$Si/$^{29}$Si	& SiO, $^{29}$SiO			&	$>11\pm{5}\phantom{001}$	&	\dots						&	\dots			&	\dots			& \dots \\
$^{28}$Si/$^{30}$Si	& SiO, $^{30}$SiO			&	$>17\pm{8}\phantom{001}$	&	\dots						&	\dots			&	\dots			& \dots \\
$^{29}$Si/$^{30}$Si & $^{29}$SiO, $^{30}$SiO	&	$1.5\pm{0.7}$				&	\dots						&	\dots			&	\dots			& \dots \\
\hline
\end{tabular}

$^a$: $^{12}$C/$^{13}$C = 4.0, $^b$:  $^{12}$C/$^{13}$C = 18.2, $^c$: $^{12}$C/$^{13}$C =2.5
\end{table*}

\subsubsection{Carbon isotopic ratios}

A number of carbon-bearing species and their corresponding isotopologues were identified in these observations: CO, HCN, HNC, CS, HC$_3$N, CN and all of their corresponding $^{13}$C isotopologues. Additionally, CCH and one of its carbon isotopologues, C$^{13}$CH was identified. The transitions of the second isotopologue, $^{13}$CCH, are outside the frequency range of these observations. 

For those species with a single isotopologue, such as CO and CS, the isotopic ratio is found by comparing the column densities of the two isotopologues. For those with multiple isotopologues, the isotopic ratio can be determined from each pair and should be equal, assuming there is no chemical fractionation towards one isotopologue. For example, HC$_3$N has three isotopologues (H$^{13}$CCCN, HC$^{13}$CCN and HCC$^{13}$CN) and the isotopic ratio may be found by dividing the column density of the $^{12}$C-containing isotopologue by each of the column densities of the $^{13}$C-containing isotopologues resulting in three $^{12}$C/$^{13}$C isotopic ratios. 

\begin{figure}
\subfigure{\includegraphics[trim= 2cm 13cm 2cm 2cm, clip=true, width=0.4\textwidth]{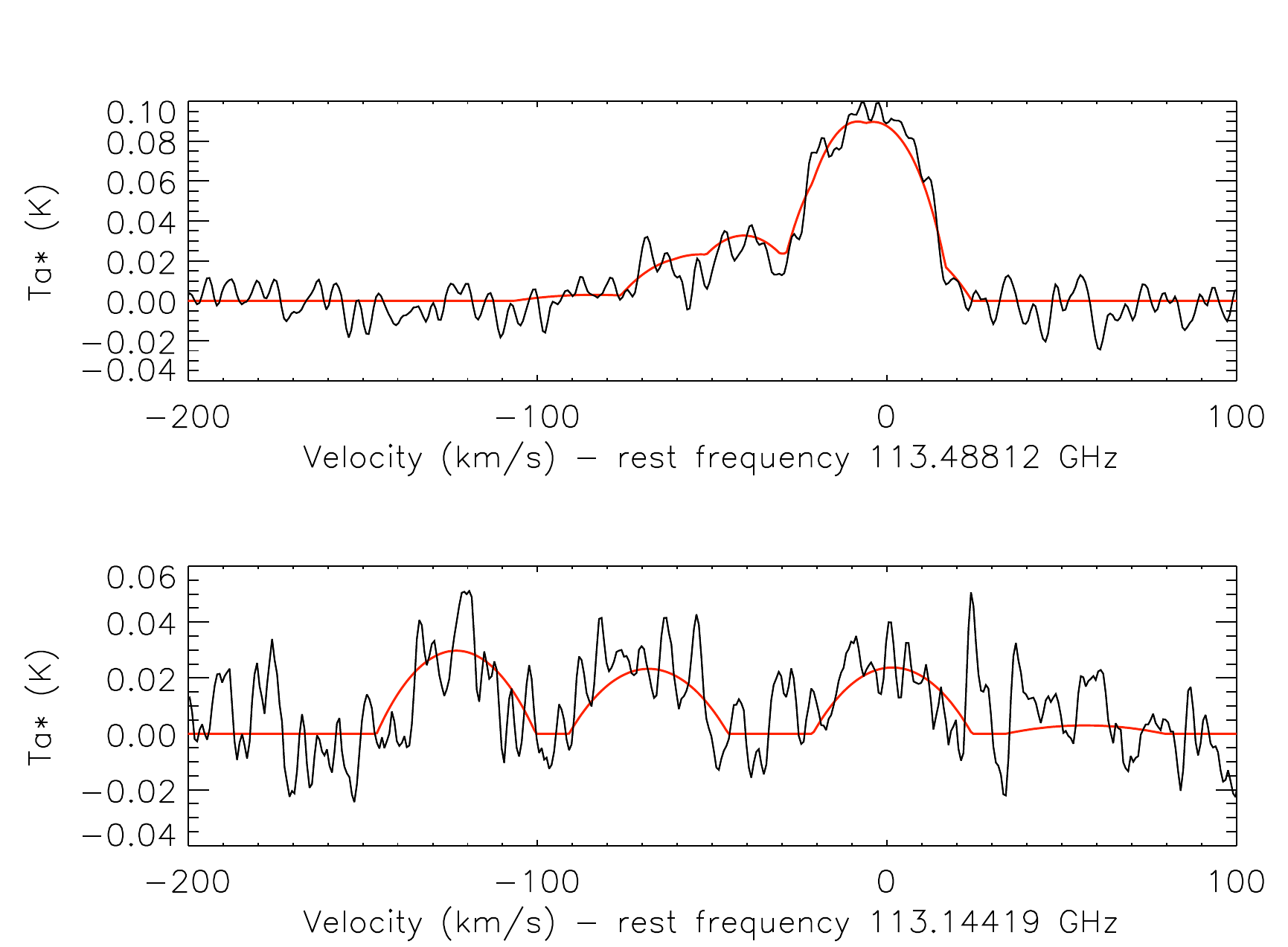}}
\subfigure{\includegraphics[trim= 2cm 13cm 2cm 2cm, clip=true, width=0.4\textwidth]{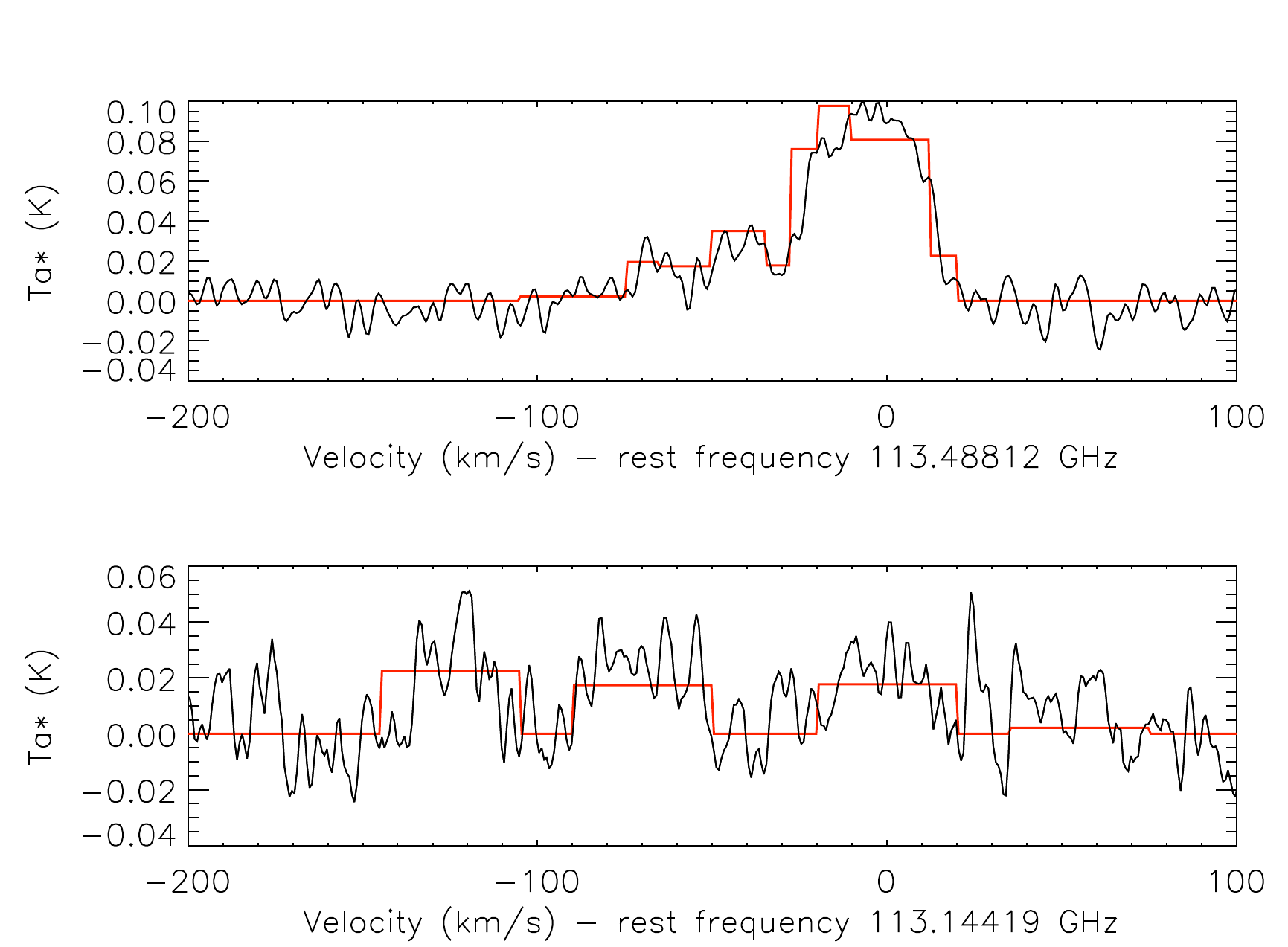}}
\caption{Figures showing example fits to the CN emission in IRAS 15194-5115. The upper two plots are optically thick fits to the data and the lower two are optically thin fits to the data.}\label{CN_fits}
\end{figure}

CO and HCN appear to be optically thick in all the AGB stars in the sample, thus the ratios found from these molecules are lower limits only and, as expected, give the lowest calculated ratios. HNC, HC$_3$N and C$_2$H appear to be optically thin from the appearance of their line profiles in IRAS 15194-5115 and their ratios agree well with one another. $^{12}$CS in some cases appeared to be on the borderline between optically thick and optically thin. In IRAS 15194-5115, the measured isotopic ratio is lower from CS than from HNC and C$_2$H, which would be expected if CS were somewhat optically thick. The hyperfine transitions of CN were, in general, blended making it difficult to identify whether the transitions were optically thick or thin. Fig. \ref{CN_fits} shows the hyperfine structure of CN with examples of both optically thick and optically thin fits to the data. The optically thick blended profile better represents the data, suggesting that the transition is optically thick. The calculated carbon isotopic ratio is lower than the ratios found from HNC and C$_2$H which also suggests that the transition is optically thick.

The ratios calculated here are lower than those found by directly evaluating the integrated intensities of the lines, without compensating for the transitions' different Einstein coefficients. For IRAS 15194-5115, for example, the integrated intensity ratios of isotopologues HNC and CS are 5.5 and 5.8 respectively.

\subsubsection{Sulphur isotopic ratios}

Three isotopologues of CS were detected in multiple sources: $^{12}$C$^{32}$S, $^{12}$C$^{34}$S and $^{13}$C$^{32}$S. A further two isotopologues were not detected in any of the sample stars: $^{12}$C$^{33}$S and $^{12}$C$^{36}$S. 

Deriving $^{32}$S/$^{34}$S can be done in two ways with the detected transitions: by comparing the integrated intensities of the $^{12}$C$^{32}$S and $^{12}$C$^{34}$S lines or by dividing the integrated intensity of $^{13}$C$^{32}$S with that of $^{12}$C$^{34}$S and multiplying by the carbon isotopic ratio of the source. 

$^{12}$C$^{34}$S, where detected, has an optically thin profile. $^{12}$C$^{32}$S, however, showed a borderline optically thick/thin profile. In IRAS 15082-4808, both methods give consistent ratios of approximately 22 for $^{32}$S/$^{34}$S. In IRAS 15194-5115, the ratios differ by a factor of $\sim1.5$, suggesting that CS is indeed optically thick in this star.

All three CS isotopologues were detected in NGC 6537. $^{32}$S/$^{34}$S was measured to be 2 and 4 from the two aforementioned methods, which is significantly lower than those of the AGB stars. However, the minor isotopologues had low signal-to-noise ratios and therefore these values should be taken with caution.

\subsubsection{Silicon isotopic ratios}

$^{28}$SiO, $^{29}$SiO and $^{30}$SiO have been detected in IRAS 15194-5115. $^{28}$SiO has a clear parabolic profile, meaning only lower limits can be derived for $^{28}$Si ratios. $^{29}$SiO and $^{30}$SiO showed square profiles and are thus optically thin. The measured $^{29}$Si/$^{30}$Si ratio (1.5) is consistent with the Solar value.

\section{Discussion}

\subsection{Emission extent}

From the observed line profiles, limits can be placed on the extent of the emitting regions. For example, the line profile of CO in IRAS 15194-5115 has a flattened parabolic profile (neglecting the contaminant emission) and $^{13}$CO has a double-peaked profile, both of which imply resolved emission. Assuming a distance of 1 kpc to IRAS 15194-5115, the beam (33\arcsec) resolves an area with radius $3.0\times10^{17}$cm, putting a lower limit on the CO emission region. The limit is dependent on the distance to the source used. Assuming a distance of 600 pc, the radius of the area subtended by the beam reduces to $1.6\times10^{17}$cm and if a distance of 1.2 kpc is assumed, this radius increases to $3.2\times10^{17}$ cm. The CO photodissociation radius reported in \citet{Woods2003} is $3.2\times10^{17}$ cm assuming a distance of 600 pc. Our lower limit value is a factor of two smaller than this value but as it is a lower limit, it does not disagree with the results of \citet{Woods2003}.

A number of other molecules, including HC$_3$N, SiO and their isotopologues, show unresolved profiles, implying that their emission region is smaller than that of CO. The upper level energies of the resolved  transitions (molecules: CO, $^{13}$CO, C$_2$H and likely CN and CS) are less than 5 K, whereas the upper level energies of the unresolved emission lines are, as expected, higher. 

\subsection{Excitation temperatures}

Rotation temperatures were determined for up to six molecules in the AGB star portion of the sample. IRAS 15194-5115 has six measured rotation temperatures. These range from 5 K for H$^{13}$CCCN to 77 K for C$_4$H. \citet{Woods2003} determine rotation temperatures for this source as part of their molecular line survey using data obtained on the Swedish-ESO Submillimeter Telescope (SEST) and the Onsala Space Observatory (OSO) 20 m telescope. Of the eight molecules that they were able to measure rotation temperatures for, three also appear in our results: H$^{13}$CCCN, HC$_3$N and SiS. Our results agree to better than 3 K (21\%) for HC$_3$N and  agree to within 1 K (15\%) for SiS and $\sim$2 K (40\%) for H$^{13}$CCCN which are all within our calculated uncertainties. Our calculated temperature for SiC$_2$ agrees with the average result obtained for their AGB star sample (24.4 K). The determined temperature of C$_3$N (31 K) has reasonable agreement with SiC$_2$. C$_4$H, however, is a little higher at 77 K.

\citet{Woods2003} have also measured rotation temperatures in the remaining two sources of our AGB sample. Our obtained value for the rotation temperatures of SiS and SiC$_2$ agree to within 5 K (50\% of our value)  and 2 K (6\% of our value) in IRAS 15082-4808 which are both within the measured uncertainties. However our calculated value for IRAS 07454-7112 is a factor of 10 higher than that measured in \citet{Woods2003}, although the uncertainties on our SiS excitation temperature in IRAS 07454-7112 are extremely large ($\pm{288}$ K). 

Differences in the excitation temperature values could be caused by the different methods used to obtain the rotation temperatures or could be an effect of the different beam sizes of the telescopes used to obtain the data: Mopra has a beam size of 36$\arcsec$$\pm3$ at 86 GHz while SEST and OSO 20 m had 57$\arcsec$ and 44$\arcsec$ respectively at the time of observation. 

HC$_3$N in IRAS 15082-4808 and both HC$_3$N and SiS in IRAS 07454-7112 appear to be optically thick. This results in the underestimation of the upper level populations and will also have an effect on the rotation temperature derived if the optical depth of the transitions varies with upper level energy.

The calculated excitation temperature of C$_4$H is unphysical (negative) in IRAS 15082-4808. This is caused by the shallow gradient of the population diagram constrained by level energies that are in relatively close proximity to one another. This temperature has therefore not been used in subsequent analyses.

\subsection{Source-averaged column densities}

Source-averaged column densities have been calculated in each source for up to 27 different molecules. Two methods were used: the population diagram method and a direct calculation for those molecules for which population diagrams were not able to be constructed. The AGB stars in the sample have higher numbers of molecular species detected on average than those of the planetary nebulae in the sample. This is typical for PNe in comparison to AGB stars due to the high levels of ionising radiation which dissociate many molecular species in the nebulae.

In the work of \citet{Ali2006}, a number of molecular column densities are predicted from gas phase models of IRAS 15194-5115. Our calculated abundances for CN, HNC, SiC$_2$, C$_3$H$_2$ and HC$_5$N agree to within a factor of two of those predicted by the models of \citet{Ali2006}. The abundance limits of molecules such as CS and HCN, whose transitions are optically thick, agree with the models of \citet{Ali2006}. Our measured abundance of SiS is two orders of magnitude lower than the predicted value. The abundance of HC$_3$N is a factor of 100 higher than the predictions of \citet{Ali2006}, but more in-line with the measured column densities of HC$_3$N in IRC+10216 by \citet{Woods2003}.

In general, the column densities of the AGB star sample are broadly similar, with differences of up to a factor of 3 between sources. Notably, IRAS 07454-7112 had no detections of C$_2$H, C$_4$H and C$_3$N, despite the column densities of the detected species being very similar to those of IRAS 15082-4808 and IRAS 15194-5115. 

\citet{Edwards2013} have studied NGC 6537 in-depth in similar frequency ranges to those observed here and use the non-LTE {\sc{RADEX}} code to analyse the data. Our column densities for HNC agree to within 20\% of that found by \citet{Edwards2013}. However, our HCN and H$^{13}$CN are higher than their calculated column densities by a factor of 4, although the ratio of the two agree to within 4\%. As the CO isotopologue regions of the spectra included contaminant absorption, it was impossible to calculate column densities for these molecules. The work of \citet{Edwards2013} does not mention suffering from contamination, although the data in this region is not shown.

\citet{Cox1992} have studied abundances of IC 4406 relative to HCN. Assuming all species are present over the same region, we can compare our source-averaged column densities to those abundances reported in \citet{Cox1992}. Our relative abundance of HNC to HCN is approximately consistent with that found in \citet{Cox1992}. However the relative abundances of CO, CN and HCO$^+$ differ from those measured by \citet{Cox1992} by factors of 2, 1.5 and 3 respectively. This may be due to the different extents of the emission regions or due to the observations having been taken from a different region of the nebula as the strength and profile of the lines has been shown to vary across the nebula.

Although both nebulae are O-rich, at least in the ionised gas, with C/O ratios of 0.79 (IC 4406, \citealp{Delgado2014}) and 0.95 (NGC 6537, \citealp{Pottasch2000}), the only oxygen containing species detected was HCO+. Species such as SO, SO$_2$ and SiO which are seen towards O-rich sources were covered by our survey, but not detected. The absence of SiO is perhaps not surprising as it is often `locked away' in grains in the outer regions of planetary nebulae (e.g. \citealp{Thronson1982}). CS, an unusual species to be present in an O-rich system, was detected towards NGC 6537. As mentioned in \citet{Woods2005}, CS can form in regions of shocked gas so this may partially explain the presence of such a molecule. Also, as the measured C/O ratio is close to 1, it is possible that at least some regions of this nebula could be carbon-rich. 

\subsection{Isotopic Ratios}

A number of isotopic ratios have been calculated for the sample, assuming each of the molecular isotopologue column density ratios is equal to the overall elemental isotopic ratio. Only lower limits could be applied to IRAS 07454-7112 due to the lack of detected $^{13}$C species, suggesting that this source has the highest $^{12}$C/$^{13}$C of the sample. Taking the H$^{13}$CCCN line (J=10-9) as an example, the rms noise in the region of this line is 0.004 K (corrected antenna temperature units) which, assuming the line is optically thin and therefore square, gives an integrated intensity of 0.113 K km/s assuming a linewidth equal to the HCCCN J=10-9 line (28.4 km/s). This results in an upper limit on the column density of $1\times10^{12}$ cm$^{-2}$ and a limit on the $^{12}$C/$^{13}$C of 38. 

IRAS 15082-4808, however, has been found to have a $^{12}$C/$^{13}$C ratio of 18.2 from CS, and significantly lower limits from HCN and CO isotopologues. This puts this carbon-rich AGB star above the threshold for a J-type star.

Both of the observed planetary nebulae are oxygen-rich. Their $^{12}$C/$^{13}$C have been measured in this work at 2.4 and 2.5 from HCN isotopologues in IC 4406 and NGC 6537 respectively. The value for NGC 6537 agrees to within 4\% of that reported by \citet{Edwards2013}. These isotopic ratios are lower than those of other planetary nebulae (e.g. \citealp{Balser2002}) as well as those of the oxygen-rich AGB stars reported by \citet{Milam2009}. 

IRAS 15194-5115 had the largest number of isotopologues detected and thus had the largest number of isotopic ratios calculated. The $^{12}$C/$^{13}$C ratios are consistent across HNC, C$_2$H and CS at 3-4. The value derived from HC$_3$N and its isotopologues is lower than the CNO cycle equilibrium value, however these are lower limits. This alters, however, if the direct method of column density calculation is used: the $^{12}$C/$^{13}$C ratio increases to $\sim4$. This is the only molecule where a significant diference is found between using the two methods and it is unclear as to why this occurs. The limits ascertained for CO, HCN are lower than these values, thus are also consistent. The $^{12}$C/$^{13}$C as derived from CN isotopologues, however, is only 2.8. The transitions of CN used are blended, with the contributions from the individual hyperfine components impossible to ascertain, thus neither the line profiles nor the hyperfine structure ratios could be examined to constrain the optical depth. It is plausible this low ratio is due to the CN transitions being optically thick or due to the assumptions made whilst disentangling the hyperfine structure not being sufficiently robust. Due to the clarity of the line profiles and robustness of the optically thin assumption, the HNC and C$_2$H ratios are the most reliable indicators of $^{12}$C/$^{13}$C in IRAS 15194-5115.

In addition to carbon isotopic ratios in IRAS 15194-5115, silicon and sulphur isotopic ratios were determined. Due to the high optical depth of $^{28}$SiO, only limits could be placed on the $^{28}$Si/$^{29}$Si and $^{28}$Si/$^{30}$Si. However, both $^{29}$SiO and $^{30}$SiO appear to be optically thin. The resulting $^{29}$Si/$^{30}$Si of 1.5 is exactly in agreement with Solar. \citet{Peng2013} measures the $^{29}$Si/$^{30}$Si in 15 oxygen-rich AGB stars and finds the sample have values that are either $\sim1.5$ or lower. Few carbon-rich $^{29}$Si/$^{30}$Si ratios are reported in the literature, due to the low intensity of the $^{29}$SiO and $^{30}$SiO emission lines. The $^{29}$Si/$^{30}$Si has been reported by \citet{Cernicharo2000} to be $1.45\pm0.13$ and by \citet{He2008} to be $1.46\pm0.11$ in IRAS 15194-5115. It is currently uncertain as to whether Si isotopic ratios are affected throughout the life of an AGB star. Discussions on this topic can be found in  \citet{Zinner2006} and \citet{Decin2010}.

$^{32}$S/$^{34}$S ratios were calculated for both IRAS 15194-5115 and IRAS 15082-4808 using two methods: directly from CS and C$^{34}$S and indirectly using the  $^{12}$C/$^{13}$C already calculated along with $^{13}$CS and C$^{34}$S. In IRAS 15082-4808, both methods yield a value of 22, the Solar value. The results from IRAS 15194-5115 from the indirect method are almost 50\% greater than that found using the direct method. This suggests that the CS line is optically thick - which is also suggested by the line profile. As both $^{32}$S and $^{34}$S are primarily produced by high mass stars in supernovae  and the $^{32}$S/$^{34}$S has remained approximately constant at the Solar ratio for the last 7 Gyr \citep{Hughes2008}, this almost-Solar ratio in both AGB stars is expected.

\subsection{Evidence for mixing processes}

\citet{Karakas2010b} has published updated stellar yields from evolutionary models. The results include the average mass fractions of various species in the stellar wind which, as is reported, are suitable for comparison with planetary nebula abundances. These models use metallicities of Z=0.02 (Milky Way metallicity), Z=0.008 (Large Magellanic Cloud metallicity), Z=0.004 (Small Magellanic Cloud metallicity) and Z=0.0001. The $^{12}$C/$^{13}$C ratio varies between 4 and 23000 depending upon initial mass and metallicity. At Solar metallicity, $^{12}$C/$^{13}$C ratios of less than ten are only found in the higher mass progenitors: M$_{\rm{star}}\geq 5$ M$_{\odot}$. 

NGC 6537 has been reported in the literature as having a core mass of 0.7-0.9 M$_\odot$ and a probable progenitor mass of 3-7 M$_\odot$ \citep{Matsuura2005}. Its C/O ratio has been found to be 0.95 \citep{Pottasch2000}. The \mbox{5 M$_{\odot}$} model of \citet{Karakas2010b} with Z=0.02 is oxygen rich (C/O = 0.7) and has a $^{12}$C/$^{13}$C ratio of 6. The final mass of the star is 0.879 M$_\odot$. These parameters have reasonable agreement with the measured values of NGC 6537, both in this work and the literature (e.g. \citealp{Edwards2013}). This model includes hot bottom burning (HBB) but no partial mixing zone (PMZ). This suggests that in NGC 6537, hot bottom burning was present, in agreement with the findings of \citet{Edwards2013}.

The carbon isotopic ratios of IC 4406 are less reliable than that of NGC 6537 due to the low signal-to-noise ratio of the $^{13}$C species transitions; \citet{Josselin2003} report $^{12}$C/$^{13}$C=23. The post-AGB mass has been estimated by \citet{Sahai1991} to be between 0.6 and 0.76 M$_\odot$, requiring a lower mass progenitor than for NGC 6537. The low-mass progenitor models (1.5-2 M$_\odot$) of \citet{Karakas2010b} at Z=0.02 result in $^{12}$C/$^{13}$C of 22 with no HBB or PMZ present and a final mass of 0.6-0.64 M$_\odot$. This implies our carbon isotopic ratio is likely a lower limit and that the star did not undergo HBB.

\subsection{J-type stars and A+B grains}

J type stars, those characterised by $^{12}$C/$^{13}$C $<10$, have been suggested as a potential origin of the SiC presolar A+B grains \citep{Zinner2006}. The A+B grains have low $^{12}$C/$^{13}$C, a trend between the Si isotopes (known as the Si MS line on a plot of $\delta^{29}$Si against $\delta^{30}$Si, with equation:  $\delta^{29}$Si$=-20+1.37\times\delta^{30}$Si, \citealp{Hoppe2010}\footnote{\smallskip$\delta^i{\rm{Si}}=1000\left(\frac{^{i}{\rm{Si}}/^{28}{\rm{Si}}}{^{i}{\rm{Si}}_\odot/^{28}{\rm{Si}}_\odot}\right)$ \citep{Decin2010}}) and more varied S isotopic ratios, but still centred on Solar.

IRAS 15194-5115 has been measured to have a low carbon isotopic ratio, of $\sim4$ from column densities of HNC and C$_2$H, in line with the values obtained from A+B grains. The silicon ratios are difficult to assess due to the $^{28}$SiO transition being optically thick. The $^{32}$S/$^{34}$S ratio is also in line with that found in A+B grains \citep{Hoppe2010, Zinner2006}. This evidence suggests that this probable J-type star could be an example of the class of object from which the A+B grains originate. 

\subsection{Uncertainty considerations}\label{uncertainties}

The uncertainties listed in Tables \ref{col_density} and \ref{isotopic_ratios} are the formal uncertainties found from propagating the calibration and measurement uncertainties through the calculations. The uncertainties on the excitation temperature have not been included in these calculations as these had to be assumed for the majority of the molecular species. In some cases these were taken as the closest temperature for which a published partition function exists to the temperature resulting from the population diagram analyses. Looking at the rotation temperatures as a whole, two outliers exists (543 K and 240 K) and the remainder lie within the range 5-77 K. Taking CO as an example: the calculated column density deviates from the value at 37.5 K by factors of 2.5 and 3.5 when the temperature is changed to 5 K and 150 K respectively. Across all molecules, the  $3\sigma$ uncertainty on the calculated column densities due to the temperature can generally be taken as a factor of 3.

A number of assumptions have been made in the analysis within this paper and these assumptions dominate the uncertainties of the results. Some of these may be quantified, others may not. The first is the assumption of optical depth. This has been tested by examining the shape of the line profiles or, when sufficiently resolved hyperfine structure exists, the lines have been fitted to accurately constrain this. Any value obtained from a transition deemed likely to be optically thick is listed as a lower limit to the column density. 

The assumption that the source fills the beam can also be tested by examining the line profiles. Some lines show resolved profiles and others show unresolved profiles, indicating that the emission regions in each case should not deviate too far from a beam filling factor of 1. Therefore the assumption that the beam filling factor is unity should not contribute significantly to the uncertainties on the calculated column densities.

The remaining assumptions are difficult to quantify, but we estimate that overall the column densities are accurate to a factor of 6. 

The isotopic ratios, though dependent upon the column densities, are expected to have a lower uncertainty as some of the effects discussed above are systematic across the isotopologues. For example, changes in the excitation temperature affect the isotopologues in a similar manner with little effect on the ratio of the two. A significant source of uncertainty emanates from optical depth assumptions: if the major isotopologue is optically thick then the measured isotopic ratio is a lower limit. 

\section{Conclusions}

Presented here are the results of a molecular line survey, covering the frequency range 80.5-115.5 GHz, carried out over a sample of five targets: three carbon-rich AGB stars and two oxygen-rich planetary nebulae. The data was obtained from the Mopra Telescope, Australia. 

68 individual transitions were identified in the data, emanating from 27 different molecular species. The hyperfine structures of C$_2$H and C$^{13}$CH were fitted according to the model of \citet{Fuller1993} to measure the optical depth and the contribution to the integrated intensity of the blended lines from the individual hyperfine components. Rotation temperatures were calculated for those species with multiple detected transitions using the population diagram method \citep{Goldsmith1999}. Source-averaged column densities were calculated using both the population diagram method and a direct calculation using a single transition. We have also detailed the method used to calculate the column densities of species with resolvable and unresolvable hyperfine structure.

From the calculated column densities, carbon, silicon and sulphur isotopic ratios have been determined, where detections allow, for the sample. This includes $^{32}$S/$^{34}$S and $^{29}$Si/$^{30}$Si ratios of the probable J-type star, IRAS 15194-5115. This probable J-type star has been shown to exhibit isotopic ratios in-line with those expected for the source of the A+B presolar grains.

\section{Acknowledgements}


We would like to thank J. Williams for taking the observations upon which this work is based and J. Greaves for discussions on hyperfine structure. The Mopra radio telescope is part of the Australia Telescope National Facility which is funded by the Commonwealth of Australia for operation as a National Facility managed by CSIRO. The University of New South Wales Digital Filter Bank used for the observations with the Mopra Telescope was provided with support from the Australian Research Council. This research was supported by the Science and Technologies Funding Council. 

\bibliography{Thesis_refs}

\appendix

\section{Spectra}

\begin{figure*}
{\centering
\subfigure{\includegraphics[trim= 2cm 13cm 2cm 2cm, clip=true, width=0.32\textwidth]{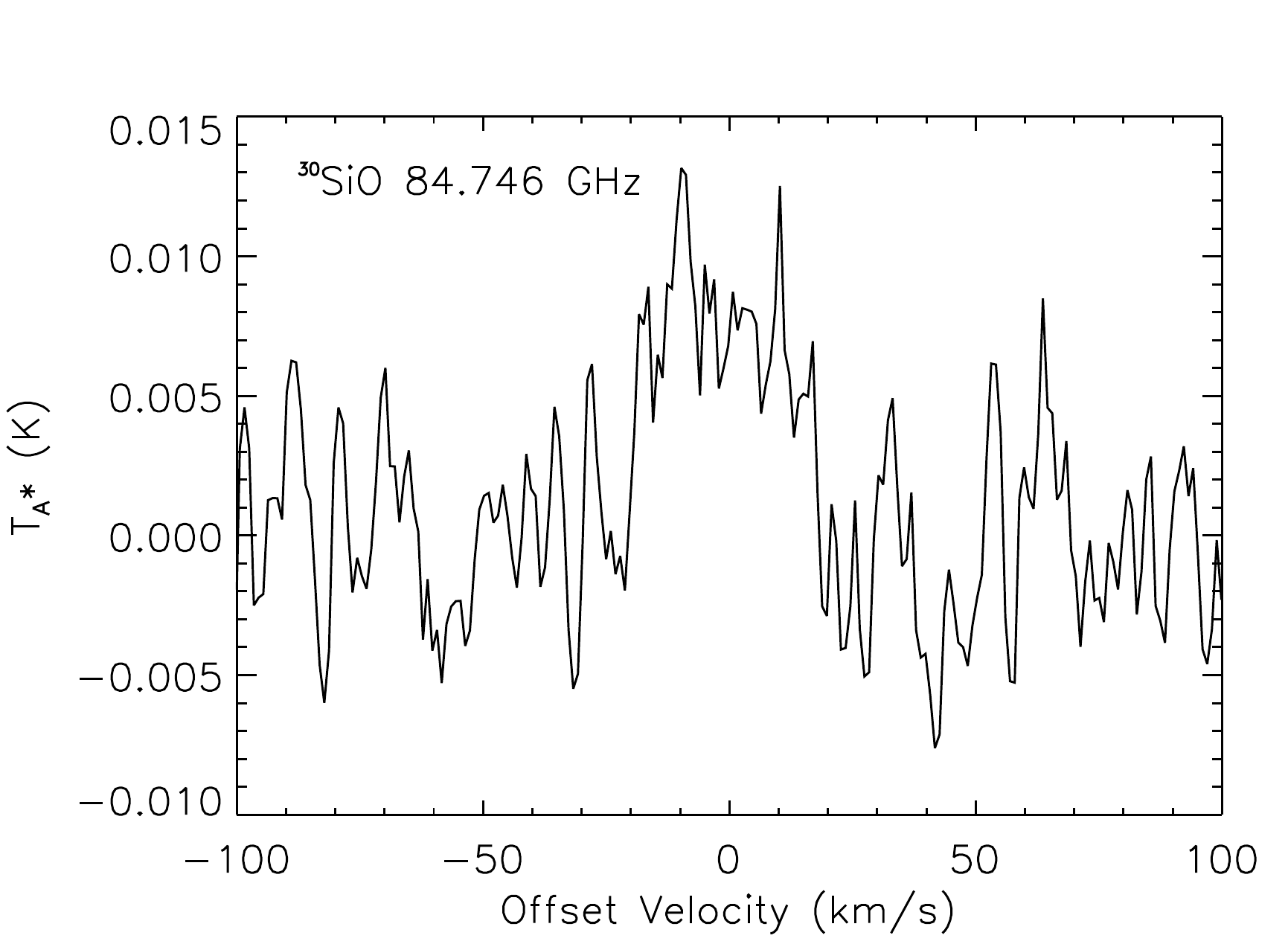}}
\subfigure{\includegraphics[trim= 2cm 13cm 2cm 2cm, clip=true, width=0.32\textwidth]{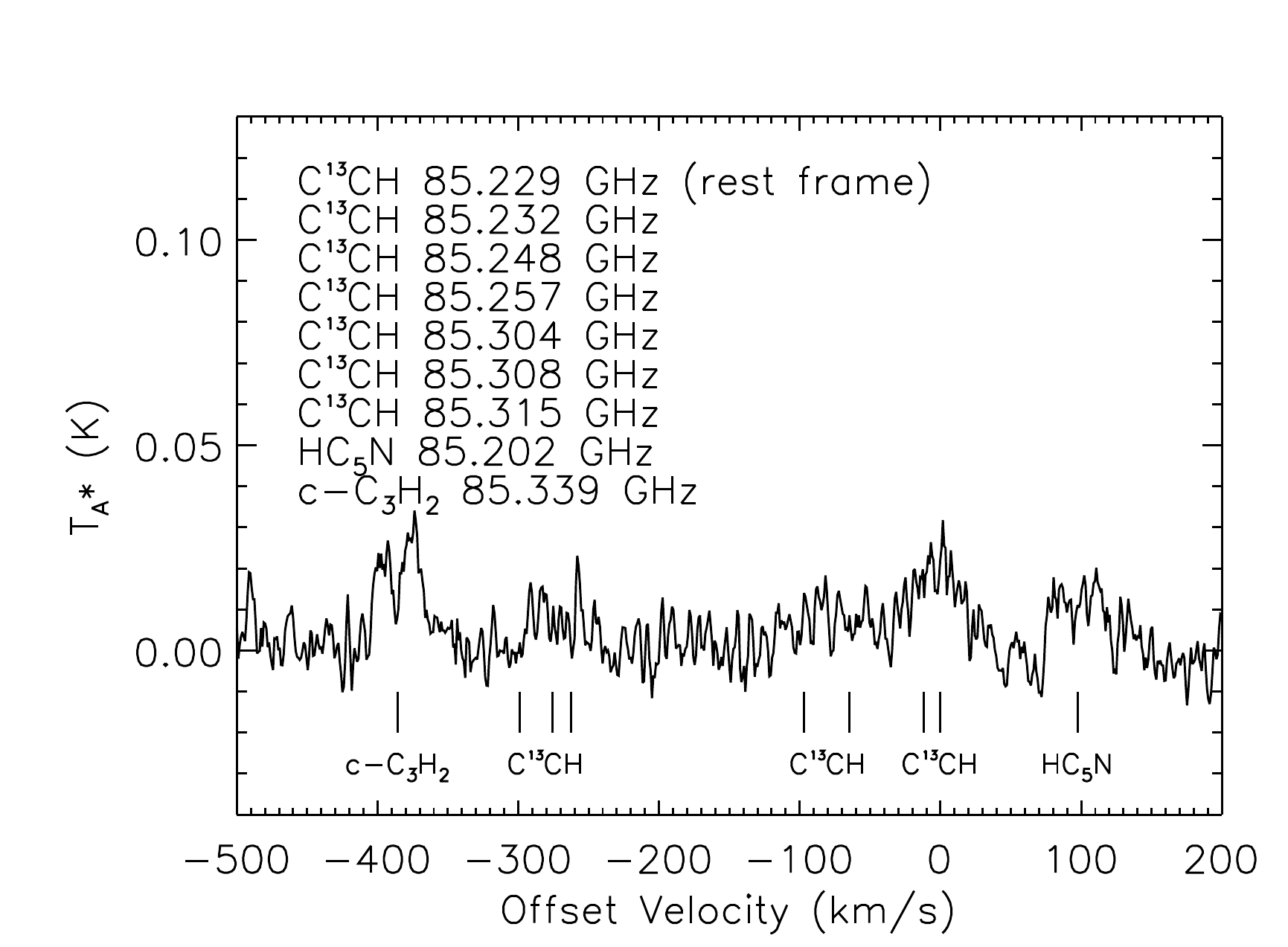}}
\subfigure{\includegraphics[trim= 2cm 13cm 2cm 2cm, clip=true, width=0.32\textwidth]{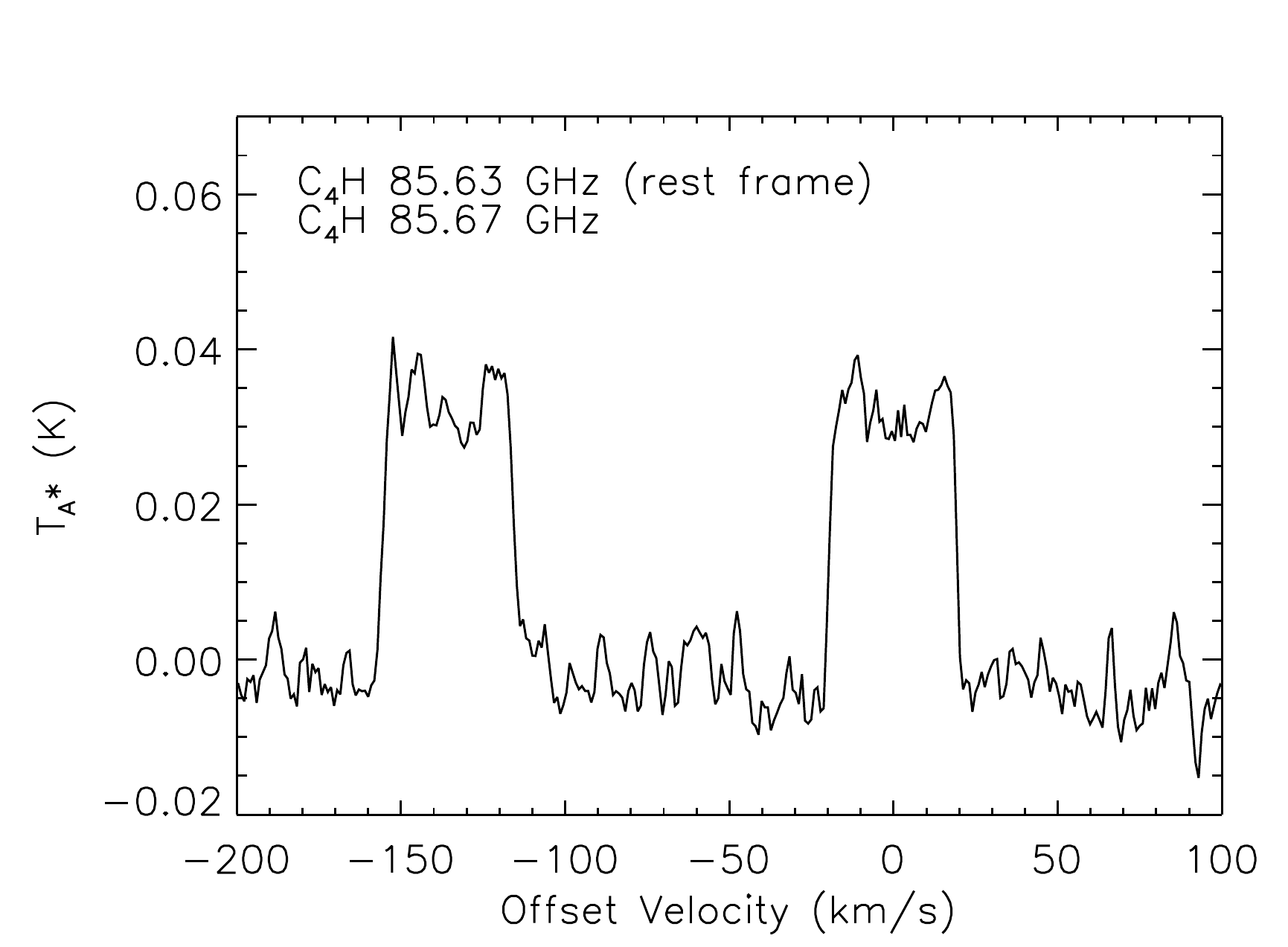}}

\subfigure{\includegraphics[trim= 2cm 13cm 2cm 2cm, clip=true, width=0.32\textwidth]{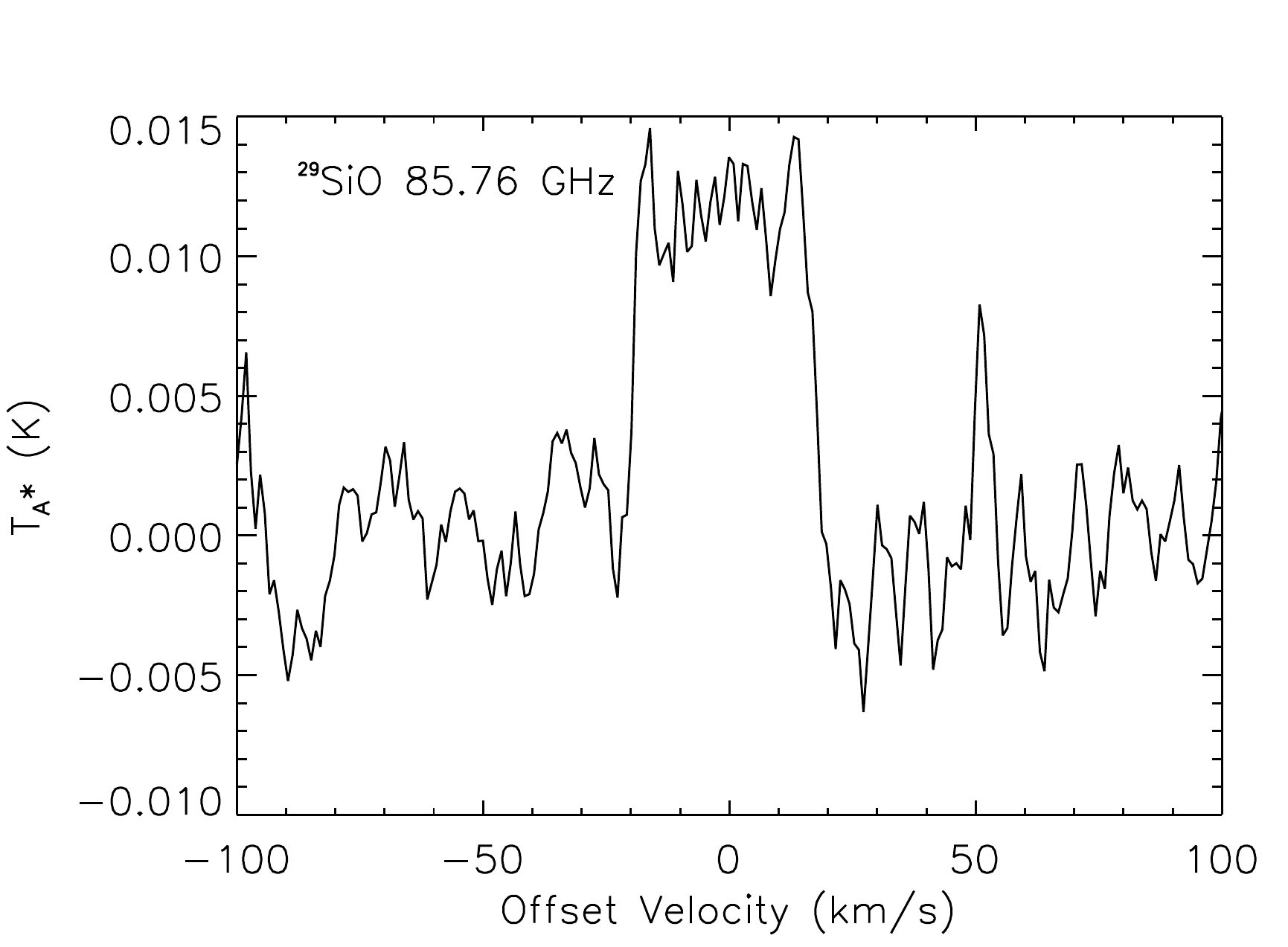}}
\subfigure{\includegraphics[trim= 2cm 13cm 2cm 2cm, clip=true, width=0.32\textwidth]{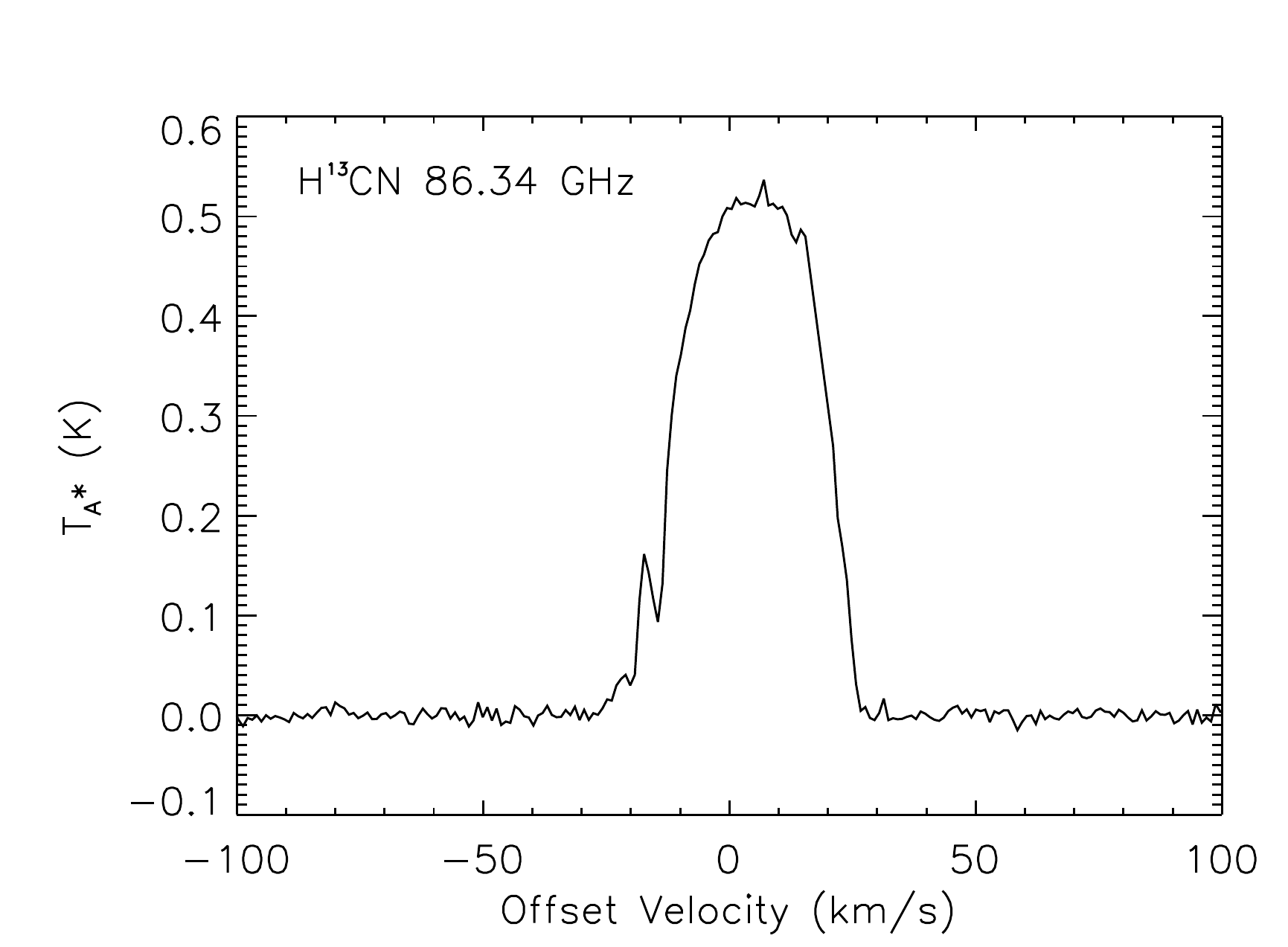}}
\subfigure{\includegraphics[trim= 2cm 13cm 2cm 2cm, clip=true, width=0.32\textwidth]{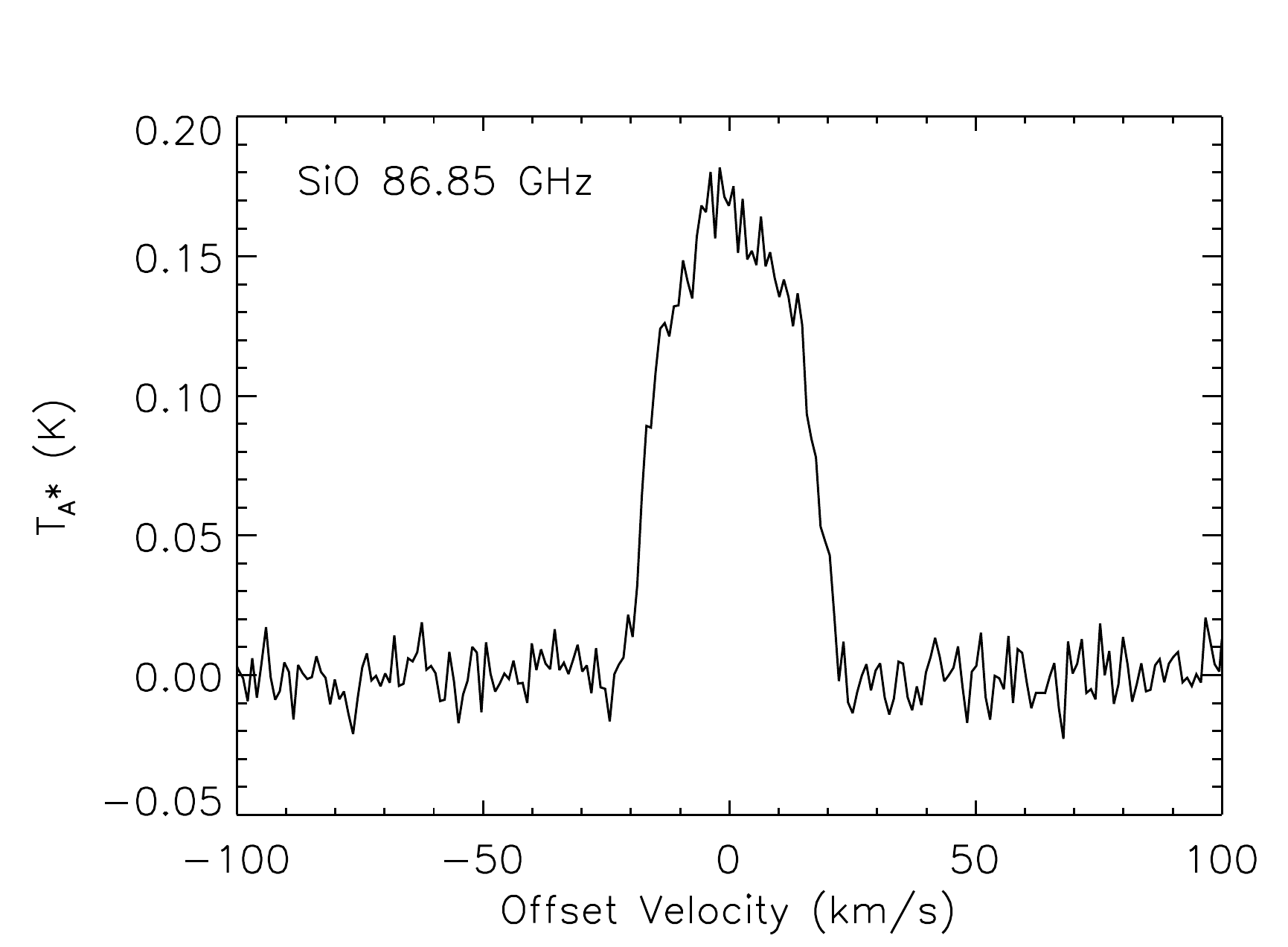}}

\subfigure{\includegraphics[trim= 2cm 13cm 2cm 2cm, clip=true, width=0.32\textwidth]{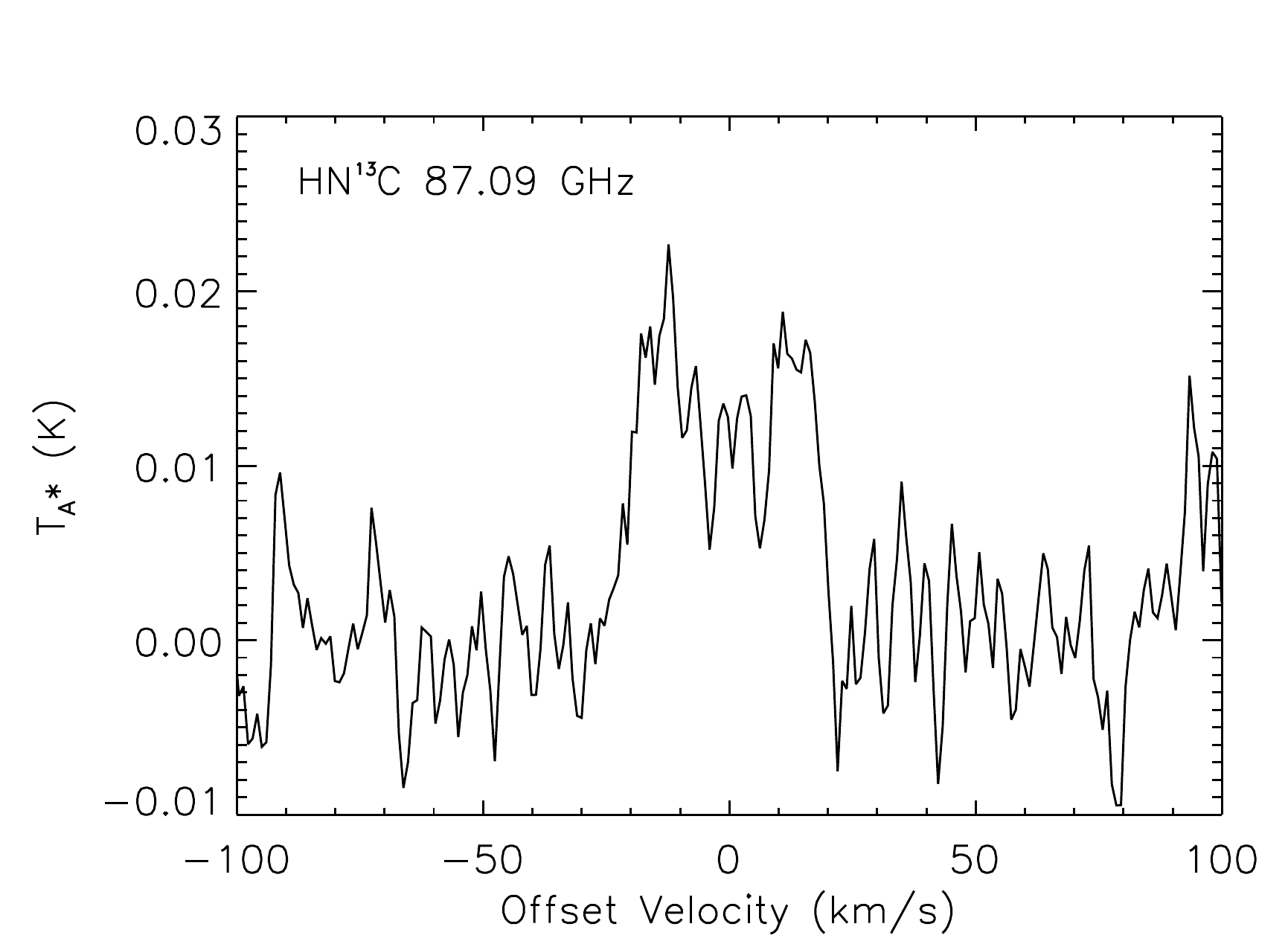}}
\subfigure{\includegraphics[trim= 2cm 13cm 2cm 2cm, clip=true, width=0.32\textwidth]{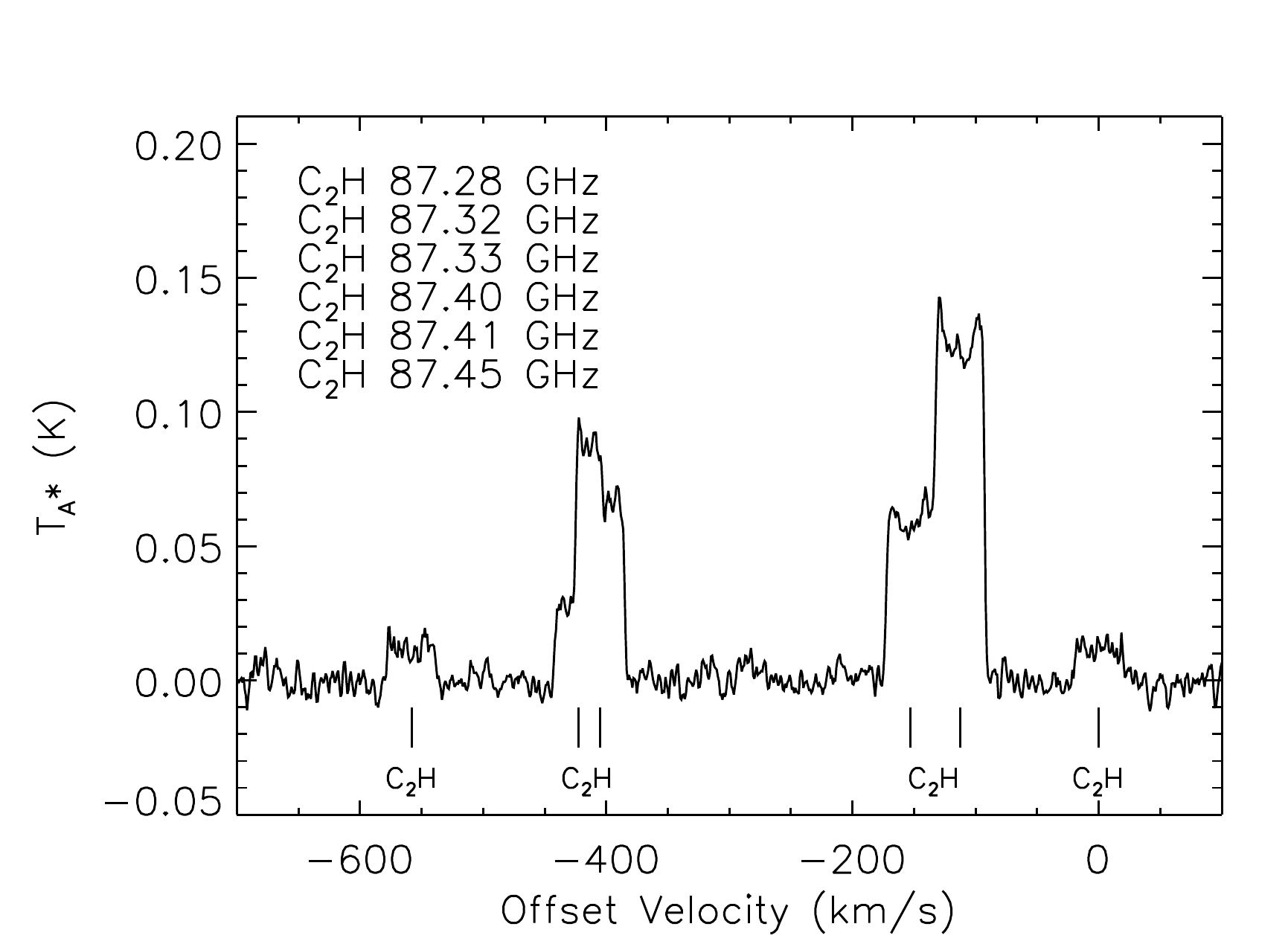}}
\subfigure{\includegraphics[trim= 2cm 13cm 2cm 2cm, clip=true, width=0.32\textwidth]{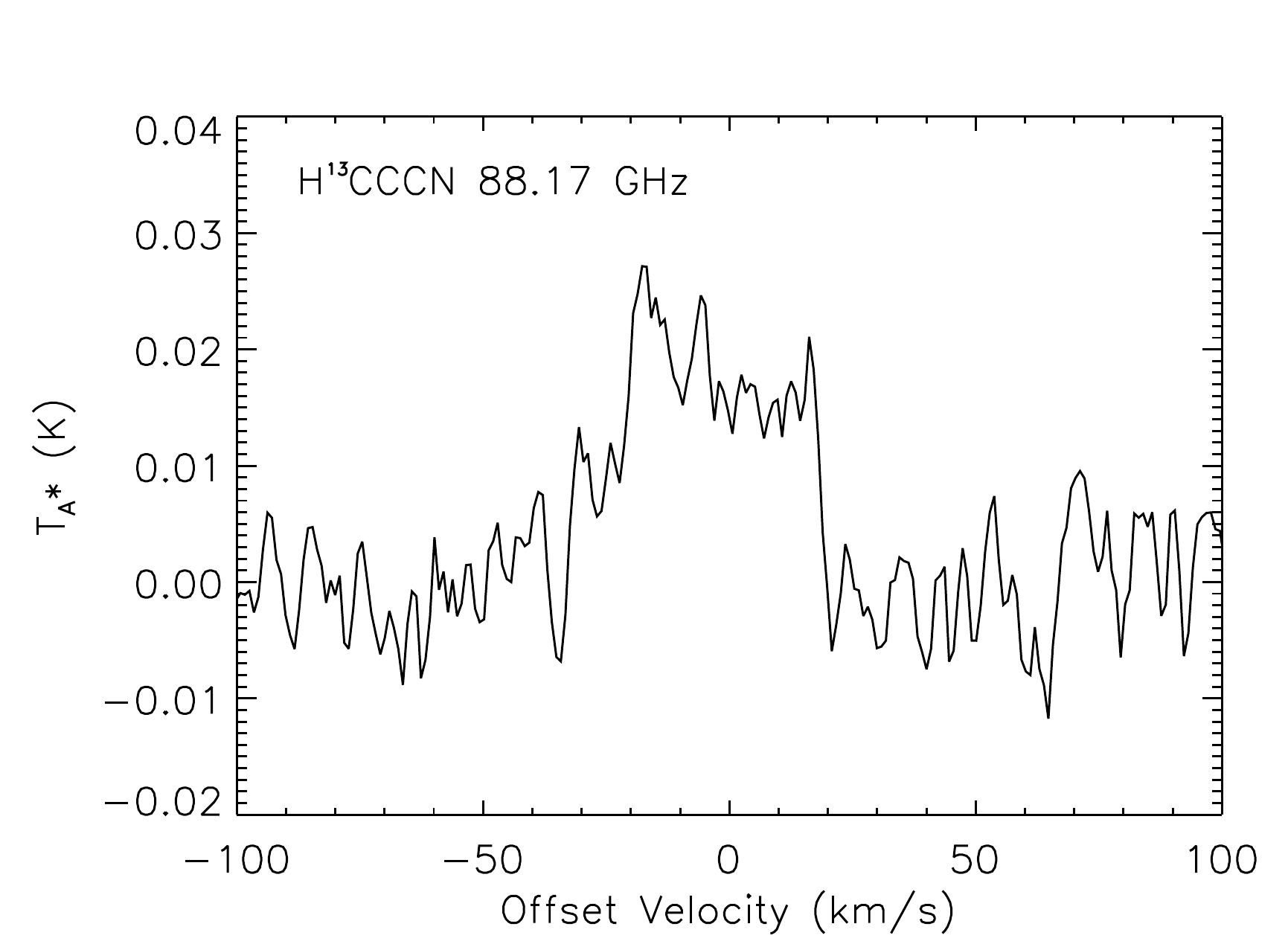}}

\subfigure{\includegraphics[trim= 2cm 13cm 2cm 2cm, clip=true, width=0.32\textwidth]{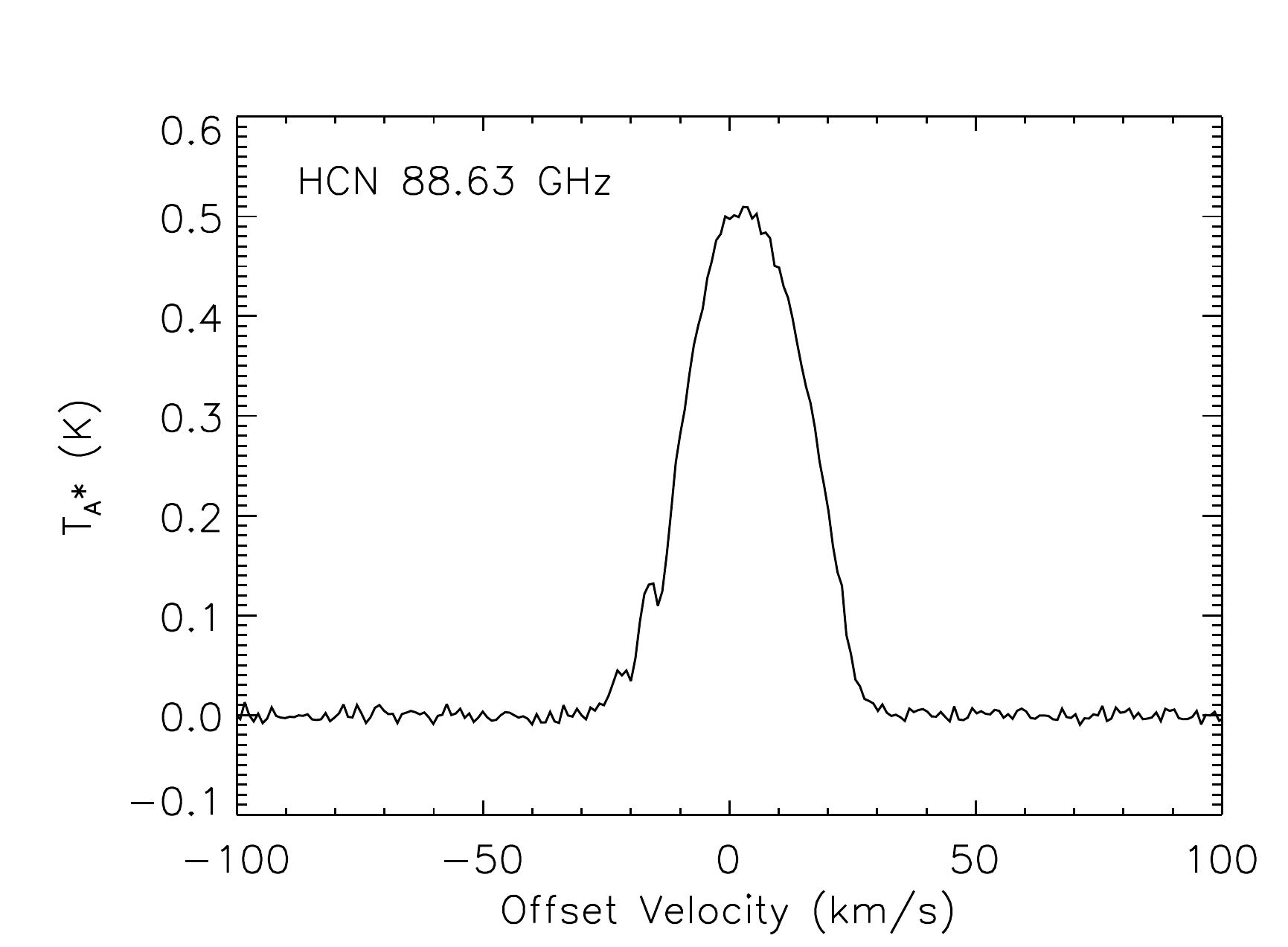}}
\subfigure{\includegraphics[trim= 2cm 13cm 2cm 2cm, clip=true, width=0.32\textwidth]{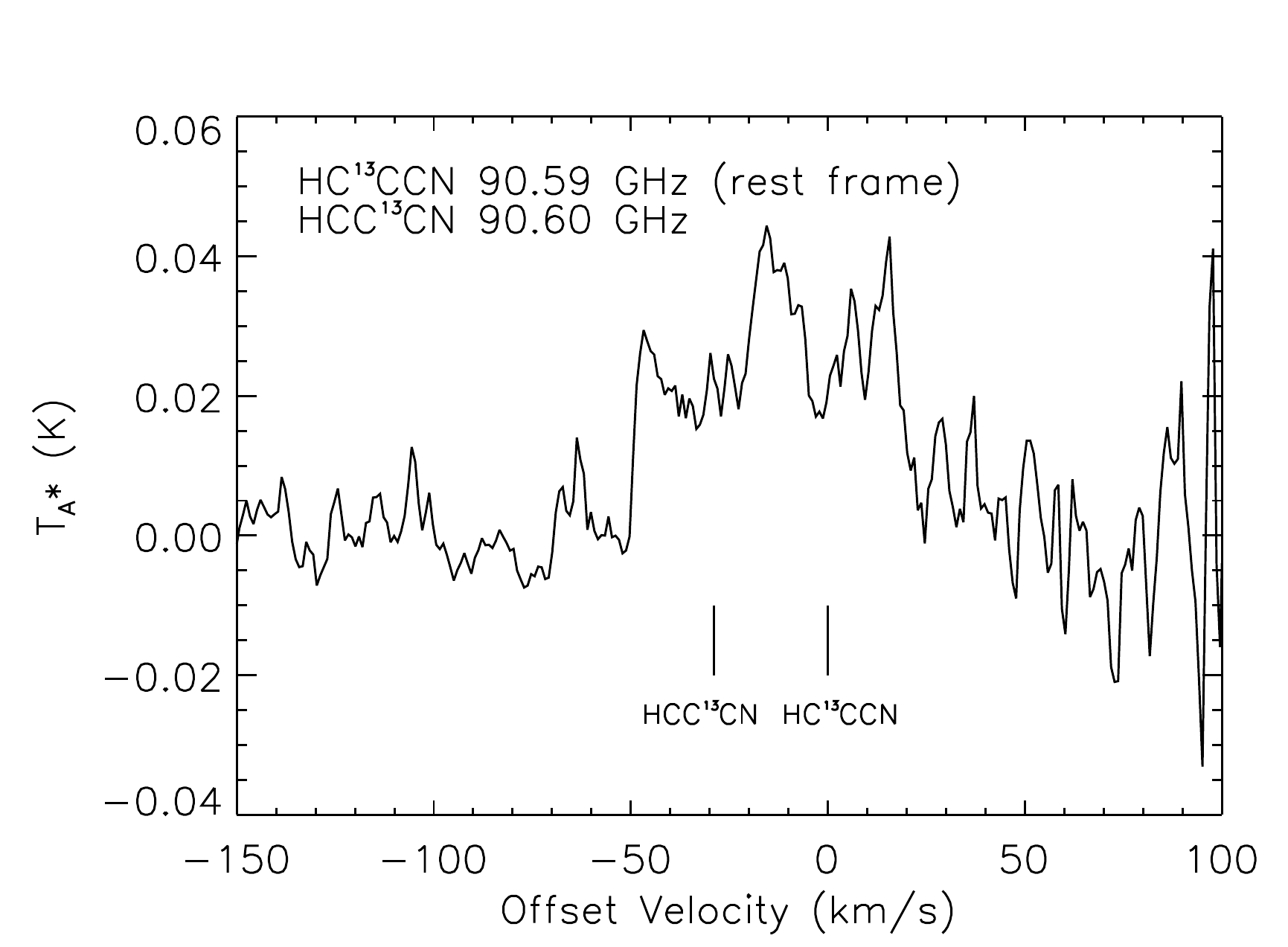}}
\subfigure{\includegraphics[trim= 2cm 13cm 2cm 2cm, clip=true, width=0.32\textwidth]{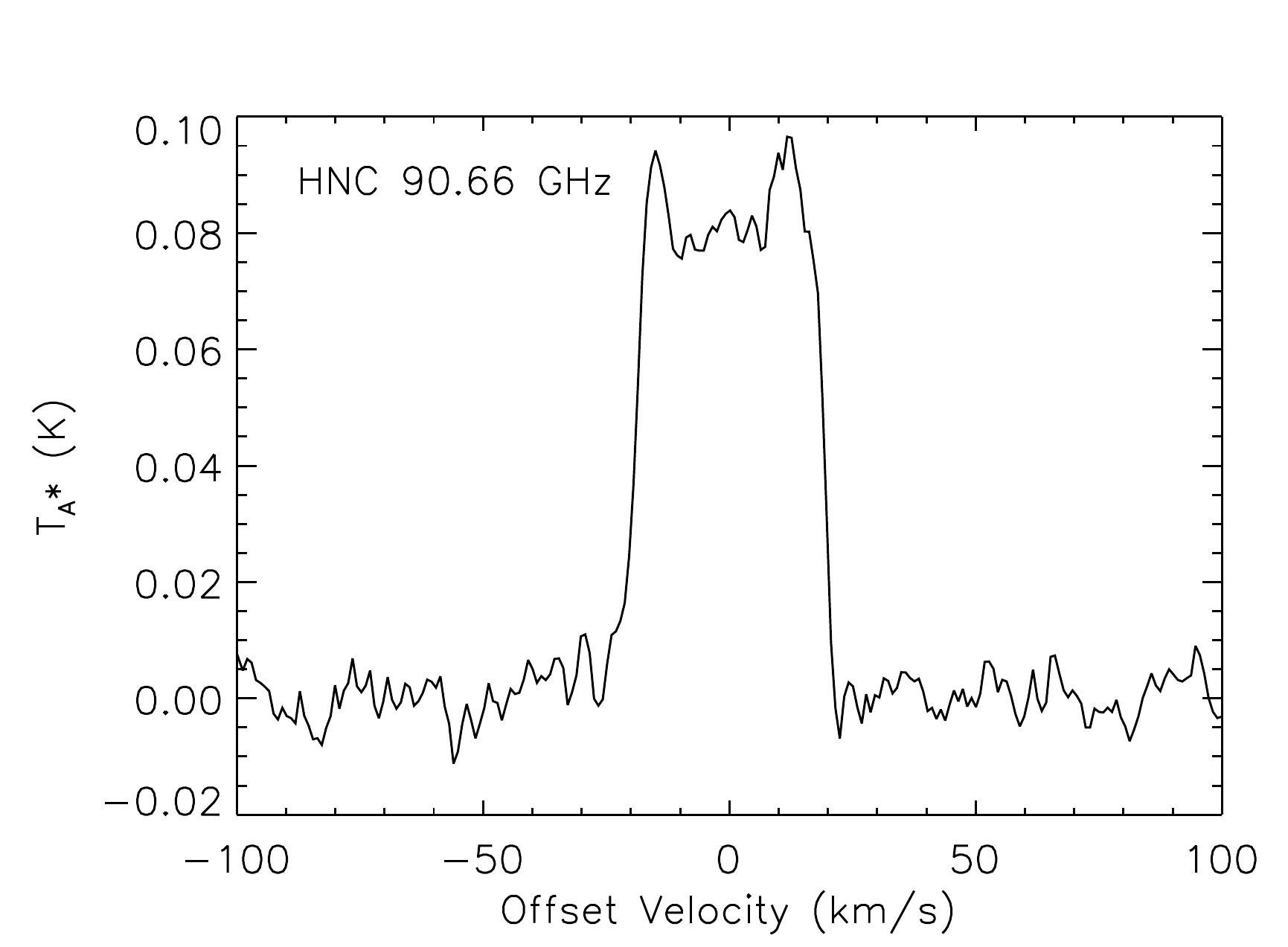}}

}

\caption{All transitions in IRAS 15194-5115. Ordinate axis intensities are in units of corrected antenna temperature, abscissa values are in units of km/s, corrected for LSR velocity of the source, taken as -15.2 km/s. CO and $^{13}$CO transitions show the stellar emission (solid line) and the interstellar contamination (dotted line).}
\end{figure*}

\begin{figure*}
{\centering
\subfigure{\includegraphics[trim= 2cm 13cm 2cm 2cm, clip=true, width=0.32\textwidth]{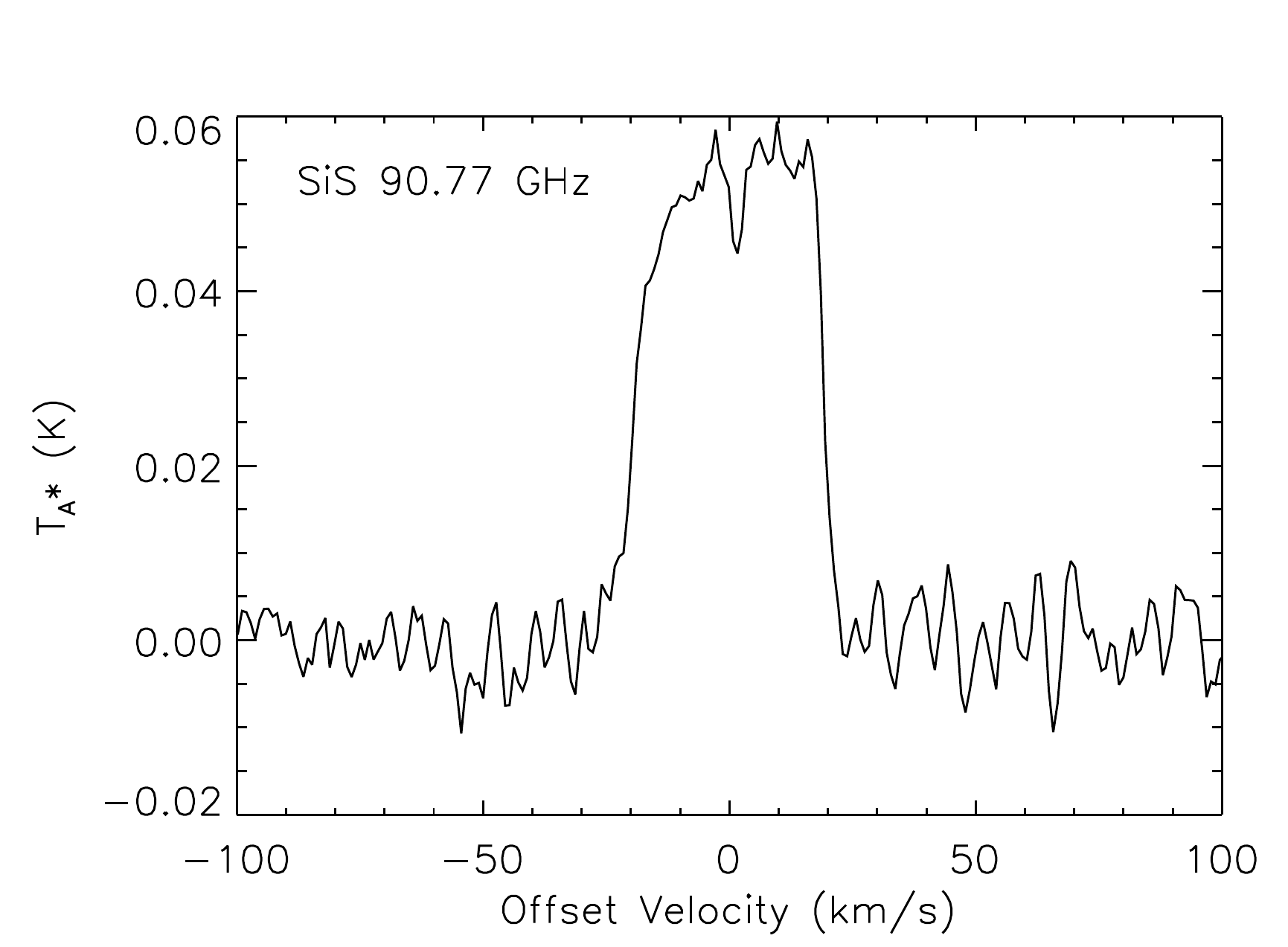}}
\subfigure{\includegraphics[trim= 2cm 13cm 2cm 2cm, clip=true, width=0.32\textwidth]{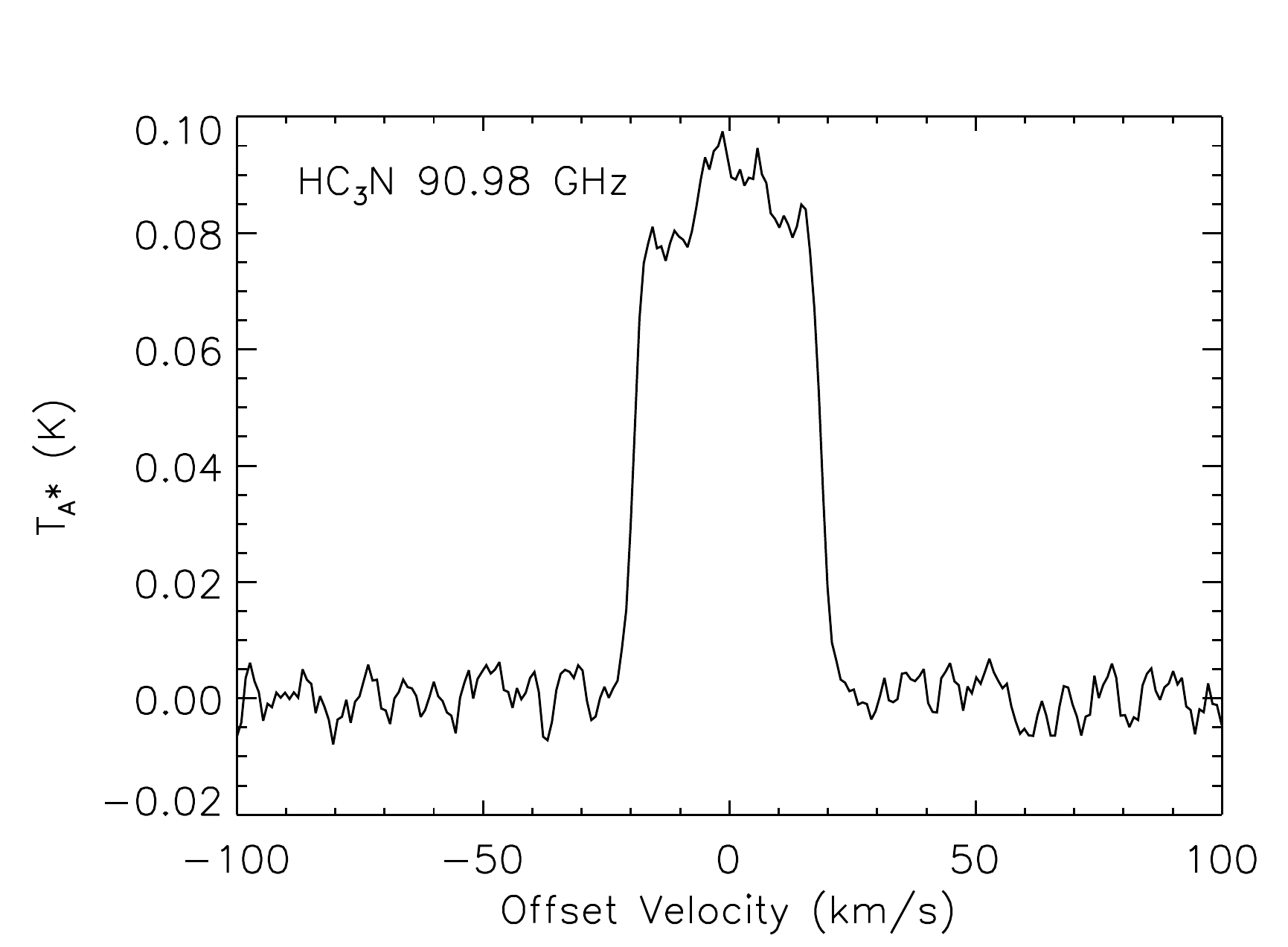}}
\subfigure{\includegraphics[trim= 2cm 13cm 2cm 2cm, clip=true, width=0.32\textwidth]{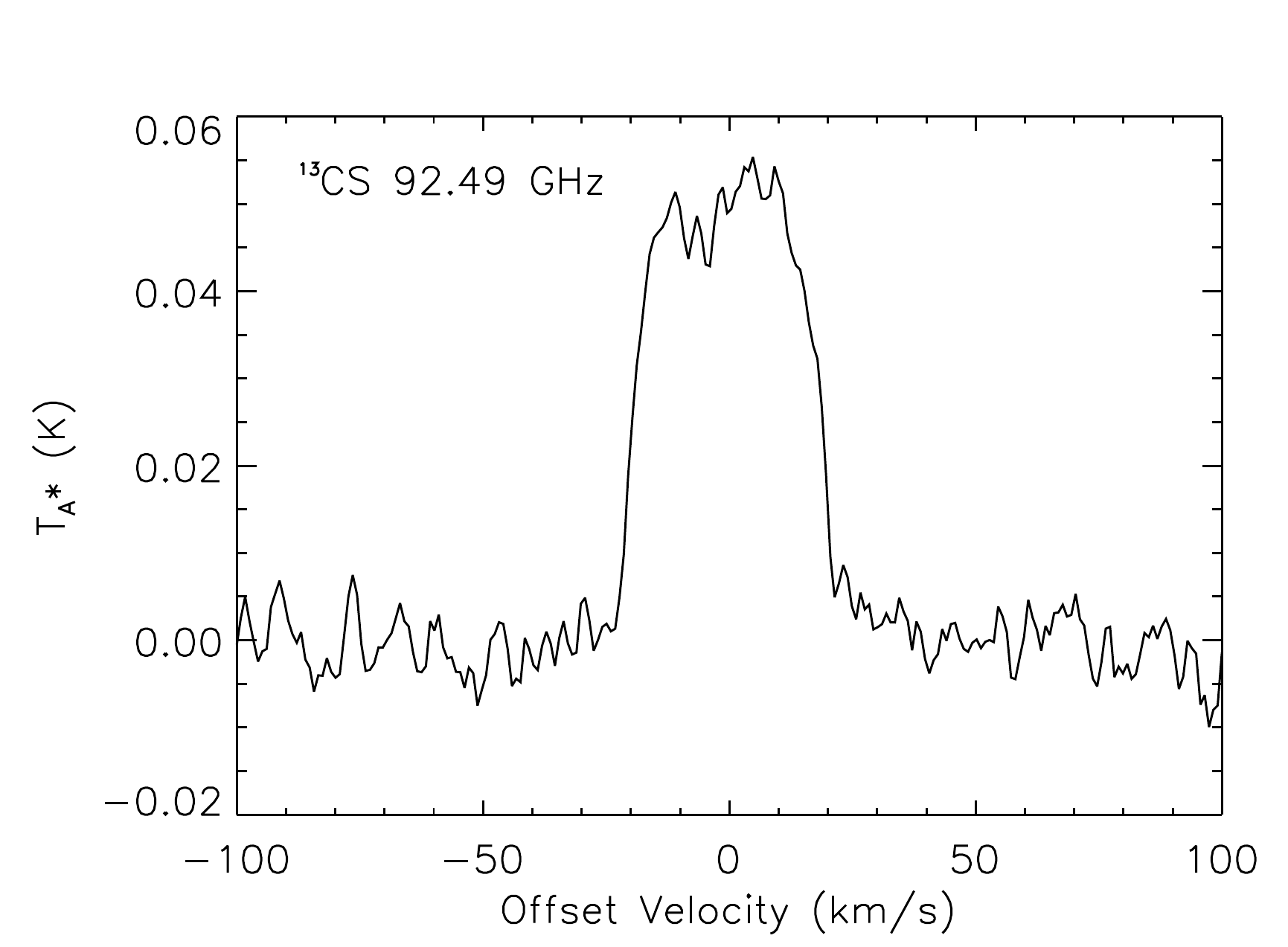}}

\subfigure{\includegraphics[trim= 2cm 13cm 2cm 0cm, clip=true, width=0.32\textwidth]{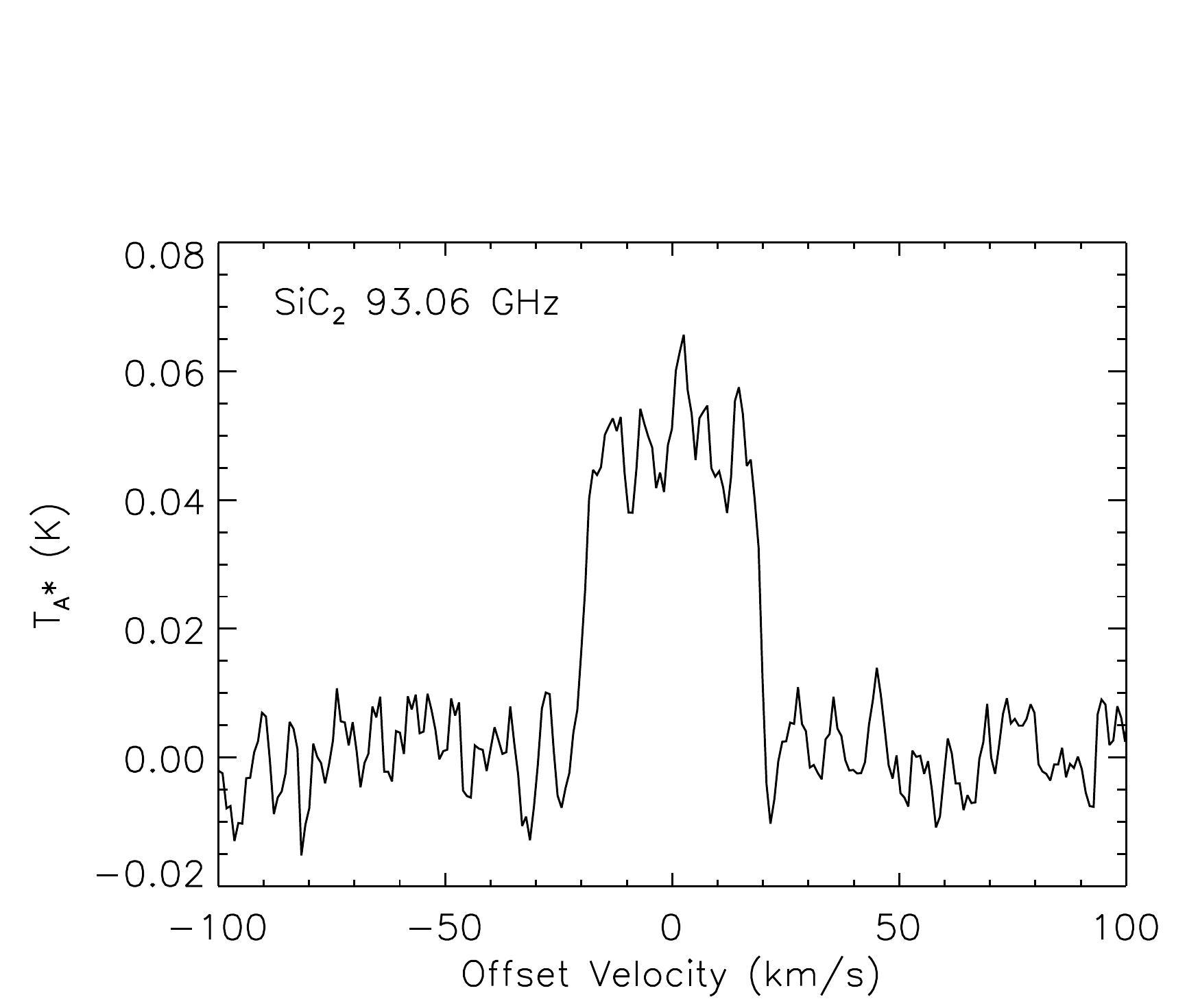}}
\subfigure{\includegraphics[trim= 2cm 13cm 2cm 0cm, clip=true, width=0.32\textwidth]{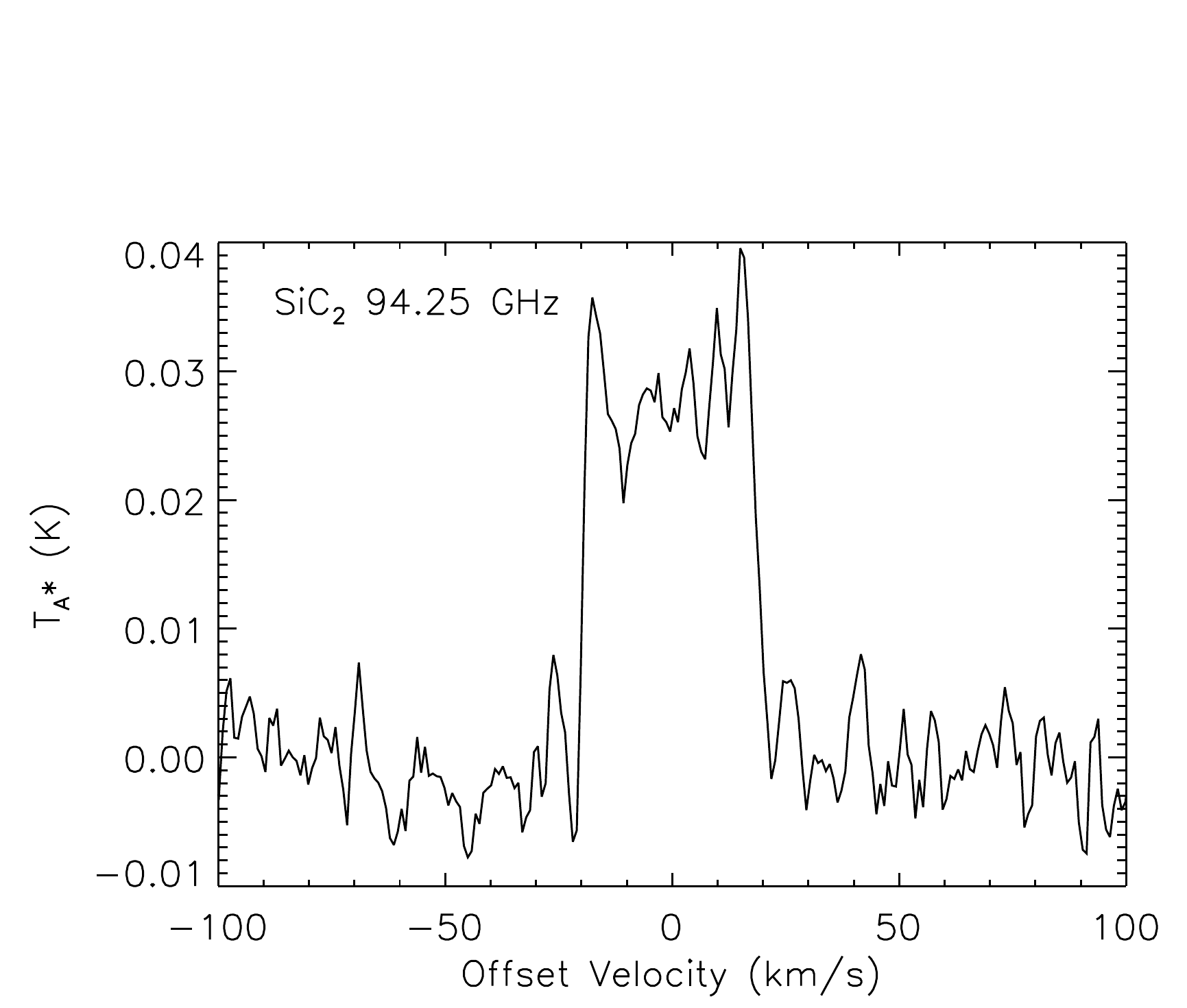}}
\subfigure{\includegraphics[trim= 2cm 13cm 2cm 0cm, clip=true, width=0.32\textwidth]{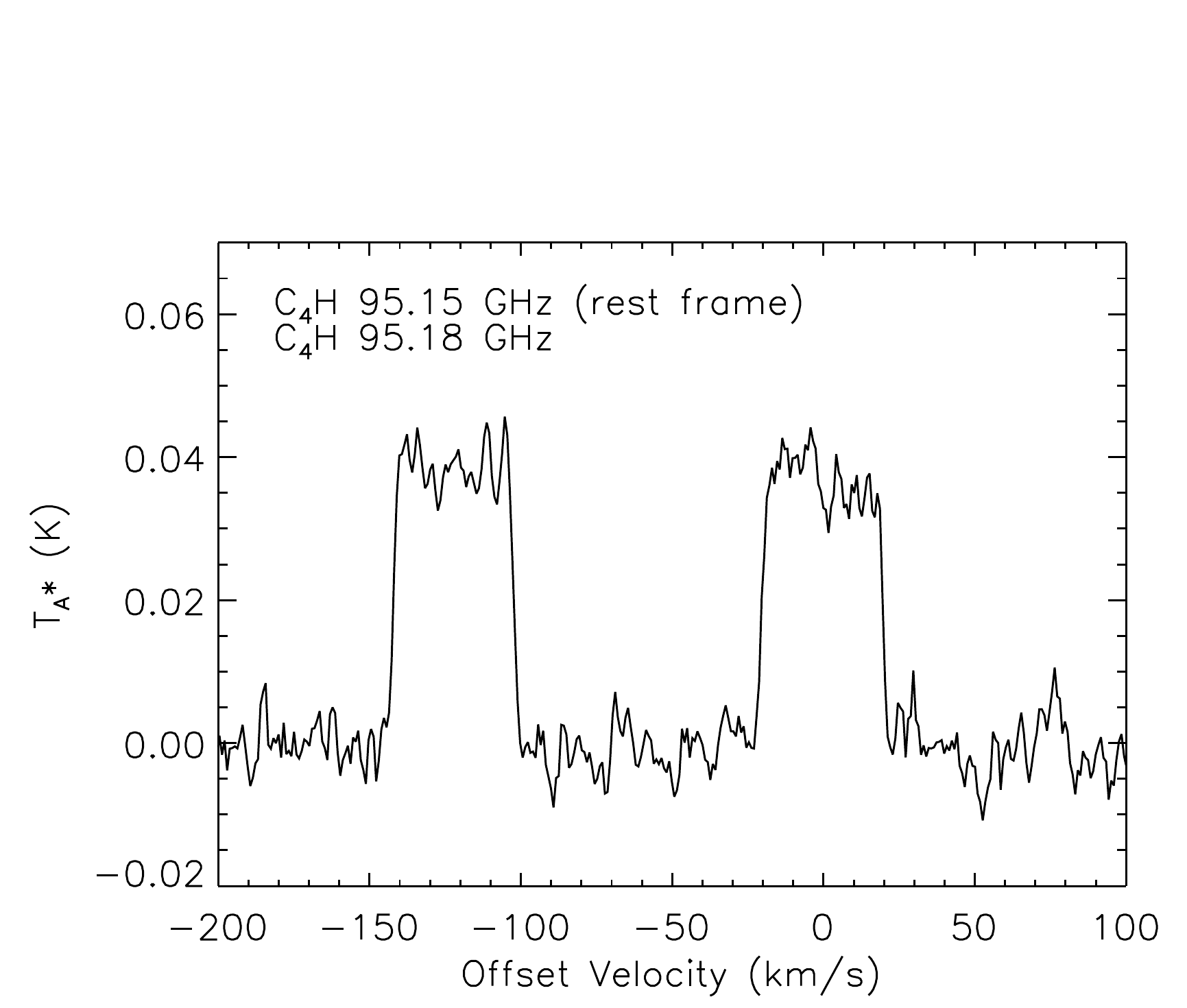}}

\subfigure{\includegraphics[trim= 2cm 13cm 2cm 2cm, clip=true, width=0.32\textwidth]{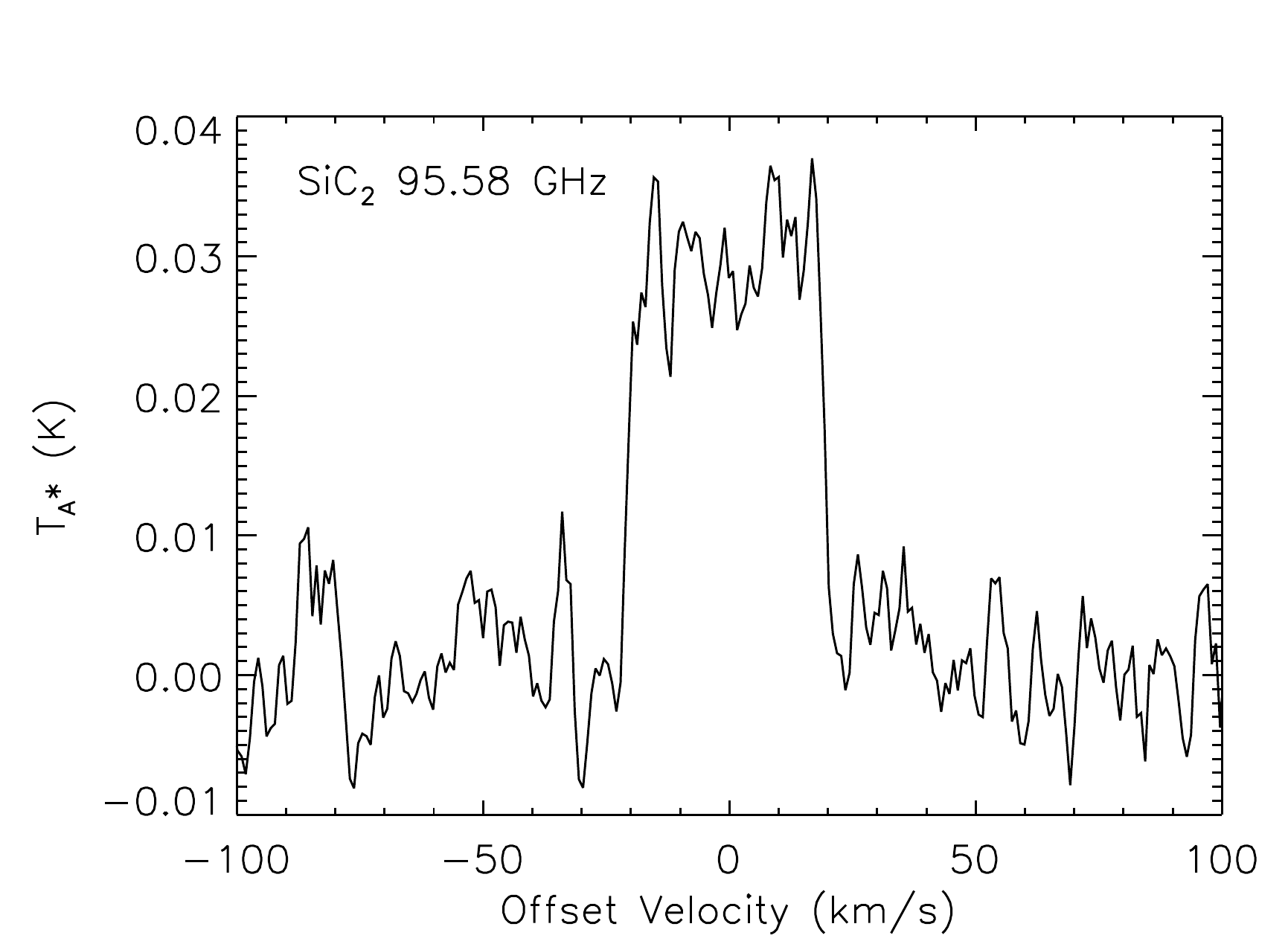}}
\subfigure{\includegraphics[trim= 2cm 13cm 2cm 2cm, clip=true, width=0.32\textwidth]{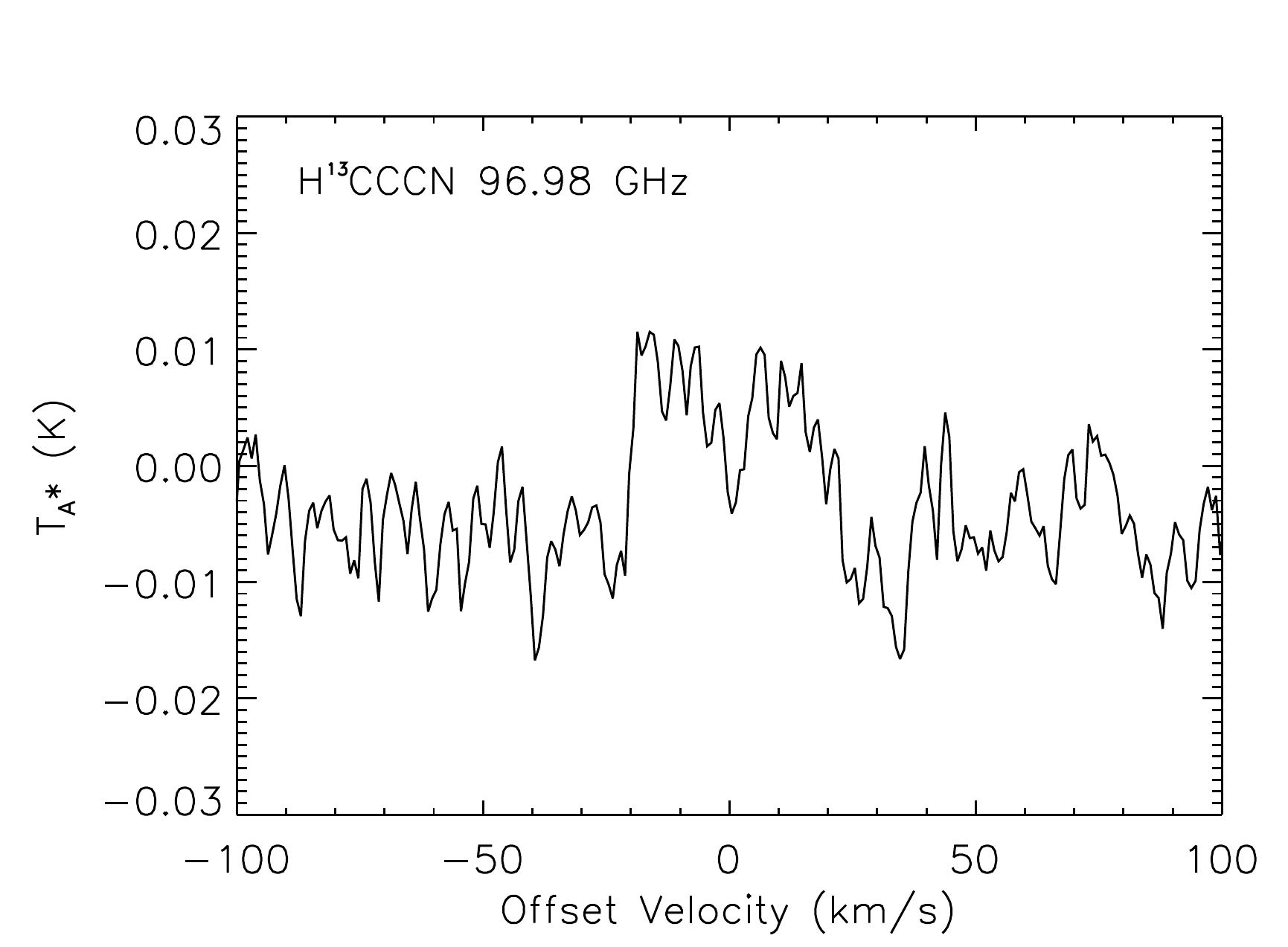}}
\subfigure{\includegraphics[trim= 2cm 13cm 2cm 2cm, clip=true, width=0.32\textwidth]{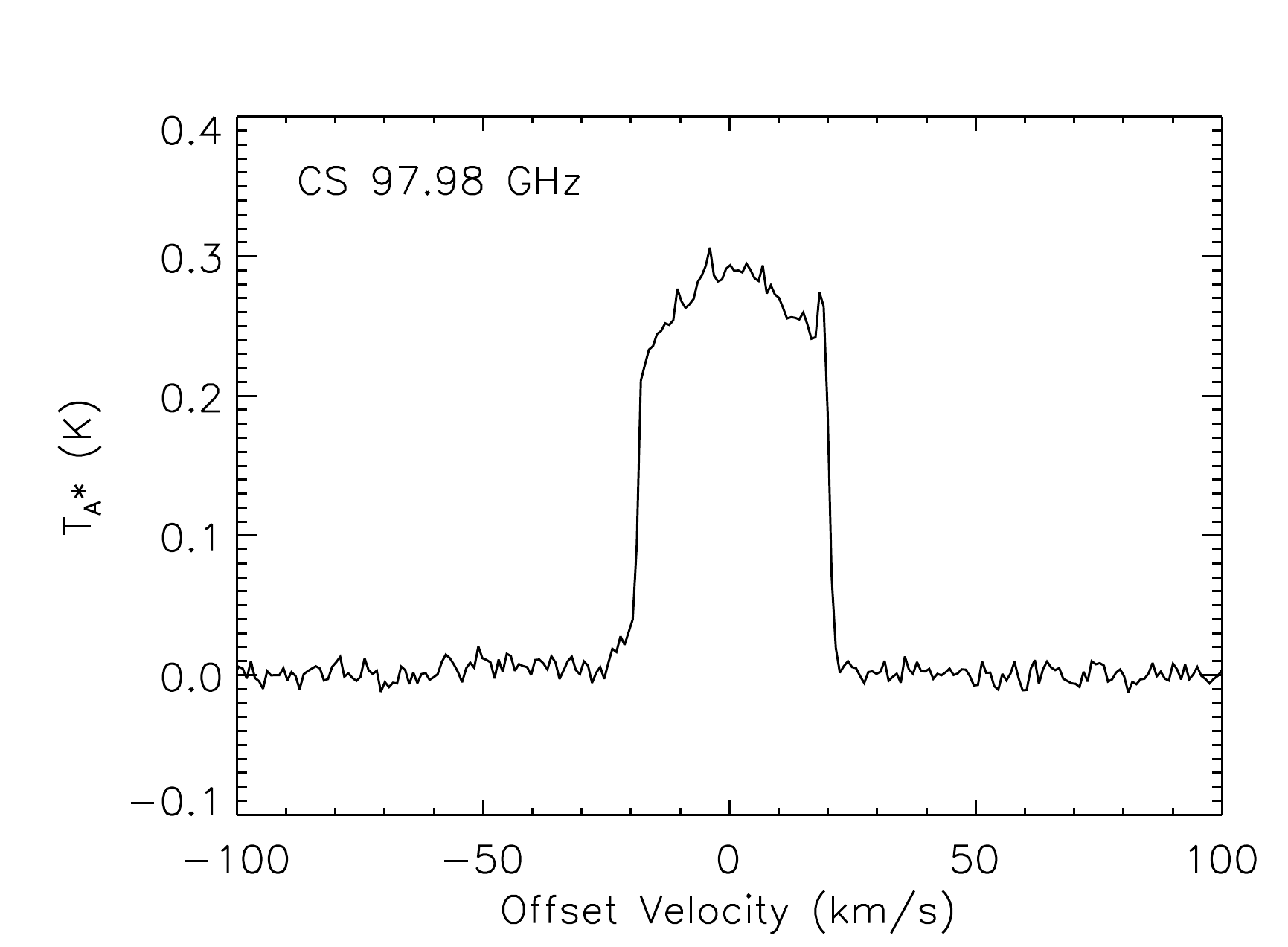}}

\subfigure{\includegraphics[trim= 2cm 13cm 2cm 2cm, clip=true, width=0.32\textwidth]{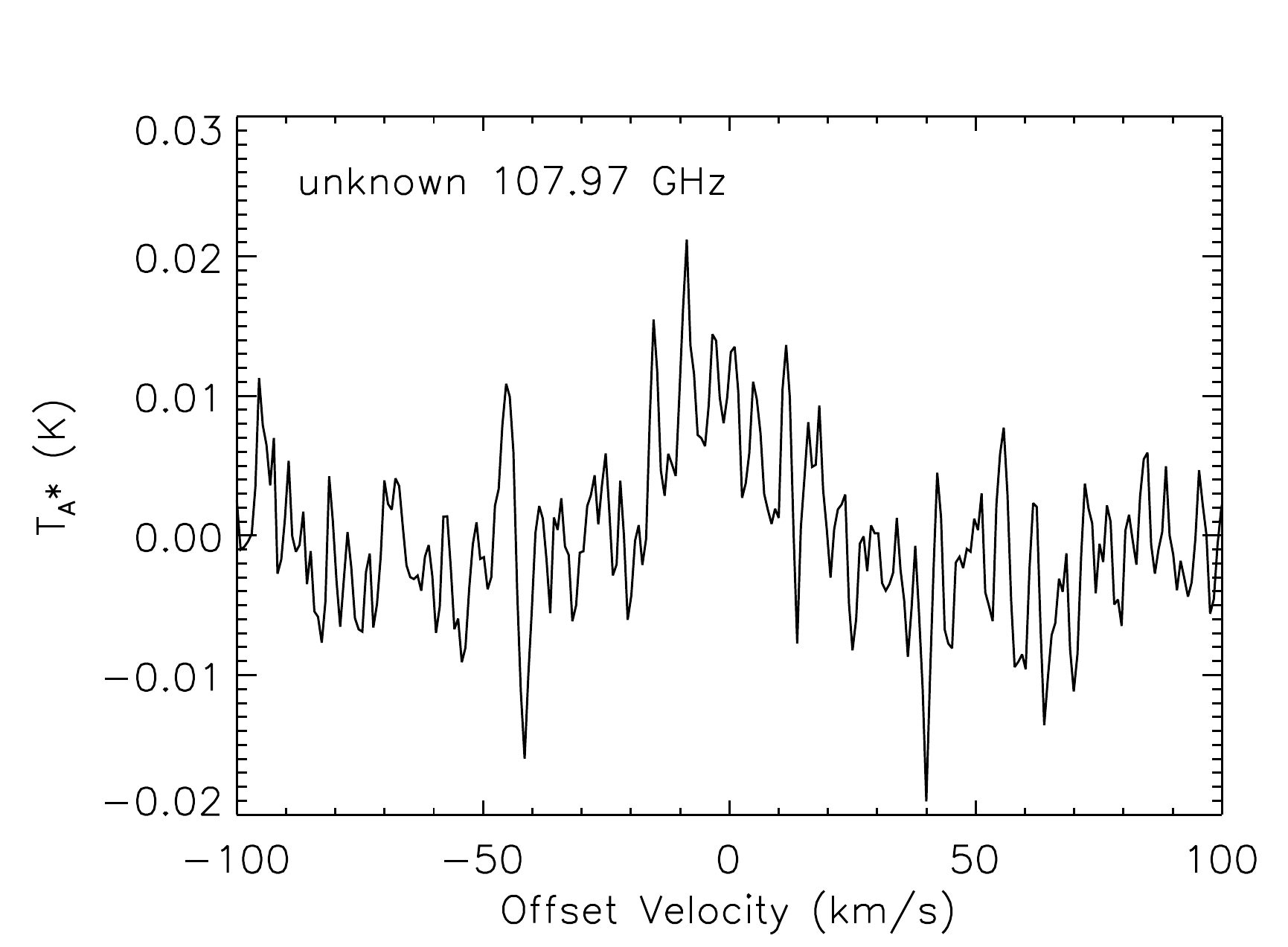}}
\subfigure{\includegraphics[trim= 2cm 13cm 2cm 2cm, clip=true, width=0.32\textwidth]{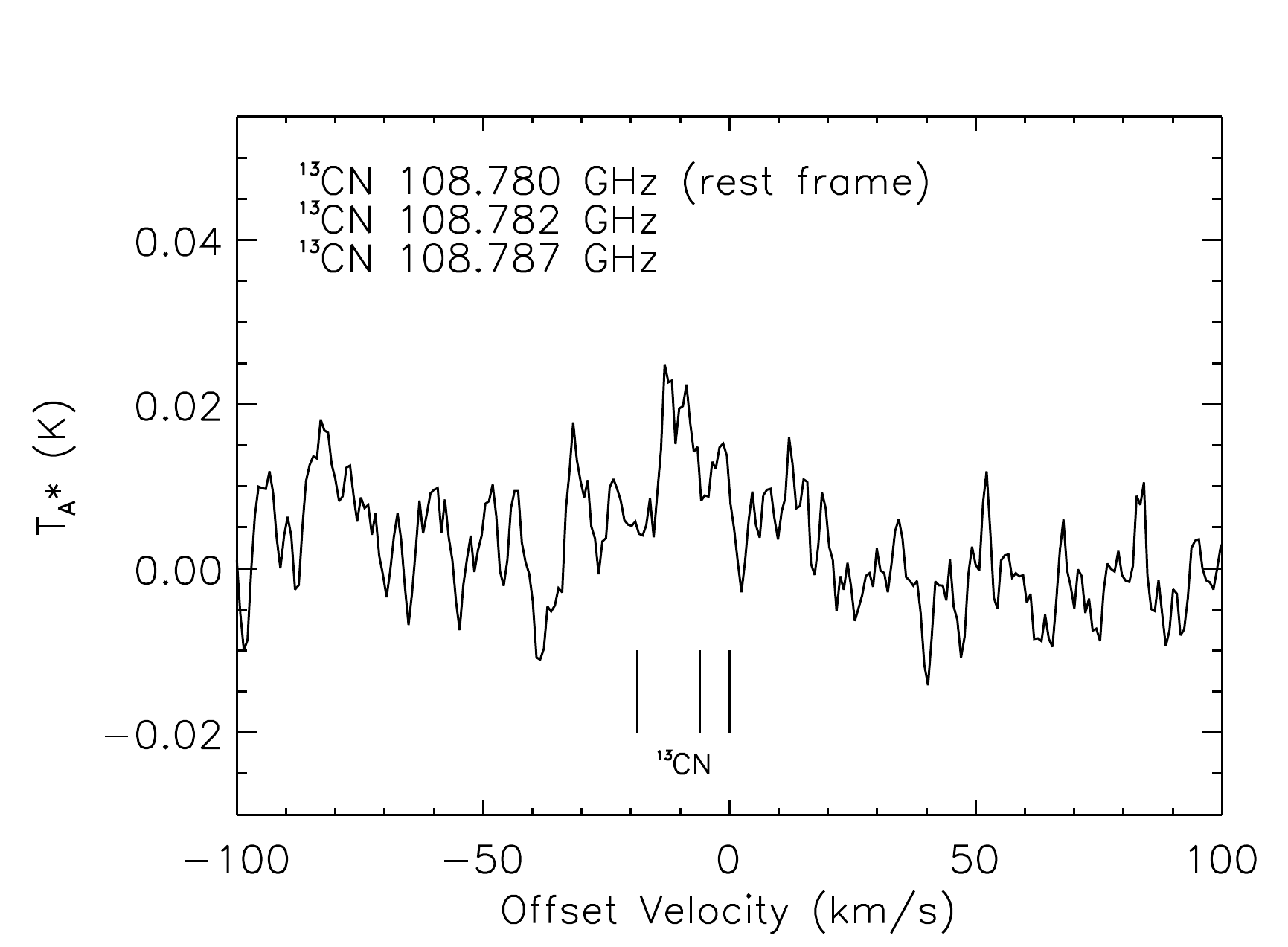}}
\subfigure{\includegraphics[trim= 2cm 13cm 2cm 2cm, clip=true, width=0.32\textwidth]{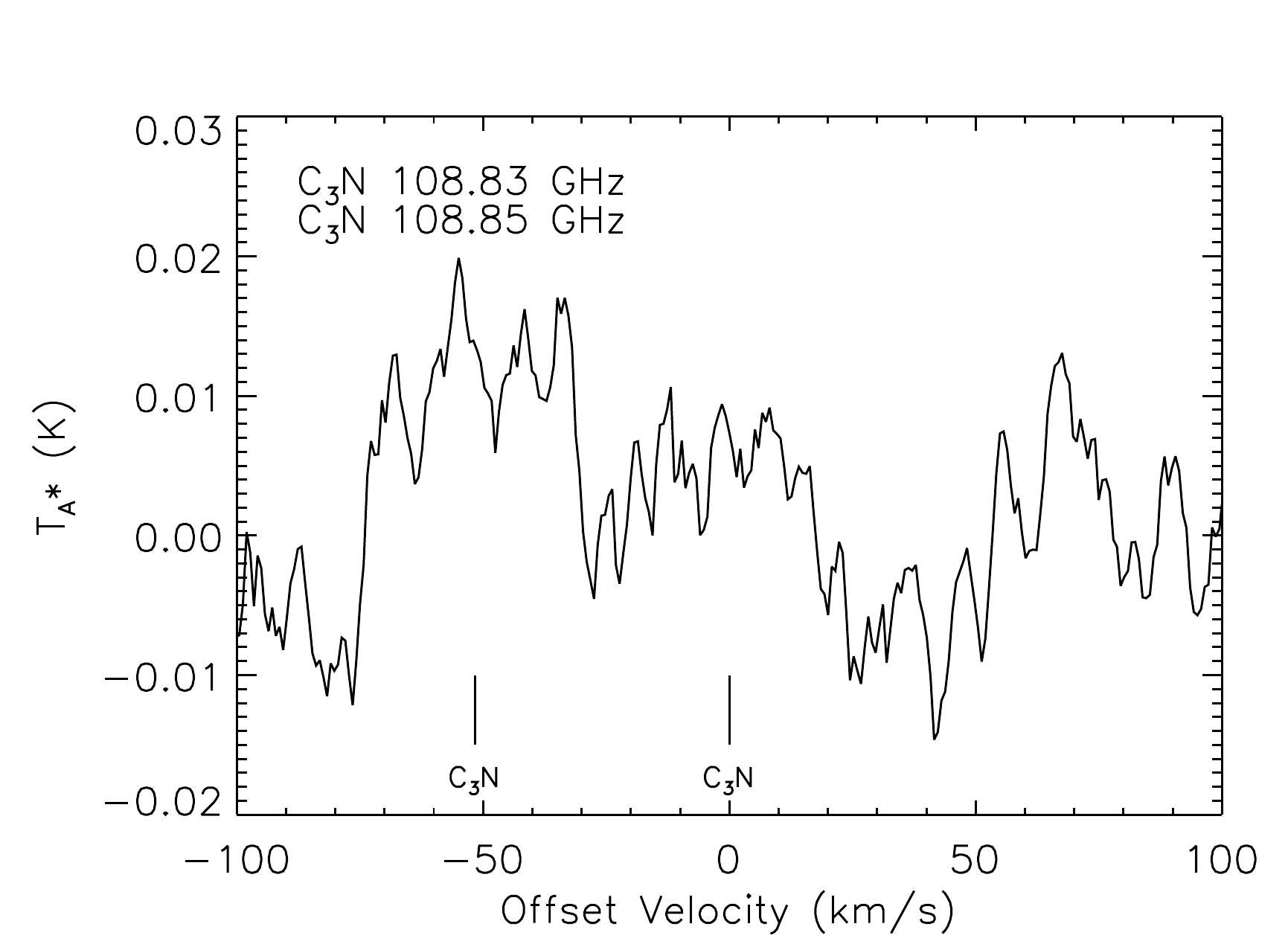}}

}
\contcaption{}
\end{figure*}

\begin{figure*}
\centering

\subfigure{\includegraphics[trim= 2cm 13cm 2cm 2cm, clip=true, width=0.32\textwidth]{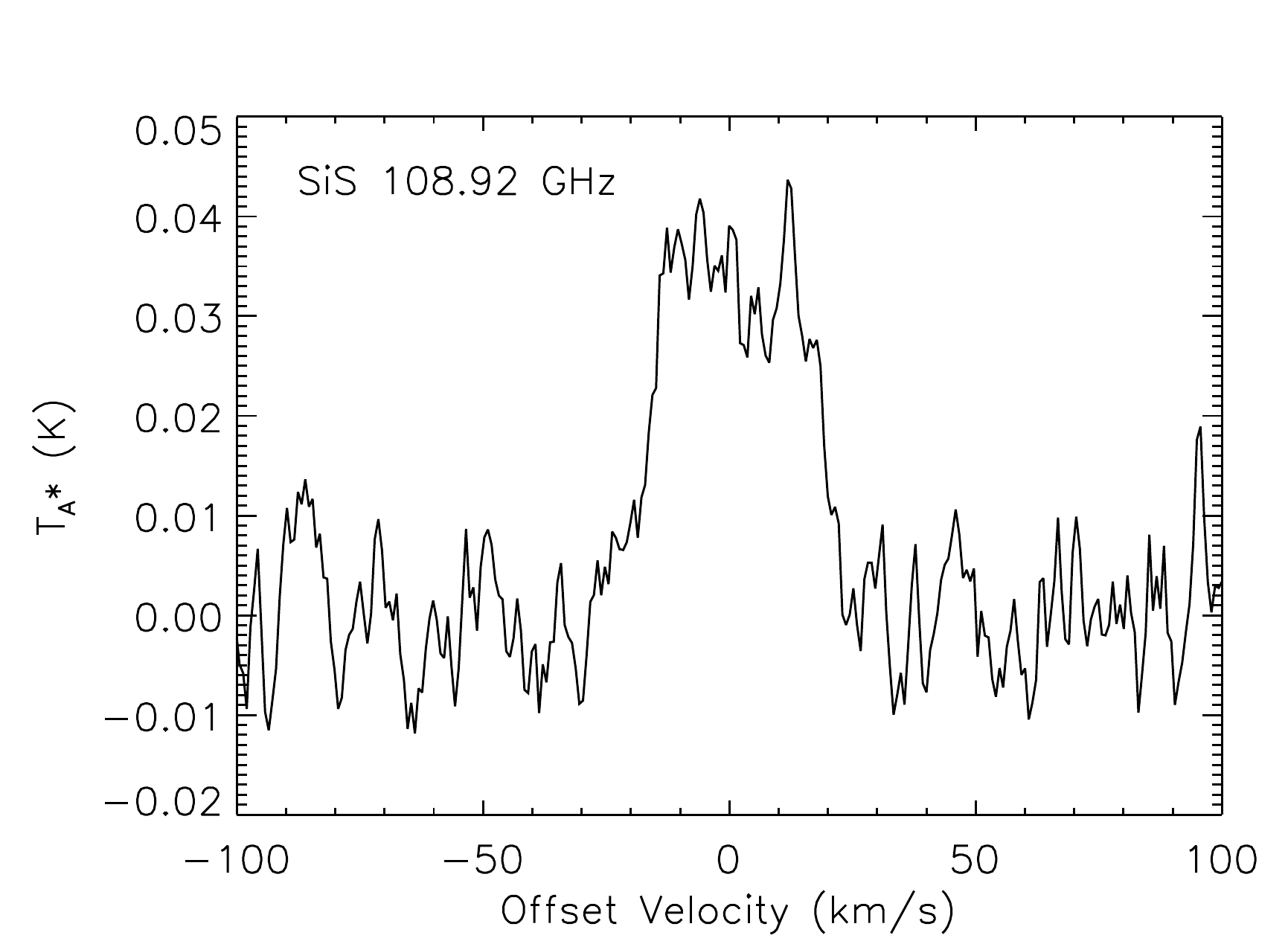}}
\subfigure{\includegraphics[trim= 2cm 13cm 2cm 2cm, clip=true, width=0.32\textwidth]{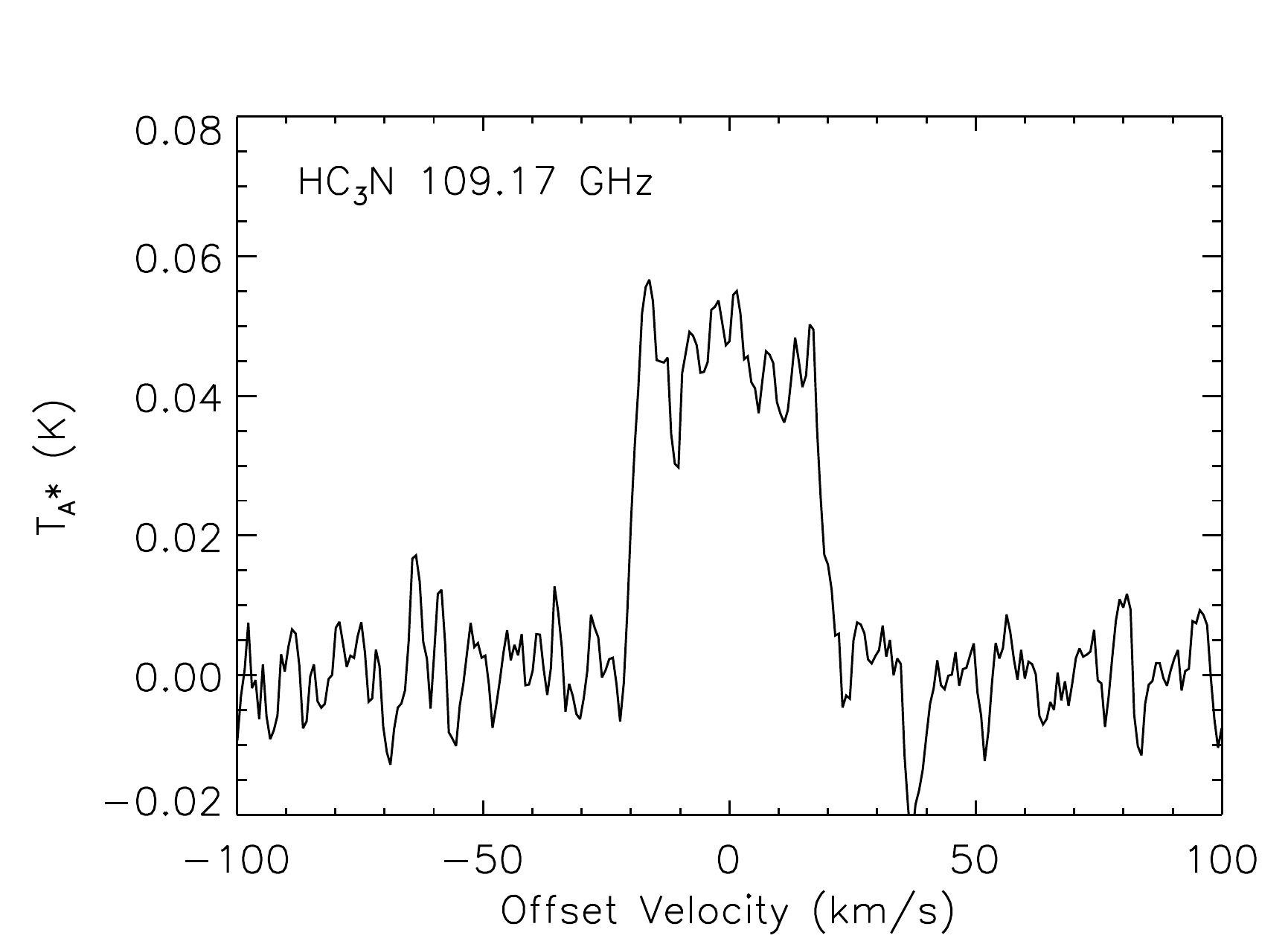}}
\subfigure{\includegraphics[trim= 2cm 13cm 2cm 2cm, clip=true, width=0.32\textwidth]{appendix_plots/15194/15194_13co.pdf}}

\subfigure{\includegraphics[trim= 2cm 13cm 2cm 2cm, clip=true, width=0.32\textwidth]{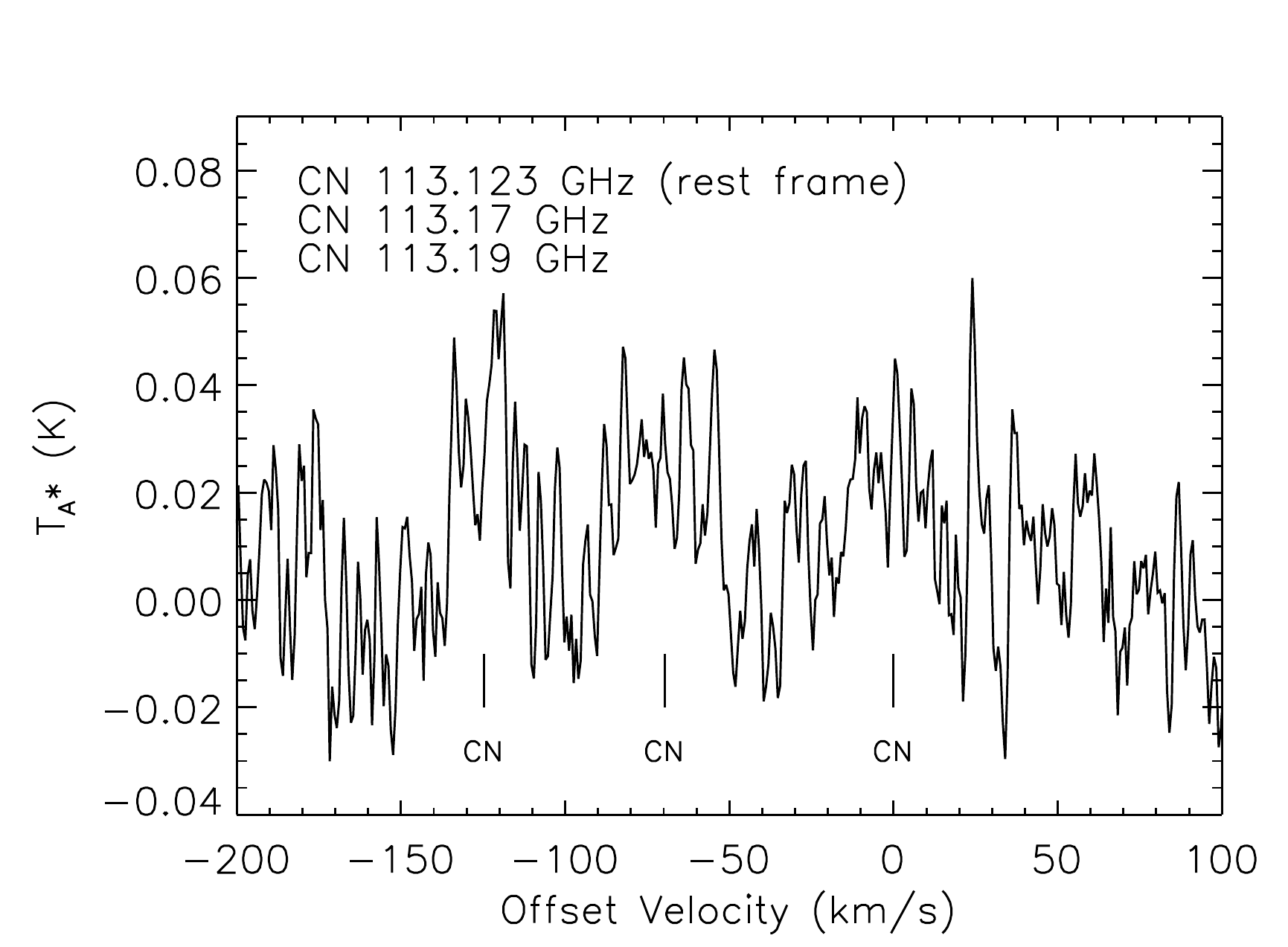}}
\subfigure{\includegraphics[trim= 2cm 13cm 2cm 2cm, clip=true, width=0.32\textwidth]{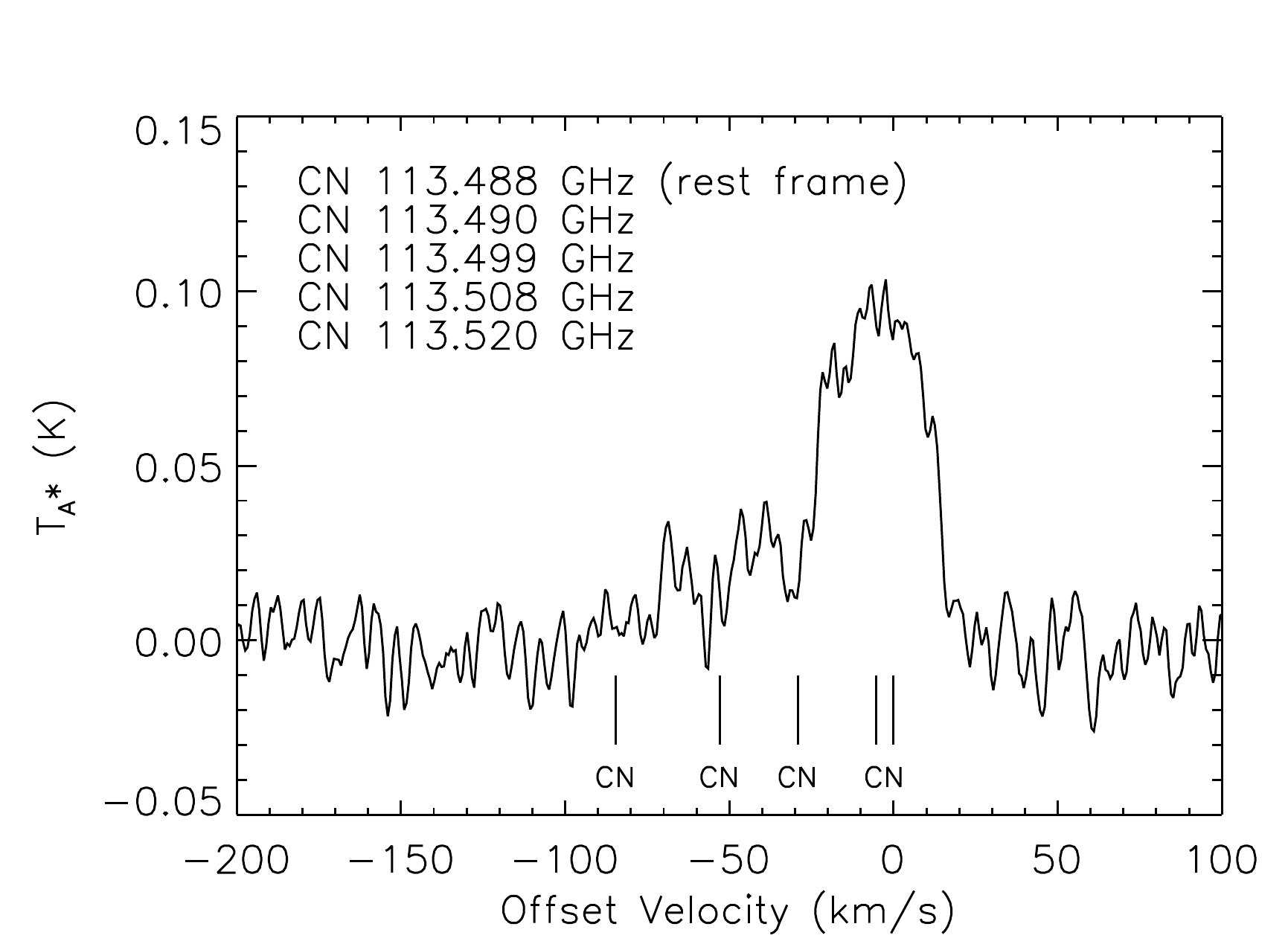}}
\subfigure{\includegraphics[trim= 2cm 13cm 2cm 2cm, clip=true, width=0.32\textwidth]{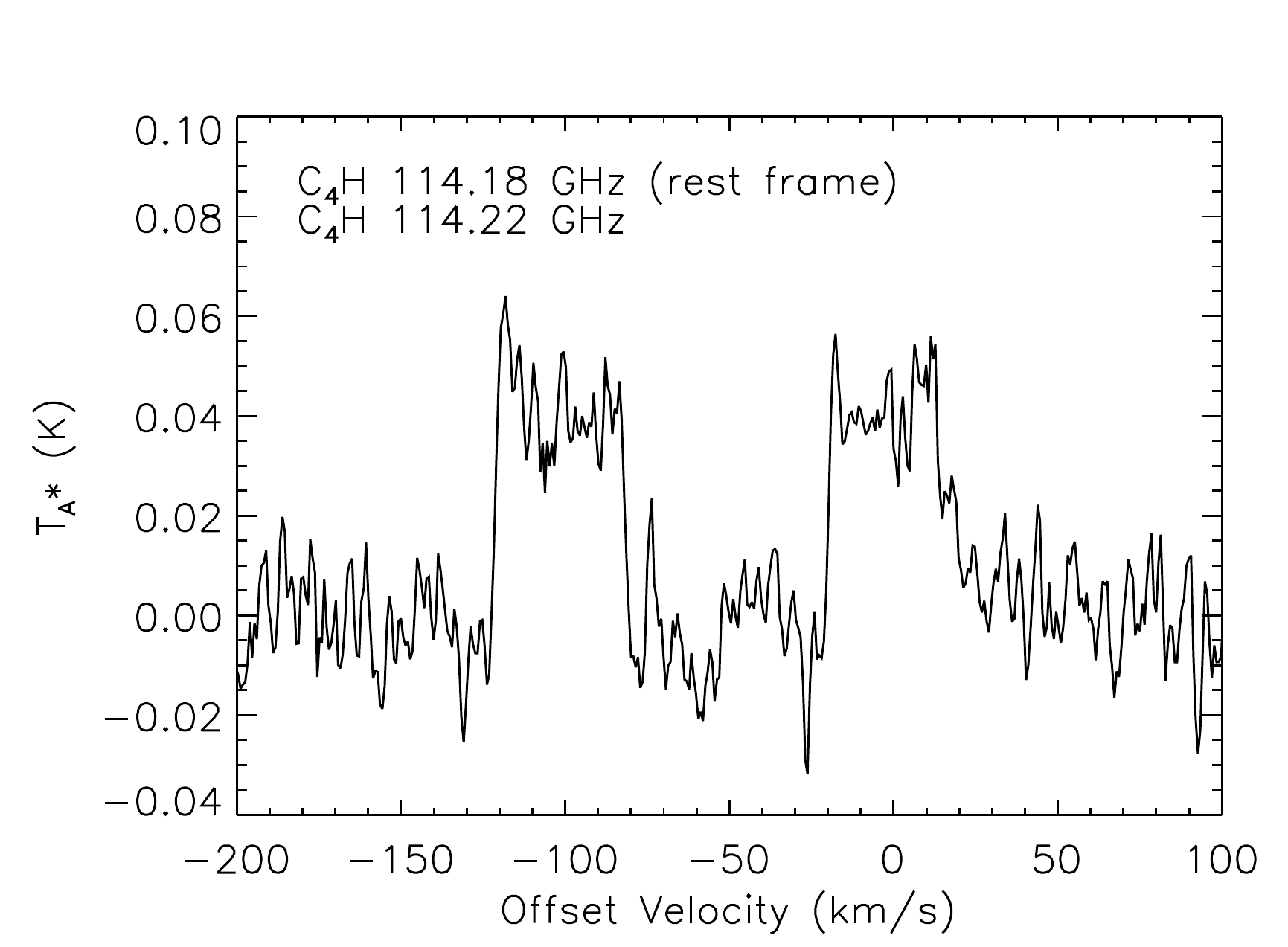}}

\subfigure{\includegraphics[trim= 2cm 13cm 2cm 3cm, clip=true, width=0.32\textwidth]{appendix_plots/15194/15194_12co.pdf}}
\subfigure{\includegraphics[trim= 2cm 13cm 2cm 3cm, clip=true, width=0.32\textwidth]{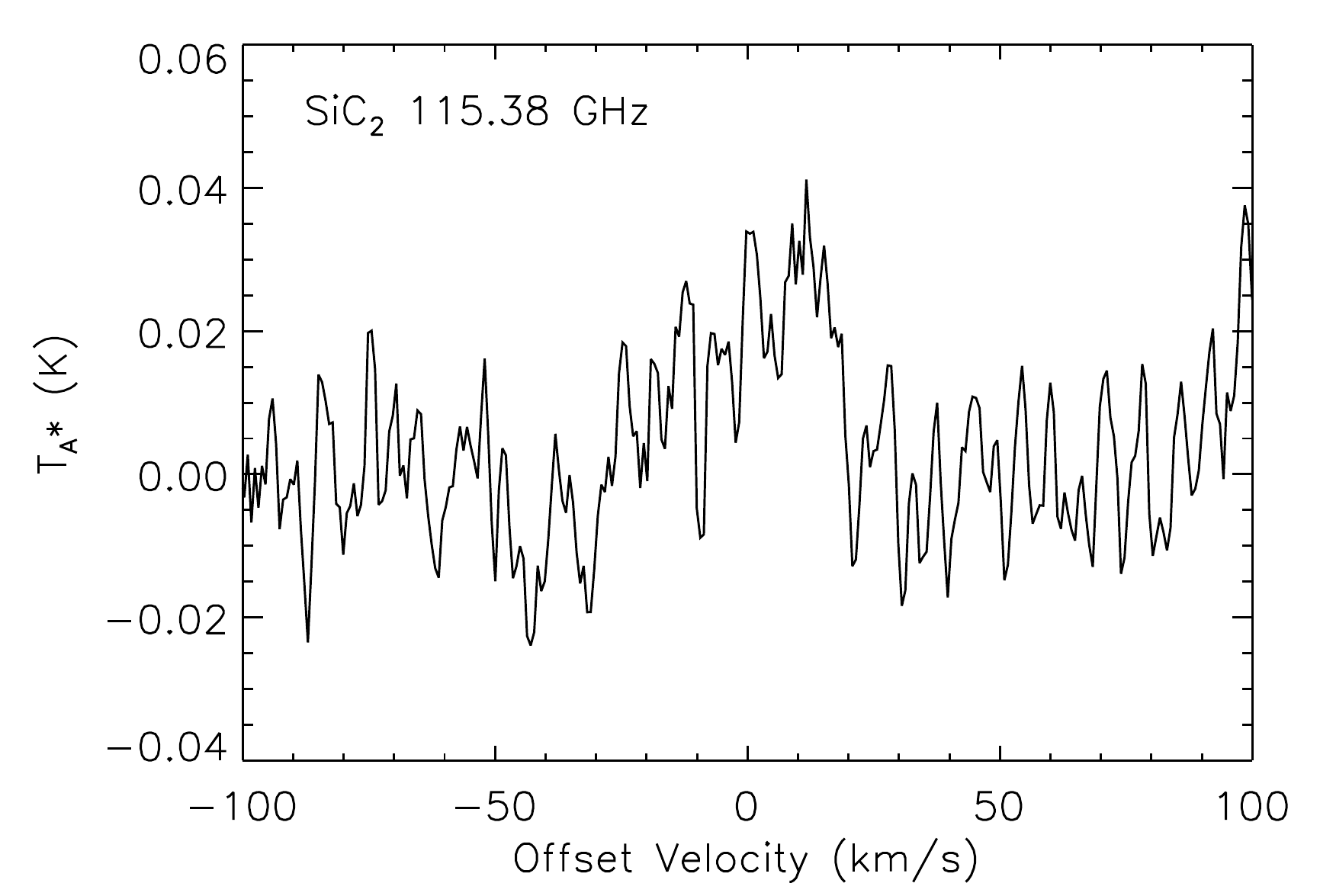}}

\contcaption{}
\end{figure*}

\begin{figure*}
{\centering
\subfigure{\includegraphics[trim= 2cm 13cm 2cm 2cm, clip=true, width=0.32\textwidth]{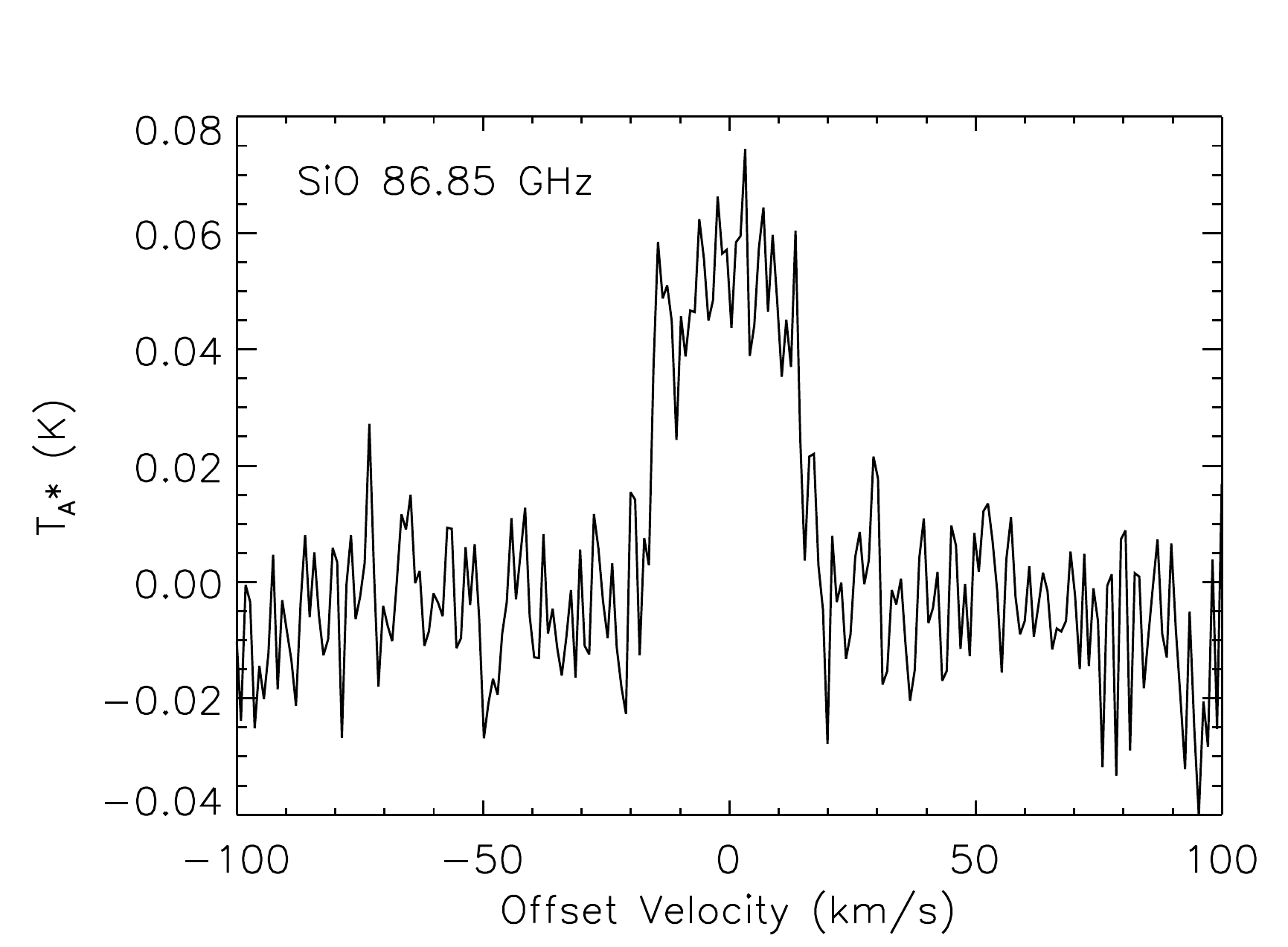}}
\subfigure{\includegraphics[trim= 2cm 13cm 2cm 2cm, clip=true, width=0.32\textwidth]{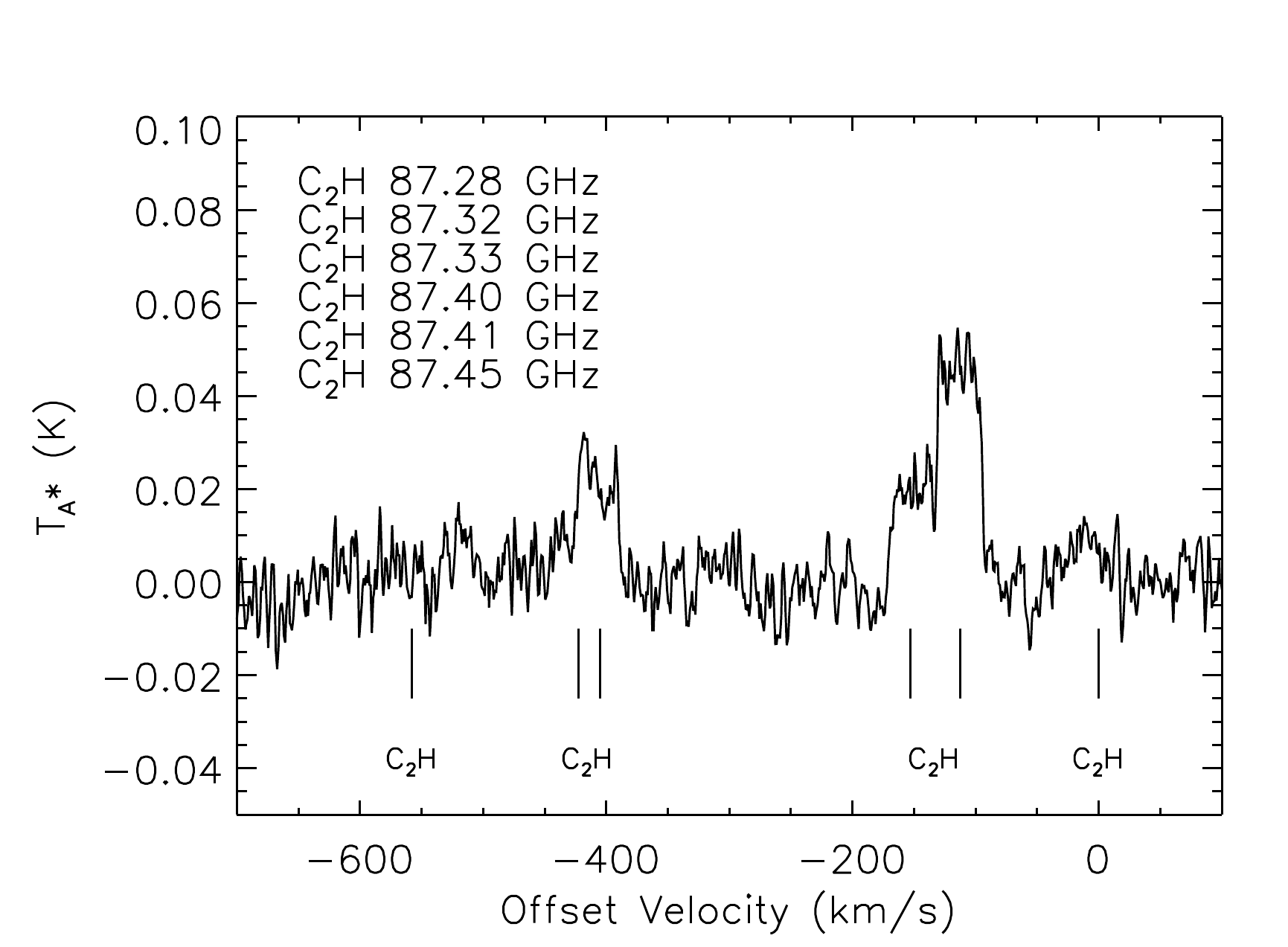}}
\subfigure{\includegraphics[trim= 2cm 13cm 2cm 2cm, clip=true, width=0.32\textwidth]{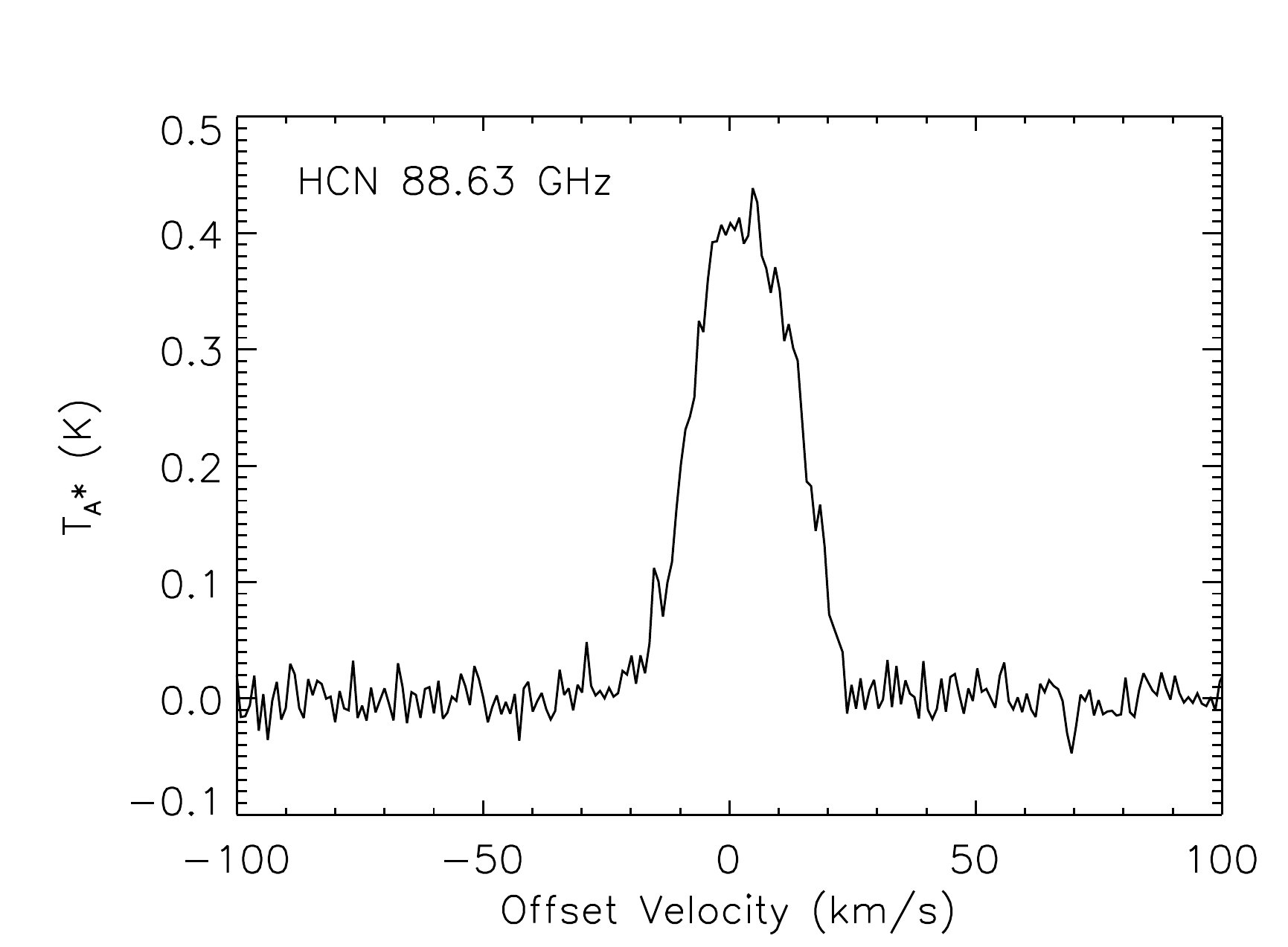}}

\subfigure{\includegraphics[trim= 2cm 13cm 2cm 2cm, clip=true, width=0.32\textwidth]{appendix_plots/15082/15082_hcn_maser.pdf}}
\subfigure{\includegraphics[trim= 2cm 13cm 2cm 2cm, clip=true, width=0.32\textwidth]{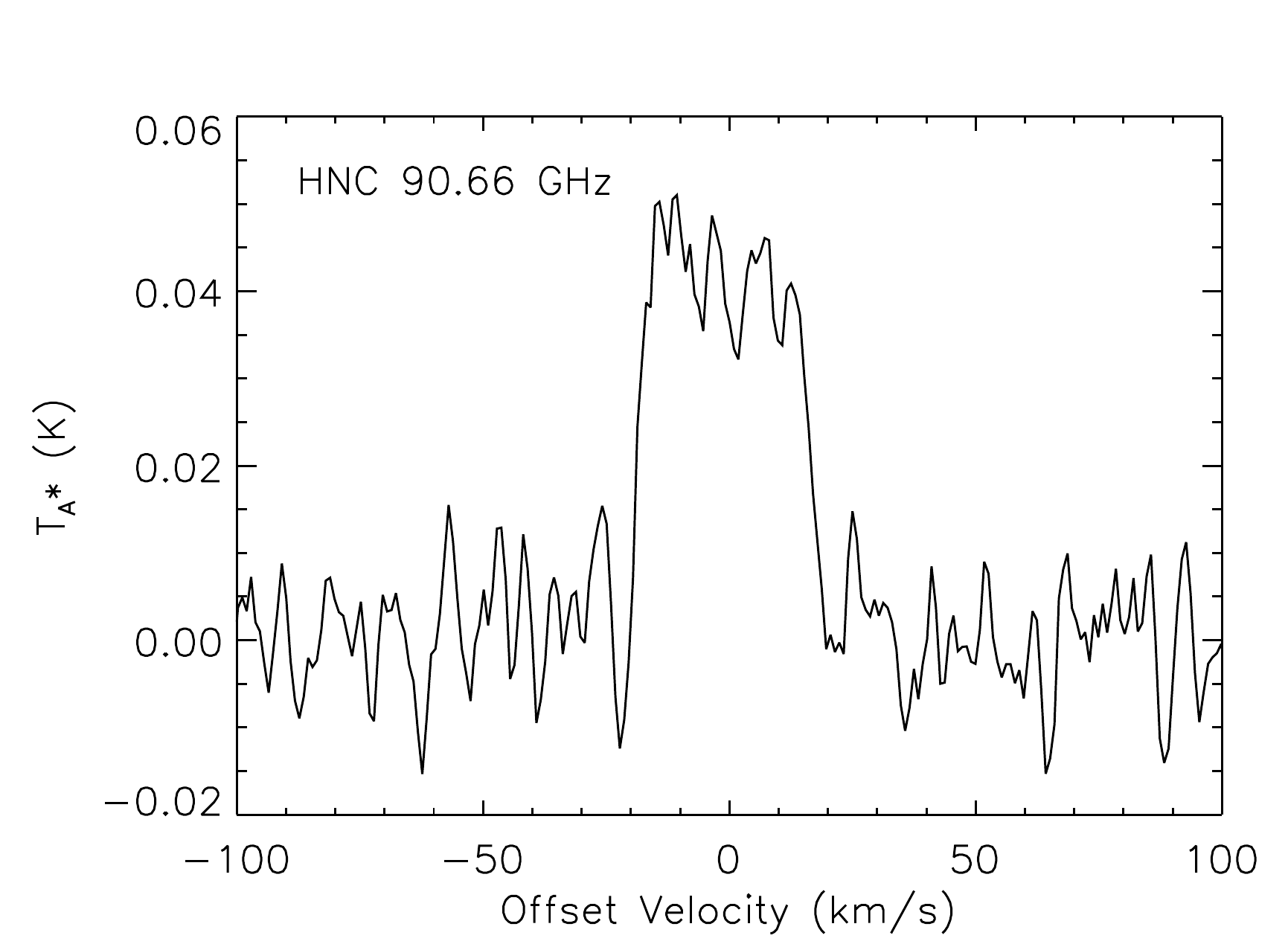}}
\subfigure{\includegraphics[trim= 2cm 13cm 2cm 2cm, clip=true, width=0.32\textwidth]{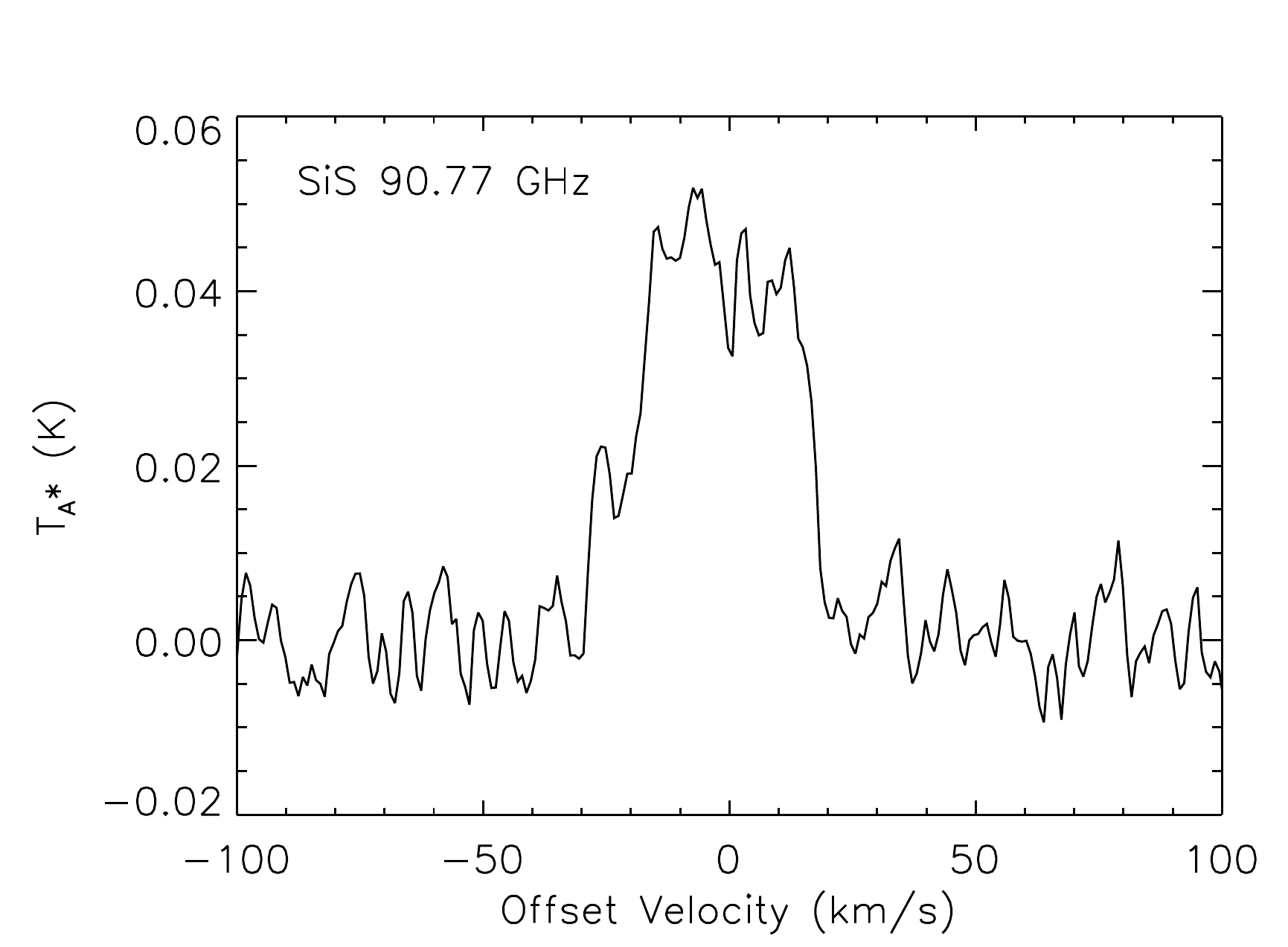}}

\subfigure{\includegraphics[trim= 2cm 13cm 2cm 2cm, clip=true, width=0.32\textwidth]{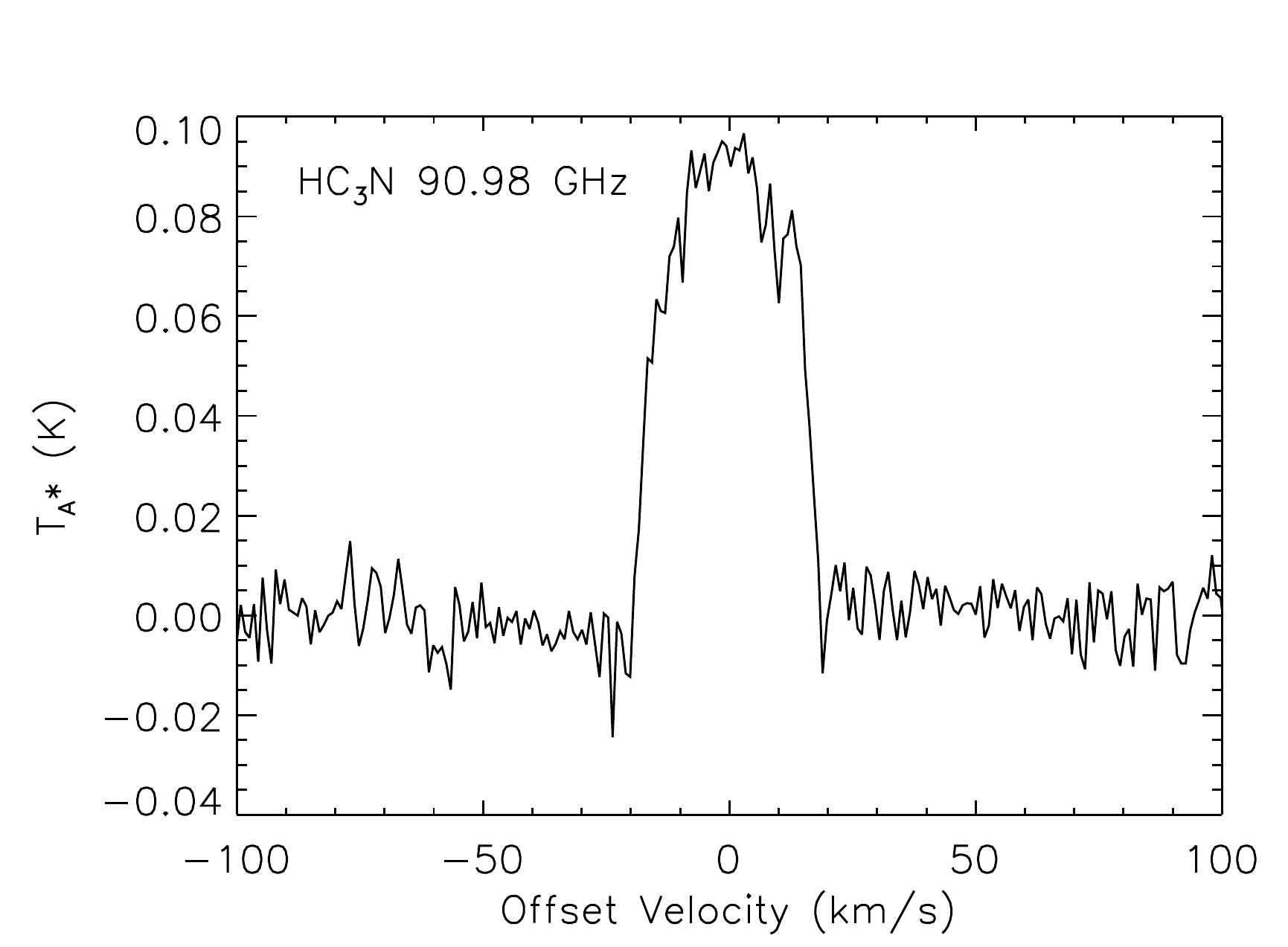}}
\subfigure{\includegraphics[trim= 2cm 13cm 2cm 2cm, clip=true, width=0.32\textwidth]{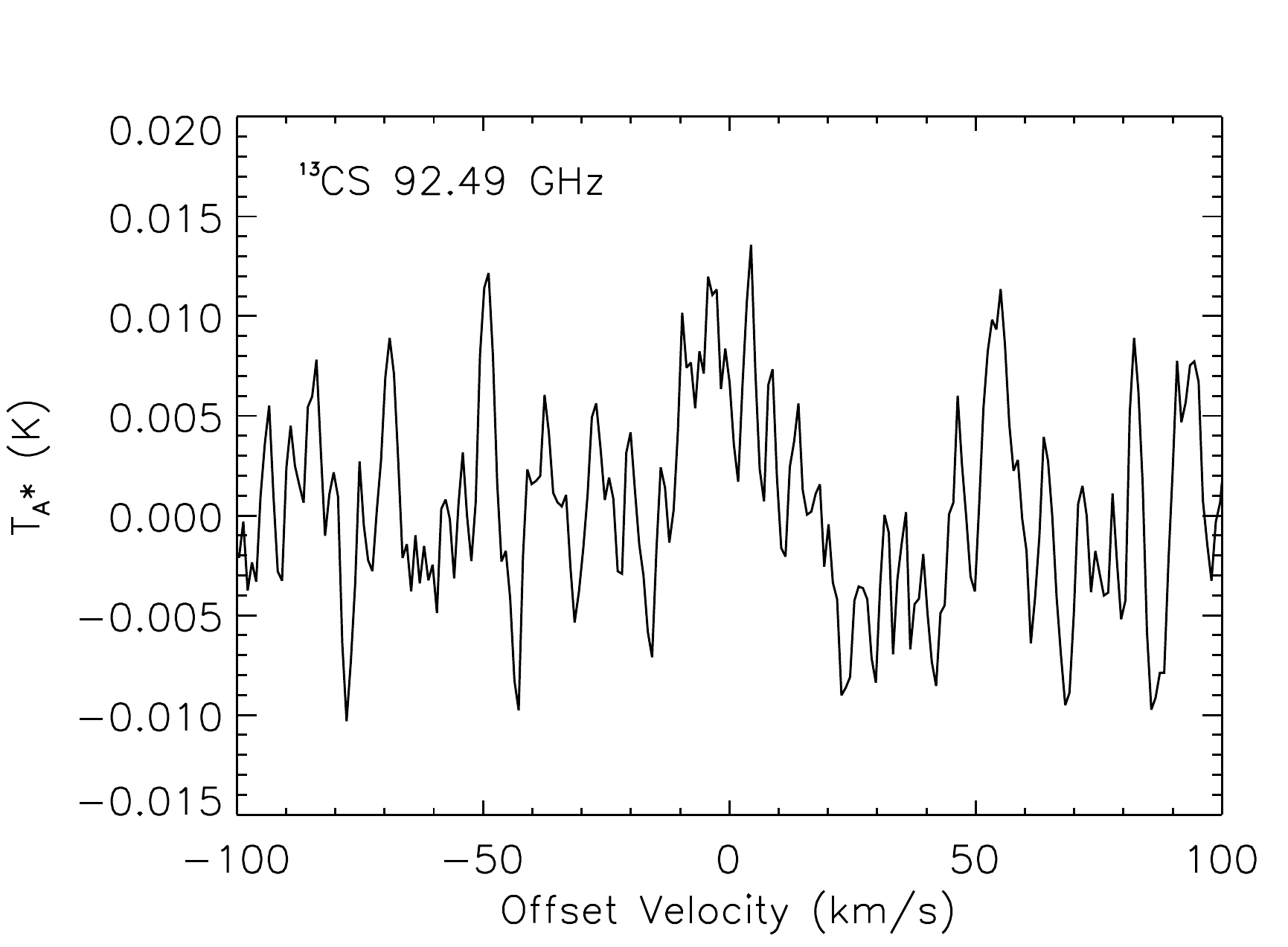}}
\subfigure{\includegraphics[trim= 2cm 13cm 2cm 2cm, clip=true, width=0.32\textwidth]{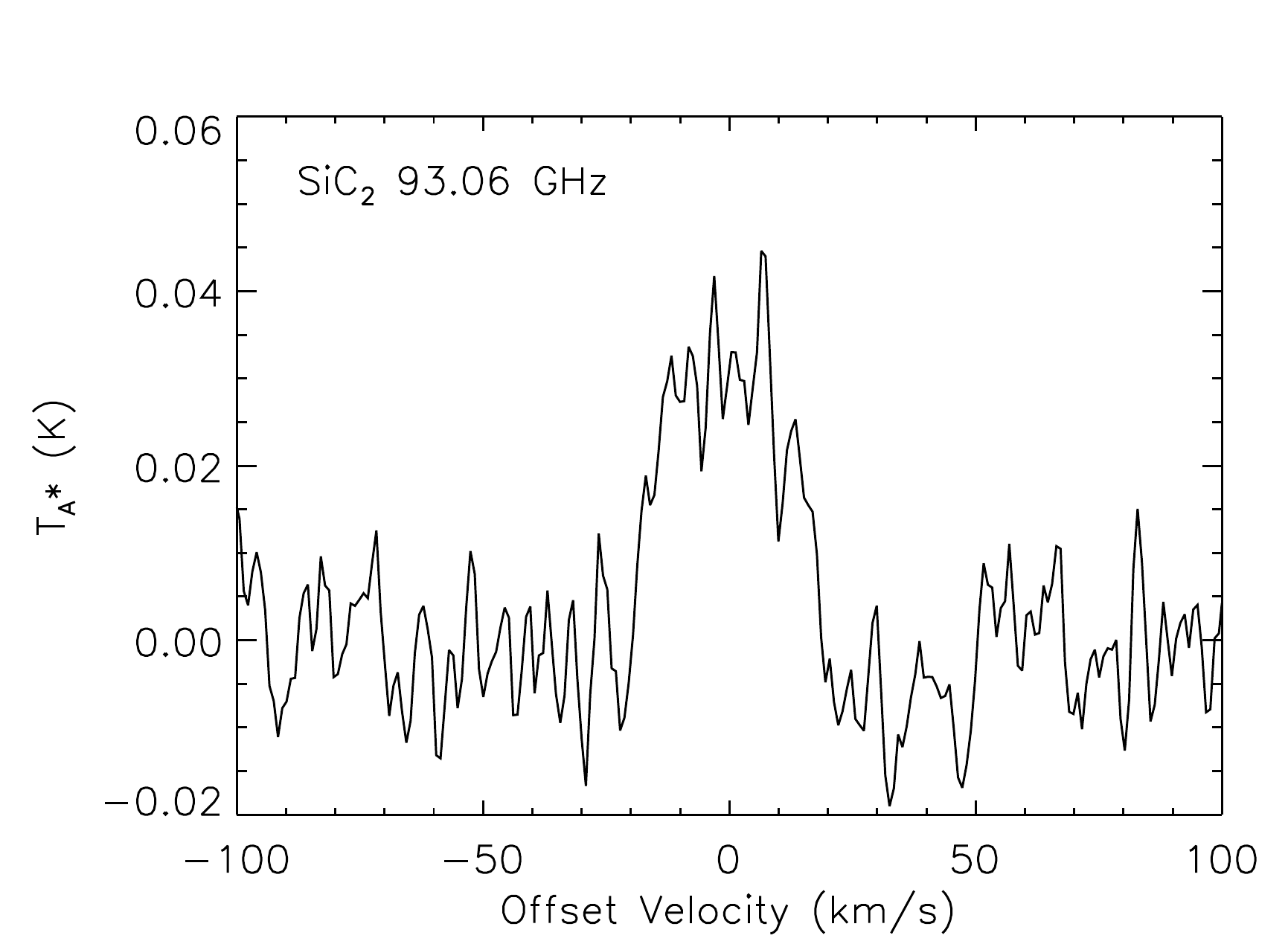}}

\subfigure{\includegraphics[trim= 2cm 13cm 2cm 2cm, clip=true, width=0.32\textwidth]{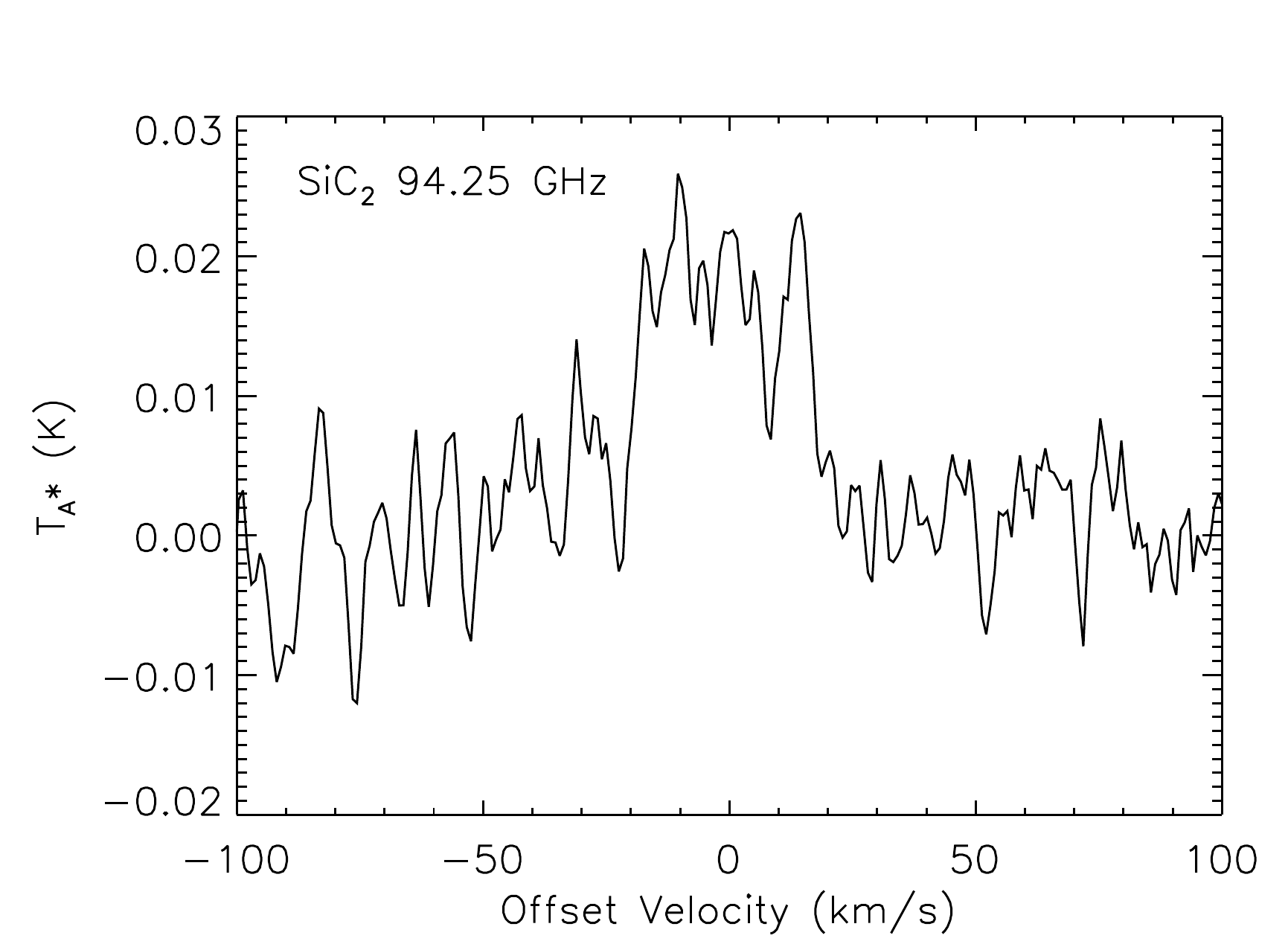}}
\subfigure{\includegraphics[trim= 2cm 13cm 2cm 2cm, clip=true, width=0.32\textwidth]{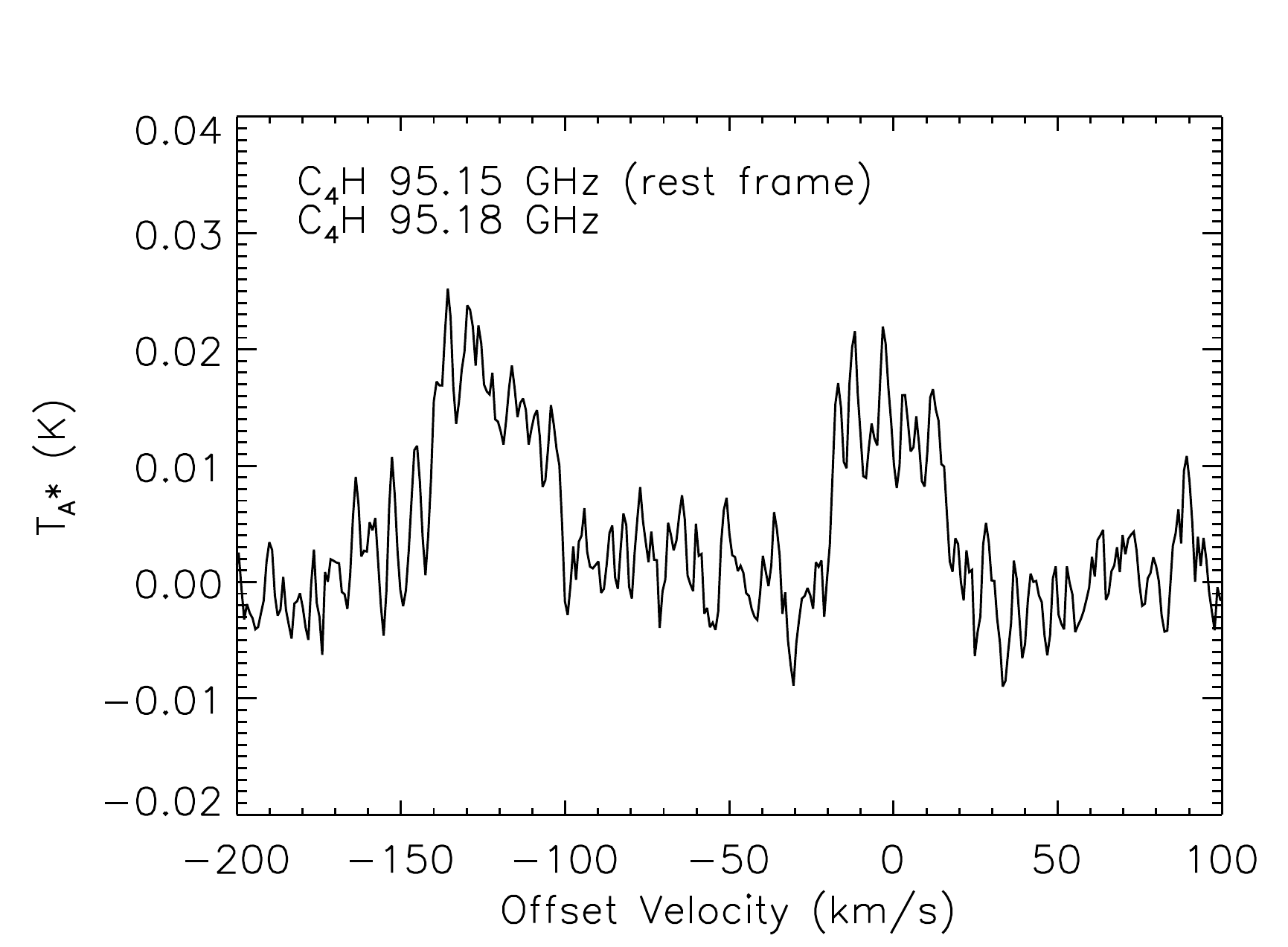}}
\subfigure{\includegraphics[trim= 2cm 13cm 2cm 2cm, clip=true, width=0.32\textwidth]{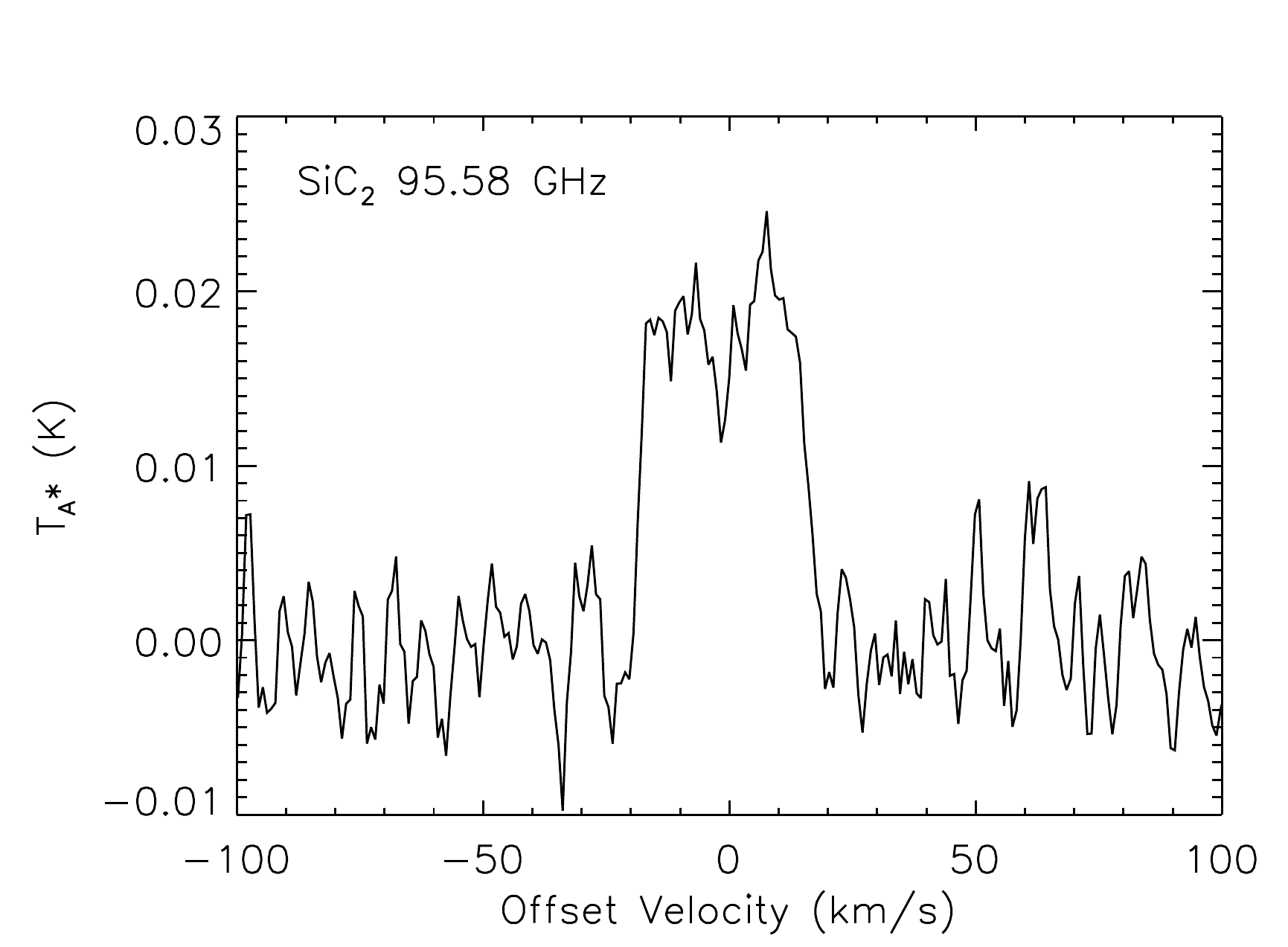}}
}

\caption{All transitions in IRAS 15082-4808. Ordinate axis intensities are in units of corrected antenna temperature, abscissa values are in units of km/s, corrected for LSR velocity of the source, taken as -3.0 km/s \citep{Groenewegen2002}.}
\end{figure*}

\begin{figure*}
{\begin{center}
\subfigure{\includegraphics[trim= 2cm 13cm 2cm 2cm, clip=true, width=0.32\textwidth]{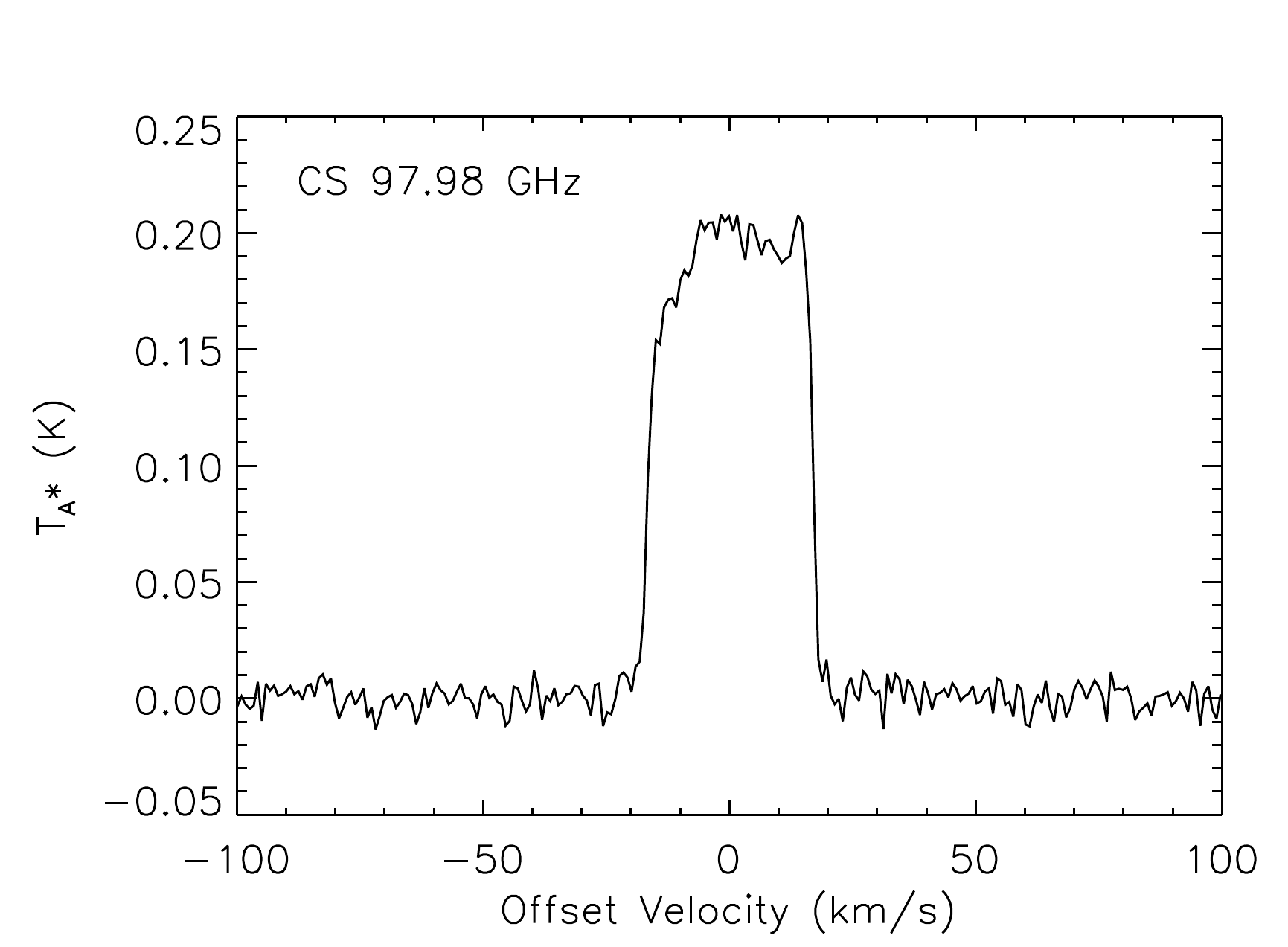}}
\subfigure{\includegraphics[trim= 2cm 13cm 2cm 3cm, clip=true, width=0.32\textwidth]{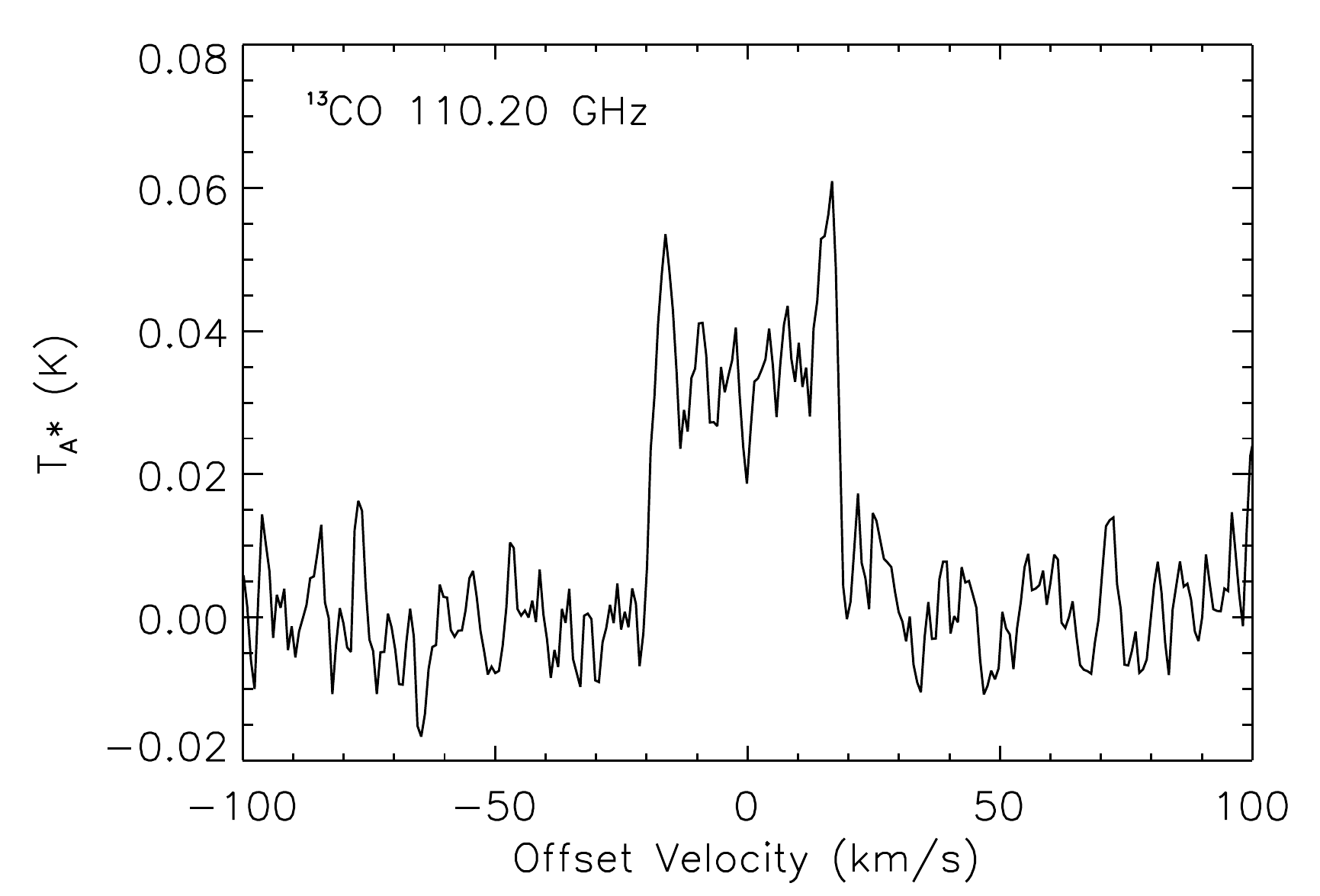}}
\subfigure{\includegraphics[trim= 2cm 13cm 2cm 3cm, clip=true, width=0.32\textwidth]{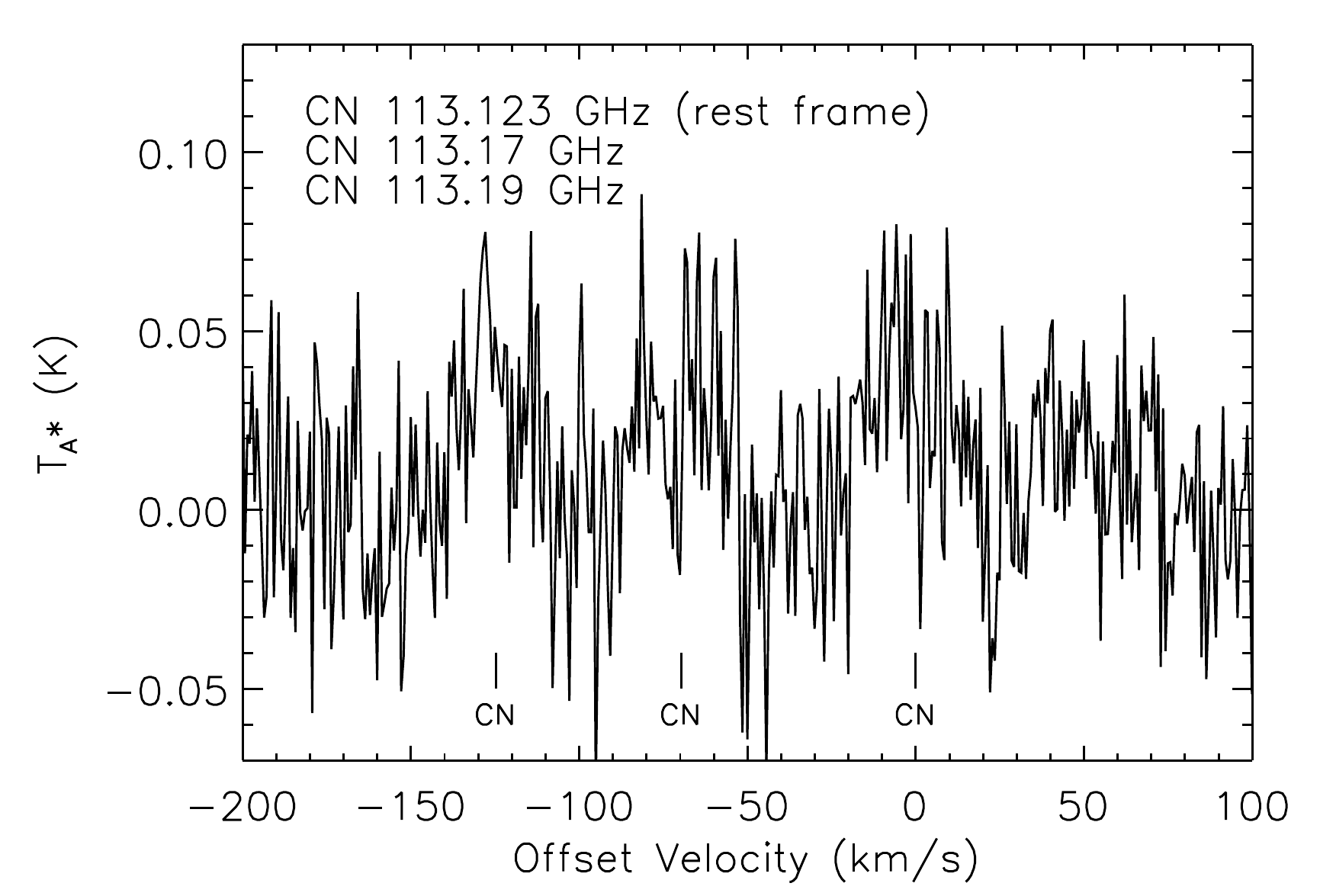}}

\subfigure{\includegraphics[trim= 2cm 13cm 2cm 3cm, clip=true, width=0.32\textwidth]{appendix_plots/15082/15082_cn-1134.pdf}}
\subfigure{\includegraphics[trim= 2cm 13cm 2cm 2cm, clip=true, width=0.32\textwidth]{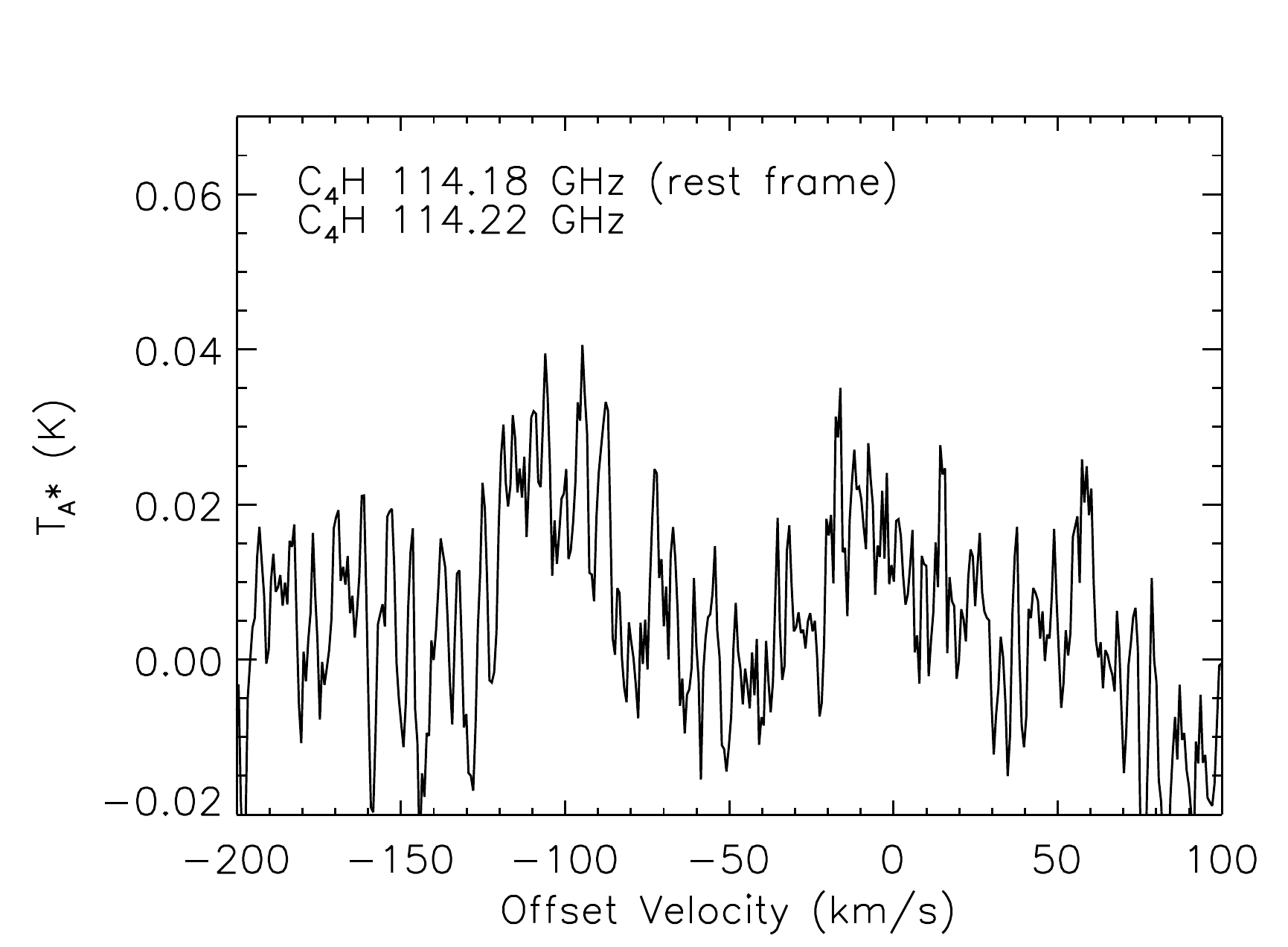}}
\subfigure{\includegraphics[trim= 2cm 13cm 2cm 2cm, clip=true, width=0.32\textwidth]{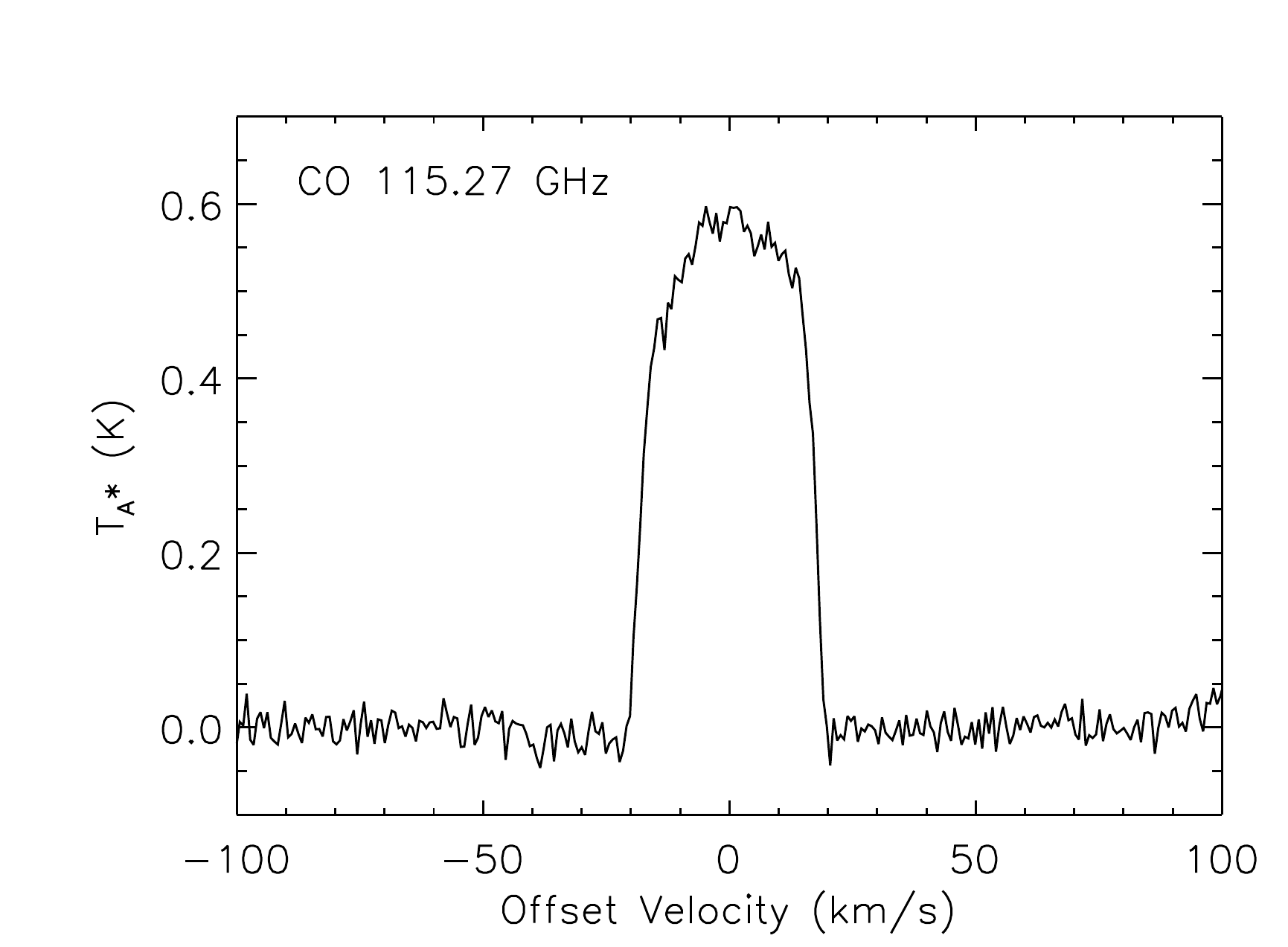}}

\subfigure{\includegraphics[trim= 2cm 13cm 2cm 2cm, clip=true, width=0.32\textwidth]{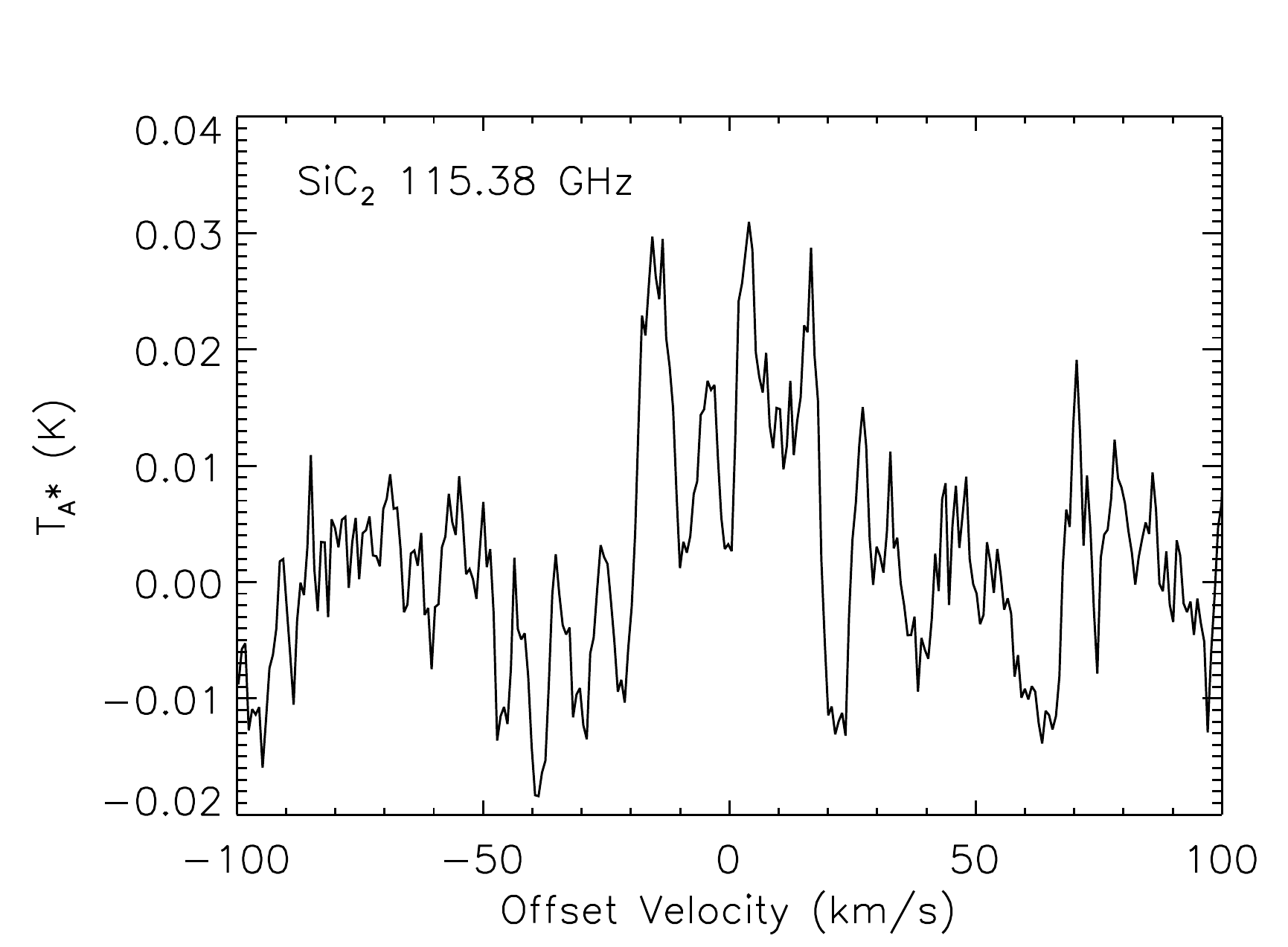}}
\end{center}
}
\contcaption{}
\end{figure*}


\begin{figure*}
{\centering
\subfigure{\includegraphics[trim= 2cm 13cm 2cm 2cm, clip=true, width=0.32\textwidth]{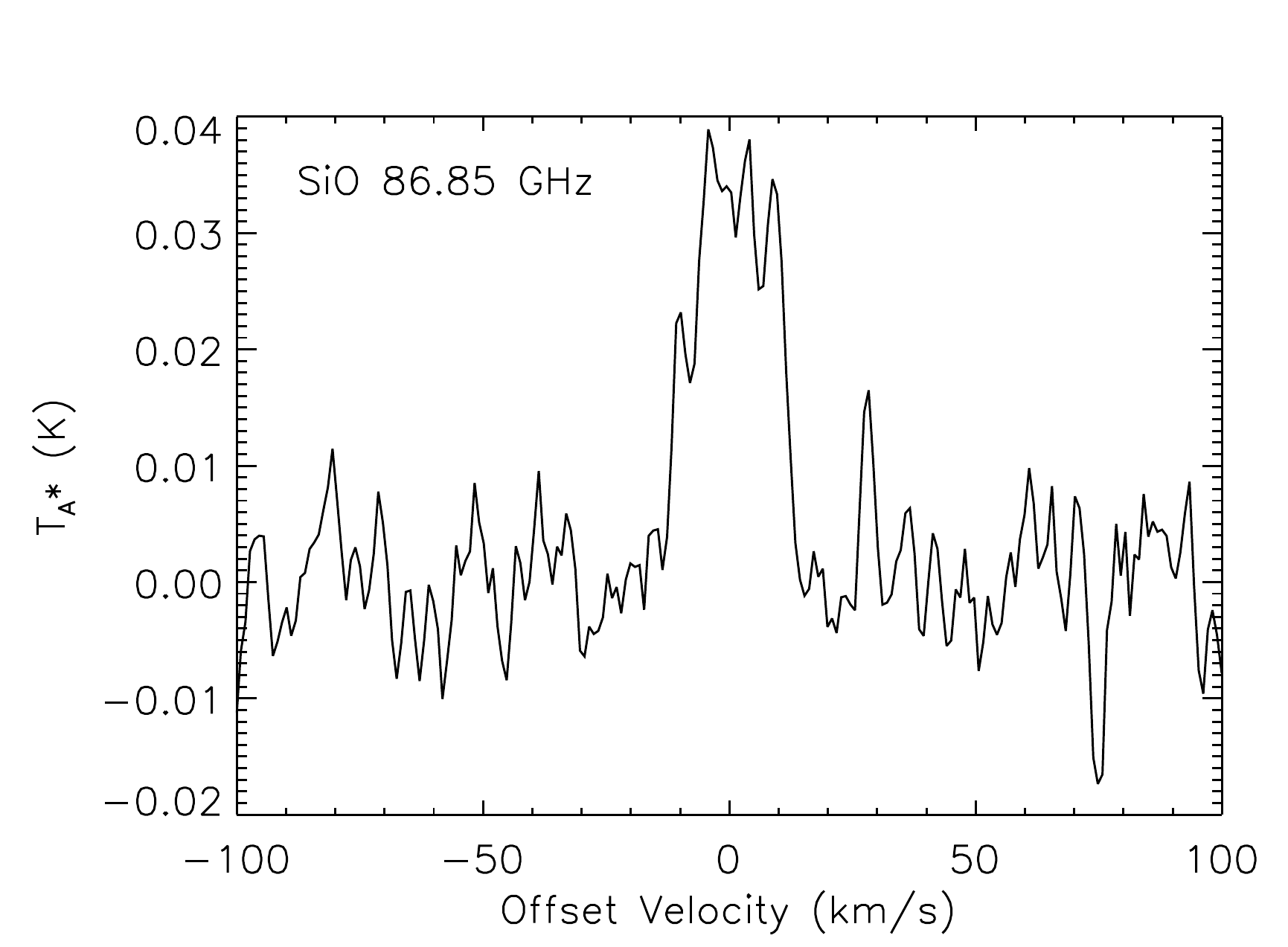}}
\subfigure{\includegraphics[trim= 2cm 13cm 2cm 2cm, clip=true, width=0.32\textwidth]{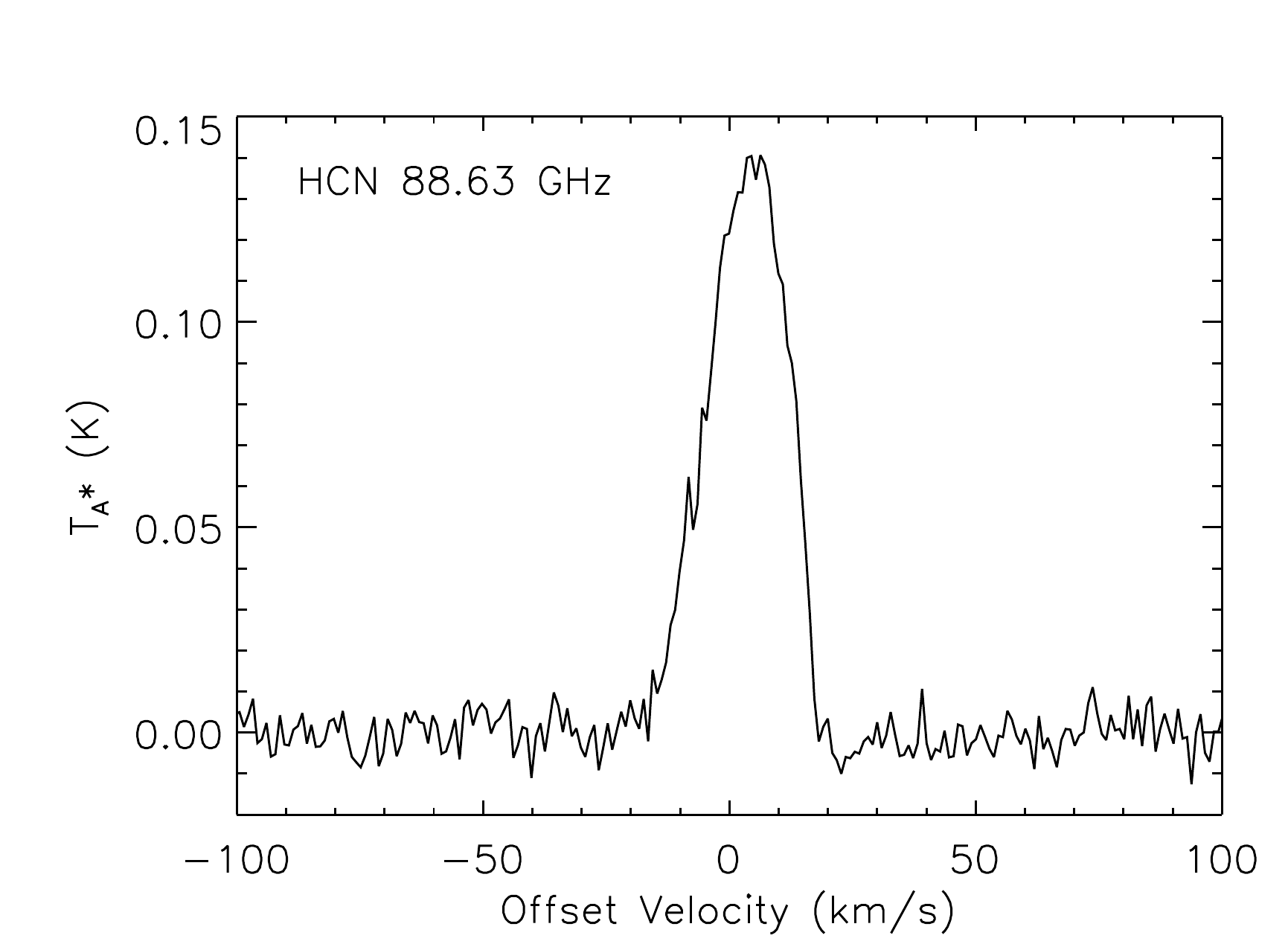}}
\subfigure{\includegraphics[trim= 2cm 13cm 2cm 2cm, clip=true, width=0.32\textwidth]{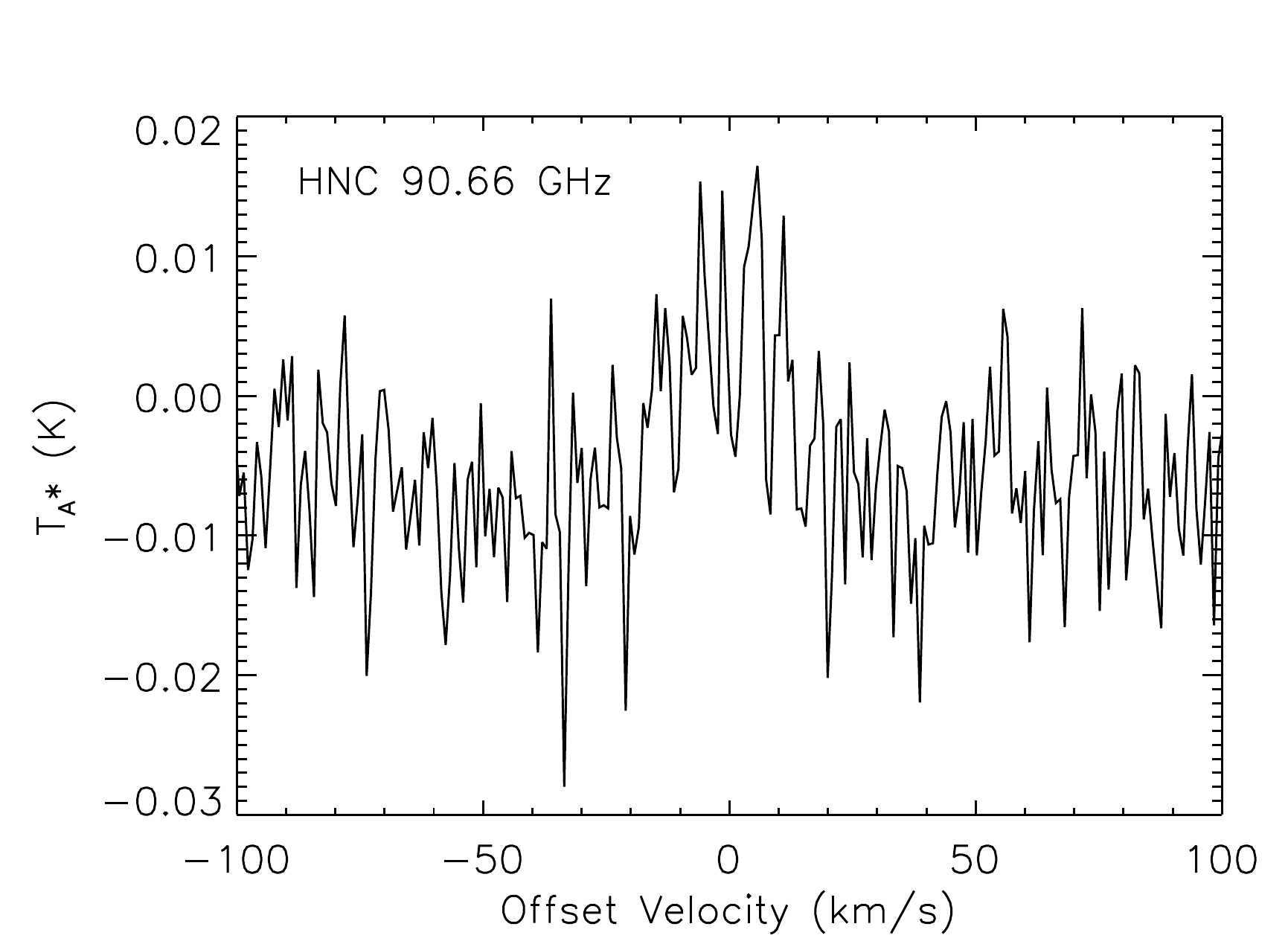}}

\subfigure{\includegraphics[trim= 2cm 13cm 2cm 2cm, clip=true, width=0.32\textwidth]{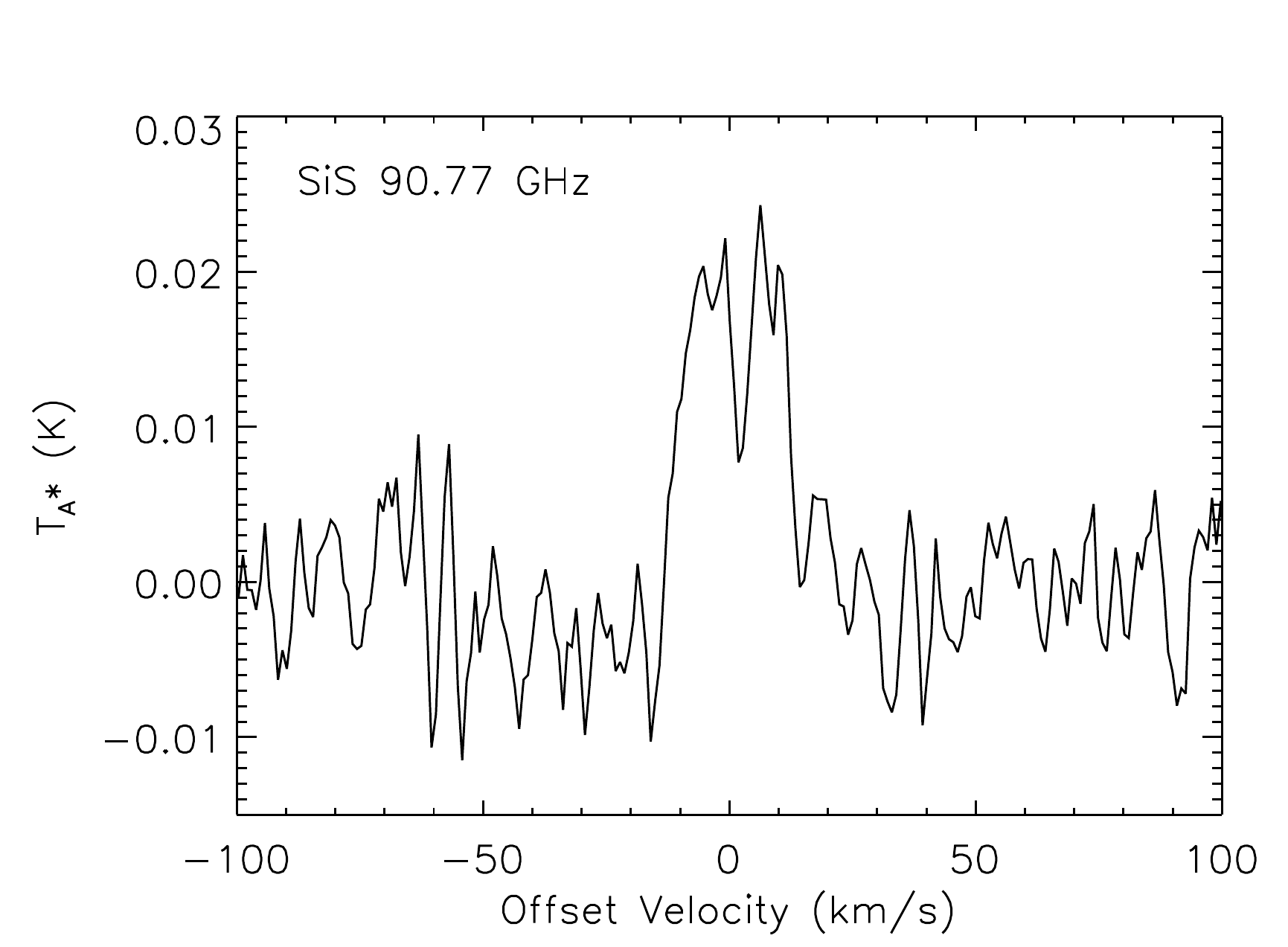}}
\subfigure{\includegraphics[trim= 2cm 13cm 2cm 2cm, clip=true, width=0.32\textwidth]{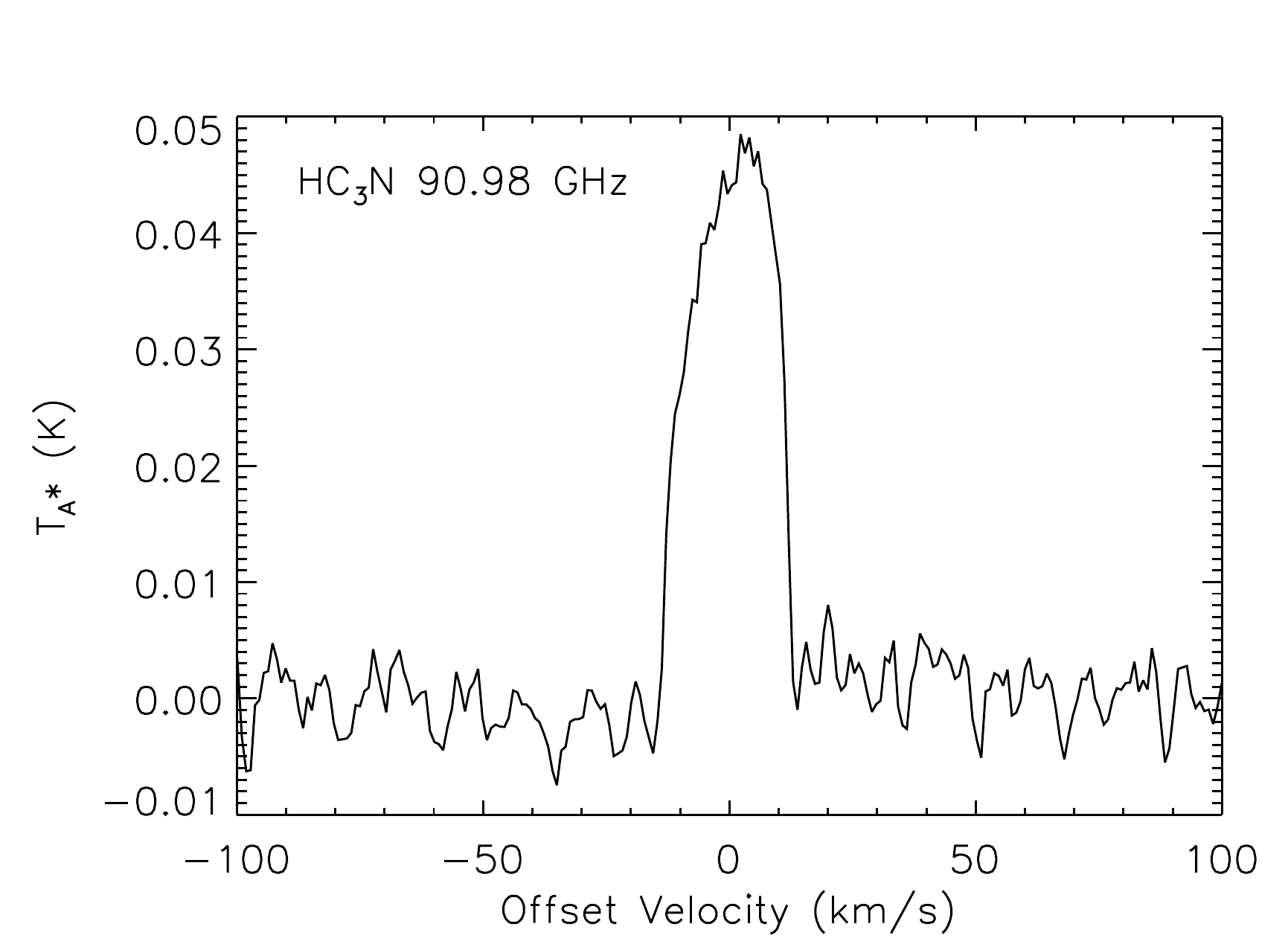}}
\subfigure{\includegraphics[trim= 2cm 13cm 2cm 2cm, clip=true, width=0.32\textwidth]{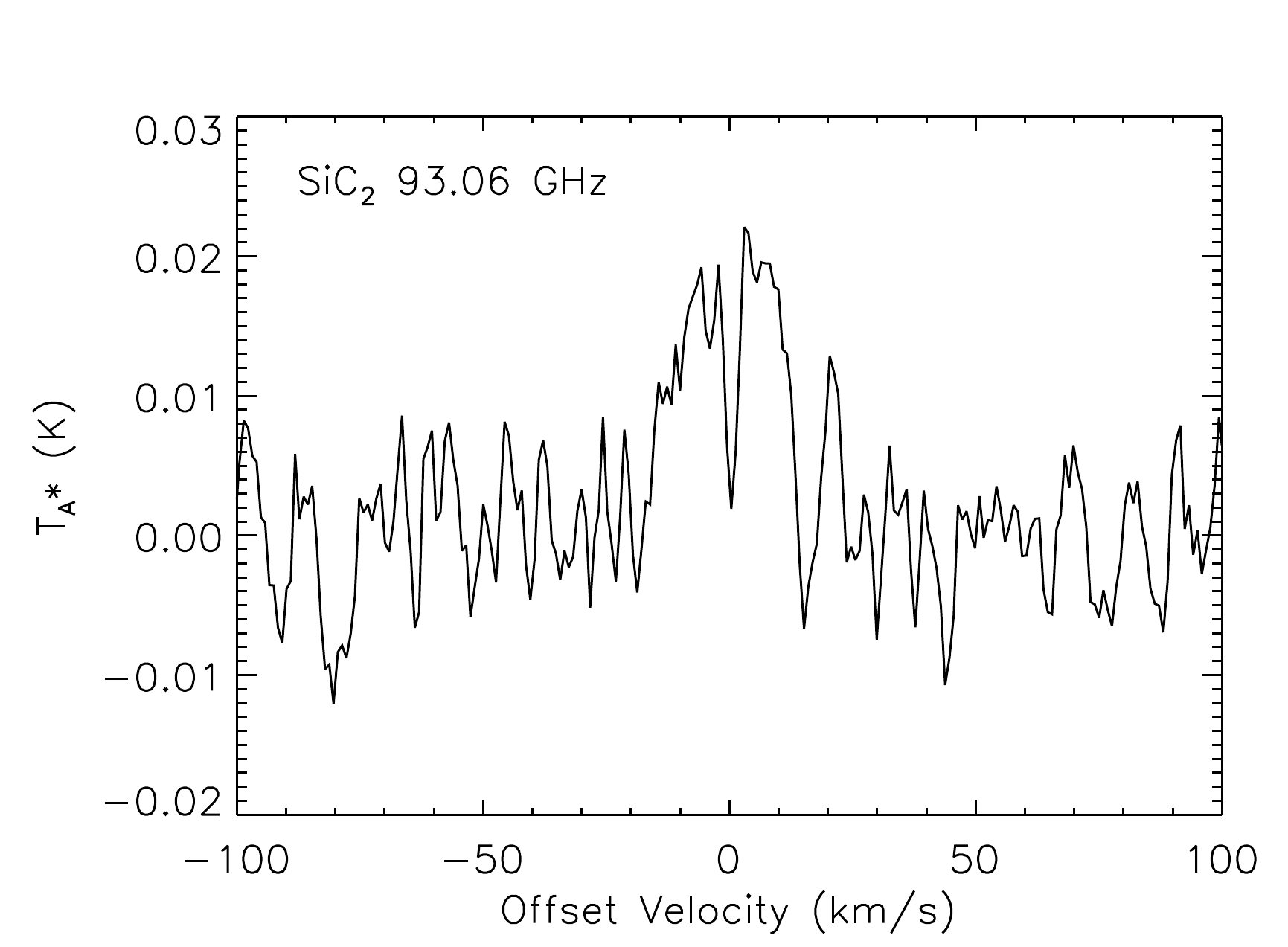}}

\subfigure{\includegraphics[trim= 2cm 13cm 2cm 2cm, clip=true, width=0.32\textwidth]{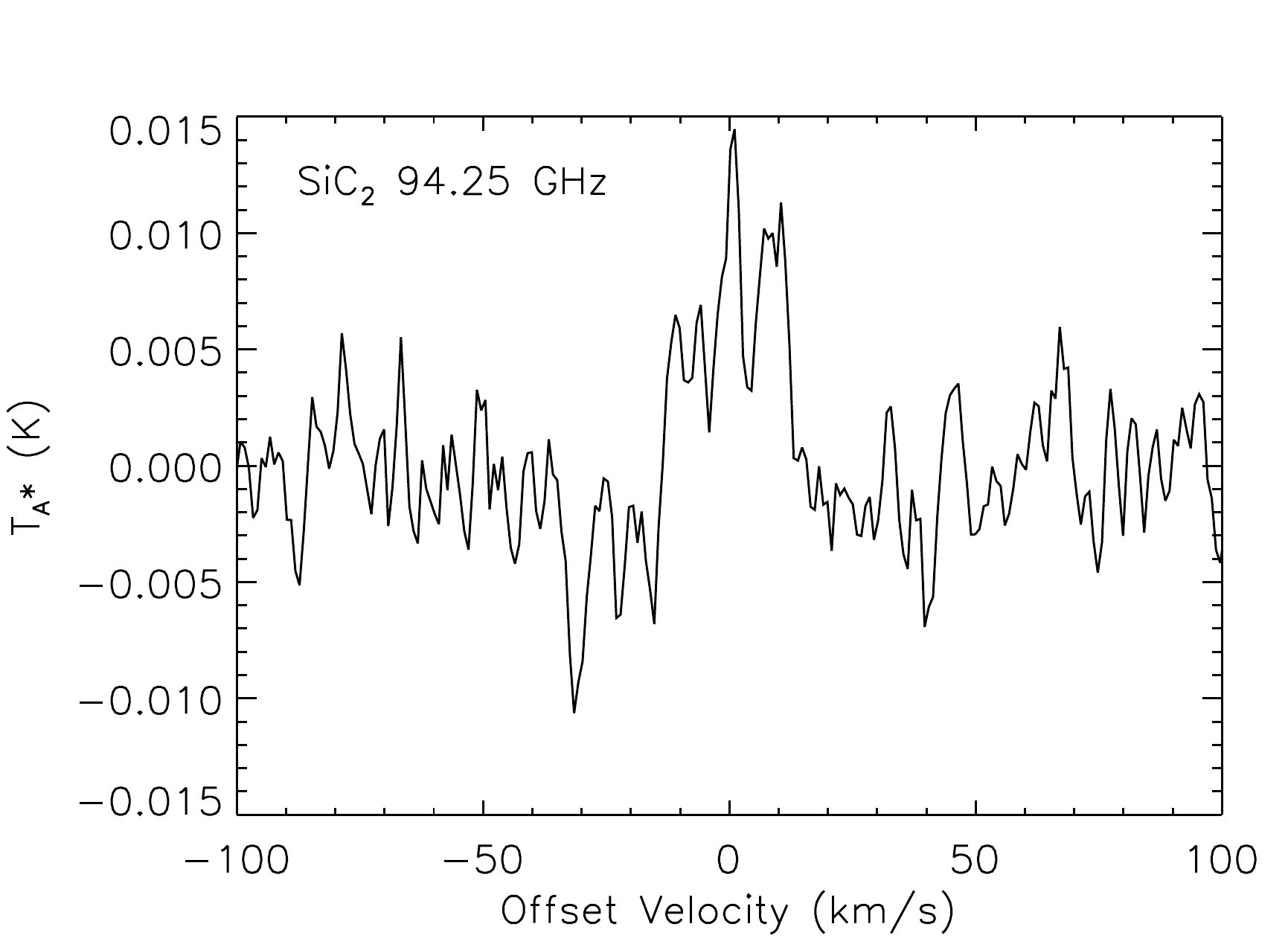}}
\subfigure{\includegraphics[trim= 2cm 13cm 2cm 2cm, clip=true, width=0.32\textwidth]{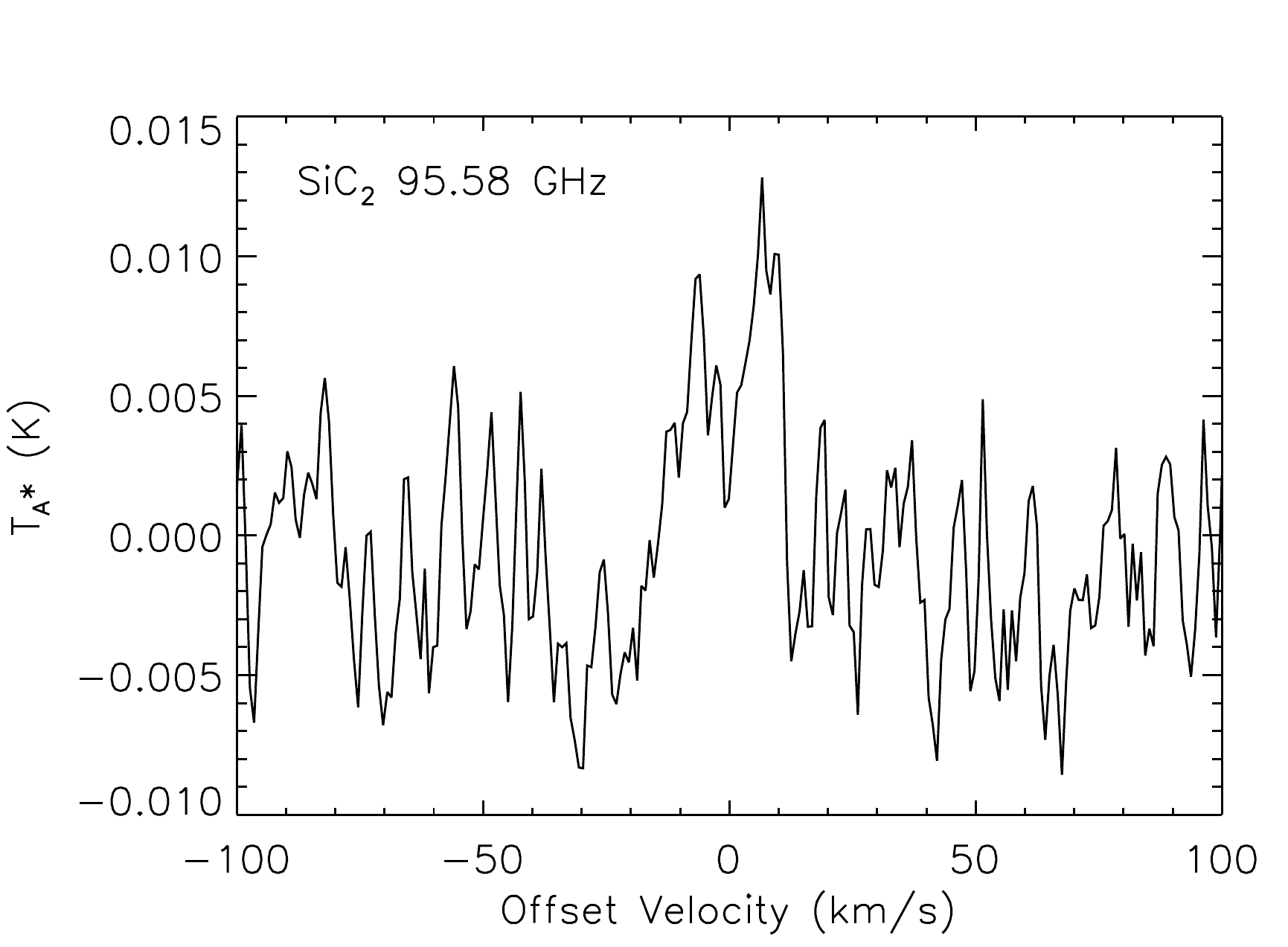}}
\subfigure{\includegraphics[trim= 2cm 13cm 2cm 2cm, clip=true, width=0.32\textwidth]{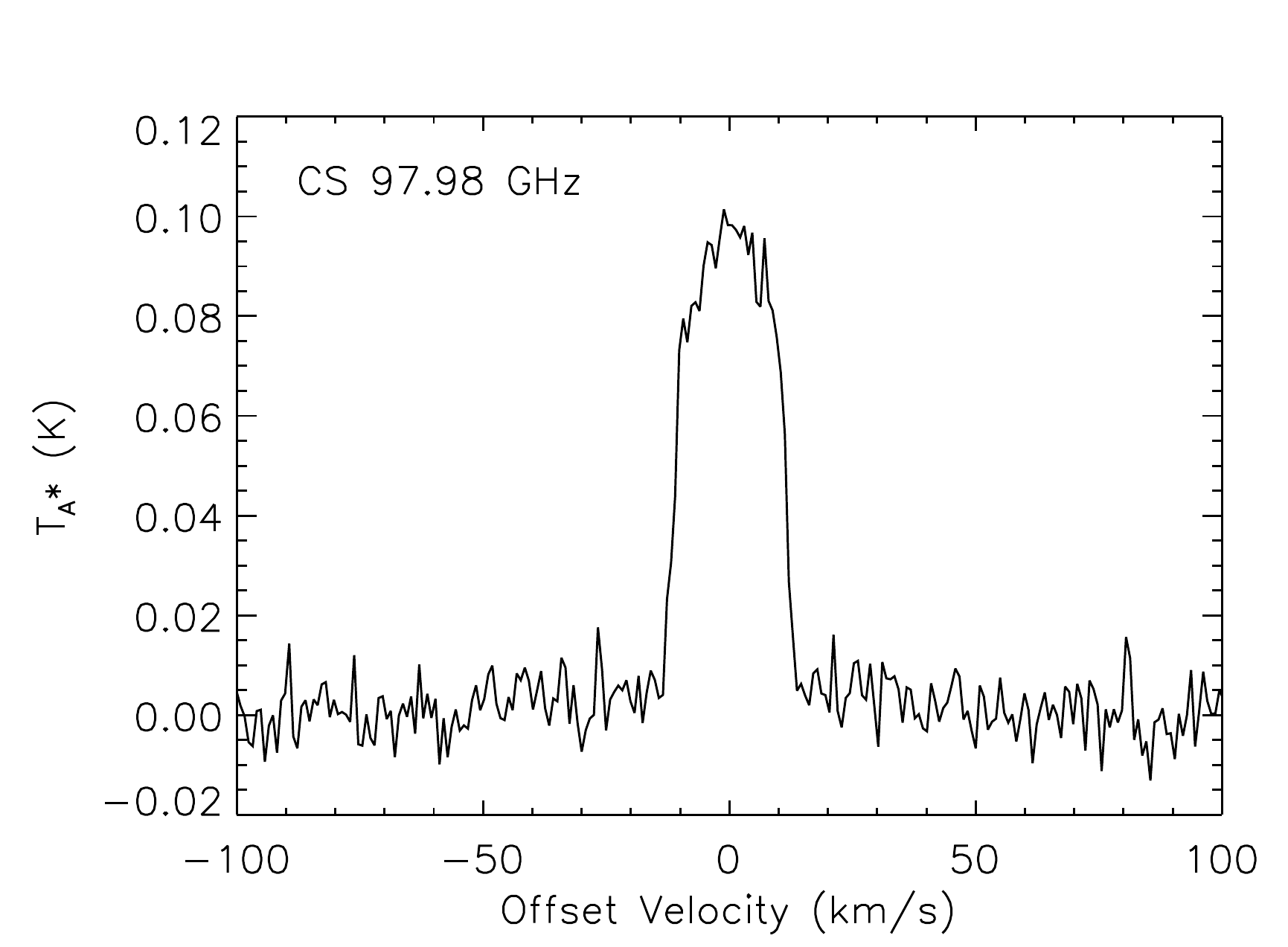}}

\subfigure{\includegraphics[trim= 2cm 13cm 2cm 3cm, clip=true, width=0.32\textwidth]{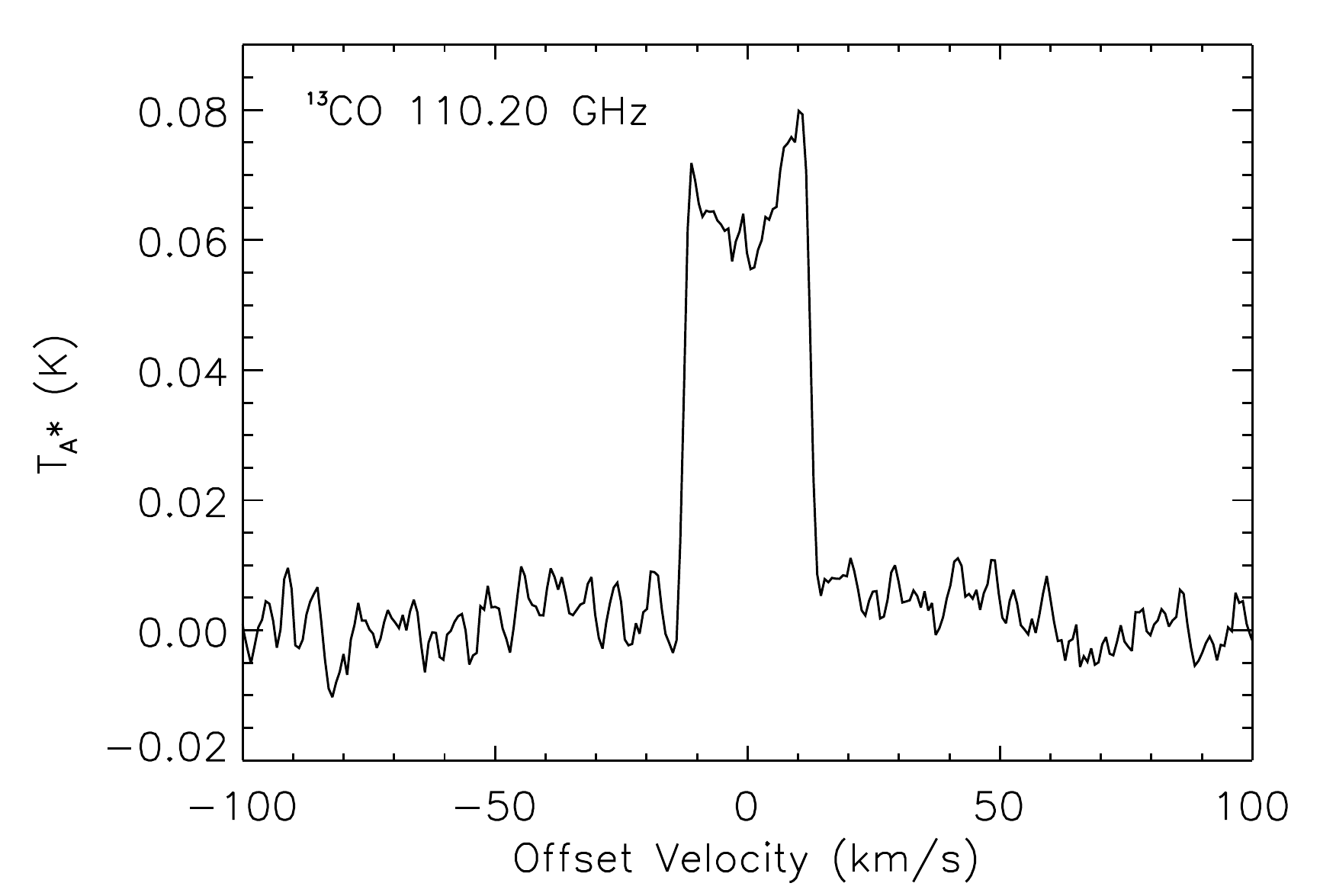}}
\subfigure{\includegraphics[trim= 2cm 13cm 2cm 3cm, clip=true, width=0.32\textwidth]{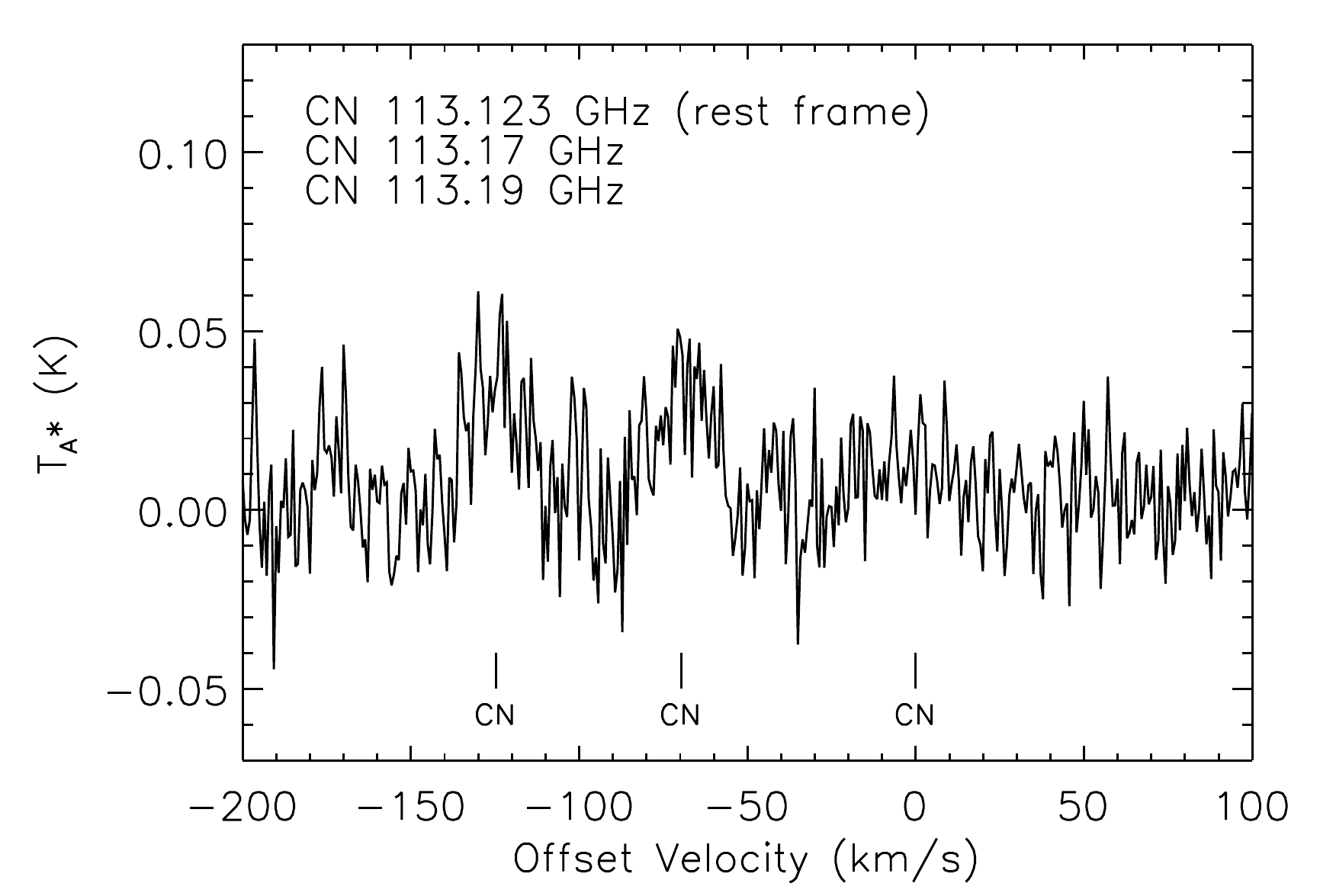}}
\subfigure{\includegraphics[trim= 2cm 13cm 2cm 3cm, clip=true, width=0.32\textwidth]{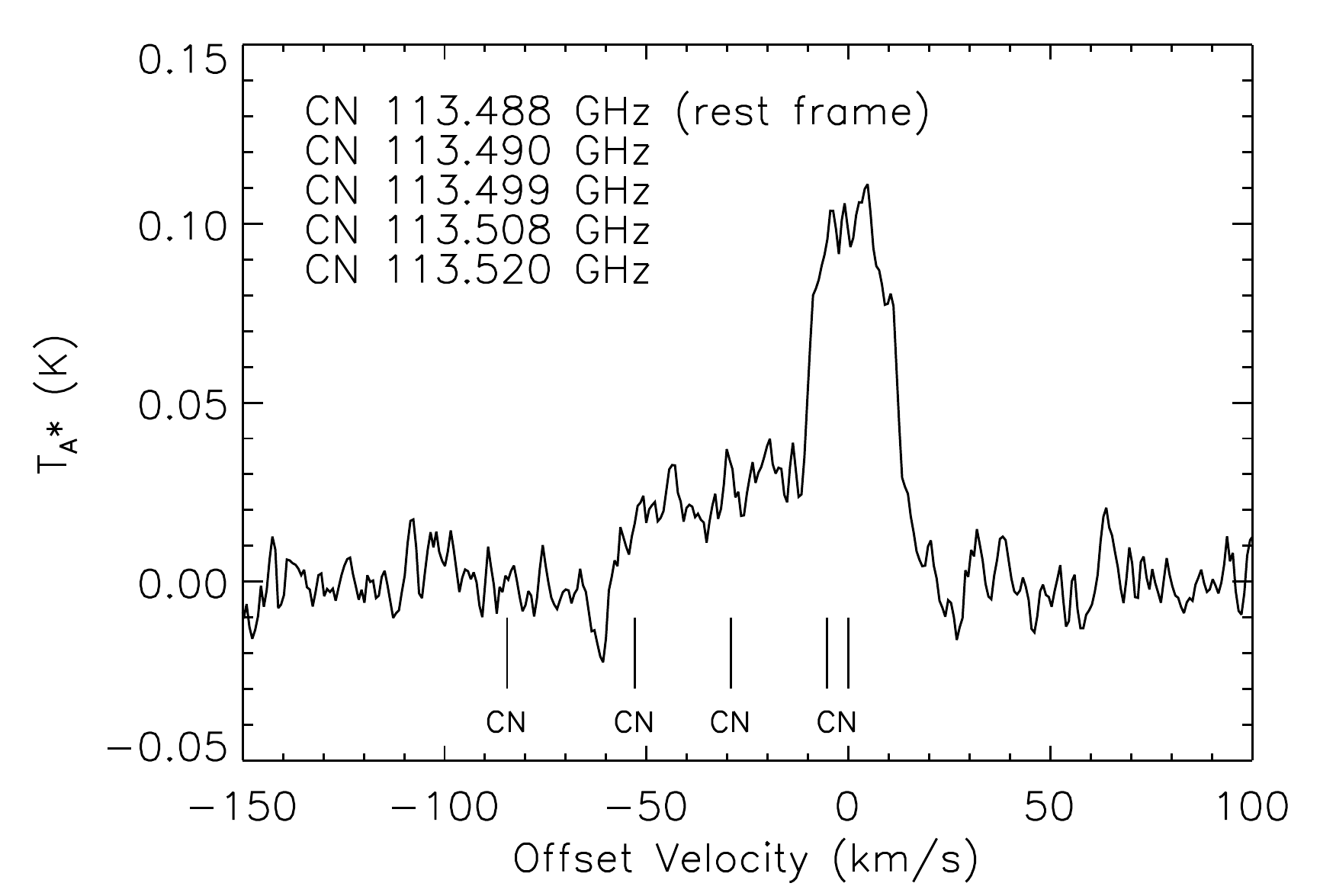}}

\caption{All transitions in IRAS 07454-7112. Ordinate axis intensities are in units of corrected antenna temperature, abscissa values are in units of km/s, corrected for LSR velocity of the source, taken as -39.8 km/s \citep{Risacher2009}.}
}
\end{figure*}

\begin{figure*}
{\centering

\subfigure{\includegraphics[trim= 2cm 13cm 2cm 2cm, clip=true, width=0.32\textwidth]{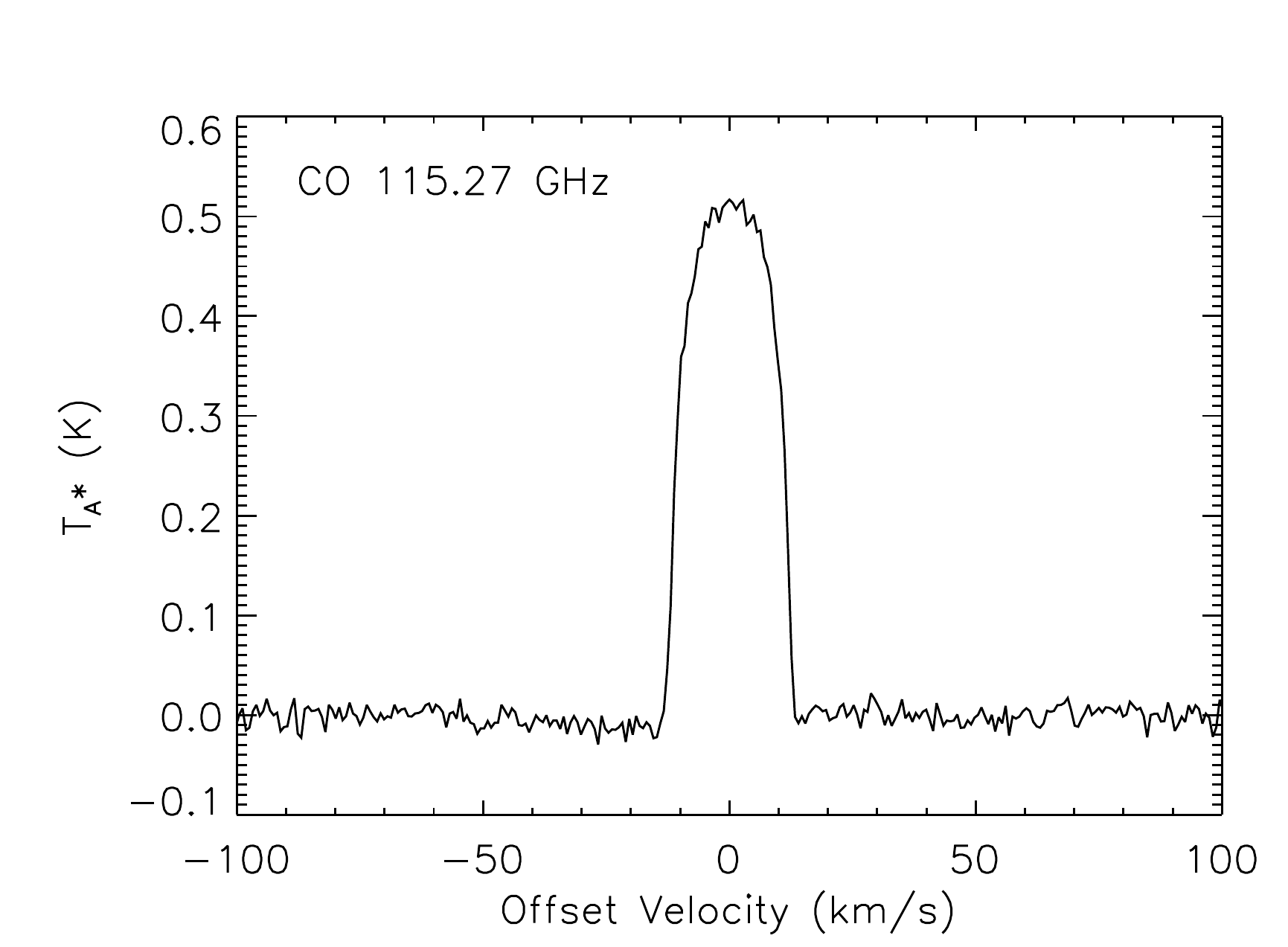}}
\subfigure{\includegraphics[trim= 2cm 13cm 2cm 2cm, clip=true, width=0.32\textwidth]{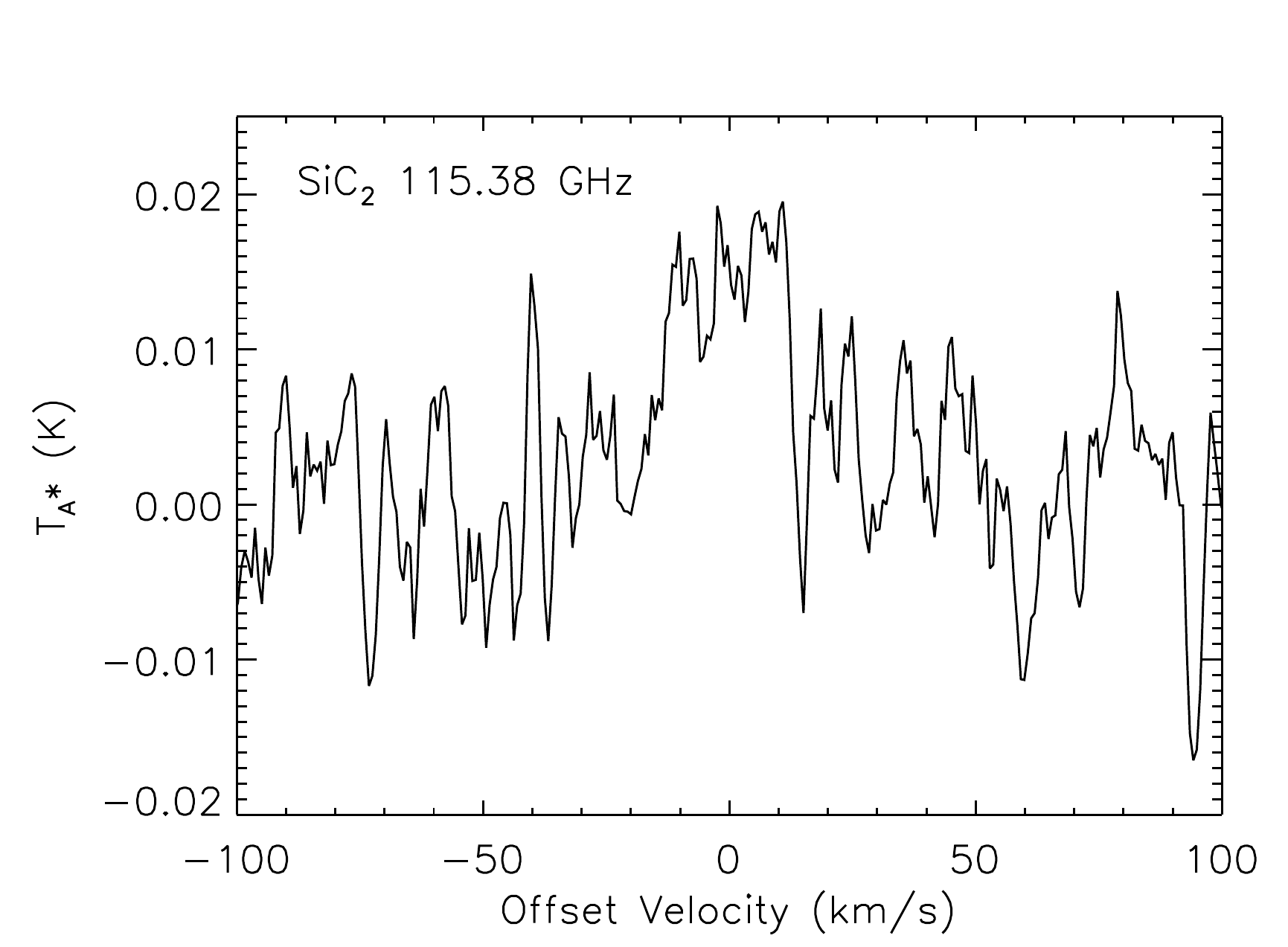}}
}

\contcaption{}
\end{figure*}

\begin{figure*}
{\centering
\subfigure{\includegraphics[trim= 2cm 13cm 2cm 2cm, clip=true, width=0.32\textwidth]{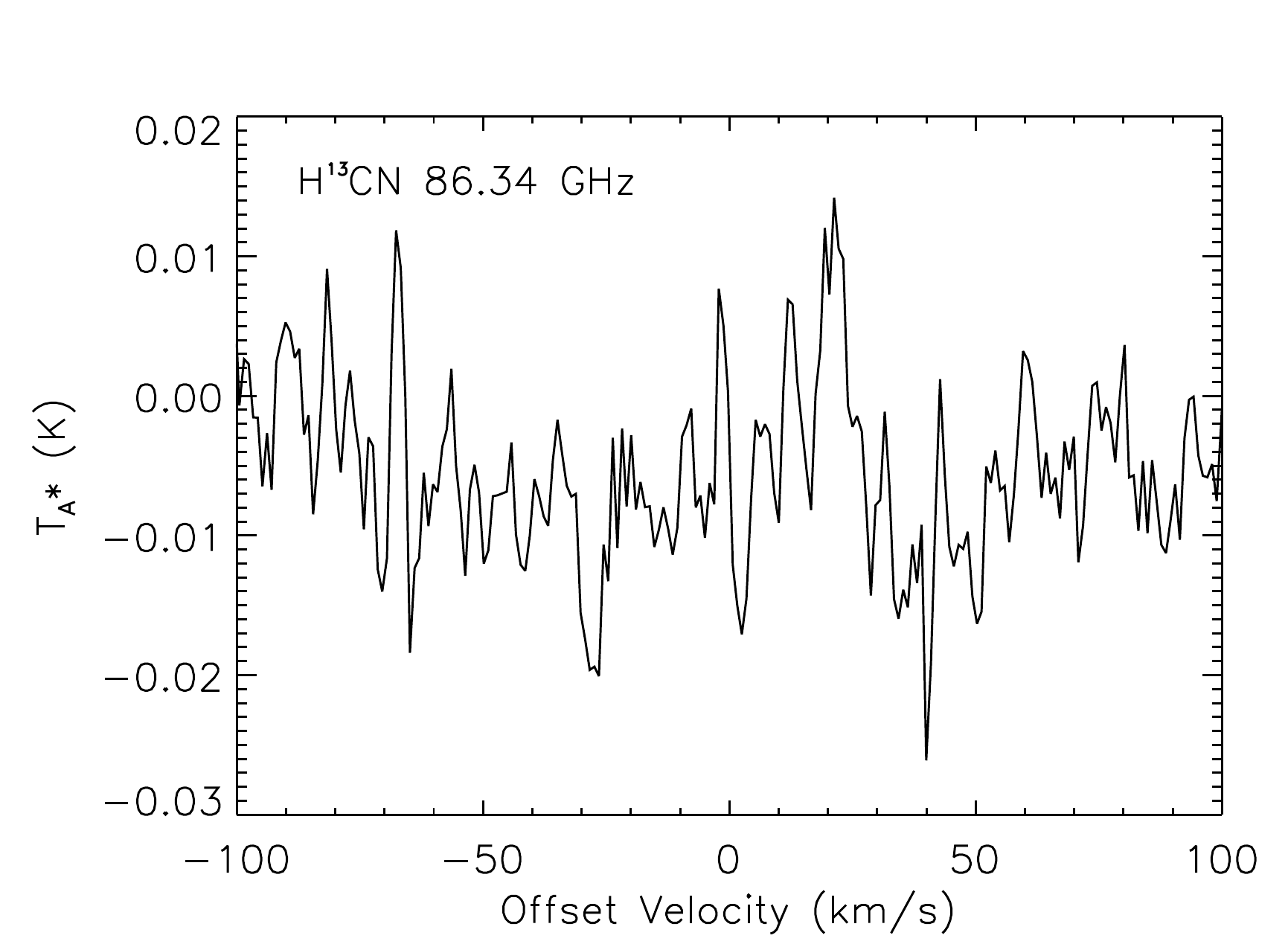}}
\subfigure{\includegraphics[trim= 2cm 13cm 2cm 2cm, clip=true, width=0.32\textwidth]{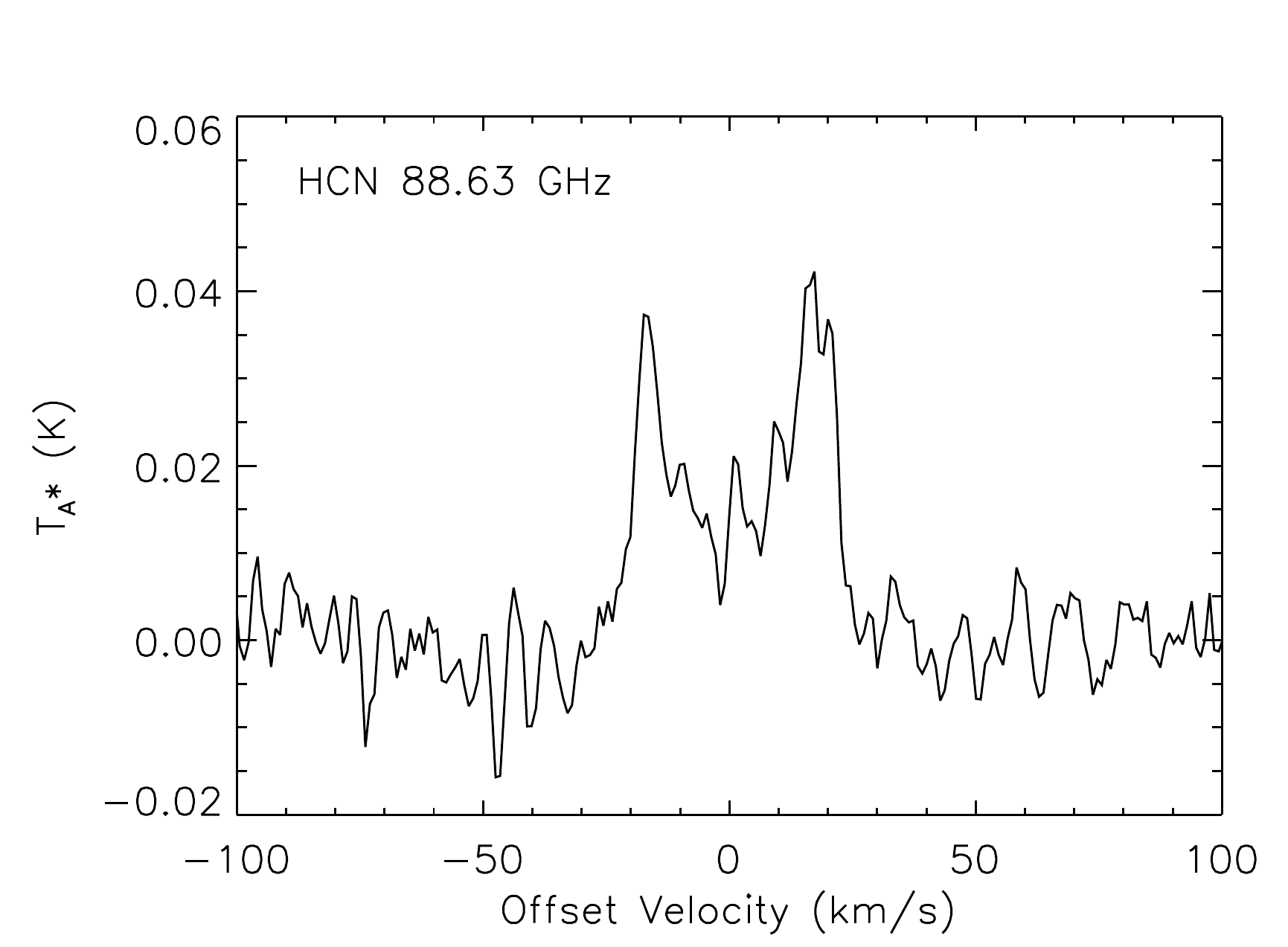}}
\subfigure{\includegraphics[trim= 2cm 13cm 2cm 2cm, clip=true, width=0.32\textwidth]{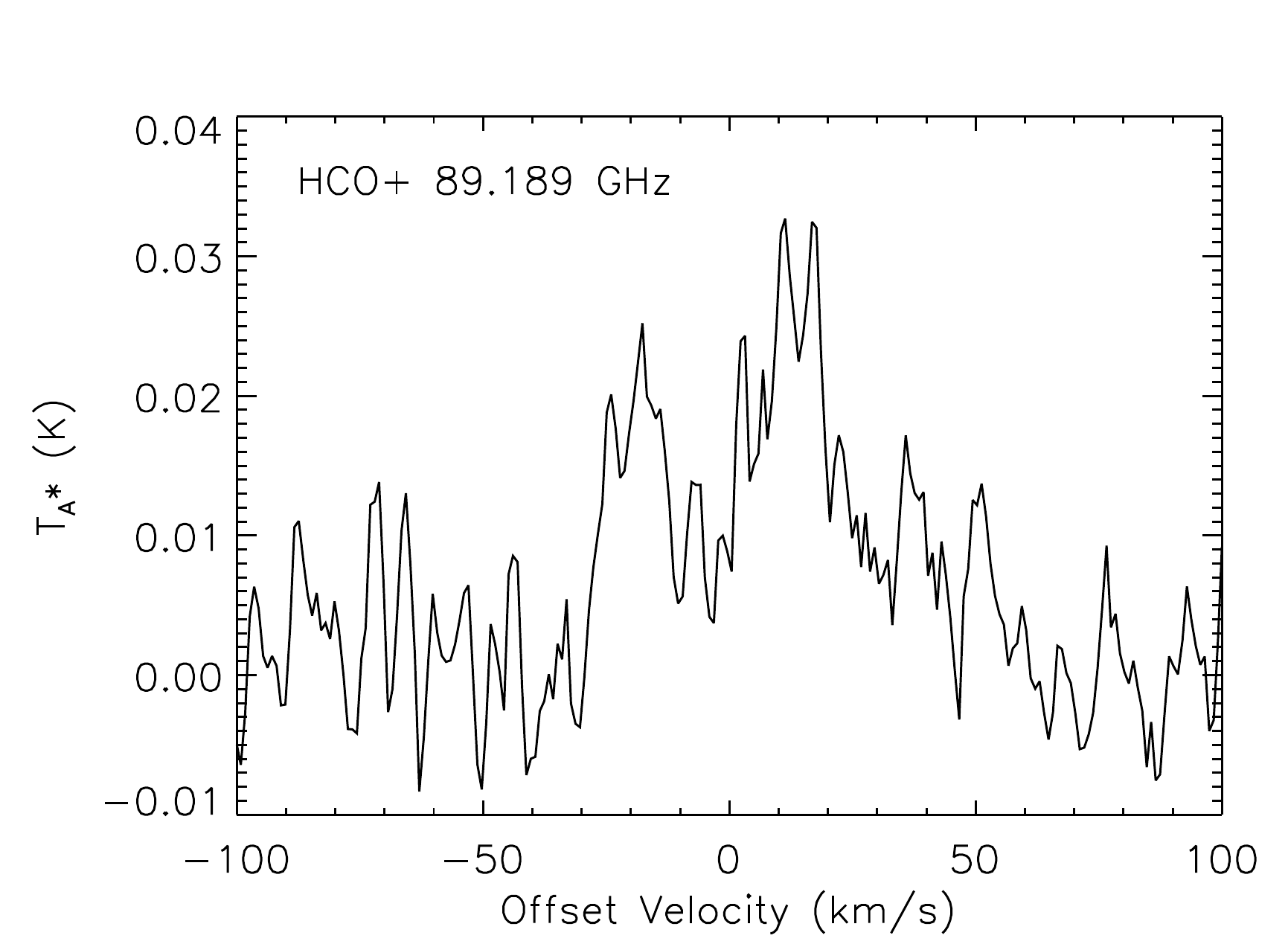}}

\subfigure{\includegraphics[trim= 2cm 13cm 2cm 2cm, clip=true, width=0.32\textwidth]{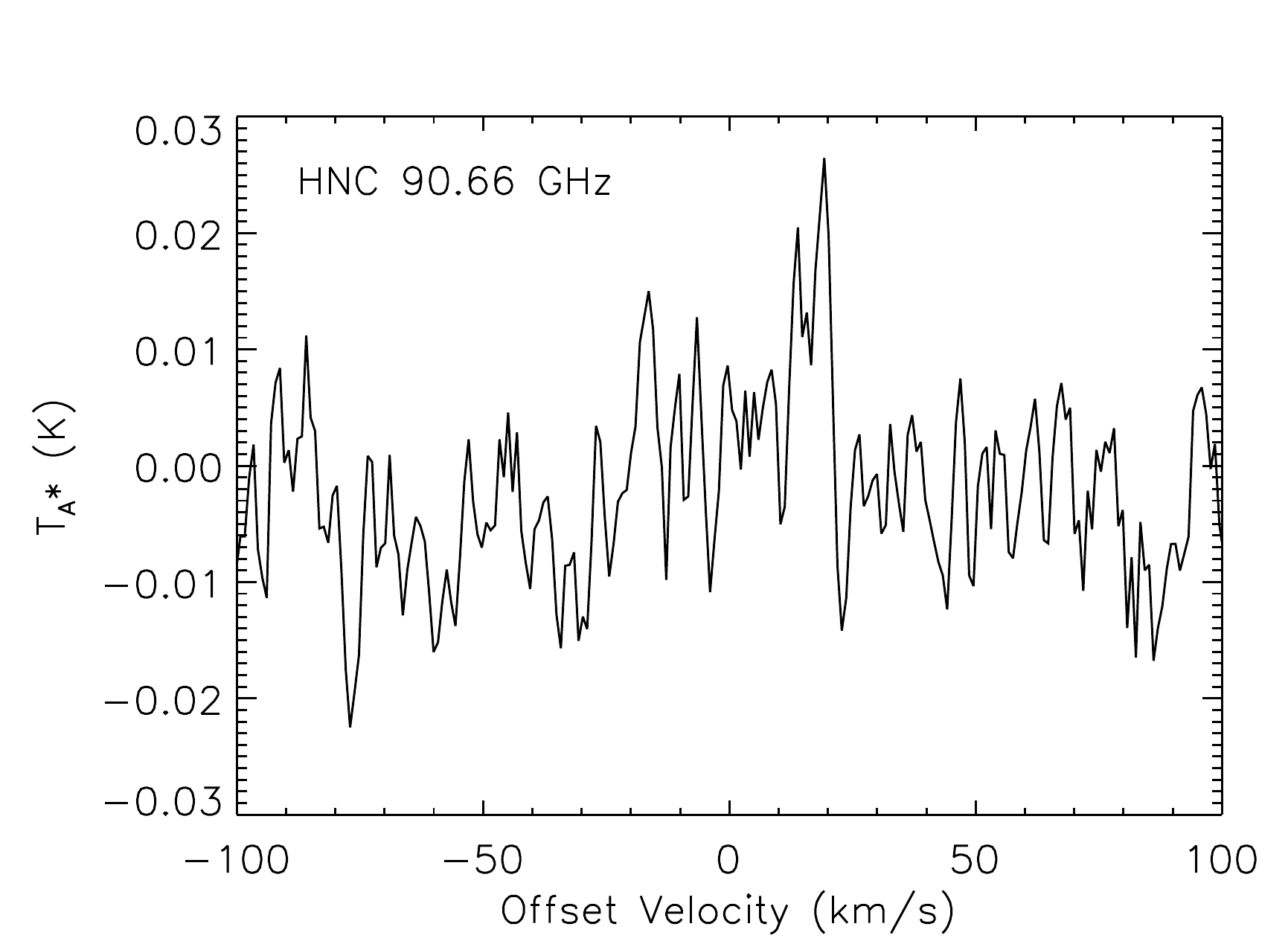}}
\subfigure{\includegraphics[trim= 2cm 13cm 2cm 2cm, clip=true, width=0.32\textwidth]{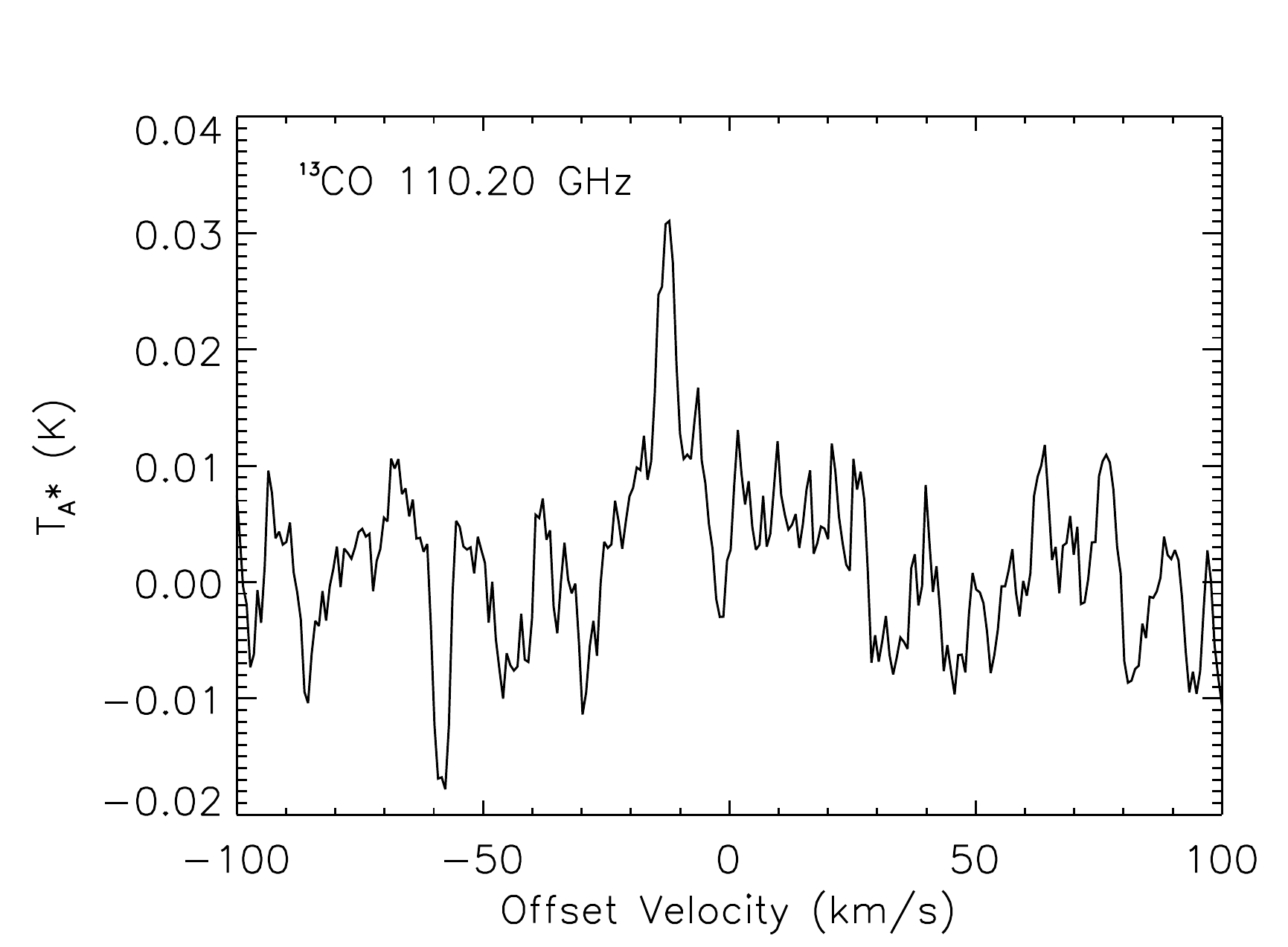}}
\subfigure{\includegraphics[trim= 2cm 13cm 2cm 2cm, clip=true, width=0.32\textwidth]{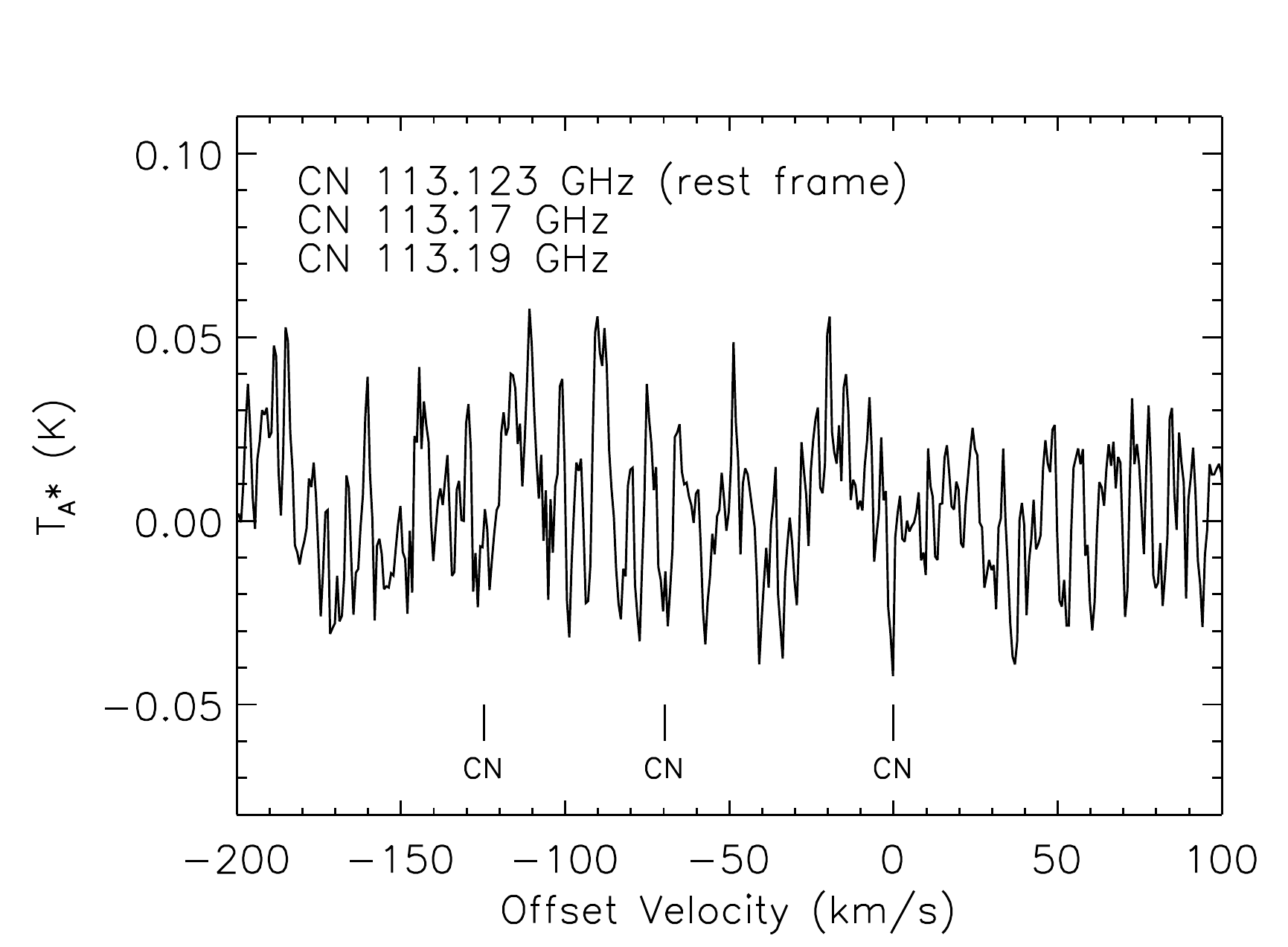}}

\subfigure{\includegraphics[trim= 2cm 13cm 2cm 2cm, clip=true, width=0.32\textwidth]{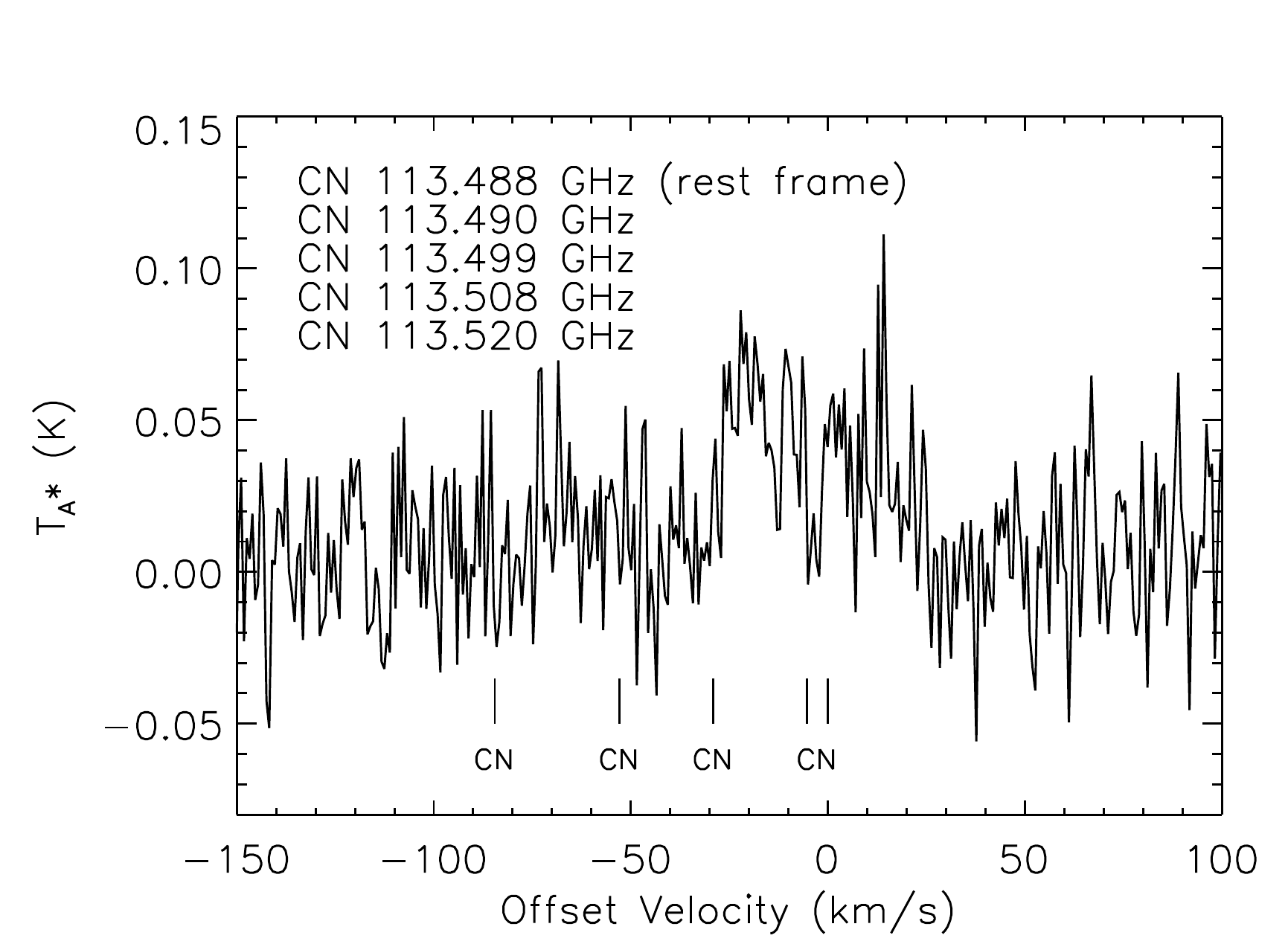}}
\subfigure{\includegraphics[trim= 2cm 13cm 2cm 2cm, clip=true, width=0.32\textwidth]{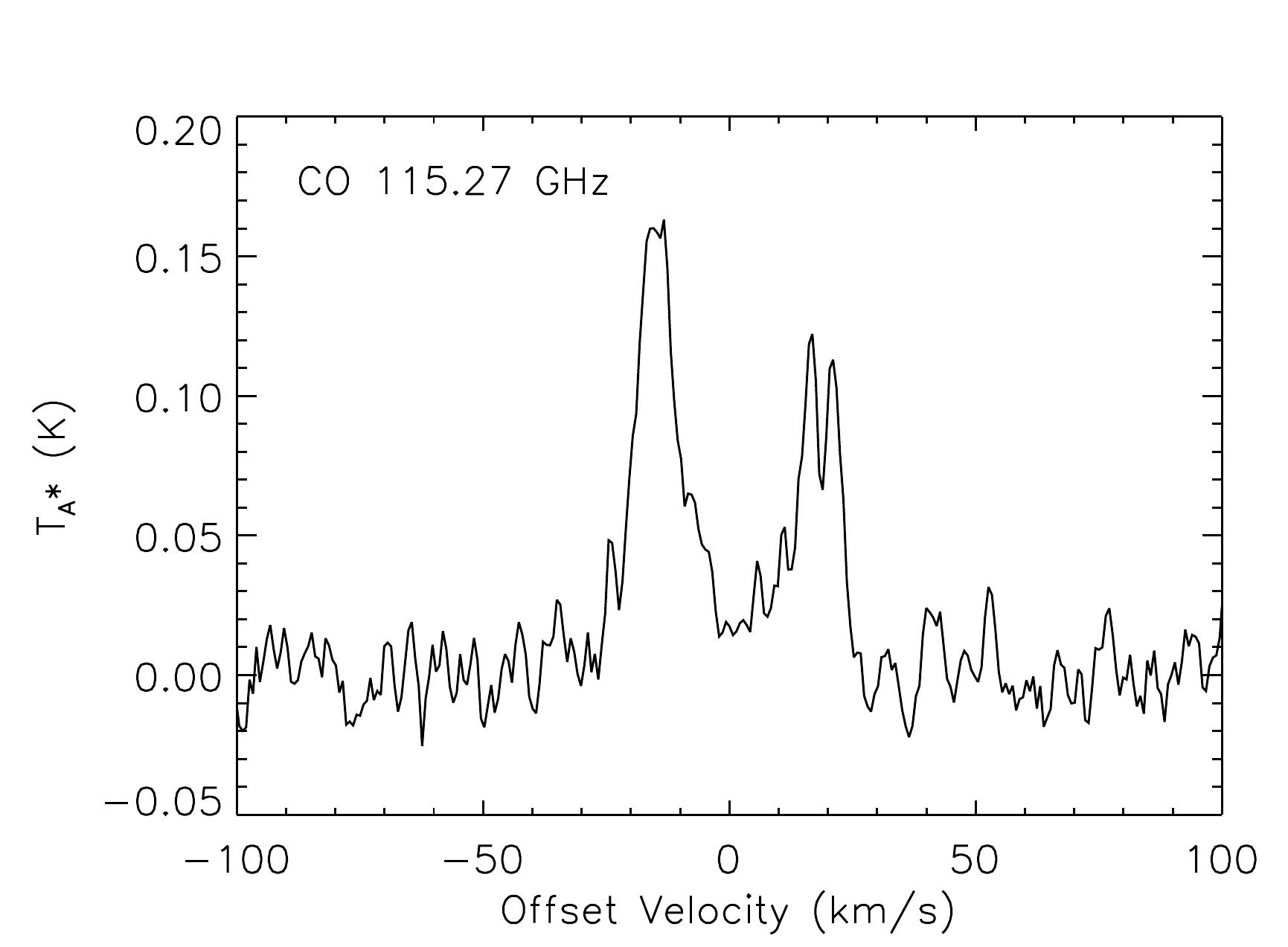}}
}

\caption{All transitions in IC 4406. Ordinate axis intensities are in units of corrected antenna temperature, abscissa values are in units of km/s, corrected for LSR velocity of the source, taken as -44.5 km/s \citep{Sahai1991}.}
\end{figure*}

\begin{figure*}
{\centering
\subfigure{\includegraphics[trim= 2cm 13cm 2cm 2cm, clip=true, width=0.32\textwidth]{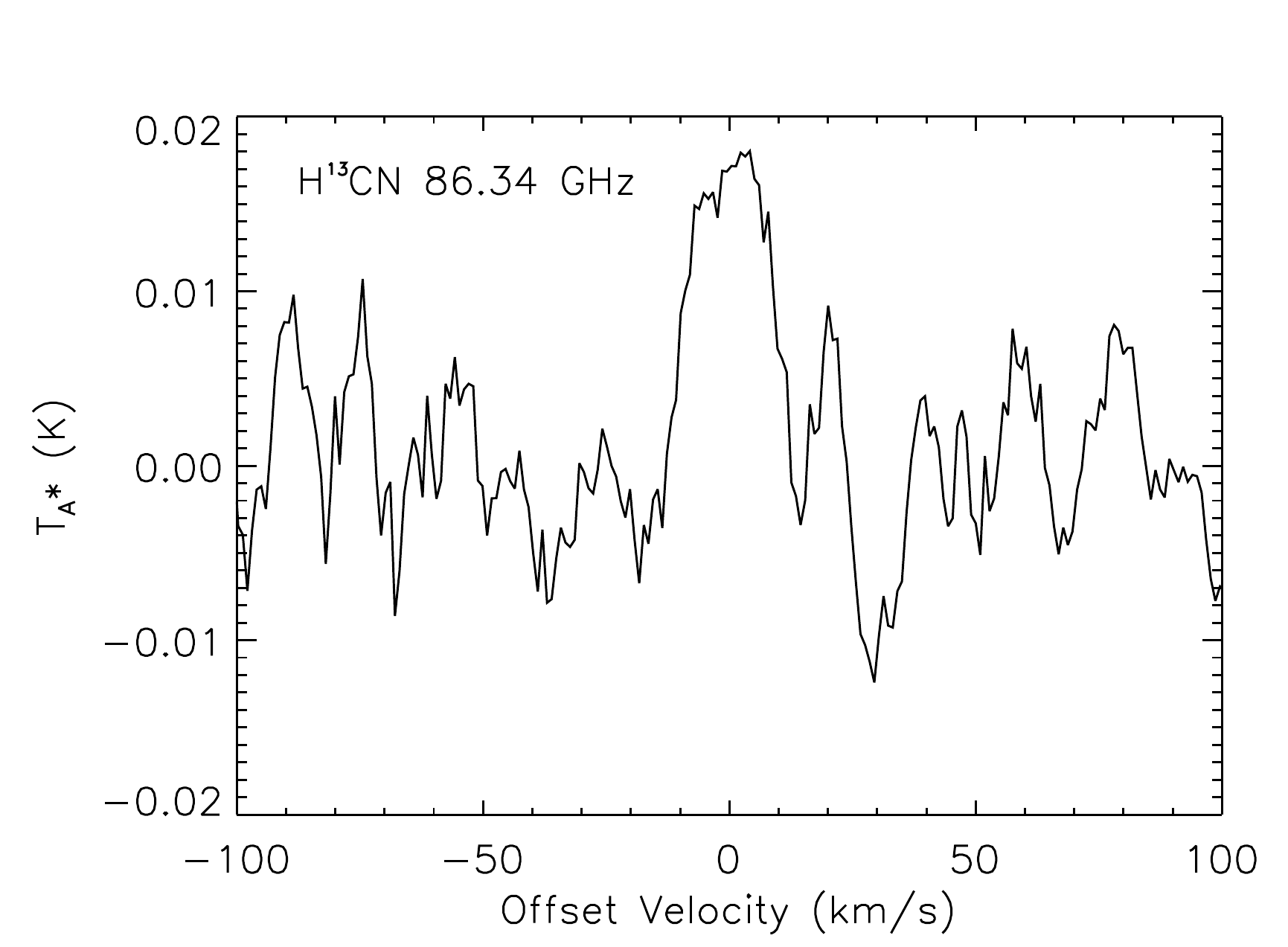}}
\subfigure{\includegraphics[trim= 2cm 13cm 2cm 2cm, clip=true, width=0.32\textwidth]{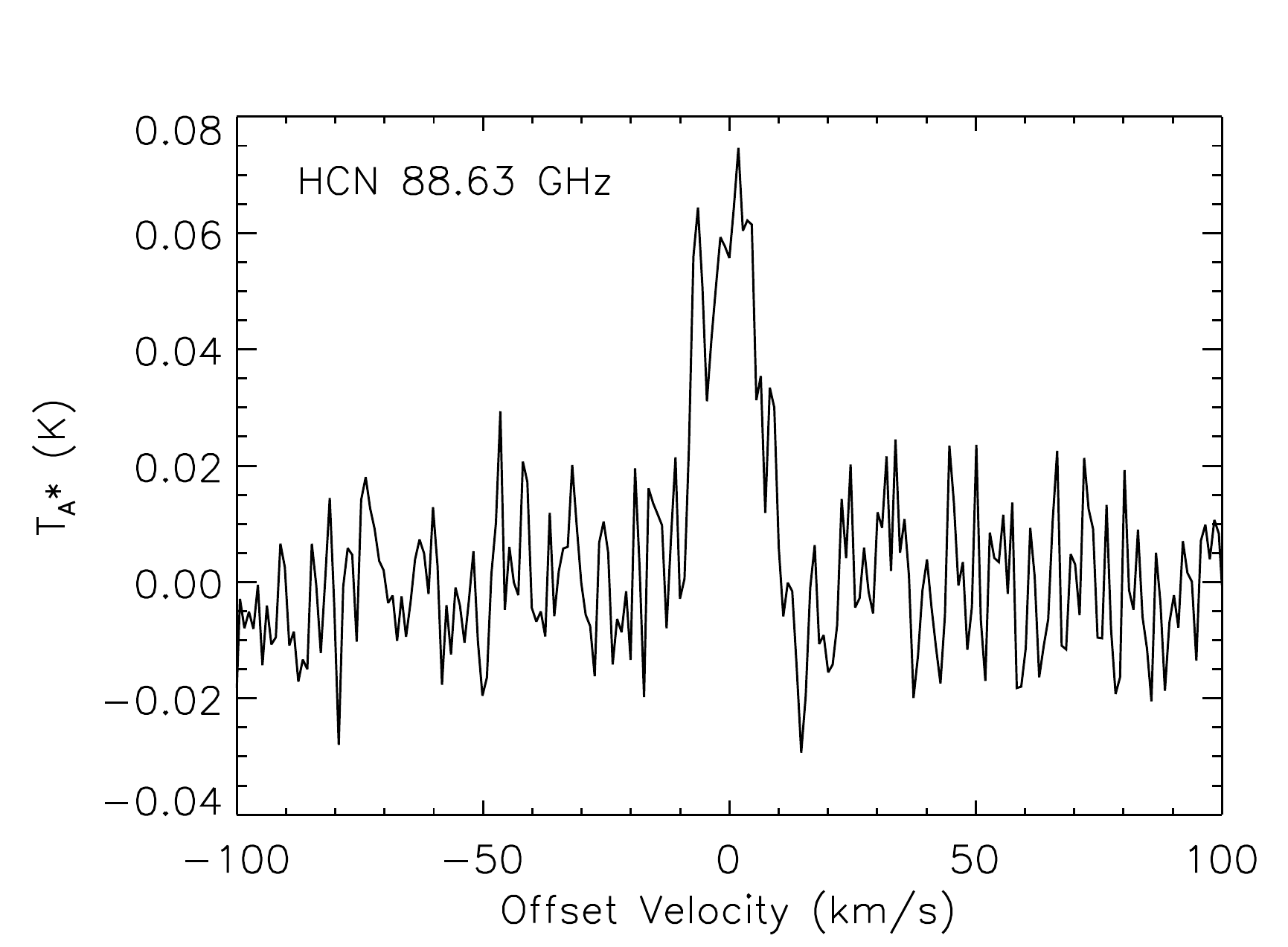}}
\subfigure{\includegraphics[trim= 2cm 13cm 2cm 2cm, clip=true, width=0.32\textwidth]{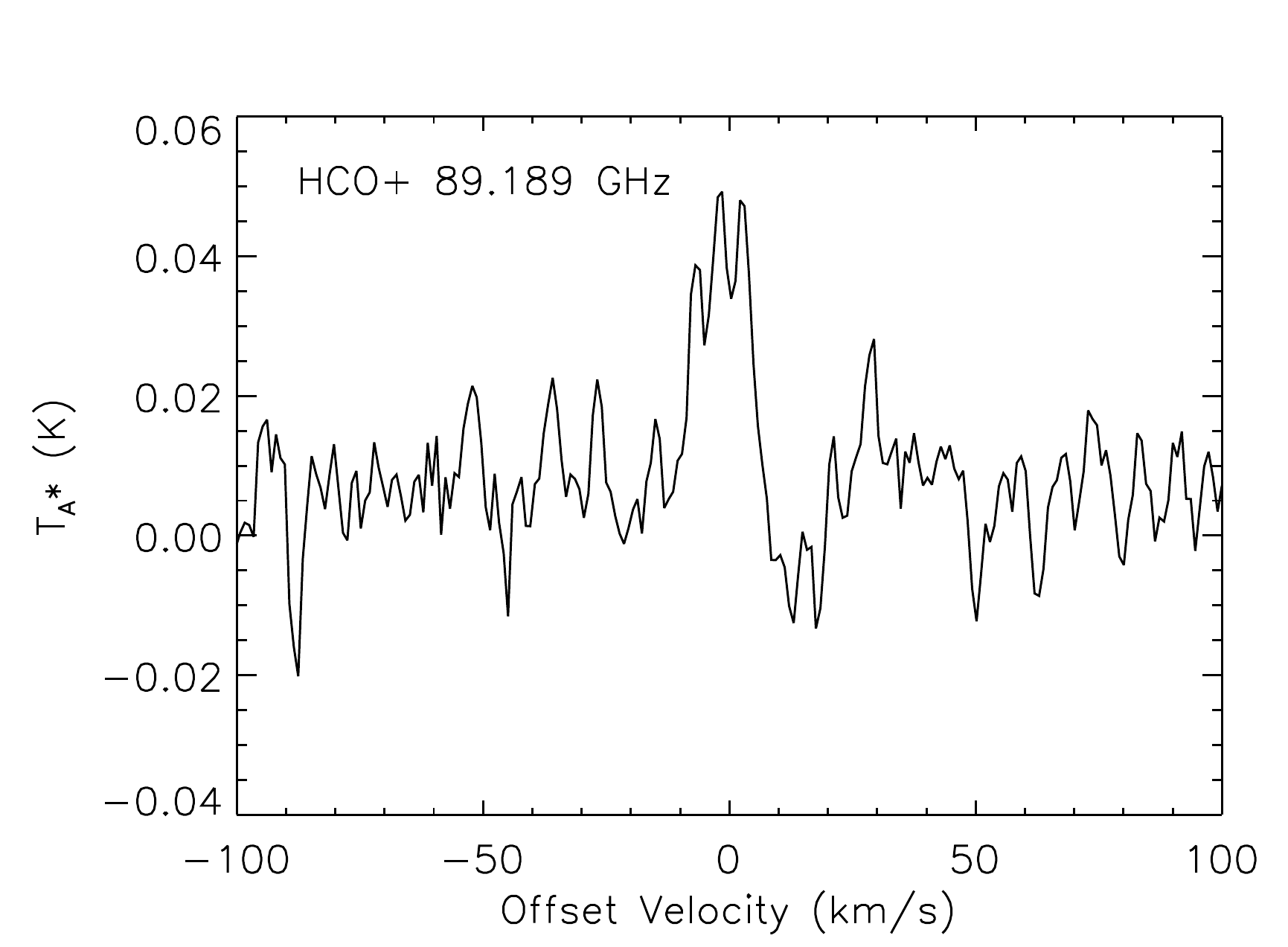}}
\subfigure{\includegraphics[trim= 2cm 13cm 2cm 2cm, clip=true, width=0.32\textwidth]{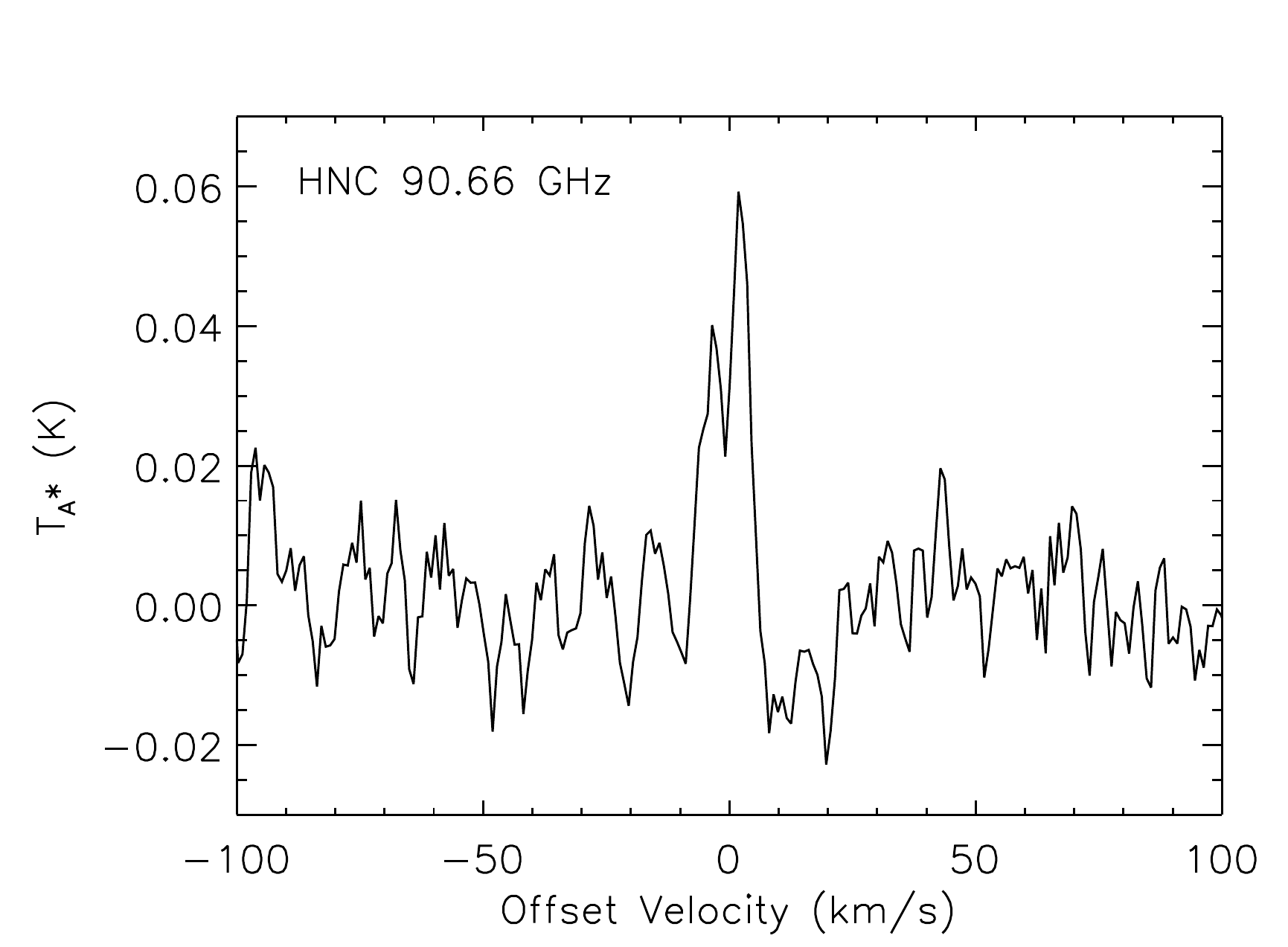}}
\subfigure{\includegraphics[trim= 2cm 13cm 2cm 2cm, clip=true, width=0.32\textwidth]{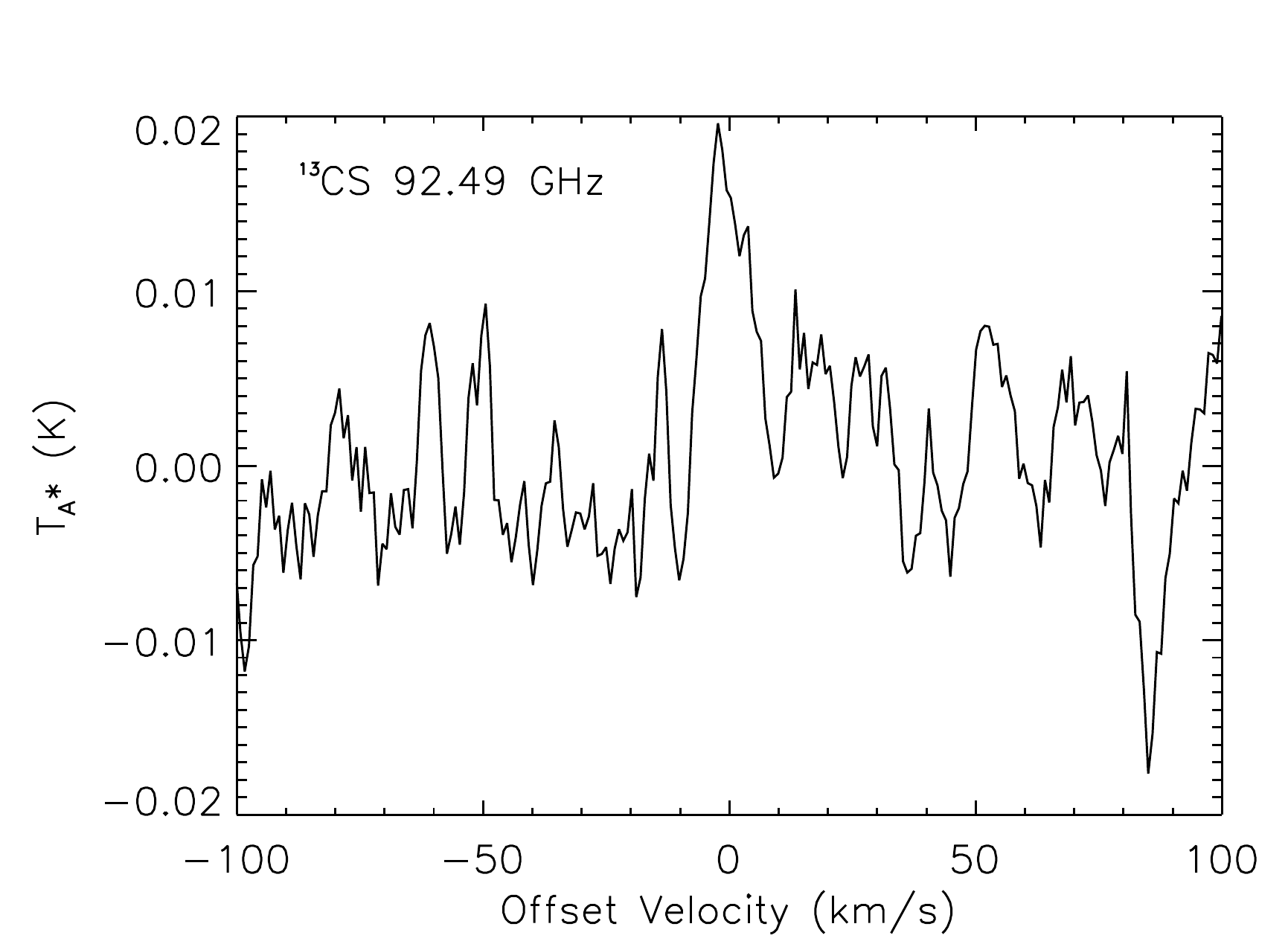}}
\subfigure{\includegraphics[trim= 2cm 13cm 2cm 2cm, clip=true, width=0.32\textwidth]{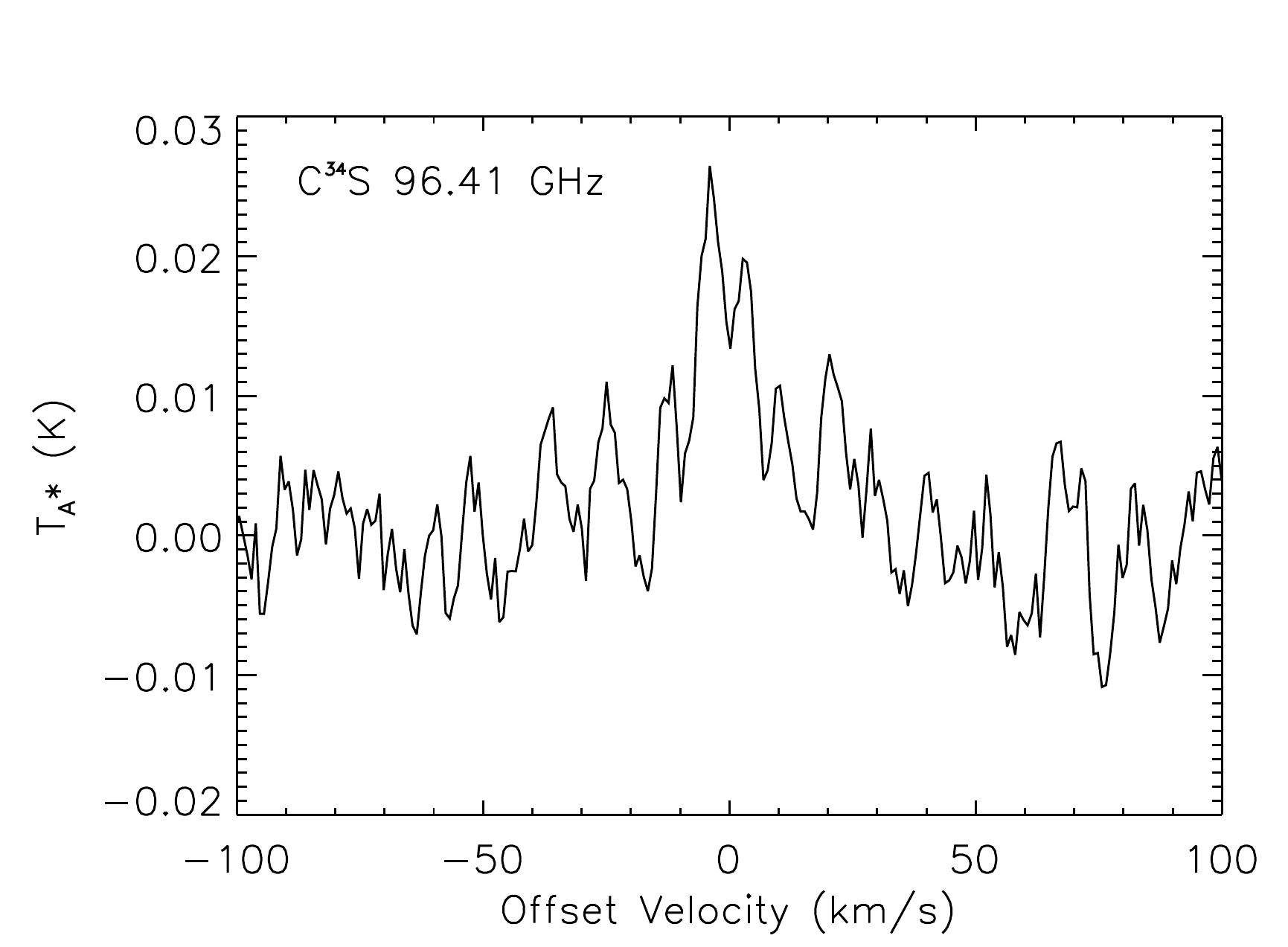}}
\subfigure{\includegraphics[trim= 2cm 13cm 2cm 2cm, clip=true, width=0.32\textwidth]{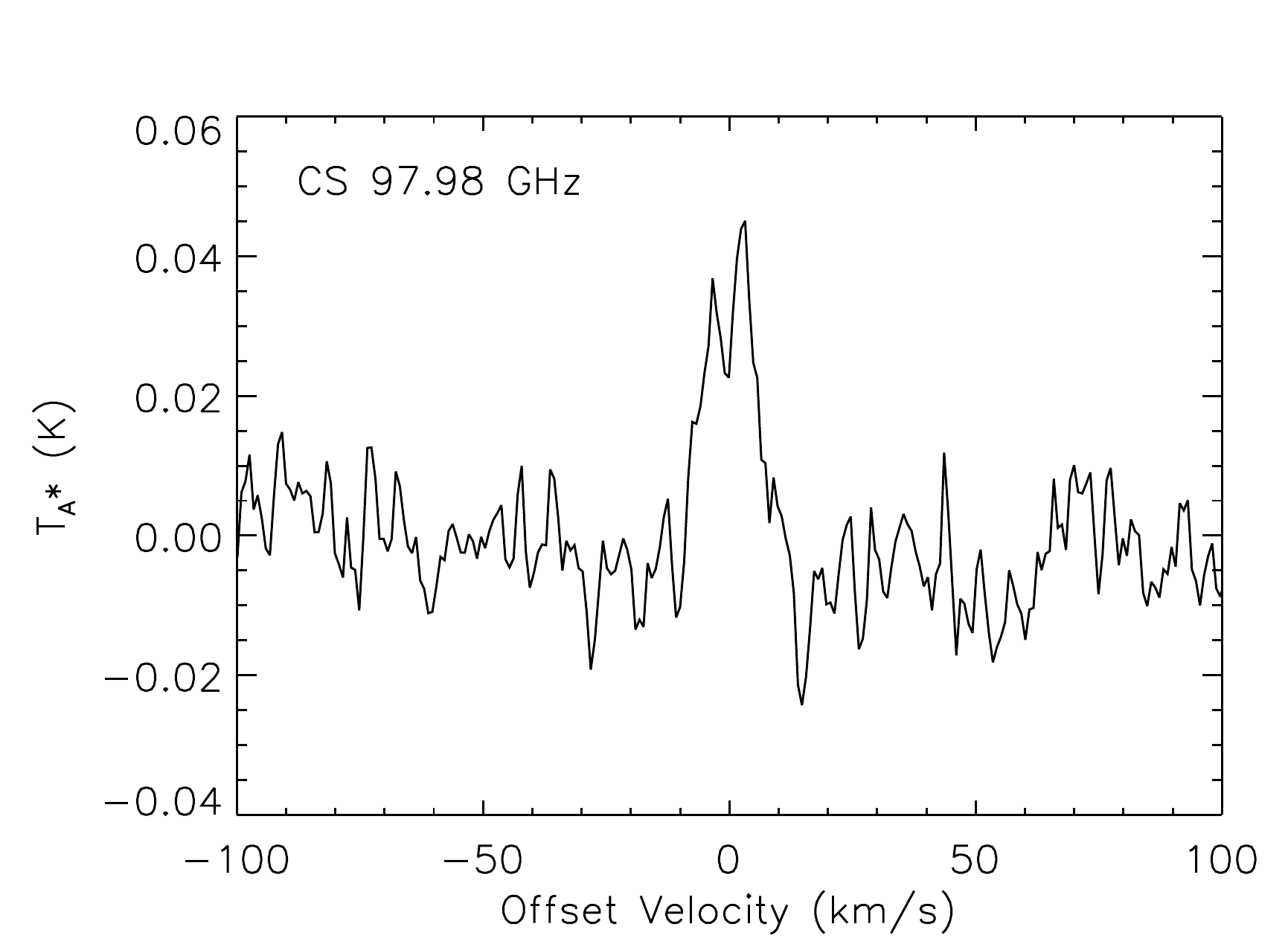}}
\subfigure{\includegraphics[trim= 2cm 13cm 2cm 2cm, clip=true, width=0.32\textwidth]{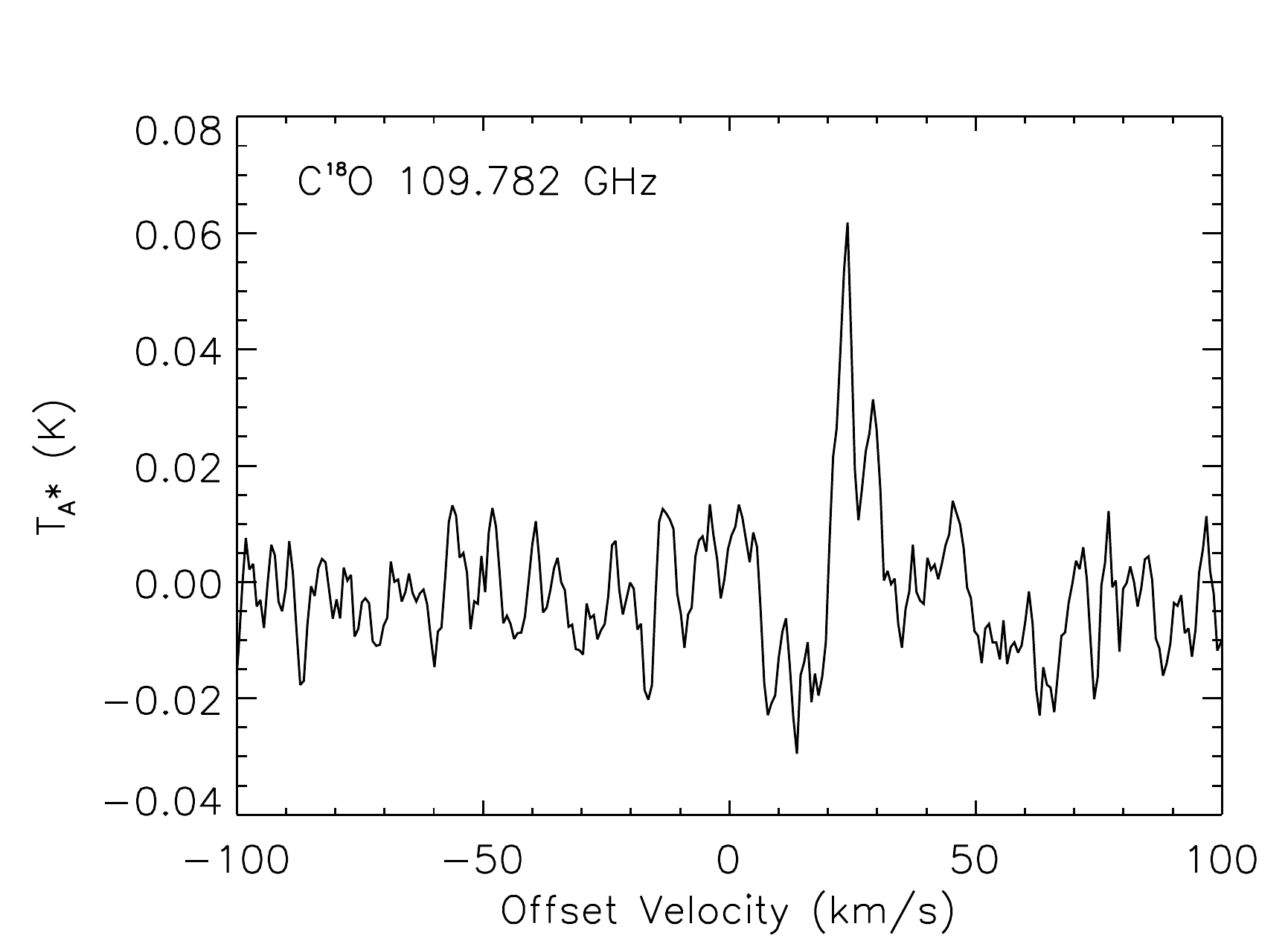}}
\subfigure{\includegraphics[trim= 2cm 13cm 2cm 2cm, clip=true, width=0.32\textwidth]{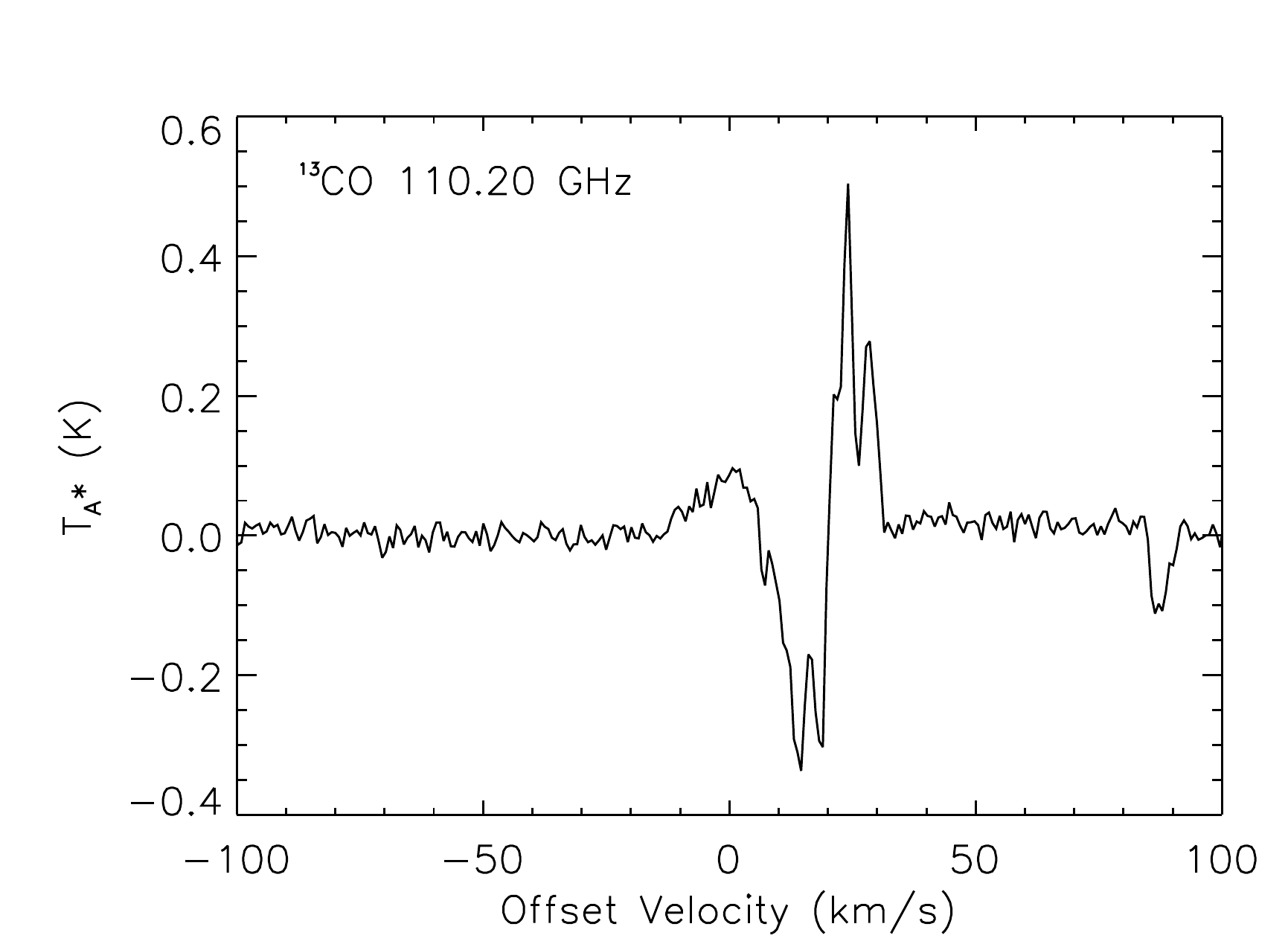}}
\subfigure{\includegraphics[trim= 2cm 13cm 2cm 2cm, clip=true, width=0.32\textwidth]{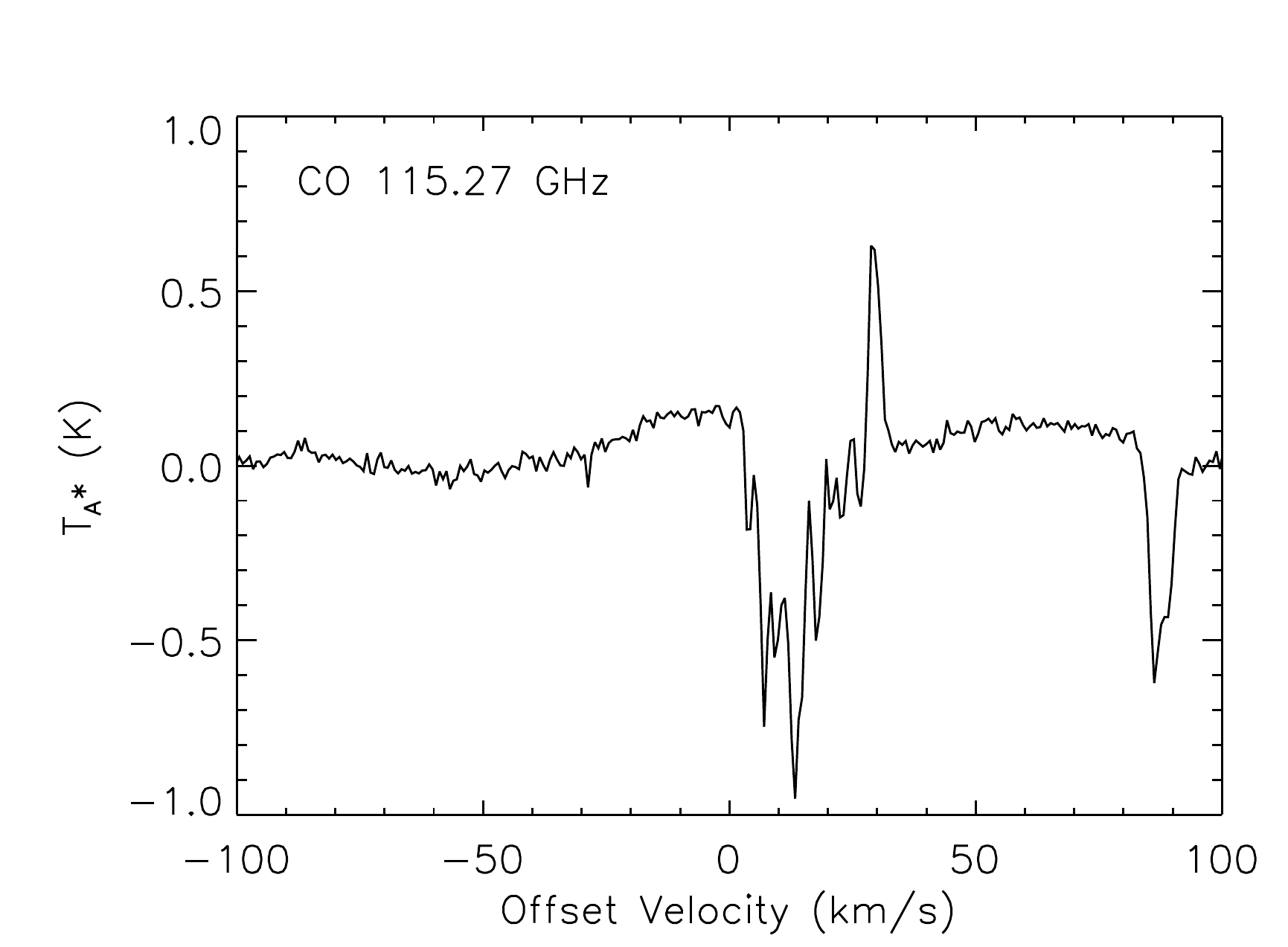}}
}

\caption{All transitions in NGC 6537. Ordinate axis intensities are in units of corrected antenna temperature, abscissa values are in units of km/s, corrected for LSR velocity of the source, taken as +10.0 km/s \citep{Edwards2013}.}
\end{figure*}



\end{document}